\documentclass[a4paper,11pt,twoside]{book}

\usepackage[hyperfootnotes=false]{hyperref}

\usepackage{mathptmx}

\usepackage{setspace}
\usepackage[a4paper,twoside,heightrounded]{geometry}

\usepackage[multiple,stable]{footmisc}

\usepackage[labelfont=bf,font=footnotesize]{caption}
\usepackage{subfigure}

\setcaptionmargin{1in}

\newcommand{\captionlabeldelim}{:}

\usepackage{fancyhdr}

\fancypagestyle{plain}{%
  \fancyhf{}
  
  \fancyhf[FLE,FRO]{\hfill\slshape\thepage}}

\makeatletter
\renewcommand{\chaptermark}[1]{\markboth{\@chapapp\ \thechapter.\ \ #1}{}}
\makeatother

\usepackage{sectsty}

\usepackage{color}
\usepackage[grey,utopia]{quotchap}

\definecolor{chaptergrey}{rgb}{0,0.4,0}

\renewcommand{\numberalign}{\raggedleft}
\renewcommand{\textalign}{\raggedleft}

\usepackage[subfigure]{tocloft}

\tocloftpagestyle{plain}

\makeatletter
\renewcommand{\cftmarktoc}{\@mkboth{\contentsname}{\contentsname}}
\renewcommand{\cftmarklof}{\@mkboth{\listfigurename}{\listfigurename}}
\renewcommand{\cftmarklot}{\@mkboth{\listtablename}{\listtablename}}
\makeatother

\renewcommand{\contentsname}{Contents}

\renewcommand{\listfigurename}{List of Figures}

\renewcommand{\listtablename}{List of Tables}

\usepackage{tocbibind}

\makeatletter
\def\@makeschapterhead#1{%
  {\parindent \z@ \centering
    \normalfont
    \interlinepenalty\@M
    \huge \bfseries \sc #1\par\nobreak
    \vskip 10\p@
  }}
\makeatother

\usepackage[toc]{appendix}

\renewcommand{\appendixtocname}{\Large\bfseries\textsc{Appendices}}

\noappendicestocpagenum

\renewcommand{\appendixpagename}{\scshape{\mdseries{Appendices}}}

\renewcommand{\appendixname}{Appendix}

\usepackage{amsmath}  

\allowdisplaybreaks   

\makeatletter
\@ifundefined{mathindent}{\relax}{\setlength{\mathindent}{0pt}}
\makeatother

\usepackage{amssymb}

\usepackage[final]{graphicx}
\usepackage{epsfig,rotating}
\usepackage{lscape}

\usepackage{natbib}

\usepackage{wrapfig} 

\newcommand{\corr}[1]{#1}    
\newcommand{\corrbox}[1]{#1}

\bibpunct[; ]{(}{)}{,}{a}{}{;}

\usepackage{upgreek}

\usepackage{array}

\usepackage{booktabs}

\usepackage{url}

\usepackage{longtable} 

\newcommand{\thesisdate}{January 21, 2013}

\makeatletter
\renewcommand*{\cleardoublepage}{%
  \clearpage
  \if@twoside
    \ifodd\c@page
    \else
      \thispagestyle{empty}     
      \hbox{}\newpage
      \if@twocolumn
        \hbox{}\newpage
      \fi
    \fi
  \fi
}
\makeatother

\usepackage{natbib}
\usepackage{aas_macros}

\usepackage{color,soul}

\usepackage{lettrine}

\usepackage[noprefix]{nomencl}
\RequirePackage{ifthen}
\renewcommand{\nomgroup}[1]{%
\ifthenelse{\equal{#1}{R}}{\item[\textbf{Roman symbols}]}{%
\ifthenelse{\equal{#1}{X}}{\item[\textbf{Constants}]}{%
\ifthenelse{\equal{#1}{Y}}{\item[\textbf{Acronyms}]}{%
\ifthenelse{\equal{#1}{G}}{\item[\textbf{Greek symbols}]}{}}}}}
\makenomenclature

\begin{document}

\frontmatter
\begin{titlepage}
\thispagestyle{empty}

{\center

\Huge
\vspace*{12pt}

\fontsize{36}{42}{\textbf{\textsc{Lucky Imaging:{\newline}Beyond Binary Stars}}}

\rm
\Large
\vspace{10mm}
Thesis submitted for the degree of \\
Doctor of Philosophy

\vspace{5mm}
by

%
\vspace{5mm}
\LARGE
Tim Staley

%
\vspace{1cm}
\hspace{-7mm}
\includegraphics[height=40mm]{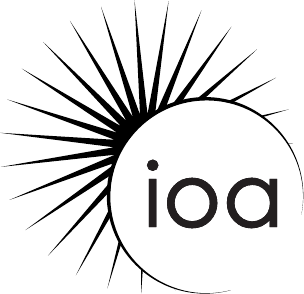} \hspace{10mm}
\includegraphics[height=40mm]{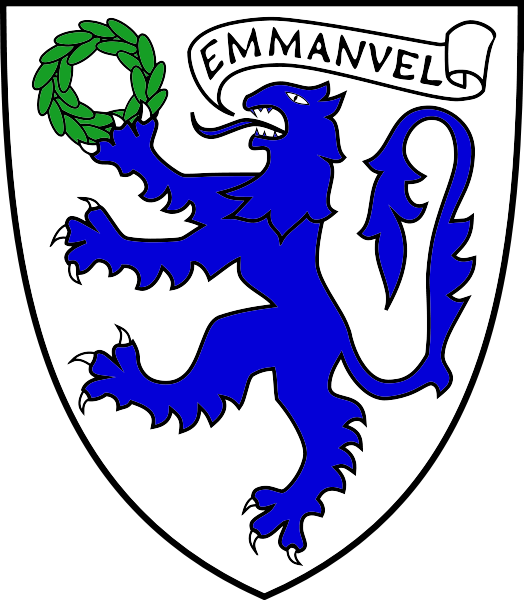} \hspace{10mm}
\includegraphics[height=40mm]{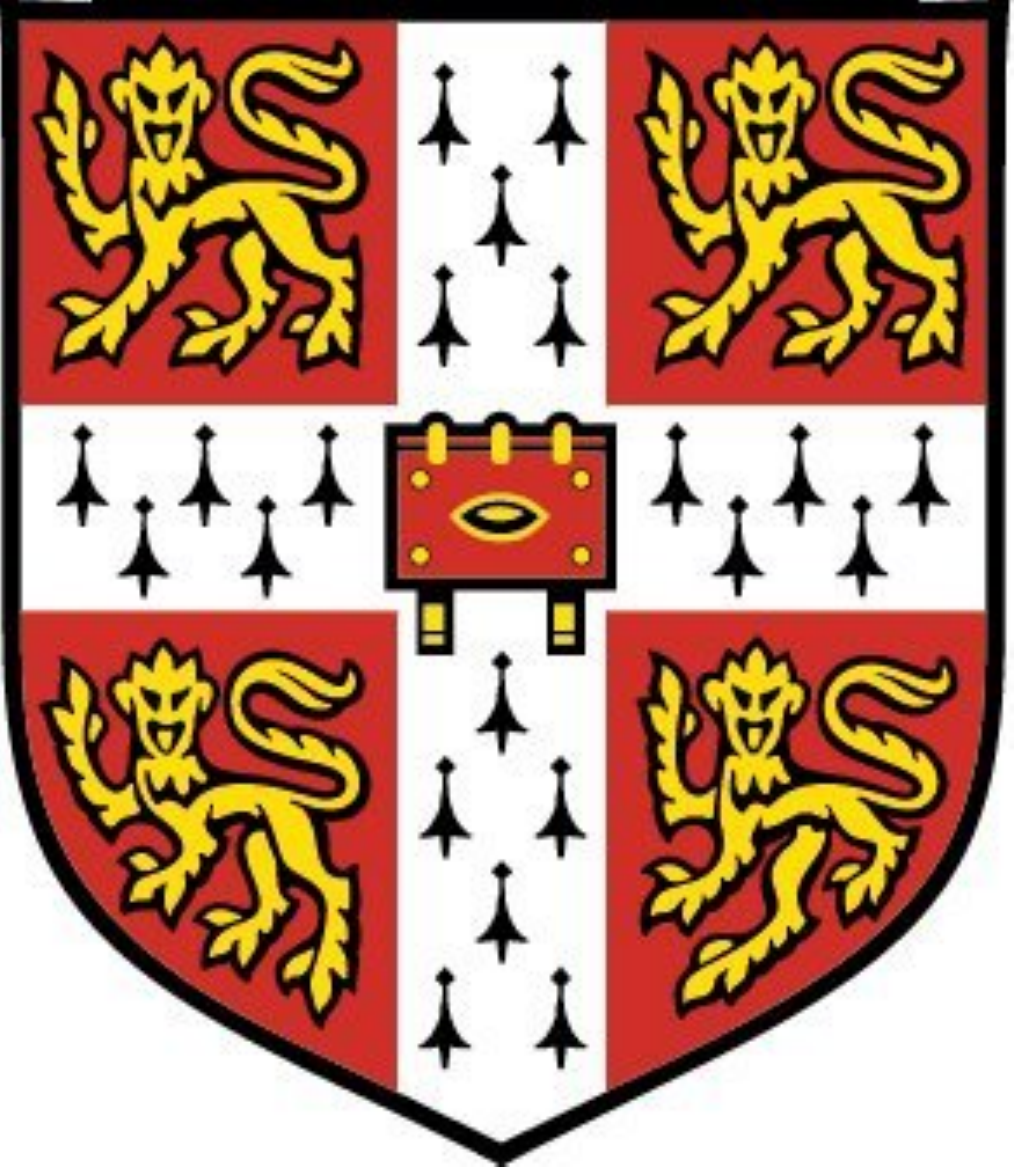} \\

\vspace{1cm}

Institute of Astronomy \\
\& \\
Emmanuel College \\

\vspace{0.75cm}

University of Cambridge
%

%
\vspace{10mm}
\thesisdate
%

}
\end{titlepage}


%


\chapter*{Declaration}
\addcontentsline{toc}{section}{Declaration}
I hereby declare that this dissertation entitled \textit{Lucky Imaging: Beyond Binary Stars} is not substantially the same as any that I have submitted for a degree or diploma or other qualification at any other University.

I further state that no part of my thesis has already been or is being concurrently submitted for
any such degree, diploma or other qualification.

This dissertation is the result of my own work and includes nothing which is the outcome of
work done in collaboration except where specifically indicated in the text.

I note that chapter~\ref{chap:intro} and the first few sections of chapter~\ref{chap:lucky_AO} are intended as reviews, and as such contain little, if any, original work. They contain a number of images and plots extracted from other published works, all of which are clearly cited in the appropriate caption.
Those parts of this thesis which have been published are as follows:
\begin{itemize}
 \item Chapters~\ref{chap:thresholding} and \ref{chap:frame_registration} contain elements that were published in \cite{Staley2010a}. However, the work has been considerably expanded upon for this document.
 \item The planetary transit host binarity survey described in chapter~\ref{chap:science} is soon to be submitted for publication.
\end{itemize}

This dissertation contains fewer than 60,000 words.

\vspace{3cm}
\parbox{10cm}{%
  \textit{Tim Staley\\Cambridge, \thesisdate}
}


\chapter*{Acknowledgements}
\addcontentsline{toc}{section}{Acknowledgements}
\markboth{Acknowledgements}{Acknowledgements}
This thesis has been typeset in \LaTeX{} using Kile
\footnote{http://kile.sourceforge.net/}
 and JabRef.
\footnote{http://jabref.sourceforge.net/}
Thanks to all the former IoA members who have contributed to the LaTeX template used to constrain the formatting. It is a dark art.

Many thanks to the Cambridge lucky imaging team. Firstly my supervisor Craig Mackay, ever on hand with sound advice and helpful input. Much hard effort has gone in behind the scenes to make LuckyCam work, and for that my thanks go to Craig, David King, Frank Suess and Keith Weller. 

Thanks also to all the administrative and student support staff at the IoA, for making my PhD experience a remarkably well organised and content one.

Thanks to my parents for their unfailing support, and all my friends in Cambridge for making my time here immensely enjoyable. Special thanks to Lindsey, for putting up with me throughout the write up, and to the windsurf club for making sure I get out enough.

I acknowledge with gratitude the support of an STFC studentship. 
This thesis is based on observations made with the Nordic Optical Telescope, operated on the island of La Palma
jointly by Denmark, Finland, Iceland, Norway, and Sweden, in the Spanish Observatorio del
Roque de los Muchachos of the Instituto de Astrofisica de Canarias. This research has made use
of NASA’s Astrophysics Data System Bibliographic Services, as well as the SIMBAD database
and VizieR catalogue access tools operated at CDS, Strasbourg, France. A few of the results
were based on observations made with the NASA/ESA Hubble Space Telescope, obtained from
the data archive at the Space Telescope Institute.

\vspace{2\baselineskip}
\noindent\textit{Tim Staley\\Cambridge, \thesisdate}

\begin{figure}[ht]
\begin{center}
\includegraphics[width=\textwidth]{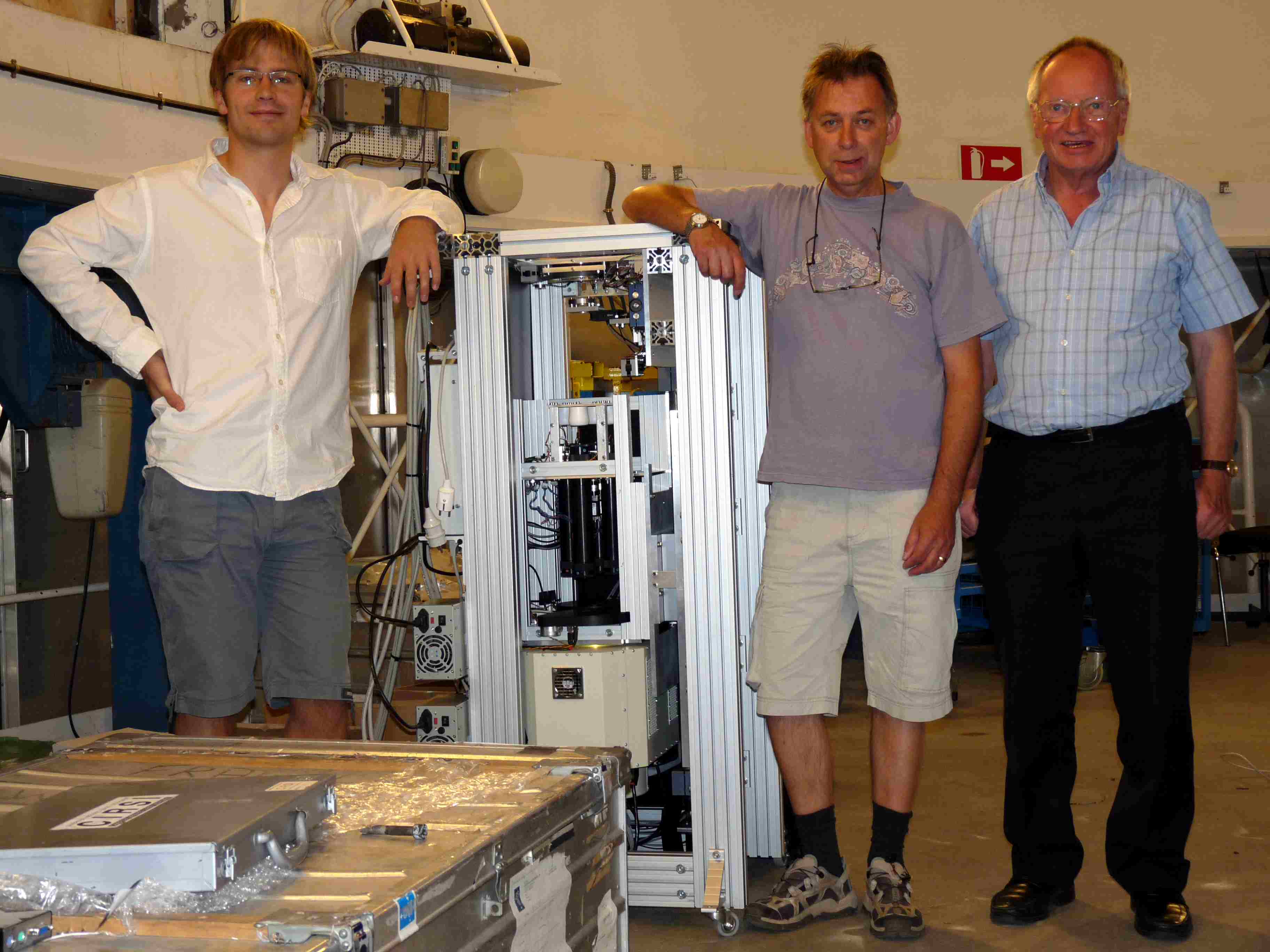}
\caption*{
The LuckyCam observing team, shortly after assembling the camera for the observing run at the Nordic Optical Telescope, La Palma, in July 2009. From left to right: myself, LuckyCam, David King and Craig Mackay.
}
\label{fig:team}
\end{center} 
\end{figure}


%

%
\newpage
\chapter*{Summary}
\addcontentsline{toc}{section}{Summary}
\markboth{Summary}{Summary}


{
\Large
\vspace{5mm}
\center
Lucky Imaging: Beyond Binary Stars
\\
\center
Tim Staley
\\
\vspace{5mm}
}

Lucky imaging is a technique for high resolution astronomical imaging at visible wavelengths, 
utilising medium sized ground based telescopes in the 2--4m class. The technique uses high speed, 
low noise cameras to record short exposures which may then be processed to minimise the 
deleterious effects of atmospheric turbulence upon image quality.

The key statement of this thesis is as follows; that lucky imaging is a technique 
which now benefits from sufficiently developed hardware and analytical techniques that 
it may be effectively used for a wide range of astronomical imaging purposes 
at medium sized ground based telescopes. Furthermore, it has proven potential 
for producing extremely high resolution imaging when coupled with adaptive optics systems on larger telescopes.
 I develop this argument using new mathematical analyses, simulations, and data from the latest Cambridge lucky imaging instrument.

\vspace{1cm}

The first half of this thesis develops new models and algorithms for 
general purpose reduction of lucky imaging data. Imaging of faint astronomical objects is achieved 
through careful calibration and analysis of the data, and utilisation of faint guide stars is improved. 
An analytic model for predicting Strehl ratio in reduced images is proposed.

The second half covers scientific results and applications, 
analysis techniques, and implementation of simulations and the data reduction pipeline. 
Results from a binarity survey of planetary transit hosts are given, 
demonstrating improved detection limits compared to previous publications.
Wide field lucky images produced from the synchronised four CCD mosaic camera 
are demonstrated and analysed for image quality across the field. Preliminary investigations of 
hybrid lucky imaging adaptive optics systems through simulation and experimental data are presented. 
Finally, the challenges of dealing with high volumes of image data in an accurate and timely fashion are covered and solutions discussed.

\vspace{2\baselineskip}
\textit{Cambridge, \thesisdate}  


\newpage
\tableofcontents


%

\newpage

\listoffigures
\addtocontents{lof}{\protect\thispagestyle{fancy}}







\mainmatter
\setcounter{equation}{0}

\chapter{Introduction}
\label{chap:intro}
In this chapter I briefly review the aspects of ground based astronomy needed to give context to the rest of this thesis. The effects of atmospheric turbulence are described and key mathematical notations set out. After defining basic terminology, the corrective techniques of adaptive optics and lucky imaging are introduced. Finally, I state my thesis proposition and summarise the following chapters.

\section{Atmospheric turbulence and light propagation}
\label{sec:atmos_effects}
Looking up at the night sky, it is easy to observe the distorting effects of the atmosphere, even on a cloudless night --- the stars twinkle. 
This phenomenon of varying intensity in the stellar light, scintillation, is one of a wider range of atmospheric optical effects known generally as ``seeing.'' Since seeing and the atmospheric turbulence that causes it play a key role in the techniques discussed here I will give a brief introduction to the models used in its description. Many of the intricacies are omitted for brevity; the reader is referred to, for example, \cite{Roddier1981} for a concise review of the relevant derivations.
\subsection{Basic turbulence modelling}
The seeing effects of turbulent flow within the atmosphere are indirect. Optical distortions are not due to the turbulent motion of the atmosphere itself, but rather due to variations in refractive index caused by temperature differences. There is an ever changing temperature gradient in the atmosphere, and seeing occurs when small pockets of air become mixed into a layer with a different temperature, causing a significant change in refractive index on small spatial scales. 
The temperature of a small volume of air equalises with its surroundings slowly compared to the mixing timescale, so we can think of the temperature as a `passive advective scalar' carried by the turbulent velocity field in much the same way as dye can be seen to follow a turbulent liquid flow (buoyancy effects being negligible). 

\begin{figure}[htp]
\begin{center}
 \includegraphics[width=1\textwidth]{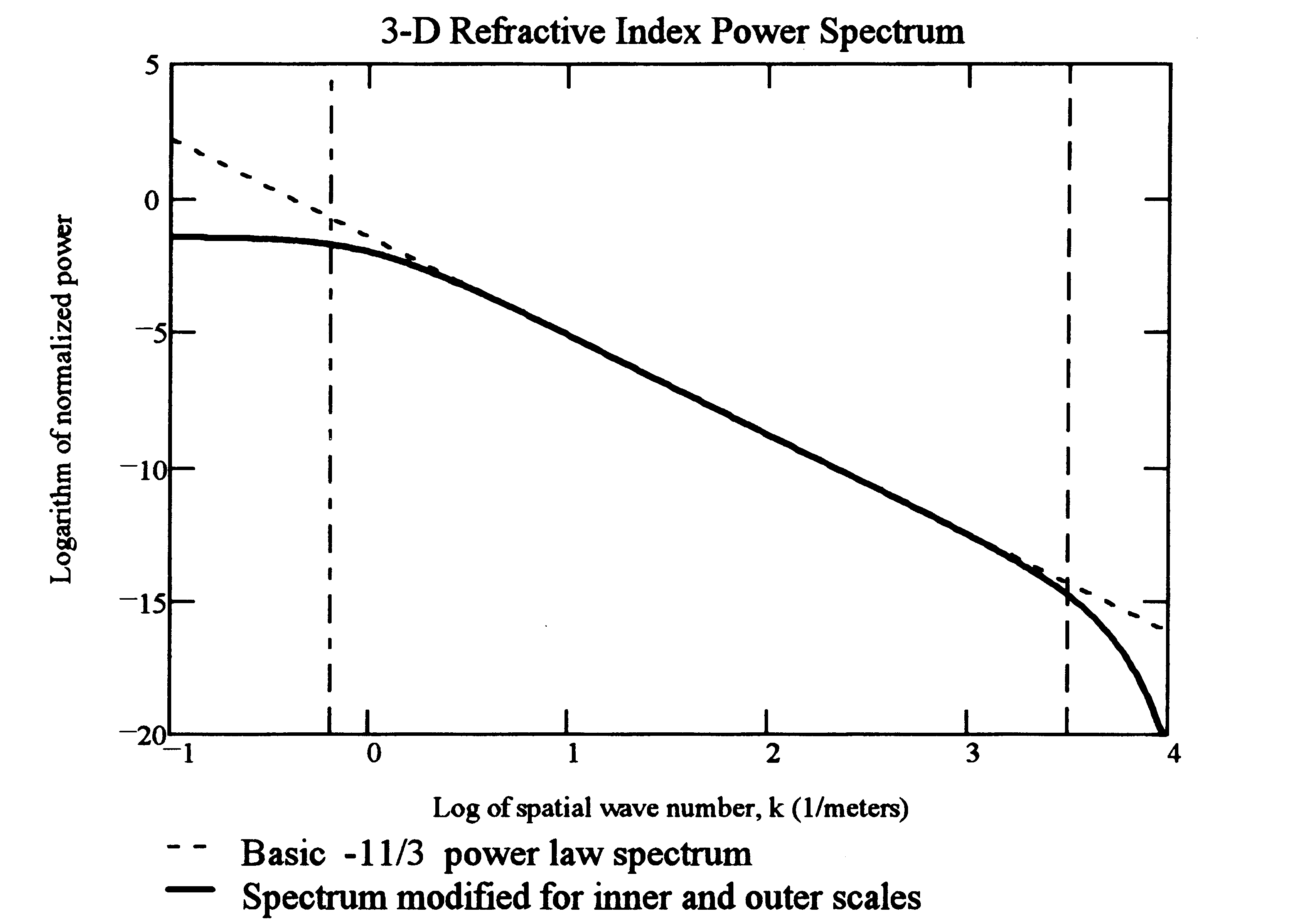}
\caption[The von Karman Power Spectrum]{The von Karman three dimensional power spectrum, which shows the Kolmogorov 11/3 slope in the inertial range, i.e. the range of scales through which the energy cascades before dissipating. The dash-dotted vertical line, left, marks an inner scale of 2mm, while the dashed vertical line to the right marks an outer scale of 10 meters. The von Karman power spectrum includes corrective terms in order to better describe the power spectrum outside the inertial range. Credit: \cite{Hardy1998}}
\label{fig:spectrum}
\end{center} 
\end{figure}

To characterise this process mathematically we first invoke the Kolmogorov (1941) scaling law for homogeneous three dimensional turbulence. The basic assumption is that energy only enters the turbulent system at the largest scales and only dissipates at the smallest due to viscosity; for example in a teacup the largest or `outer' scale might be the stroke length of a stirring spoon. We further assume that between these large and small scales, in what is known as the `inertial range,' kinetic energy is conserved.%
\footnote{Note that the assumptions of Kolmogorov's 1941 theory do not account for energy stored as or converted into pressure fluctuations, which are sometimes significant.}
The energy propagates down to the inner scale through a series of subdividing vortices in what is known as a Richardson cascade (after the fluid dynamicist Lewis Fry Richardson). 
This allows a swift dimensional analysis arriving at the power law for the spectral energy density in three dimensions:
\begin{equation}
 \Phi(\mathbf{k}) \sim |\mathbf{k}|^{-11/3}
\end{equation}
where $\Phi$ is the spectral energy density, and $k$ is the wavenumber corresponding to a spatial scale $r$, with the relation $k = 2\pi / r$.
This is the ubiquitous `Kolmogorov power spectrum,' which has been verified for the velocity field in a wide range of turbulent fluids, at scales in the inertial range (Figure~\ref{fig:spectrum} illustrates the drop off at the inner and outer scales). 
Kolmogorov also introduced the use of the structure function as a tool for studying turbulence. Random variables such as velocity are often not stationary, that is to say the underlying mean value of the random process may be slowly varying. We use the structure function, denoted $D_f$ and defined as
\begin{equation}
 D_f(\tau)=<[f(t+\tau) - f(t)]^2>
\end{equation}
i.e. the mean square difference between two locations, to look at the variations on different scales. The Kolmogorov structure function for the velocity field in three dimensional homogeneous turbulence is 
\begin{equation}
 D_v(r)={C_v}^2r^{2/3}
\end{equation}
where $C_v$ is a parameter depending on the energy of the turbulence.

\cite{Obukhov1949} and \cite{Corrsin1951} independently suggested that a passive advective scalar should have the same structure function and power spectrum as the velocity field. 
This assumption is somewhat flawed (see Section~\ref{sec:intermittency}), but the structure function largely satisfies the standard Kolmogorov power spectrum, and this has proven to be an adequate hypothesis for much modelling of astronomical seeing.

\subsection{Image propagation}
The next step is to consider the interaction of the star light with this turbulence. Taking the classical view of light as oscillations in a complex scalar field we can consider incident monochromatic light with representation \begin{equation}
 \psi_0(\mathbf{ r},t)=Ae^{i(\phi_0 + 2\pi \nu t + \mathbf{ k} \cdot \mathbf{ r} )}
\end{equation}
which experiences perturbations due to the varying refractive index of the turbulence and emerges as
\begin{equation}
 \psi _p (\mathbf{r}) =(\chi_a(\mathbf{r})e^{i\phi_a(r)} )\psi_0
\end{equation}
for any instantaneous observation.
Here $\chi_a$ is the fractional change in the amplitude, A, due to atmosphere, and $\phi_a$ is the wavefront perturbation, which evolve as the turbulent atmospheric temperature structure changes.

Since we are considering `clear air turbulence'
\footnote{i.e. without clouds partially absorbing the light, or any other change in atmospheric transmission.} 
the varying refractive index of the atmosphere only directly affects the \emph{phase} of the light; amplitude variations are a secondary effect caused by the diffraction of light as it propagates between the turbulent layer and the observer. For the purposes of modelling a good astronomical site at longer wavelengths it is usually appropriate to neglect diffraction and scintillation effects. This is called the geometric propagation model, as it effectively models bundles of rays
\footnote{Vectors representing paths of light propagation.} 
whose wavefronts are advanced or retarded by the turbulent phase perturbations collectively referred to as the `phase screen.'

\cite{Tatarski1961} showed that for an atmospheric refractive index $n$, which has turbulence induced variations with structure function 
\begin{equation}
D_n(\mathbf{r})={C_n}^2 |\mathbf{r}|^{2/3}
\end{equation}
the structure of the atmospheric phase screen $\phi_a$ takes the form 
\begin{equation}
D_{\phi}(\mathbf{r}) \varpropto \int_0^z  C_n^2(z) dz |\mathbf{r}|^{5/3},
\label{eq:phase_structure_func}
\end{equation}
where $C_n^2(z)$ is the structure function parameter, varying with height $z$ in accordance with different intensities of turbulent variation in refractive index.
This integral can be rewritten in terms of a single parameter; the Fried coherence length $r_0$ \citep{Fried1965}:
\begin{equation}
 D_\phi(r) = 6.88 {\left( \frac{r}{r_0} \right) }^{5/3},
\end{equation}
which describes how the mean square phase difference between two points scales with the distance between them. 
The Fried coherence length represents the diameter of a telescope which will be almost unaffected by the seeing
\footnote{Specifically, the Fried coherence length defines the aperture of a telescope with the same optical transfer function as the atmospheric effects. See \cite{Roddier1981} for more detail.}
, and varies with the wavelength of observation as $\lambda^{6/5}$. As the ratio between telescope diameter and Fried length increases the point spread function that would ideally be an Airy disc splits up into multiple bright spots known as speckles, created by interference between light from regions across the telescope aperture within which the phase differs by around 1 radian, sometimes referred to as `coherent cells.' This ratio, $D_{tel}/r_0$, is a useful dimensionless number when comparing the severity of atmospheric turbulence for different telescope sizes, observing wavelengths, and prevailing atmospheric conditions. This phase disturbance scaling law has been verified by observation over a range of scales, with deviations as the outer scale of the turbulence is approached. This characterisation of the phase screens can be used as a model to simulate the effects of atmospheric turbulence upon astronomical telescopes, as covered in chapter~\ref{chap:lucky_AO}. 

\section{Characterising wavefront aberrations and image quality}
\label{sec:aberration}
\begin{figure}[htp]
 \includegraphics[width=\textwidth]{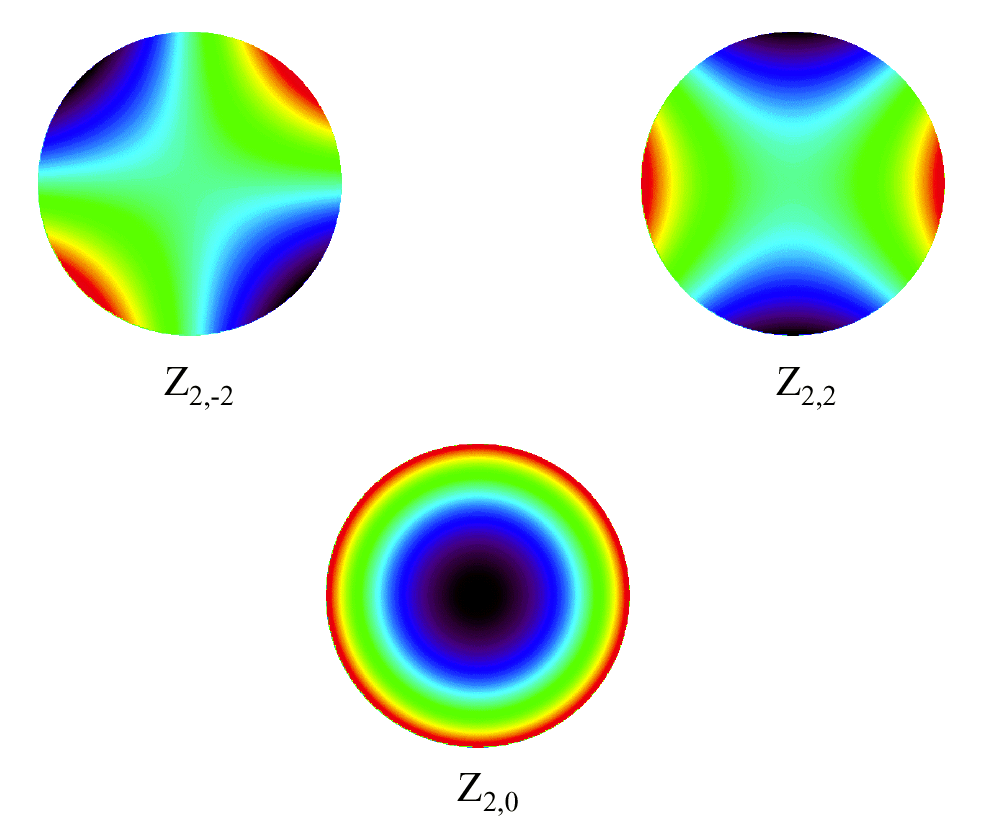}
\caption[Zernike modes]{Colour plots showing the shape of Zernike modes for astigmatism and defocus.}
 \label{fig:zernike}
\end{figure}
Since terms describing wavefront perturbations and image quality are often of use when discussing astronomy in the presence of atmospheric turbulence, it is worth introducing them for the unfamiliar reader before proceeding.

\subsubsection{Zernike modes}
In the same way that sine and cosine functions can be used to represent a signal composed of many different frequencies via a Fourier transform, a surface representing the phase of a non-planar wavefront entering a telescope pupil can be broken down into orthogonal components. A suitable set of functions for performing such a deconstruction are the Zernike modes \citep{Noll1976}. The low order Zernike modes correspond to the lowest frequency changes across the wavefront, and due to the Kolmogorov power spectrum these large scale changes always have the greatest amplitude (on average). Therefore when performing modal correction --- i.e. trying to correct for each Zernike mode in turn --- the largest improvements come from correcting the low-order terms first. Hence the terms `low order' and `high order' adaptive optics. The first Zernike term is piston, important for comparing phase across two interferometry apertures but irrelevant for a single aperture. The next two terms are tip-tilt, which together create 
an overall slope of the wavefront - this slope causes the point spread function (PSF) to move about in the focal plane. Higher order terms cause all manner of distortions, well known to manufacturers of optics and telescopes. Figure~\ref{fig:zernike} illustrates the wavefront shape for the astigmatism and defocus modes.

\subsubsection{Point Spread Function}
When describing the image quality of an optical system, the point spread function is usually the main focus of discussion. This is the light intensity distribution as measured in the focal plane, for an unresolved source (i.e one well approximated by an infinitesimal `point' source of light). Note that there are subtleties in describing the PSF, for example one may choose to include or neglect the effects of pixellation and sampling. These distinctions are covered briefly in chapter~\ref{chap:frame_registration}.

\subsubsection{Strehl ratio and full width at half maximum}
While Zernike modes describe phase perturbations directly, they do not describe the effect of atmospheric perturbations upon the image quality, and indeed the PSF. There are two ubiquitous measures for these effects. 

The first is the Strehl ratio, defined as the ratio between the peak intensity of the observed PSF, and that which would be measured if no phase perturbations were present and the optical system were performing perfectly. \corr{To determine this ratio we require the total source flux, in order to estimate the ideal peak intensity. Accurate measurements of Strehl ratio therefore require both accurate photometry and a well sampled image so that the peak intensity may be accurately estimated.} The ideal peak intensity is often well approximated by using an Airy disc model for the ideal image, although this may not be the case if significant support structures are present or a segmented primary mirror with large inter-segment spacing is used.

The second is full width at half maximum (FWHM). As the name suggests, this enumerates the angular width of the PSF at half the peak intensity. The implicit assumption is that the PSF is axisymmetric; while often not exactly true this is usually a reasonable approximation to the half maximum radius.

Many other measures such as full width at half enclosed flux (FWHEF) and encircled energy (EE) at a given radius exist, but Strehl ratio and FWHM are the most commonly used.

\subsubsection{Seeing width}
Another commonplace term in ground based astronomy is seeing width. This is simply the FWHM of a long exposure image obtained via conventional imaging at a large telescope. The atmospheric effects upon the PSF make this effectively independent of the telescope aperture for $D_{tel}/r_0$ of, say, 10 or greater. The seeing width $\epsilon$ is dependent upon observing wavelength through the relation \citep{Tokovinin2002}:
\begin{equation}
 \epsilon = \frac{0.98\lambda}{r_0}
\end{equation}

For use as a general metric of seeing conditions $\epsilon$ is usually quoted for a reference wavelength of 500nm. 

\subsubsection{Wavelength dependence}
\corr{It is worth mentioning that metrics of image quality which refer to characteristics of the PSF, such as Strehl ratio, FWHM, seeing width, and isoplanatic angle (see below) are all strongly dependent upon the observation wavelength. This is due to two factors. Firstly, propagation path differences caused by atmospheric turbulence or imperfect optics equate to different fractions of different wavelengths. For example, a 250nm optical path difference may be quite detrimental at 400nm (blue light) but will have little effect at 2200nm (`K band' infrared). Secondly, the ideal PSF also varies accordingly, such that longer wavelengths will have a wider, shallower ideal PSF. The resulting combined effect is usually a much improved Strehl ratio when observing at longer wavelengths.}

\corr{A corollary is the fact that since $r_0 \propto \lambda^{6/5}$, the seeing width actually decreases as the observation wavelength is shifted to longer wavelengths according to the equation:}
\begin{equation}
 \epsilon \propto \frac{1}{\lambda^{1/5}}
\end{equation}
\corr{for a given severity of atmospheric turbulence, when in the seeing limited regime (i.e. $D_{tel}/r_{0} \gg 1$). }

\section{Adaptive optics}
The most widespread method of attempting to overcome atmospheric turbulence effects in astronomy is adaptive optics. A brief recap of some key concepts and issues in this significant subfield of optical engineering is appropriate here, as it lends context to the discussion of lucky imaging to come.
\begin{figure}[htp]
 \includegraphics[width=\textwidth]{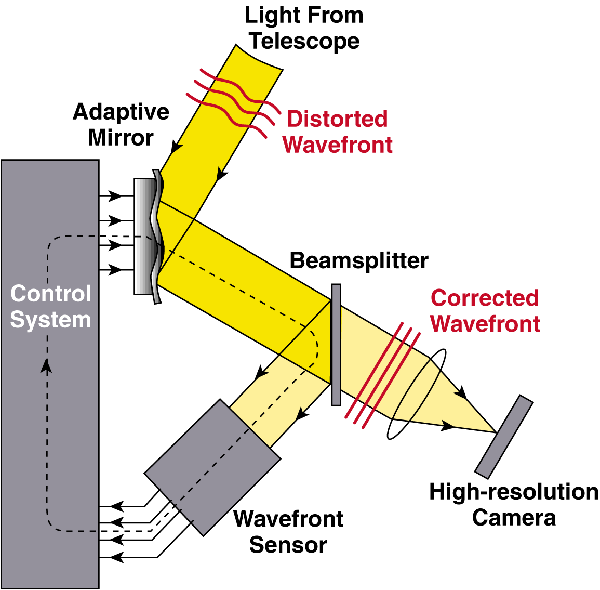}
\caption[AO Schematic]{A schematic of a basic adaptive optics system. Credit: Claire E. Max, UCSC}
 \label{fig:ao_schematic}
\end{figure}

A full introduction to adaptive optics (AO) is beyond the scope of this document, and the reader is referred to the excellent books of \cite{Tyson2000} for an overview and \cite{Hardy1998} for a full review. Figure~\ref{fig:ao_schematic} serves as an illustration of the general premise. In short, a portion of the incoming light is directed via a beamsplitter to some form of wavefront sensor. From the sensor, data is passed to a control system which estimates the perturbations to the wavefront. The control system then calculates appropriate commands to send to the corrective component, a deformable mirror. By using the deformable mirror to alter the path lengths for different sections of the wavefront a correction to the phase perturbations can be applied. The deformable mirror is placed before the beamsplitter in the optical train, creating a closed loop of feedback and correction. The aim is to perfectly counteract the ever changing atmospheric phase perturbations, resulting in a planar wavefront and thus 
achieving the full resolution and light concentration potential of the telescope. 

Note that the wavefront sensor requires a bright source in order to sense the phase perturbations with good signal to noise ratio, in a time span shorter than the typical evolutionary timescale of the atmospheric turbulence. The easiest way to achieve this is to observe a single bright star, termed the ``guide star.'' In practice, sensing and correcting the atmospheric phase perturbations with the required accuracy and speed is a challenging feat of engineering, and even with idealised systems there are fundamental limits to the performance. 

\subsection{Field of view and isoplanatic angle}
\label{sec:AO_fov_intro}
Since the phase perturbations of star light are accrued during propagation through the atmosphere, they depend on exactly which section of atmosphere the light wave passed through. As a result, correcting for phase perturbations for one source will generally only provide partial correction for sources separated by a significant angular width. For widely separated sources the phase perturbations may even be near independent, such that applying corrections for one increases the phase variance of the other.

In order to quantify this effect with a single number we use the isoplanatic angle. This is defined as the angular separation between the guide star and a secondary source at which Strehl ratio is decreased by a factor of $1/e$. 
\corr{As with Strehl ratio, the isoplanatic angle is strongly dependent upon the wavelength of observation. At shorter wavelengths, the path differences observed in light from sources of a given angular separation becomes apparent at much smaller angles. Hence observations at shorter wavelengths entail a smaller isoplanatic angle.}

Typically the isoplanatic angle of a standard AO system is small compared with the field of view --- for example, measurements at the Paranal observatory range from around 2.5 -- 4 arcseconds \citep{Sarazin2002} at 500nm. Even supposing total sky coverage (see below) this is a serious limitation, particularly for surveys or large extended objects. There are two developments of adaptive optics which seek to enlarge the isoplanatic patch, multi-conjugate adaptive optics (MCAO) and multi-object adaptive optics (MOAO). Both techniques use multiple guide stars to probe the atmosphere in different directions and so tomographically reconstruct a 3D model of the turbulence structure. Multi-conjugate systems then use 2 or more deformable mirrors conjugated to the optical heights of the turbulent layers, to correct for each layer in turn. In this way the phase correction is arranged to be different for different parts of the focal plane --- so each target within the corrected field of view has the phase disturbance of 
its own propagation path corrected for, enlarging the effective isoplanatic patch.

There is currently one MCAO demonstrator on sky, with the acronym MAD, at the ESO Very Large Telescope facility (VLT). \cite{Gullieuszik2008} report J band imaging with a mean FWHM of 0.15 arcseconds across a 45 by 45 arcsecond region, when the estimated seeing FWHM at J band was 0.52 arcseconds. This is an impressive accomplishment, and suggests that imaging of a Hubble space telescope resolution and field of view is now becoming available from the ground in the J and K bands. However, the sky coverage issues described below still apply. Target acquisition is also quite slow, with a typical loop closing period of 20-40 minutes on MAD. The system relies on favourable turbulence conditions for best performance, where most of the $C_n^2$ profile is confined to only 2 layers, one at ground level and the other at 8.5km. Future MCAO systems could have more conjugated mirrors and variable conjugation heights, but this reduces light throughput and increases complexity.

Multi-object adaptive optics is a proposed technique mainly applicable to integral field unit spectroscopy \cite[see for example][]{Ass'emat2007}. The idea is to use laser tomography and many deformable mirrors to correct the phase disturbances separately for each spectrography target in the field. Since it is impractical to perform wavefront sensing for all the targets this requires open loop wavefront correction. \corr{Open loop correction presents a significant technical challenge, because it requires phase corrections to be made accurately without any direct measurements of the current state of the deformable mirrors --- so their positioning must be correct first time.} However it is not so reliant as MCAO on the turbulence being highly confined to few layers. 

The widest field of view likely to be achieved is with ground layer adaptive optics (GLAO). This attempts to correct only for the layer of turbulence closest to the telescope, which may often contribute half or more of the integrated  $C_n^2$ profile depth. This gives a wide isoplanatic angle because the propagation paths for different targets are close at this height. GLAO is mainly aimed at survey and spectrography programs, as simulations suggest that typical FWHM improvement is a comparatively modest 0.1 arcsecond better than seeing width, but applies over a field of up to 49 by 49 arcminutes \citep[For a low order AO system on an 8m telescope,][]{Andersen2006}. As the number of guide stars increases the potential improvement under bad seeing conditions becomes larger. Such a system has been proposed for the Gemini observatory.

\subsection{Sky coverage}
\label{sec:ao_sky_coverage}
One of the major obstacles to widespread use of adaptive optics has been sky coverage. The requirement for a bright enough guide star in the vicinity of the target to be observed significantly limits the regions of sky to which the technique can be applied. This has led to the development of artificially generated ``laser guide stars'' (LGS). For easy differentiation, a bright star used for guiding is generally referred to as a ``natural guide star'' (NGS).

The current generation of adaptive optics systems have limiting natural guide star magnitudes in the R band varying from 4th to 14th magnitude, with curvature wavefront sensors providing lower order wavefront information but working with sources up to 4 magnitudes fainter than Shack-Hartmann sensors \citep{Racine2006}. A generous isoplanatic patch size of 40 arcseconds (in K band, say) then implies sky coverage of about 5\%.

\begin{figure}[htp]
 \includegraphics[width=\textwidth]{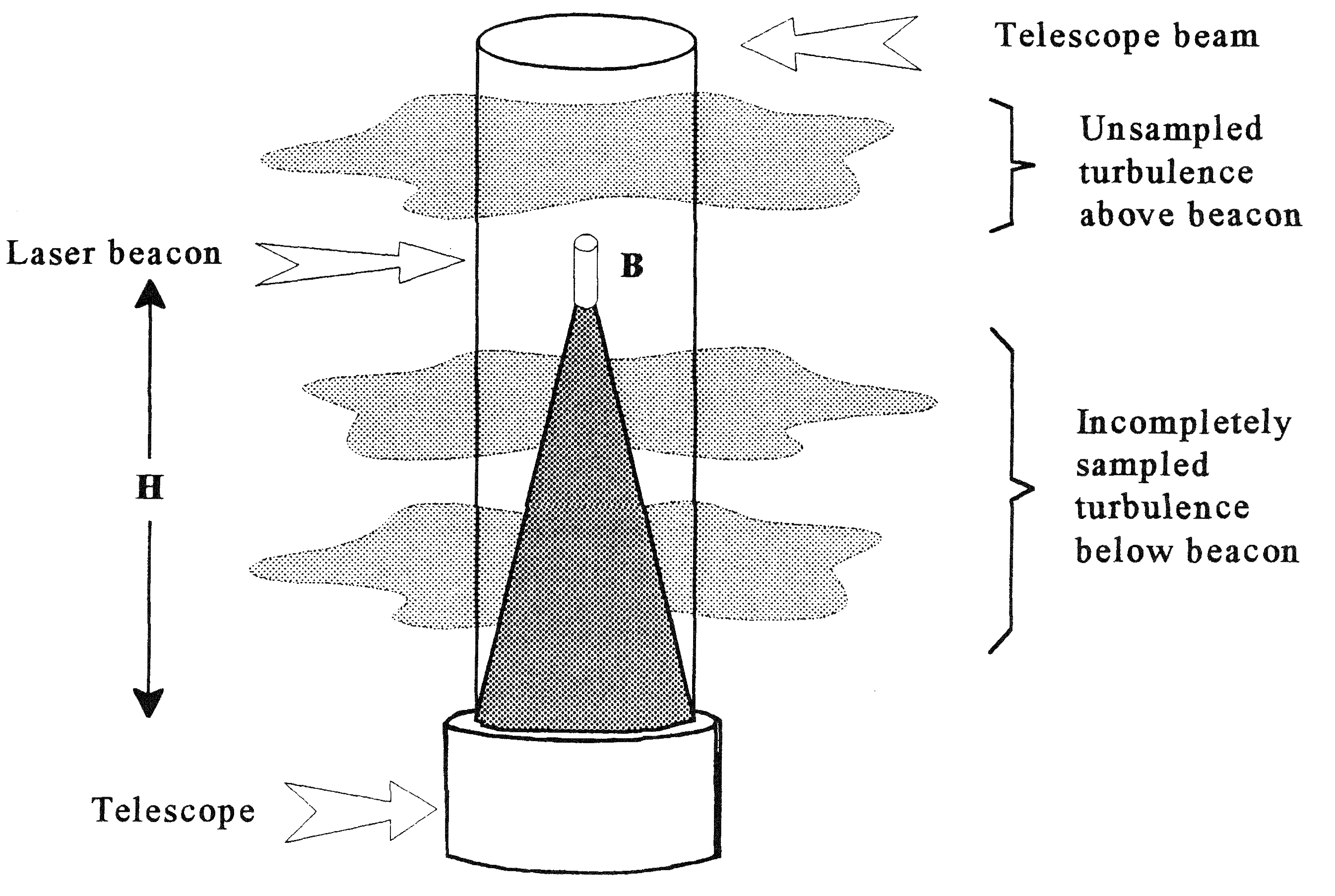}
\caption[The cone effect of LGS-AO]{Rayleigh backscatter laser guide stars suffer from the `cone effect.' Credit: \cite{Hardy1998}}
 \label{fig:cone}
\end{figure}

The AO community has sought to overcome this severe target constraint by developing laser guide stars that can be pointed anywhere in the sky; these come in two varieties, Rayleigh scattering and Sodium excitation lasers. Rayleigh back scattering laser guide stars are a maturing technology in terms of reliability and cost, to the level that a low-cost robotic LGS-AO system is being developed for 1.5m class telescopes \cite{Law2008a}. However, there are several problems associated with Rayleigh guide stars, as reviewed in \cite{Devaney2007}:
\begin{itemize}
 \item No tip-tilt determination: Tip-tilt information cannot be recovered from an LGS because the refraction of the beam on the way out cancels with the refraction on the way back. The usual solution is to observe an NGS simultaneously. Since the NGS is only being used to recover lowest order information it has less severe requirements than a full AO guide star --- for example at the VLT this tip-tilt star must be brighter than 17th magnitude in V band and within about 60 arcseconds of the target \citep{Davies2008}. This requires an additional beamsplitter in the optics, and is an intrinsic problem with all laser guide stars.

 \item Focal anisoplanatism or cone effect: The typical Rayleigh scattering height of 8-12 km presents problems for high order AO. The less than infinite propagation height means that the laser wavefronts are slightly curved, introducing focal anisoplanatism, and the highest layers of atmospheric turbulence are not probed (see Figure~\ref{fig:cone}). This problem increases in magnitude with telescope size.
 \item Spot elongation: Away from the axis of propagation the elongation of the laser scatter column can be seen, and this extension of the source makes it harder to use with a wavefront sensor. Again this problem gets worse as telescope size increases. 
\item No fly zone: High power laser guide stars are required to employ aircraft spotters and switch off the beam if there is any chance of an encounter. 
\end{itemize}

Due to the cone effect and spot elongation it becomes impractical to use a Rayleigh scattering laser guide star as telescope size increases. Sodium lasers excite the layer of atmospheric sodium at around 95km, and so ameliorate the cone effect and to a lesser extent the spot elongation. However, they come with their own problems --- sodium frequencies are currently only widely available through use of dye lasers, which are inefficient and unreliable compared to solid state.  For example, operation of the Keck system requires an extra observing assistant and a laser technician. Solid state sodium lasers are in development however, so wider application may not be many years away. The larger size of sodium lasers usually prohibits them from being placed on-axis of the main telescope, and so the beams have to be piped by fibre optics, introducing another technical challenge. Finally, even if the equipment works as intended the sodium layer varies in height and density,%
\footnote{\corr{The sodium layer owes its existence to meteor showers, resulting from the sodium atoms released as they burn up in the atmosphere. Hence the abundance varies according to the seasonal meteor density.}}
so that if conditions are bad it may only produce a faint fluorescence. In conclusion, sodium lasers are certainly an improvement, but not a panacea.

\subsection{Observation wavelengths}
The current generation of adaptive optics only work well at longer wavelengths where the atmospheric phase perturbations are a smaller fraction of the observing wavelength, or equivalently, $D_{tel}/r_0$ ratios are lower. Typically observations are made in the infra-red J,H and K bands, with Strehl ratios ranging from $\sim$0.4--0.7 in K and $\sim$0.2--0.3 in the shorter J band \citep{Racine2006}. As a result, 10 metre class AO assisted telescopes such as Keck and the VLT typically achieve imaging resolution with a FWHM of order $\sim$0.1 arcseconds --- on a par with, but not exceeding, the resolution of the Hubble Space Telescope (of course this data is complementary, since it provides information from a different wavelength regime). It is worth noting that tests of next generation adaptive optics systems are promising \citep[e.g.][]{Esposito2010} and the situation may improve once the new systems come online, but certainly these figures are likely to remain valid for most observatories in the near future.

To summarise, adaptive optics is now a relatively successful tool for high resolution ground based observing with infra-red wavelengths, but under significantly restricted conditions. The engineering complexities presented mean that AO systems are expensive, and require highly skilled personnel in order to operate and maintain them. This will remain the case as the systems evolve and increase in complexity in order to overcome some of the limitations of more basic systems.

\section{Lucky imaging}
\label{sec:lucky}
\subsection{A very brief history}
The origin of lucky imaging is credited to \cite{Fried1978}, who gave one of the first quantitative analyses of the subject in his paper ``Probability of getting a lucky short-exposure image through turbulence'' --- giving rise to the technique's name. It is of historical interest to note the earlier work of R.E. Hufnagel (whom Fried cites in his own work). Hufnagel proposed similar methods and derived the result that lucky imaging achieves the best results when the ratio $D_{tel}/r_0 \sim 7$ \citep{Hufnagel1966}, but was not widely published.%
\footnote{More information on Hufnagel's work may be found at:
\newline http://www.ast.cam.ac.uk/research/instrumentation.surveys.and.projects/lucky.imaging/references}
However, the technique was not feasible with the detectors available at the time, and so its development into an observing tool has taken over 20 years.

The basic concept of lucky imaging is simple. The atmospheric seeing is random; as such there are moments in time when the seeing is significantly better or worse than average. The coherence time associated with the seeing is on the order of tens of milliseconds, so to pick out the very best moments requires a camera that can record frames on a similar timescale. 

With the continued development of CCDs throughout the 1980s and 1990s the technical challenges started to become tractable; although detectors were still either sensitive, but too slow to freeze the atmospheric changes \citep[e.g.][]{Nieto1987}, or fast but noisy --- a good example of this is \cite{Dantowitz2000}, in which the authors produced high resolution images of Mercury (a very bright source, often brighter than 0th magnitude) on the Palomar 60 inch telescope at 60 frames per second with digital video equipment. \cite{Baldwin2001} used a specialised CCD at frame rates of up to 185 frames per second to prove the concept of lucky imaging with stellar sources on the 2.5m Nordic optical telescope, but were still limited to 6th magnitude stars by readout noise. Development of the electron multiplying CCD cameras described in chapter~\ref{chap:EMCCD_calibration} finally combined the desired properties of fast frame rates and low read noise, enabling application of lucky imaging to a wide range of targets.

%
%

Another technological advance key to lucky imaging is the ever increasing performance of computer systems, both in terms of processing power and data throughput, since lucky imaging involves a fairly sizeable amount of data processing. After data acquisition, each frame is cleaned of detector artefacts and analysed in order to assess the severity of the atmospheric turbulence at that moment (detailed in chapter~\ref{chap:frame_registration}). The best frames are then aligned to correct the effects of atmospheric tip-tilt shifting the image in the focal plane, and co-added via a Drizzle algorithm \citep{Fruchter2002}. As more frames are combined, signal to noise ratios slowly improve and faint stars become visible above the background noise. 

\subsection{Pros and cons}
The key limitation of lucky imaging is that the the technique is only feasible when telescope aperture size and atmospheric observing conditions result in $D_{tel}/r_0$ ratios of around 7 or less, \corr{as detailed in table~{\ref{tab:fried_selection_table}}. For higher ratios the chance of obtaining good frames diminishes rapidly, following the formula} {\citep{Fried1978}}:  
\begin{equation}
 Prob \approx 5.6 \exp \Big[-0.1557(D_{tel}/r_0)^2\Big]
\end{equation}

The observation wavelengths are restricted by the availability of sensitive high speed detectors, the sensitivity of current electron multiplying CCDs declining as the wavelength moves from the visible into the infra red. For typical seeing at good astronomical sites this limits the technique to application on telescopes of 3m diameter at most, resulting in images with FWHM of order $\sim$0.1 arcseconds. However, lucky imaging fills the niche of high resolution imaging in the visible wavelengths where adaptive optics is ineffective, the only other option being to obtain space based observations. %

\corrbox{Future developments in Mercury-Cadmium-Tellurium infrared detectors \citep{Rogalski2005,Rothman2009}
may change this situation, 
enabling lucky imaging in the infrared as a cheap alternative to adaptive optics on medium sized telescopes. However, a proper feasibility study considering current technology is beyond the scope of this thesis.}

\begin{table}
\begin{center}

\begin{tabular}{|c|c|}
\hline 
\hline 
 $D / r_0$ & Probability \\
\hline
\hline
 2 & 0.986 $\pm$ 0.006 \\
 3 & 0.765 $\pm$ 0.005 \\
 4 & 0.334 $\pm$ 0.014 \\
 5 & $(9.38 \pm 0.33) \times 10^{-2}$\\
6 & $(1.915 \pm 0.084) \times 10^{-2}$\\
7 & $(2.87 \pm 0.57) \times 10^{-3}$\\
10 & $(1.07 \pm 0.48) \times 10^{-6}$\\
15 & $(3.40 \pm 0.59) \times 10^{-15}$\\
\hline 
\hline
\end{tabular}
\caption[Probability of getting a lucky short exposure]{Reproduced from \cite{Fried1978}: Fried's original estimates of the frequency of good quality short exposures (i.e. those short exposures with an RMS wavefront perturbation of 1 radian or less) as a function of the ratio between telescope aperture size and the atmospheric coherence length, $D/r_{0}$. The estimates were made by Monte Carlo evaluation of the complex probability integral describing the wavefront as a superposition of Karhunen-Loeve eigenfunctions --- these are similar to the Zernike modes but statistically independent for turbulence with a Kolmogorov power spectrum. Fried's estimate that lucky imaging would work best around $D/r_{0}=7$ has proved remarkably accurate \citep[see e.g.][]{Tubbs2003}.}
\label{tab:fried_selection_table}
\end{center}
\end{table}

Compared to adaptive optics, lucky imaging is a relatively low cost tool for high resolution imaging. While an EMCCD camera may cost a few tens of thousands of dollars, the price of a full adaptive optics system may number in the millions. Operation could also be made relatively simple and reliable, since no moving parts are involved. 

Another advantage is that zero target acquisition time (after telescope slew) means lucky imaging can make efficient use of observing time, for example several hundred binary targets were observed over a few nights with a queue observing routine during a February 2008 observing run \citep{Law2008}.

Finally, isoplanatic angles appear to be larger than for adaptive optics, typically 30 arcseconds in average to good seeing. The faint guide star limit is also considerably better than for adaptive optics, with previous limiting estimates of approximately 16th magnitude in SDSS \textit{i'} band \cite{Tubbs2003,Law2006} (more on this in chapter~\ref{chap:frame_registration}). The combination of these factors results in much better sky coverage when compared to natural guide star AO.

\subsection{Current status of lucky imaging techniques}
\subsubsection{Conventional lucky imaging}
Standard lucky imaging is now a proven technique for providing diffraction limited imaging at telescope diameter to Fried coherence length ratios of up to $D_{tel}/r_0 = 7 $, for example in the SDSS \textit{i'} band on the 2.5m Nordic Optical Telescope \citep{Law2005,Lodieu2009}. Electron multiplying CCDs capable of high frame rates are becoming more widely available, and other teams have successfully implemented their own routine usage lucky imaging systems which are often employed for projects surveying binarity \citep{Hormuth2008a,Oscoz2008}. Early EMCCD detectors were only available in small pixel formats, but larger devices are now available with a format of $\sim\!\!1\mathrm{k}^2$, increasing the field of view. Noise artefacts present in earlier cameras are now much less troublesome. 

\subsubsection{Lucky imaging enhanced adaptive optics}
In July 2007 the Cambridge lucky camera was tested behind the `PALMAO' adaptive optics system on the 200 inch Hale telescope at Palomar observatory \citep{Law2008}. PALMAO usually operates in K band, giving Strehl ratios of up to 0.5 \citep{Troy2000}. Using the lucky camera, observations were made at wavelengths of 500nm, 710nm, and 950nm. At 710nm lucky selection improved the FWHM from 75 to 40 milliarcseconds, and improved the Strehl ratio from 0.06 to 0.15, demonstrating that lucky imaging can greatly improve the imaging performance of an adaptive optics system at shorter wavelengths. Investigation of the decorrelation time for the PALMAO data set suggests the lucky camera may need to run at or above the operating frequency of the AO system to achieve best performance. There is some discussion of the future potential of such systems in chapter~\ref{chap:science}, and methods for modelling and predicting their performance are covered in chapter~\ref{chap:lucky_AO}.

\section{New hardware: The quad-CCD LuckyCam}
\label{sec:luckycam_2009}
\begin{figure}[htp]
\begin{center}
 \includegraphics[width=0.95\textwidth]{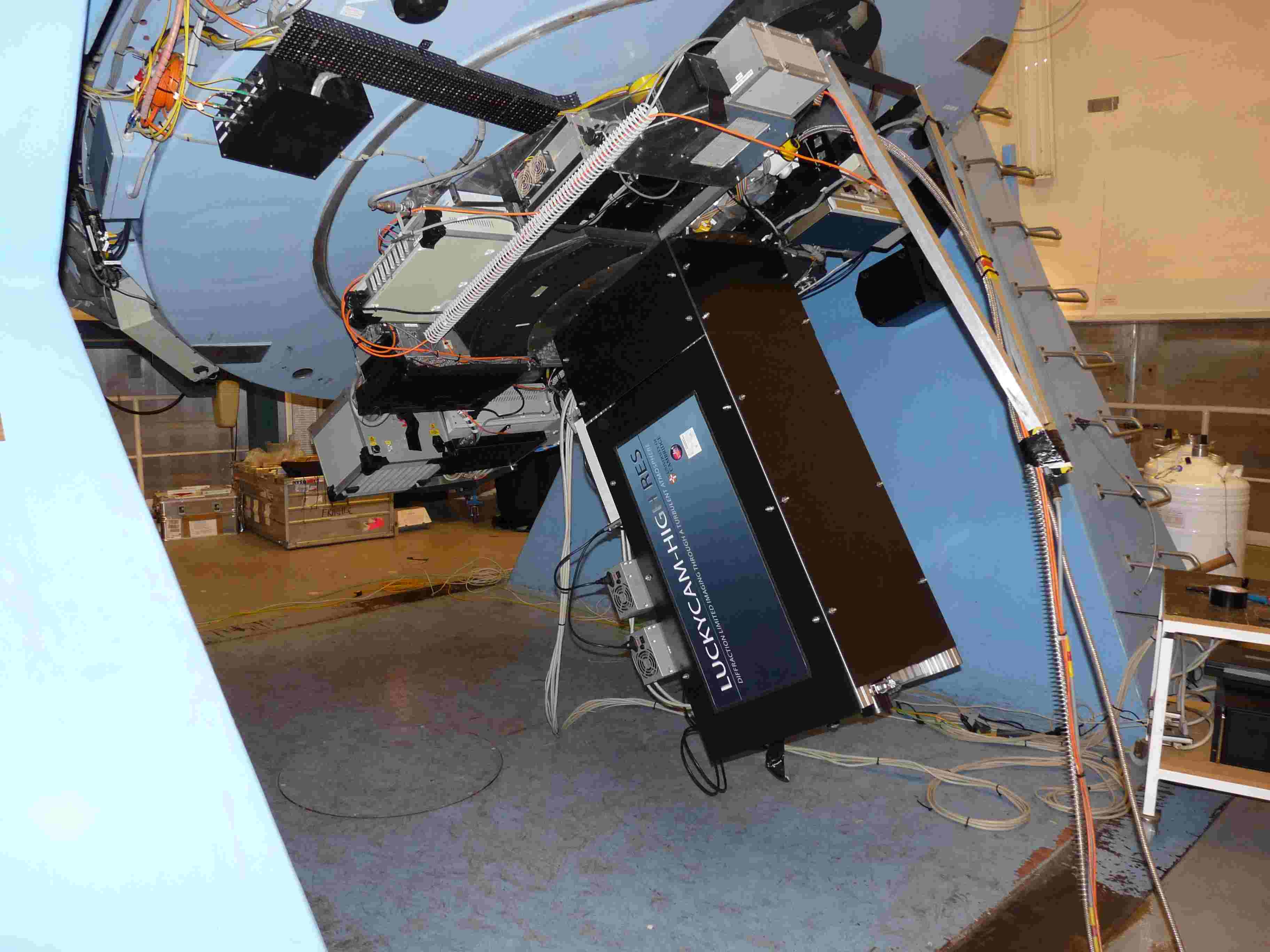}
\caption[LuckyCam 2009 on the NOT]{`LuckyCam' in the July 2009 configuration, mounted upon the 2.5m Nordic Optical Telescope. The camera housing cleared the floor with a scant few millimetres to spare.}
 \label{fig:luckycam2009_img}
\end{center}
\end{figure}
\corrbox{
A large portion of this dissertation is based upon data taken with a new design of lucky imaging camera, recorded during a July 2009 observing run at the Nordic Optical Telescope, at the Observatorio del Roque de los Muchachos, La Palma (illustrated in  Figure~{\ref{fig:luckycam2009_img}}). The 8 night observing run was undertaken by Craig Mackay, David King and myself, with remote software support from Frank Suess.
}

\begin{figure}[p]
\centering
\subfigure[Schematic of camera layout.]{
	\includegraphics[width=0.8\textwidth]{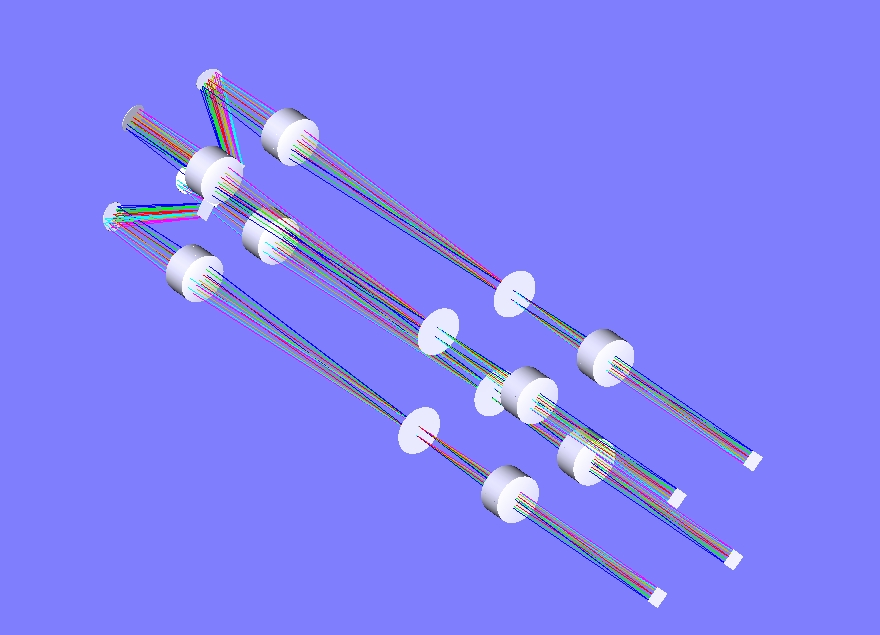}
	\label{subfig:schematic}
}
\subfigure[The `pyramid' mirrors used to split the field of view.]{
	\includegraphics[width=0.45\textwidth]{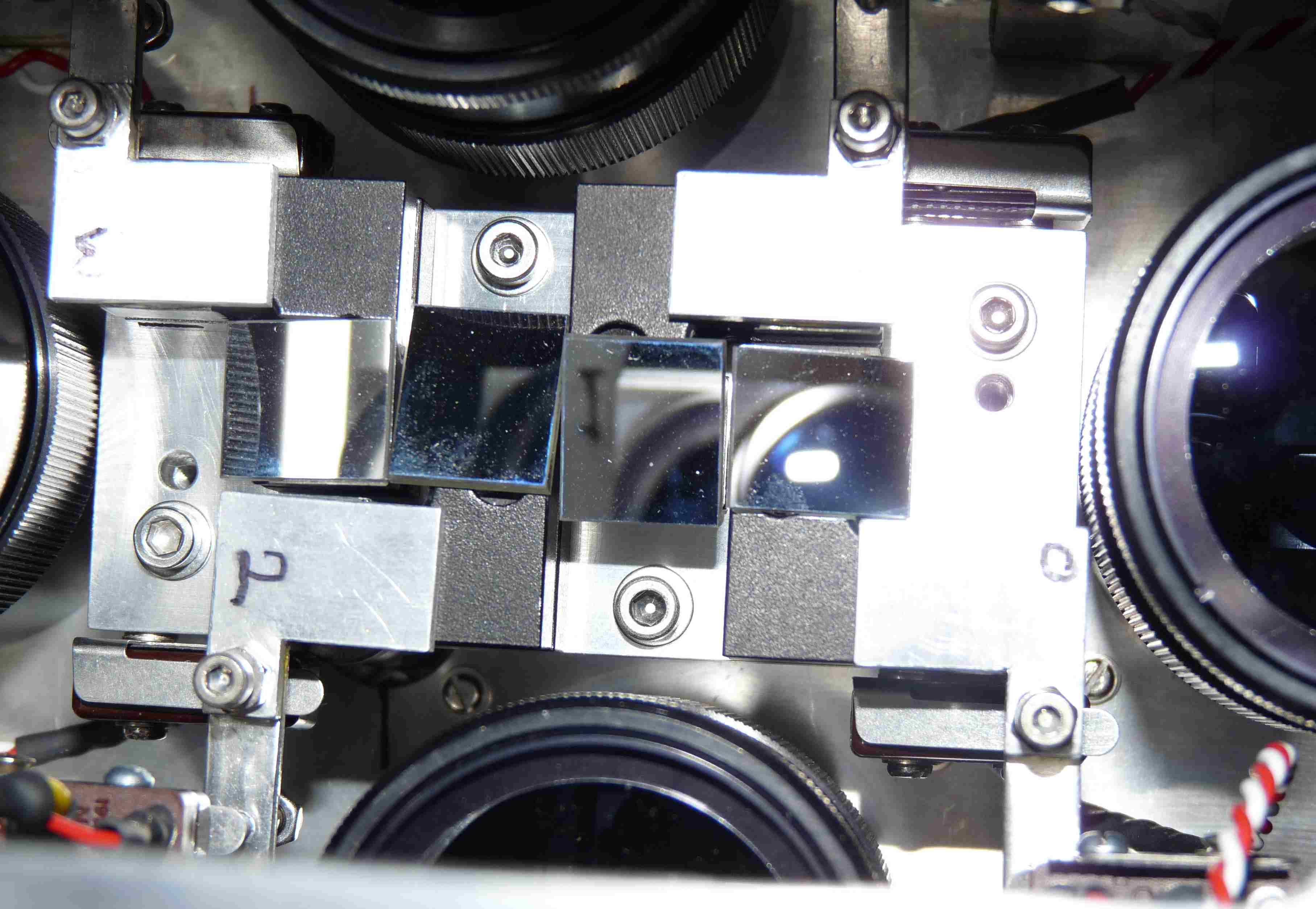}
	 \label{subfig:pyramid}
}
\subfigure[CCDs viewed through the dewar entry apertures.]{
	\includegraphics[width=0.45\textwidth]{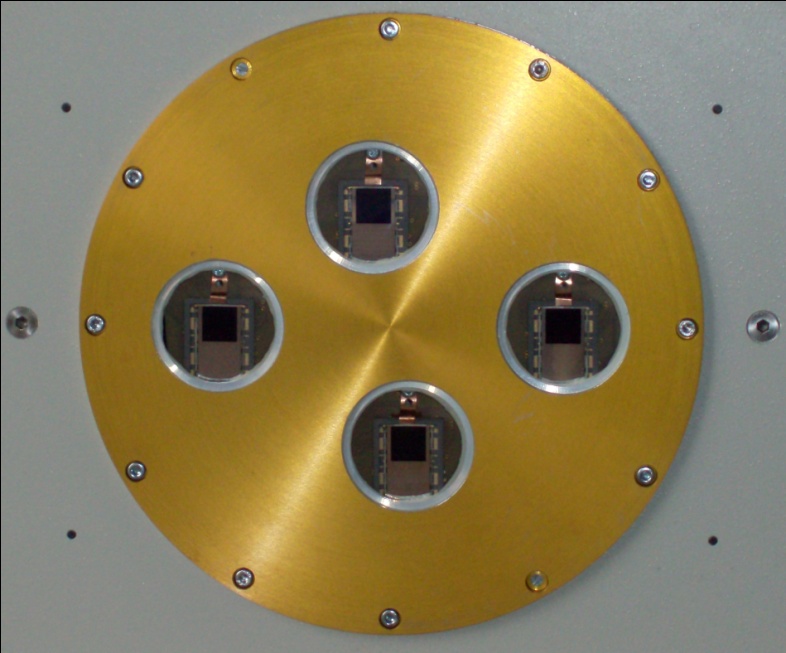}
	 \label{subfig:dewar}
}
\caption[LuckyCam 2009 camera schematic]{\subref{subfig:schematic}: A simplified schematic of `LuckyCam' in the July 2009 configuration. Light enters through the front of the camera (upper left in picture) and the focal plane is split into sub-fields by four mirrors angled in different directions. 
These are referred to as the `pyramid' mirrors, although the mirrors are actually arranged in a row \subref{subfig:pyramid}. 
The light is then reflected back along four parallel collimation tubes, before being refocused onto the detectors in the dewar \subref{subfig:dewar}. 
This design gives a mosaic of closely spaced subfields, while allowing sufficient space around the individual CCDs for readout electronics. 
}
 \label{fig:luckycam2009_schema}
\end{figure}

\begin{figure}[ht]
\centering
\subfigure[A side on view of the whole camera.]{
	\includegraphics[height=0.4\textheight]{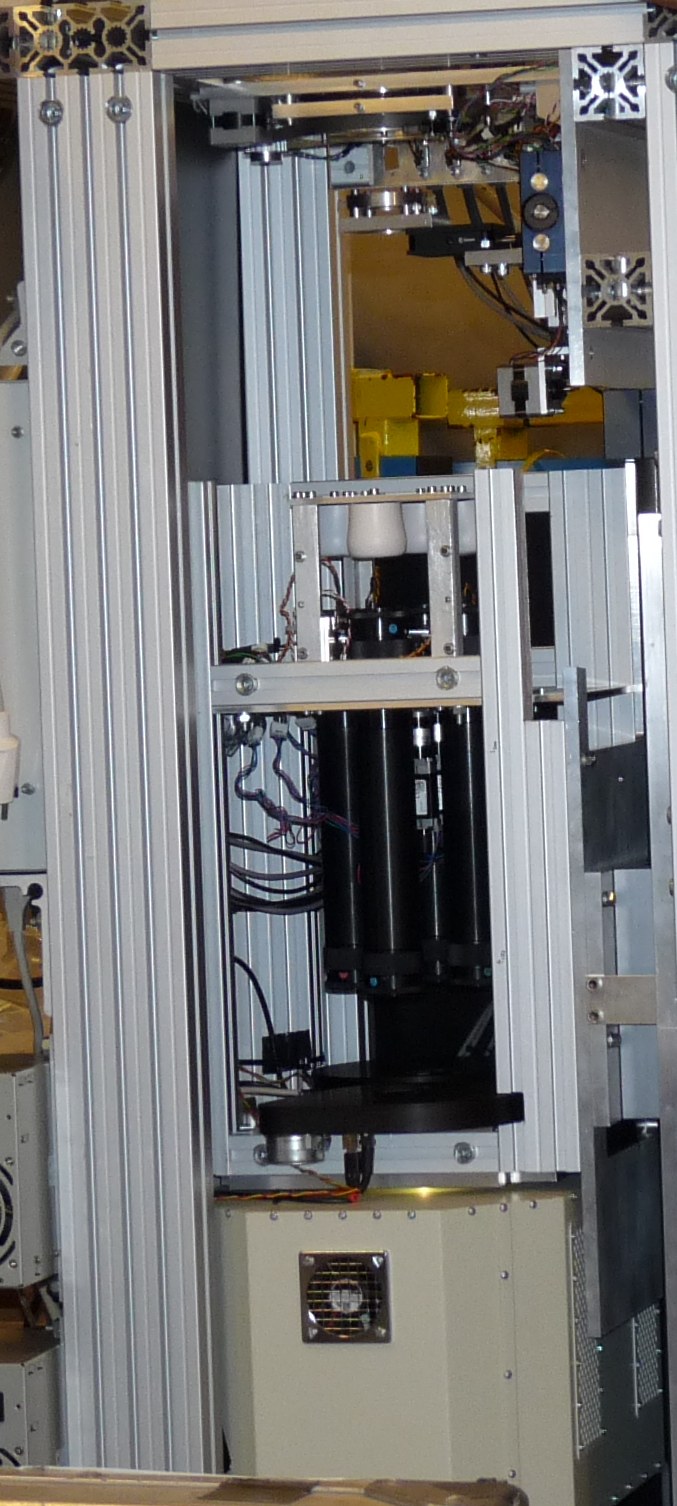}
	 \label{subfig:full_camera}
}\hfil
\subfigure[Close up of the mosaic optics.]{
	\includegraphics[height=0.4\textheight]{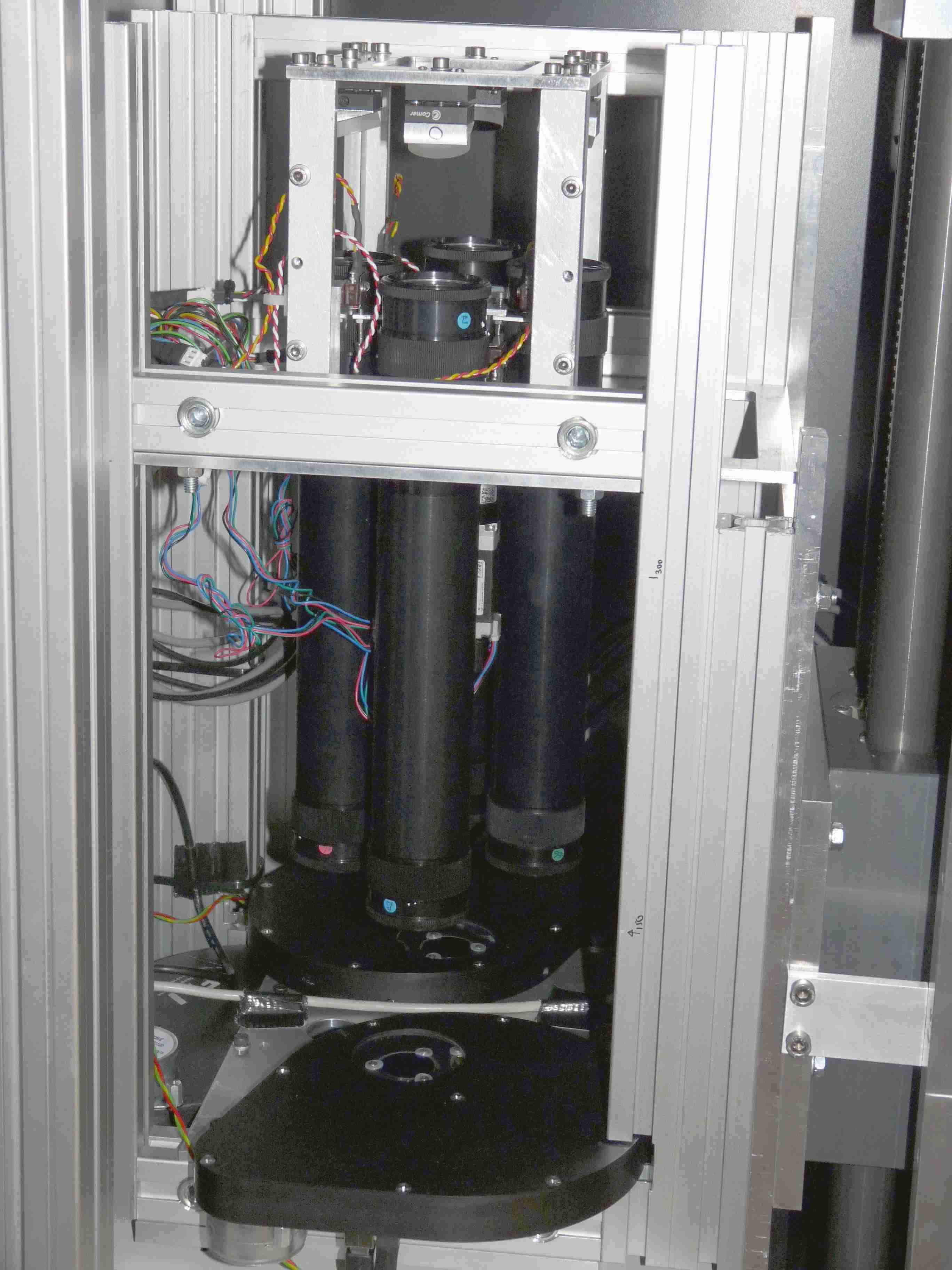}
	 \label{subfig:camera_close_up}
}
\caption[LuckyCam 2009 internals]{A side view of the LuckyCam internals, as configured for the 2009 observing run. In both images, the camera is pointing upwards (so light enters from the top). The left image shows the foremost optical components at the top of the image --- a lens wheel, which changes the focal length (and hence pixel size / field of view), and also an achromatic dispersion correction unit. The lower two thirds show the dewar and mosaic optics. These are mounted on a sliding plate which can be moved up and down using a worm drive mechanism, to adjust the focal length (which allows use of the different lenses).

The right image shows a close up of the mosaic optics. At the top the image the four pick-off mirrors can be seen (the central pyramid mirrors are obscured from view by the collimation tubes). At the bottom of the image two of the four filter wheels are visible.

}
 \label{fig:luckycam2009_internal}
\end{figure}

\corrbox{
One of the aims of the trip was to thoroughly test the new camera, which utilises 4 synchronised EMCCDs running in parallel. 
Each EMCCD is a $\sim\!\!1\mathrm{k}\times1\mathrm{k}$ pixels format, resulting in a total mosaic of $\sim\!\!1\mathrm{k}\times4\mathrm{k} $ pixels, with a frame rate of $\sim\!\!20$Hz. 
The main reason for adding this complexity is to achieve a wider field of view whilst retaining good sampling of the image, although there are secondary benefits such as improved dynamic range and enabling the use of multiple filters simultaneously (by using different readout configurations and different filters for each detector, respectively).
The layout of the camera for the July 2009 run is presented in  figures~\ref{fig:luckycam2009_schema} and \ref{fig:luckycam2009_img}, while the detector properties are detailed in table~\ref{tab:luckycam_props}.
The data acquisition and analysis requirements for the camera are considerable, and the software and techniques I developed to deal with the data are covered in chapter~\ref{chap:data_reduction}. 
}

\pagebreak[4]
\section{Statement of thesis}
The proposition of this thesis is twofold; firstly that lucky imaging hardware and data reduction techniques are now sufficiently mature that it has become a viable tool for a wide range of astronomical science applications that benefit from high resolution imaging at visible wavelengths. Secondly, coupled with adaptive optics technology it has the potential for extremely high resolution imaging that will open up entirely new areas of investigation.

 \begin{table}
\begin{center}

\begin{tabular}{|c|cc|}
\hline
\hline 
 & Red & Near-Infrared \\
 & 700--1000nm & 1200--2200nm \\
\hline \hline 
0.3 arcsec resolution & Tip/tilt & Tip/tilt \\
\hline
0.1 arcsec resolution, & Lucky imaging & NGS AO  \\
Strehl ratio $>$ 0.2 & (speckle imaging) & (speckle imaging)  \\
 & (interferometry) & (interferometry)  \\
\hline
0.1 arcsec resolution, & Lucky imaging & LGS AO  \\
using faint guide stars & & \\
\hline 
\hline
\end{tabular}
\caption[Ground imaging capabilities]{Lucky imaging provides ground imaging capabilities not otherwise available. Tip/tilt refers to an ``image stabilization system,'' effectively an AO system which only corrects for image motion in the focal plane due to the tip-tilt component of the phase perturbations. Speckle imaging and interferometry are listed in brackets because they are usually used for specialised purposes, where a model of the target object is already known or suspected (e.g.. searches for binary stars). Reproduced from \cite{Law2007}. }
\label{tab:ground_img_capabilities}
\end{center}
\end{table}

Lucky imaging as a generally applicable tool for astronomy is desirable, because it provides information that is difficult or impossible to obtain with other ground based systems (table~\ref{tab:ground_img_capabilities}), and will soon be unobtainable from space, too. 
The Hubble Space Telescope, now producing excellent images with the latest set of electronics, will not receive any further servicing missions now the shuttle fleet has retired, and will almost certainly be decommissioned in the next 5-10 years. The James Webb Space Telescope, if it flies, will produce images of inferior resolution in the visible. Adaptive optics technology is advancing, but correction in the visible is still poor even with the very latest cutting-edge systems, and only works over a small field of view. Lucky imaging provides a cheap, viable alternative for obtaining high resolution images in the visible to complement the high resolution infra-red imaging that is becoming more widespread.

When I began my PhD project, lucky imaging was a proven concept, but limited in scope of application. The strenuous requirements for high frame rates necessitated relatively small fields of view, and challenging detector noise characteristics and non-uniformities made imaging of very faint targets difficult. As a result the technique had been successfully applied to close binaries of similar magnitude, but not much else.

A large portion of my research has focused upon overcoming these issues, and demonstrating the potential applications made possible as a result. 
The pace of hardware development has been rapid, and the lucky imaging camera deployed in 2009 ran with a mosaic field of view of approximately $1\mathrm{k}\times4\mathrm{k} $ pixels, resulting in a factor of 16 increase in critically sampled field of view, and correspondingly in data rates, compared to the  $512\times512$ pixel squared system described in \cite{Law2007}. The larger field of view  greatly improves both the number of observable lucky imaging targets, and the efficiency as a potential survey tool. The calibration and reduction algorithms have also been substantially developed and refined, such that when coupled with reduction techniques developed specifically for EMCCDs, imaging of very faint targets is now a proven possibility. %
When coupled with adaptive optics, the unprecedented resolution makes possible entirely new scientific investigations.

\section{Chapter summaries}
The rest of this thesis is structured as follows:

Chapter~\ref{chap:EMCCD_calibration} details the models and calibration techniques required to get the best results from electron multiplying CCDs. Chapter~\ref{chap:thresholding} then uses those models to examine thresholding schemes to improve signal to noise ratio at low light levels. Implementation and application of the thresholding schemes to real data is covered.

Chapter~\ref{chap:frame_registration} reviews some basic elements of sampling and image processing theory, and presents improvements to the lucky imaging frame registration procedure, which enables more effective use of faint guide stars. An analytical model to predict the Strehl ratio in reduced lucky images is proposed.

Chapter~\ref{chap:science} covers the scientific applications of lucky imaging. An image analysis procedure for stellar binarity surveys is presented, along with new faint companion detections to planetary transit host stars, which significantly expand the range of contrast ratios in binary stars resolved with lucky imaging. A case is made for lucky imaging as a generally applicable high resolution imaging tool that may often provide an adequate substitute for Hubble Space Telescope data. Finally, a possible application of lucky-imaging-enhanced adaptive optics is discussed.

Chapter~\ref{chap:lucky_AO} reviews the theory of Monte Carlo simulations for modelling adaptive optics and lucky imaging systems. Preliminary performance investigations of a hybrid lucky imaging adaptive optics system are presented.

Finally, chapter~\ref{chap:data_reduction} describes the software and techniques I developed during my PhD for dealing with the large datasets and novel data reduction problems that lucky imaging presents.

\chapter{Calibration of electron multiplying CCDs}
\label{chap:EMCCD_calibration}
``The only uniform CCD is a dead CCD'' \citep{Mackay1986}. 
Due to imperfections in manufacturing causing slightly different pixel areas, readout electronics varying their characteristics during a frame read, and so on, all CCDs have at least slight non-uniformities in their pixel response to incident light. To avoid introducing errors in scientific data it is important to calibrate and correct for these. The electron multiplying CCDs used in the Cambridge lucky imaging camera require some specialized calibration routines.
Some elements of these calibration procedures were introduced in \cite[see Section~\ref{sec:previous_emccd_cals} below for specifics]{Law2007}, however considerable effort has gone into developing more accurate and comprehensive methods, so I present these in some detail. 

As an aside, where possible I shall adopt the same notations in this chapter as those of \cite{Tulloch2011}, in the hopes of achieving some amount of uniformity across the literature.

\section{The physics of electron multiplying charge-coupled devices}
\label{sec:emccd_background}
Conventional charge couple devices or CCDs are widely used as detectors for astronomy due to their linear response characteristics, high quantum efficiency, wide dynamic range, and ever increasing pixel array sizes. The reader is referred to \cite{Howell2000} and \cite{Tulloch2011} for relevant reviews, but it will be useful to recall the fundamentals of CCD and EMCCD operation here.

Photons incident upon a pixel element of a CCD may excite electrons from the silicon substrate if the photon energy is of the same order as the electron excitation band gap. The freed electrons (known as photo-electrons due to their origin) are then held in place by a potential well due to the voltage at electrodes within each pixel known as gates. When the exposure time has elapsed, the accumulated charge packets of electrons are shifted across the CCD by cycling the voltage values of neighbouring pixel gates through multiple phases
\footnote{Specifically, conventional CCDs use 3 voltage phases. EMCCDs use 2 or 4 phases during parallel charge transfer, and 3 phases during serial transfer through the readout register.}
 (a process known as clocking). Eventually all electron packets are passed through a component called the `charge amplifier', which converts the charge into a voltage. Finally the voltage is passed to an analogue-to-digital unit (ADU) for digitization, and the resulting integer value is recorded.

The conversion of the photo-electrons to voltage by the charge amplifier introduces additive noise (``readout noise''), such that the inferred pixel values will be spread in a Gaussian distribution about the true value representing the photo-electrons captured. As a result, even at slow readout speeds (tens of kHz pixel rates) typically a CCD experiences readout noise of RMS equivalent to a few photo-electrons. Lucky imaging requires frame rates of 20Hz or higher, resulting in pixel readout rates greater than 20MHz for a large CCD. At these rates a conventional CCD may experience read noise of tens or hundreds of electrons (e.g. \citet{Tubbs2003} cites readout noise of 50-60 electrons at 5.5MHz pixel rates). This level of noise makes it impossible to perform lucky imaging on all but the brightest stars with any useful level of signal to noise when utilizing conventional CCDs.

Fortunately, electron multiplying CCDs (EMCCDs, also known as low-light-level or LLLCCDs) \citep{Jerram2001, Mackay2001} offer a solution. In an EMCCD an additional multi-stage ``serial register'' is inserted into the circuit prior to the readout components. This is a series of gates similar to those used for the pixel grid of a conventional CCD, but with one of the three phases operating at much higher voltage than would usually be applied, typically around 40v compared with the 12v applied for conventional parallel transfer\citep{Mackay2010}. The large electric fields generated accelerate the electrons sufficiently to give a significant probability of impact ionisation at each stage --- effectively releasing an extra electron and so multiplying the charge in a stochastic manner. The probability of ionisation at each stage is small, but using many stages in series it is possible to achieve a high degree of gain in the average signal level, albeit in a stochastic manner. The mean gain may be altered by 
making small adjustments to the voltage, and is usually set to many times the average readout noise variation. As a result, the previously insurmountable read noise now equates to a mere tenth or twentieth of a photo-electron signal, with electron multiplication factors of one or two thousand.

\section{Basics of CCD calibration }
\label{sec:CCD_cal}
When calibrating conventional CCDs it is customary to take a handful of calibration frames, some with shutters closed to calibrate internal detector effects \corr{(detailed below)}, and some `flat fields' with uniform illumination to calibrate the detector response to light. 
These are then averaged and subtracted to infer the bias pedestal, dark current, and flat field \corr{(pixel light response). All these quantities may vary somewhat from pixel to pixel, so they are recorded as pixel maps} and used for data reduction as follows \citep{Howell2000}:
\begin{equation}
 \text{Reduced Object Frame} = \frac{\text{Raw Object Frame} - \text{Bias Map} -\text{Dark current Map}}{\text{Flat Field Normalisation Map}}
\end{equation}

Here the bias \corr{map} records variation in the `bias pedestal', an offset introduced by the readout electronics which has the effect of ensuring all raw pixel values are positive \corr{(and hence digitizable)}. The dark current \corr{map} corrects for internally generated signal, usually dominated by thermal emissions within the CCD, which would otherwise be mistaken for signal from photo-electrons. \corr{Bias and dark current maps are calibrated by taking frames with different exposure lengths. The bias calibration frames are recorded with short exposure times to minimize dark current contributions. These short exposures can be analysed in conjunction with dark current calibration frames of different exposure times, to estimate the pixel value contributions from bias pedestal and dark current.}

The flat field is a normalising pixel map which corrects for different levels of sensitivity to incident light - these may be due to intrinsic variations in pixel sensitivity, and also extrinsic factors such as partial obscuration by dust in the camera lens.

\corrbox{Note that there is some ambiguity in the nomenclature with regards to calibration frames. It is common practice to take at least a handful of each sort of calibration frame, and then combine them (usually via their median pixel values) to create averaged frames for use in data reduction. However, the raw and median frames are often referred to interchangeably for brevity (the median of the bias frames simply being called the `bias frame,' for example). In this thesis I shall take care to refer to the raw exposures as e.g. `bias frames,' and the final version for use in data reduction as the `bias map.'}

Even for a conventional CCD, these calibrations are only valid for a certain length of time. Varying operating temperatures may alter the bias characteristics or thermal emission levels, camera lenses get cleaned and then slowly accrue dust again. Typically such variations occur at a significant level on timescales of hours or longer.

With EMCCDs, things are a bit more complex. Not only are there extra variables which must be accounted for, some of these effects alter from observation to observation, and potentially even from frame to frame. Since the Cambridge group EMCCD camera is custom built, distinguishing and calibrating the various different effects is a task that must be carefully undertaken with each new camera configuration. 

Both spatial and temporal variations of the CCD response must be corrected for as far as possible on a \emph{frame by frame}  basis --- i.e. prior to any further data reduction --- for several reasons. First, it enables accurate determination of the gain since it stabilises the bias pedestal, resulting in a cleaner histogram from which to estimate the EM gain (cf. Section~\ref{sec:gain_cal}). Second, if pixel thresholding photon-counting techniques are to be applied (cf. chapter~\ref{chap:thresholding}), then the data must be debiased before thresholding, otherwise a uniform threshold level will result in non-uniform fractions of photon-events passing the threshold across the detector. Third, the `drizzle' algorithm (Figure~\ref{fig:drizzle_schema}) by definition combines values from different pixels, by default assigning them equal weights and resulting in some loss of information and correlated noise - it is therefore desirable to ensure that the input data truly are as uniform as possible before drizzling.
 Finally, the lucky imaging data has potential for analysis of high-speed variation in sources as explored in Section~\ref{sec:fast_photometry}, which obviously requires that detector corrections are applied to every frame.

The rest of this section details the various phenomena which must be considered for accurate EMCCD operation, and methodology to calibrate and correct for any non-uniformities. 
Unless stated otherwise, all results refer to full frame readout of the Cambridge EMCCD detector (1072 by 1040 pixel e2v CCD201) used in the summer 2009 observing run at a frame rate of $\sim$21Hz, i.e. 26MHz pixel rate.

\section{Motivation and comparison with previous work on EMCCD calibration}
\label{sec:previous_emccd_cals}
\cite{Law2007} introduced some methods for calibration and reduction of lucky imaging data taken with EMCCDs, addressing the need for re-calibration of the column-to-column bias pattern with each run, and a method for estimating the electron multiplication gain by fitting a straight line to the pixel histogram. While reasonably effective, the accuracy of the algorithms used was not investigated. 

Further work in this area was needed for a number of reasons. Firstly, the most recent camera uses a different model of EMCCD, with slightly different noise and uniformity characteristics, which took some work to determine and correct. Secondly, lucky imaging observations have previously been largely focused upon binary detection. Only relative photometry was performed, usually in a small region of the detector. The larger format EMCCDs and the multi-EMCCD mosaic camera now in use allow application of lucky imaging to a wider range of targets, but require more precise calibration of detector noise and  non-uniformities for accurate relative photometry across the CCD. Furthermore, inter-CCD relative photometry requires accurate determination of the EM gain for each CCD. Finally, for faint objects we can further increase photometric accuracy using pixel value thresholding techniques, but determination of the optimal threshold level requires a detailed knowledge of the CCD noise characteristics (see chapter~\
\ref{chap:thresholding}).

In summary, to achieve better accuracy in the reduced images, I have:
\begin{itemize}
 \item Refined the bias determination algorithms, and tailored them to the current CCD model (Section~\ref{sec:bias_pedestal}).
 \item Introduced calibration steps for internally generated signal (Section~\ref{sec:internal_signal}).
 \item Tested and refined the EM gain calibration algorithms, and developed algorithms to further calibrate detector effects such as EM register clock induced charge (sections~\ref{sec:internal_signal}, \ref{sec:gain_cal}).
\end{itemize}


\section{Probability distribution models for EMCCDs}
\label{sec:pixel_PDFs}
Conventional CCDs are typically calibrated using a handful of frames. In the case of the lucky imaging the stochastic signal multiplication of the EMCCDs introduces an extra source of variance, and the fast frame rate produces many frames in a short time, so it makes sense to analyse many thousands of frames. This enables us to sensibly make use not only of the average or median pixel values, but to go further and model the pixel value distributions, varying model parameters to fit the full histograms of pixel values obtained. First, we need a model.

A simple model of the EMCCD consists of two stages, the first representing the stochastic EM gain process. 

Before proceeding, we must deal with a notation issue. \cite{Tulloch2011} introduce the notation of using $g_A$ to denote the ``avalanche multiplication gain'' or ``EM gain,'' the dimensionless average multiplication factor between electrons input to the serial register, and electrons output. They use the subscript to distinguish this from the conventional ``\emph{system} gain,'' denoted $g_s$, as measured with conventional CCDs, where system gain refers to the number of photoelectrons represented by 1 ADU, and so has units of photoelectrons / ADU. They also introduce the term $g_{s0}$ to denote the actual ratio between electrons at the input to the readout electronics, and ADU output. The same convention is followed here.

\begin{figure}[htp]
\begin{center}
 \includegraphics[width=1\textwidth]{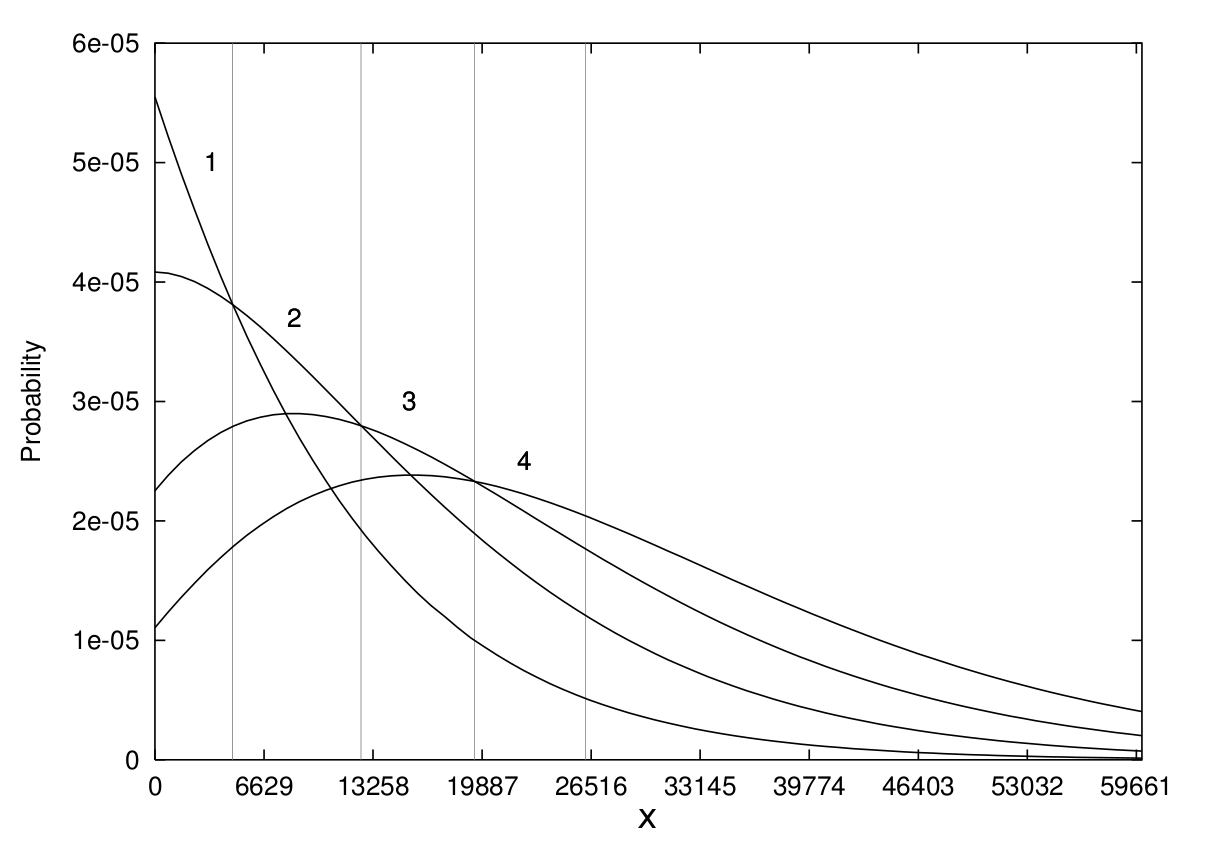}
\caption[EMCCD serial register PDF curves.]{Probability distribution function curves for the number of electrons output, x, from an EMCCD serial register, given 1, 2, 3 or 4 input photo-electrons \corr{(denoted $n$ in  equation}~\ref{eq:serial_register_pdf}), \corr{for an avalanche gain $g_a=6629$}. Vertical lines denote the output thresholds at which different numbers of input photo-electrons are most likely. Due to the overlap of the output PDFs, the number of input photo-electrons can never be certain for any given sample. Reproduced from \cite{Basden2003}.
}
\label{fig:emccd_serial_register_PDFs}
\end{center}
\end{figure}

For relatively high gain and low photon flux, a histogram of the number of electrons output from the serial multiplication register, x, for n photo-electrons input, over many samples, is well modelled by the probability distribution (PDF):
\begin{equation}
p(x | n) = \frac{x^{n-1}e^{-x/g_A}}{(g_A)^n(n-1)!}
\label{eq:serial_register_pdf} 
\end{equation}
which has a mean of $ng_A$ and a variance of $ng_A^2$ \citep{Matsuo1985,Basden2003}. At high photon flux levels the register output probability distribution function is better modelled by a Gaussian of the same mean and standard deviation. Figure~\ref{fig:emccd_serial_register_PDFs} illustrates the resulting PDF curves. 

For a single photo-electron input, the EM register output distribution simplifies to an exponential decay curve. At low light levels where the chance of 2 or more photo-electrons being captured by a single pixel is negligible, we may consider two distribution components, representing pixels which have captured 0 or 1 photo-electrons. The second stage of this model is simply addition of Gaussian readout noise. Under the assumption that no photo-electrons captured results in zero charge at readout, then we may model the EMCCD output distribution as follows. If we denote the serial register output probability distribution for a single photo-electron input as $f(x)$ with:
\begin{equation}
f(x) = \left\{
 \begin{array}{rl}
   0 & \text{if } x < 1\\
   \dfrac{e^{-x/g_a}}{g_a} & \text{if } x \geq 1
 \end{array} \right.
\end{equation}
 then we would expect a histogram of pixel values to approximately conform to the following distribution where $\phi$ is the Gaussian distribution, $r$ is the standard deviation of the readout values, $b$ is the bias pedestal, and $*$ represents convolution:

\begin{equation}
\label{eq:low_flux_simple_PDF}
 P(x) \quad = \quad  P(N_{phot}=0) \times \frac{1}{r}\phi \left( \tfrac{x-b}{r}\right) \qquad  + \qquad P(N_{phot}=1)\times \left[ f(x-b)*  \frac{1}{r}\phi \left(\tfrac{x-b}{r}\right) \right]
\end{equation}

---that is, a sum of a Gaussian readout noise distribution representing pixels which have captured no photo-electrons, and an exponential distribution due to single photo-electron events, likewise convolved with the Gaussian readout noise.

\subsection{Clock induced charge}
\label{sec:CIC}
\begin{figure}[htp]
\begin{center}
 \includegraphics[width=1\textwidth]{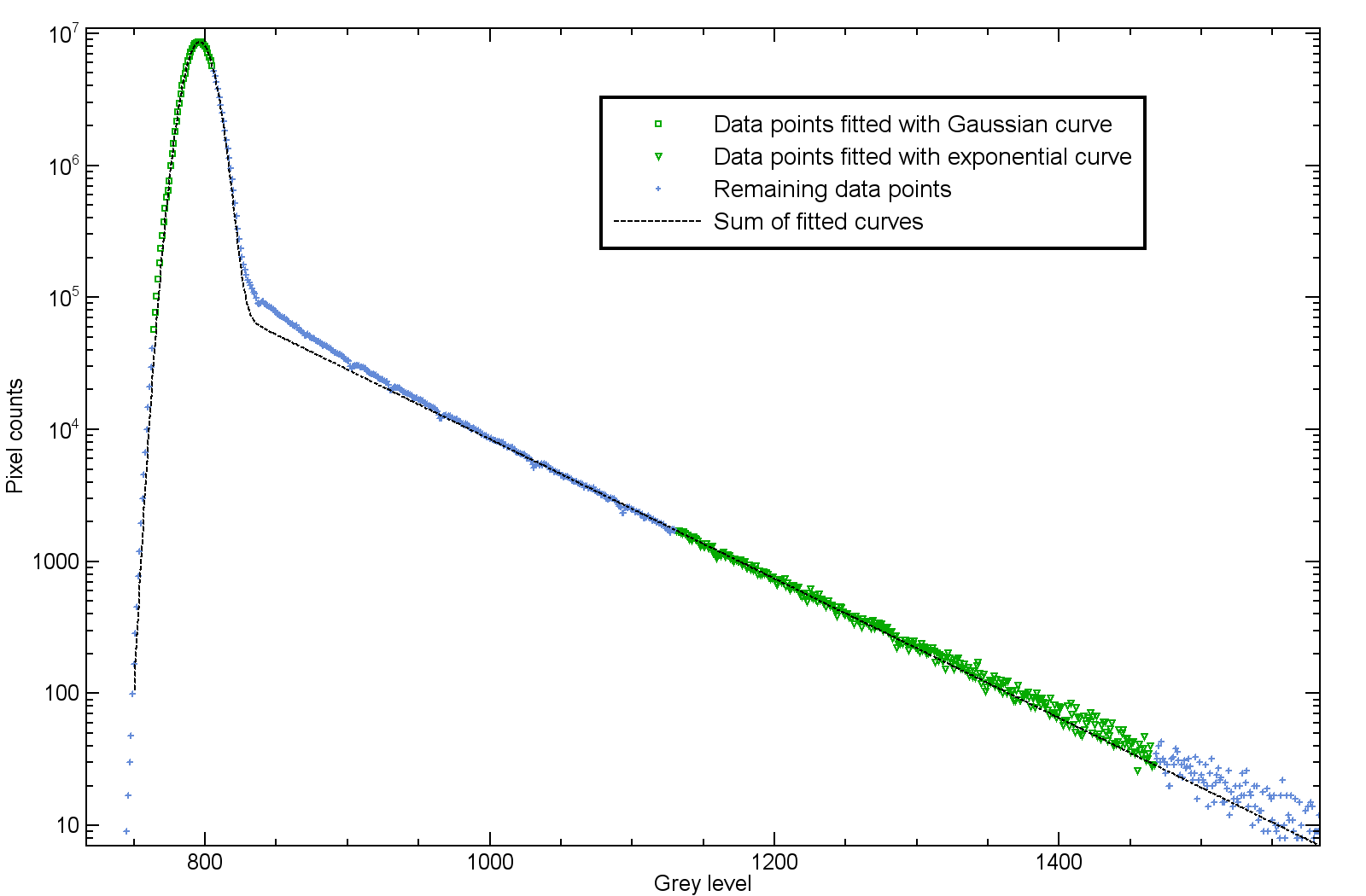}
\caption[Calibration Frames Histogram]{A histogram of pixel values in 3000 \corr{bias} frames (with a low signal due to internally generated signal) taken with the Cambridge CCD. Square and triangle points denote data selected by the algorithm for parameter fitting (described in text). Dashed line denotes sum of fitted models. Note excess around the knee of the curve due to serial register CIC electrons as described in Section~\ref{sec:internal_signal}.}
\label{fig:emccd_histogram}
\end{center}
\end{figure}

A histogram obtained from real EMCCD data without incident light, but with a small signal level due to internally generated signal (described in Section~\ref{sec:internal_signal}),  is displayed in Figure~\ref{fig:emccd_histogram}. For most of the pixel range the data is well fitted by the model of equation~\ref{eq:low_flux_simple_PDF}, the exception being around the knee of the curve where the data lies above the model. This excess of low pixel values is due to an effect not included in our simple model --- clock induced charge. Clock induced charge (CIC) is an important source of noise, consisting of internally generated electrons released by clock transitions during the CCD readout process. 

\begin{figure}[htp]
\begin{center}
 \includegraphics[width=0.8\textwidth]{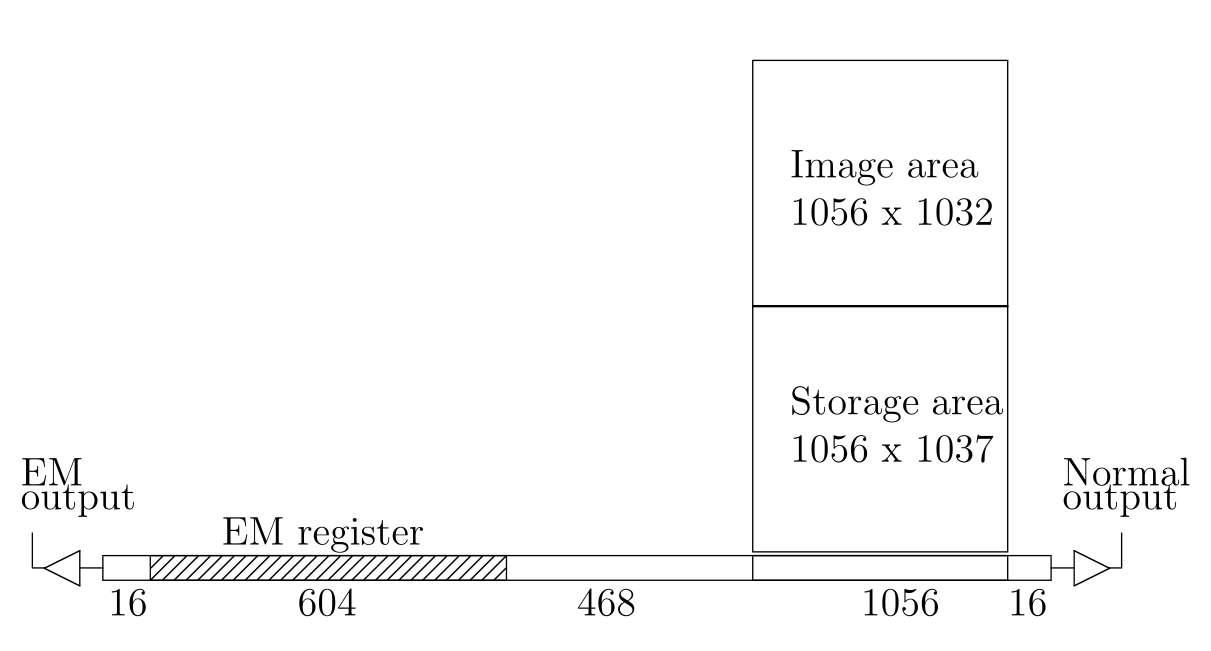}
\caption[CCD201 Schematic]{A schematic diagram of the E2V CCD201 layout, reproduced from \cite{Tulloch2011}. The readout process consists of several stages. First there is a simultaneous vertical transfer of the entire frame into the storage area (``frame transfer''). Next, the storage area is vertically transferred row by row into the serial register, where the pixels are transferred horizontally through the EM register and to the digitization components. There is a small chance that CIC electrons may be generated during any of these transfers. CIC originating prior to the EM register is indistinguishable from photo-electron signal.
}
\label{fig:CCD201_schema}
\end{center}
\end{figure}

\begin{figure}[htp]
\begin{center}
 \includegraphics[width=0.8\textwidth]{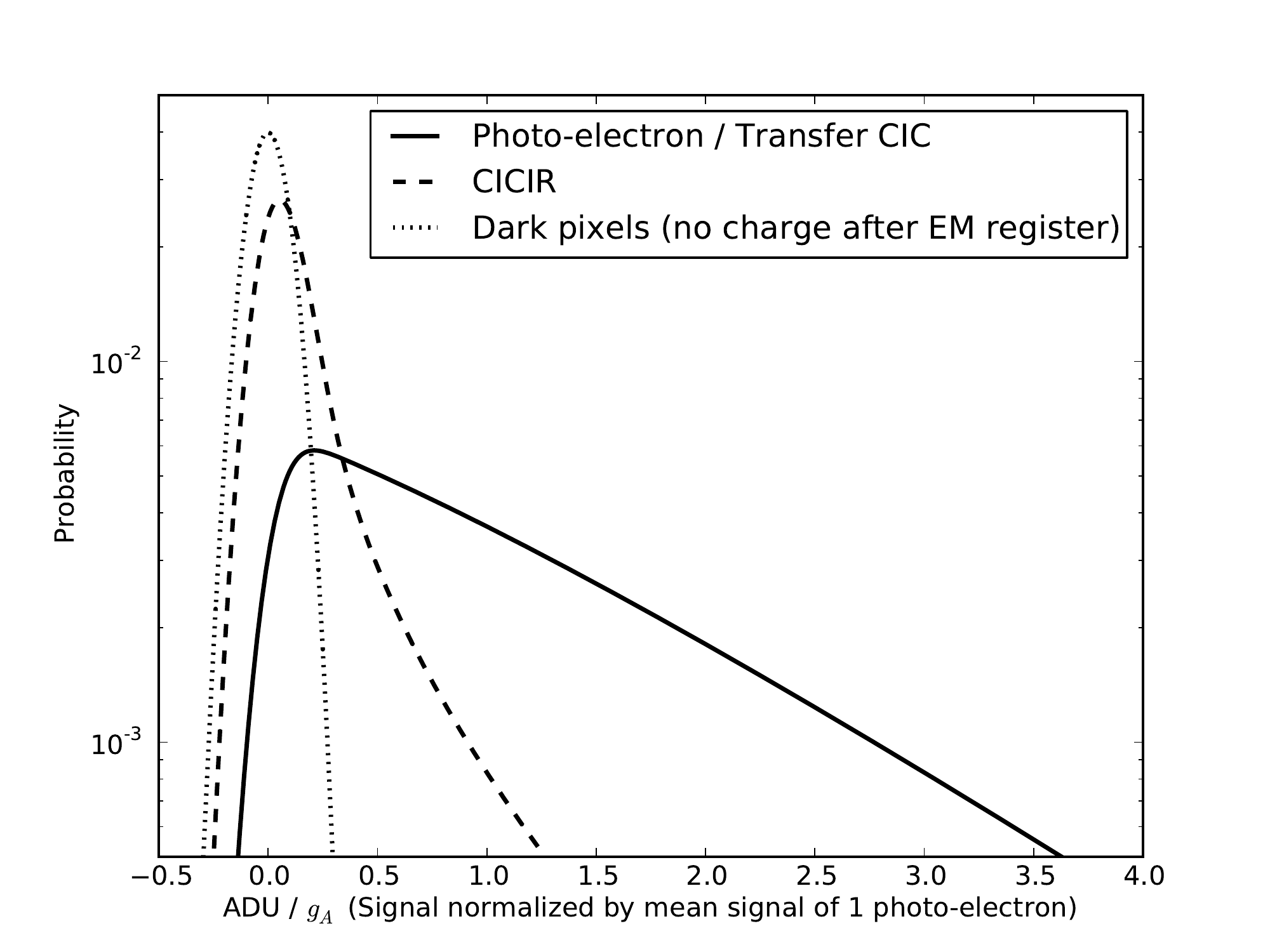}
\caption[EMCCD model PDF components]{An accurate model of the EMCCD output PDF must include a component to represent CIC originating in the EM register (CICIR). This particular plot represents a distribution with a combined signal (light level plus transfer CIC) prior to the EM register of 0.1 photo-electrons per pixel per frame, and CICIR event rate of 0.05 per pixel per frame. All components have been convolved by the Gaussian distribution of the readout noise. The improvement in fit to real data is displayed in Figure~\ref{fig:CICIR_fit}.}
\label{fig:hist_model_components}
\end{center}
\end{figure}

We may consider CIC in two categories. I shall refer to stray electrons freed during the transfer of charge across the CCD prior to the EM register (see Figure~\ref{fig:CCD201_schema}) as ``transfer CIC.'' 
Transfer CIC events are indistinguishable from photo-electron events, since they originate before the EM gain register and so experience the same stochastic gain process. The signal due to transfer CIC should be calibrated using bias frames; and subtracted from on-sky data if a true estimate of the sky background is required. 
CIC generated during transfer through the EM register will on average experience a lower gain, since it passes through fewer amplification stages. I shall refer to these events as CICIR (CIC in register). These account for the poorly fitted region of histogram in Figure~\ref{fig:emccd_histogram}. We may model the CICIR distribution by assuming an equal probability of a CIC event occurring at any stage in the EM register. Then the distribution will be a superposition of many different exponential decay curves, each representing single input electrons passing through different numbers of amplification stages, and therefore undergoing a different mean avalanche gain. The mean signal contribution from one of these events will be a factor of $1/ln(g_A)$ smaller than that of a photo-electron \citep{Tulloch2011}. Figure~\ref{fig:hist_model_components} illustrates the relative contribution of the CICIR component needed to accurately model the EMCCD output data.

\section{Histogram analysis algorithms}
\subsection{Bias determination at low pixel counts for the purpose of single frame bias estimation}
\label{sec:bias_algos}
When reducing EMCCD data, we often need a way to estimate the bias pedestal from a fairly small pixel events sample, perhaps a thousand recorded pixels.%
\footnote{\corr{For example, when using a faintly illuminated region of an on-sky dataset to track bias pedestal drift.}}
We desire such an algorithm to be both extremely robust to noisy data, and also computationally efficient since it must be run at least once per frame of data to deal with bias drift (described in Section~\ref{sec:bias_drift}). A naive approach is to simply use the mode of the histogram, however when applied to raw data composed of integer ADU values this only estimates the bias offset to the nearest integer, and is quite susceptible to noise in the histogram, which may cause the bias estimate to ``jump around,'' varying rapidly and unrealistically. \cite{Law2007} employed the tenth percentile as a metric to track bias drift which does not suffer from this noise induced variation, however not only is the sort computation required moderately time consuming, it is a biased estimator at best, consistently underestimating the bias pedestal with an offset that varies depending on the distribution of the readout noise. 

After some experimentation \corr{with fitting routines}, I settled on a thresholded centroid algorithm as an efficient, robust and accurate estimator of bias pedestal for tracking bias drift. The weighted mean of all histogram bin central values with bin counts above 80\% of the maximum count is used, and this proves very effective.

\subsection{Electron multiplication gain}
\label{sec:gain_cal}
The Cambridge lucky imaging camera incorporates controls for on-the-fly adjustment of the EM serial register voltage, and hence control over the EM gain. This is useful as it allows the user to make adjustments based on the current field of view, \corr{avoiding saturation of brighter sources whilst maintaining the best signal to noise allowed by the dynamic range for faint objects imaged on the same detector}. 

One key advantage of using a synchronised multi-CCD mosaic camera is that the gain may be controlled independently across the CCDs. Careful positioning of the mosaic field of view often makes it possible to observe a bright guide star on one CCD at low EM gain, while observing faint sources nearby at much higher EM gain with a second CCD --- effectively giving extremely wide dynamic range.

The cost of this flexibility is that the gain becomes an extra variable to be carefully calibrated. If we wish to perform inter-CCD relative photometry across a mosaic field of view, then the gain setting of each individual CCD must be accurately measured so that we may normalise the images prior to comparison.

As described in~\cite[pp. 43-46]{Law2007}, the EM gain is most easily calibrated using a histogram of pixel events from a low flux area of a field of view. The alternative method is to switch the EM gain off for comparison using calibrator stars, but this is time consuming and largely unnecessary.
\citeauthor{Law2007} used a fit to the straight line region (on a log plot) of the exponential decay curve of the serial register output histogram for a single input photo-electron (fig.~\ref{fig:emccd_serial_register_PDFs}), with the fitting region determined either manually or based on goodness-of-fit.

To determine the EM gain and other detector characteristics from the pixel histogram, I developed a two stage, fully-automated fitting routine. In the first stage, the bias pedestal, EM gain, and light level are estimated by fitting models to the histogram sections highlighted in Figure~\ref{fig:emccd_histogram}\corr{(the light level is estimated by comparing the relative integrated areas under the Gaussian readout hump and the tail of photo-electron events)}. This initial fit takes less than 0.01 seconds to compute.  Optionally, a fit is then made to the whole histogram using a fully featured distribution model with parameters corresponding to bias pedestal, readout noise, EM gain, light level, and serial register CIC event frequency. The full fit takes around 1.5 seconds to compute.

\begin{figure}[htp]
\centering
\subfigure[EM gain estimation using simple histogram model]{
	\includegraphics[scale=0.6]{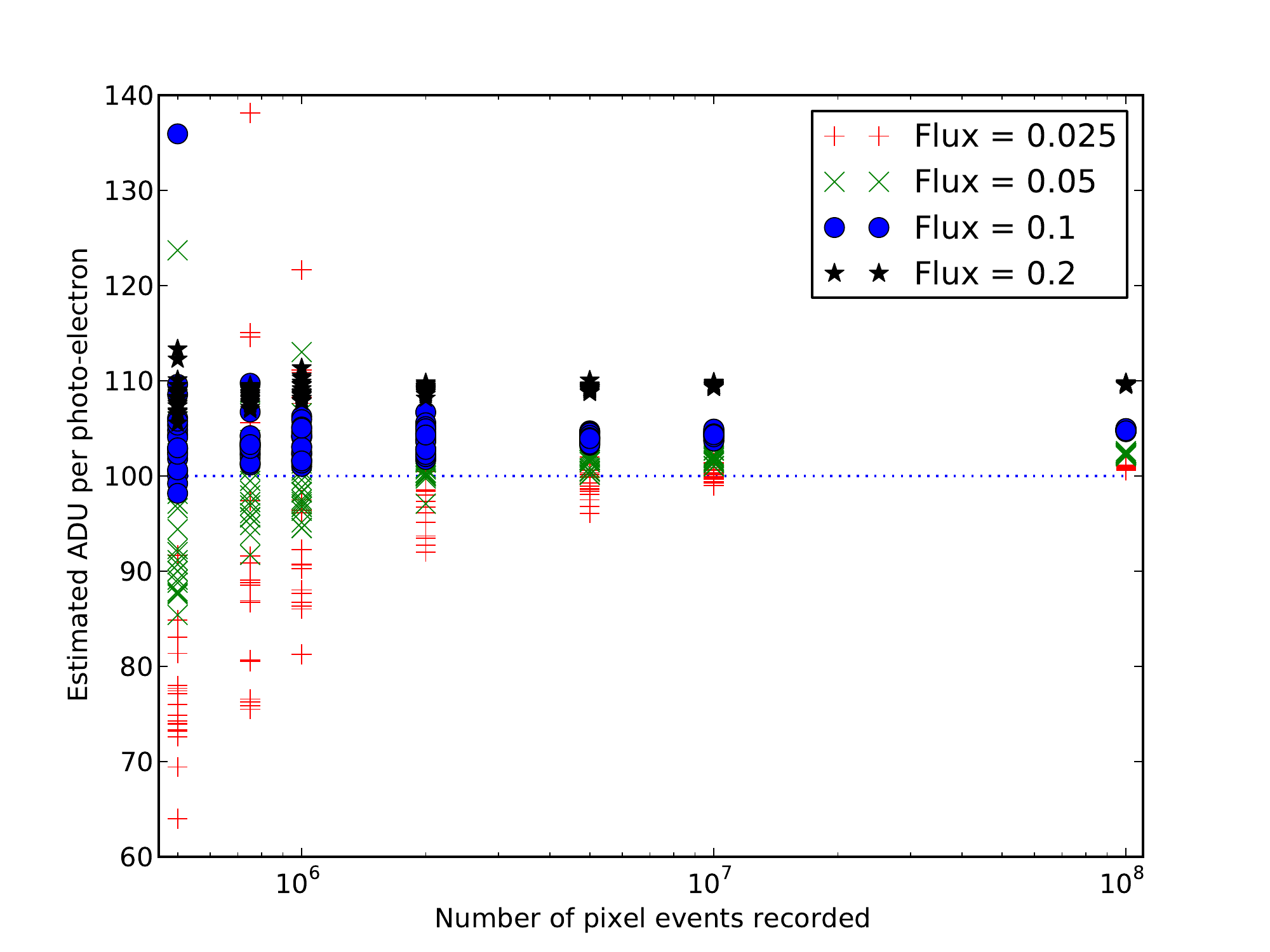}
	\label{subfig:gain_convergence_init}
}
\\
\subfigure[EM gain estimation using full multi-component histogram model]{
	\includegraphics[scale=0.6]{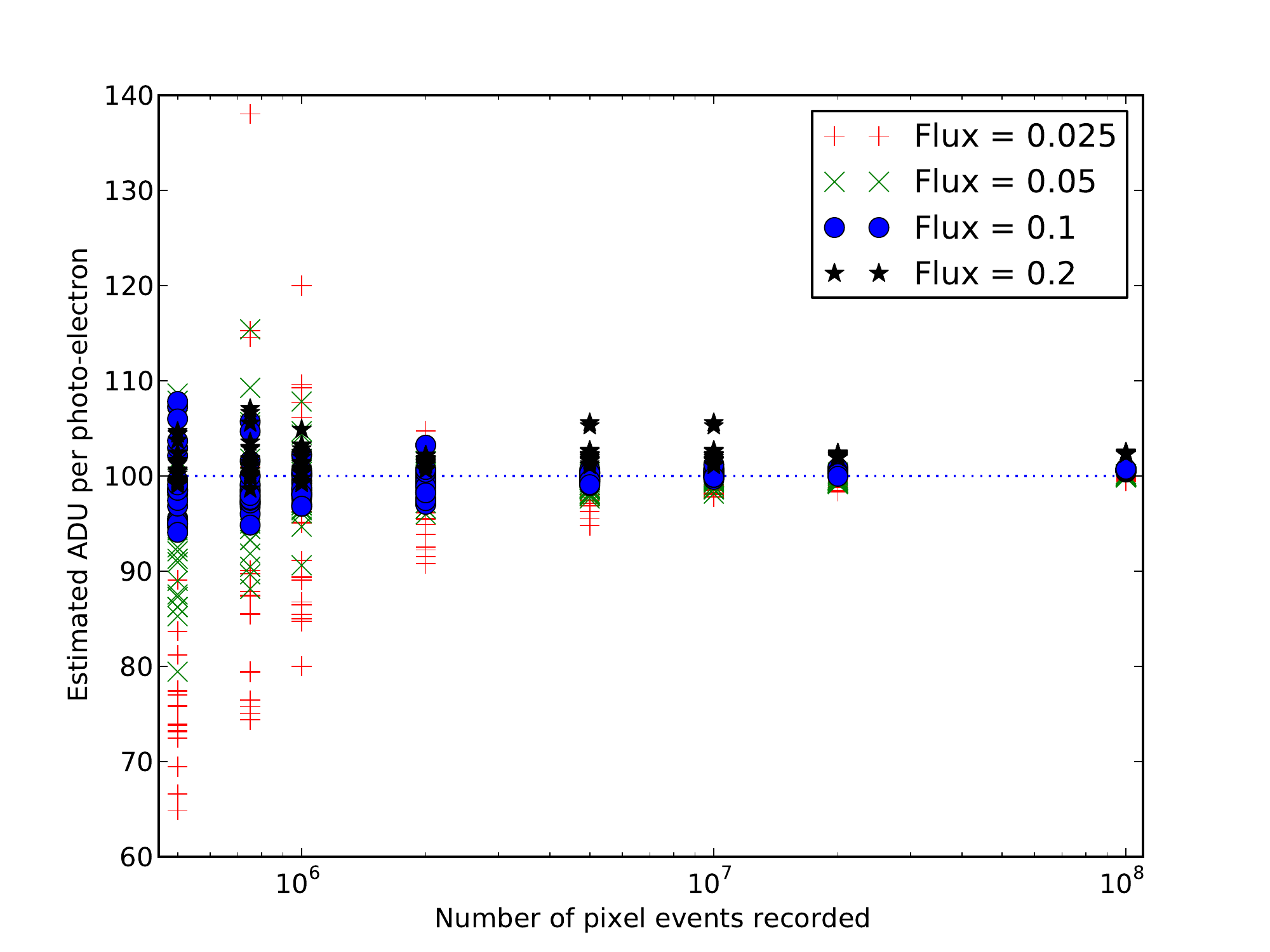}
	 \label{subfig:gain_convergence_fixed}
}

\caption[Testing EM gain estimation algorithms]{Testing electron multiplication gain estimation algorithms against simulated data. System gain $g_s=\frac{1}{100}$, so that a single photo-electron produces a typical data value of 100 ADU. Various light levels in photo-electrons per pixel per frame are plotted (denoted ``flux'' in legend for brevity). \subref{subfig:gain_convergence_init}: A simple histogram model which does not account for double photo-electron events results in systematic over-estimation of the EM gain. \subref{subfig:gain_convergence_fixed}: Full model gives better accuracy. Further details in text.}
\label{fig:gain_convergence}
\end{figure}

I undertook some investigations into the algorithm performance, to verify that the estimation algorithm did not introduce any systematic errors, and also determine how the accuracy improved as the histogram converges to the underlying distribution with increasing number of recorded pixel-events. 

I tested the code against simulated data generated with the Monte-Carlo methods described in Section~\ref{sec:EMCCD_sim}. The results are plotted in Figure~\ref{fig:gain_convergence}. While the code produced consistent results, initially a systematic over-estimation of the EM gain was evident, especially at higher light levels. This corresponds to pixel events where 2 photo-electrons are captured, which raises the number of pixel events in the tail of the histogram. Adding an extra component to the model which accounts for the 2 photon pixel events largely corrects this error. Alternatively, if the high speed single stage fitting procedure is preferred, it appears that the gain estimate can be corrected to give reasonably good accuracy simply by computing the relative frequency of 1 and 2 photon pixel events --- for example, if the light level is 0.2 photons per pixel per frame, then the EM gain will be overestimated by a factor of:
\begin{equation}
\frac{P(N_{phot}=2)}{P(N_{phot}=1)} = 0.1;\qquad N_{phot}\sim Poisson(\lambda=0.2)
\end{equation}
The simulations suggest at least $10^7$ recorded pixels are required to give a reasonable estimate of the gain, with larger samples giving better accuracy, though this is partially dependent on the light level. This is of interest for reasonably crowded fields where much of the frame has light levels too high for EM gain estimation, suggesting that if a background region of 100 by 100 pixels may be selected, then 1000 or more frames of data will give a reasonable estimate. 
 
\section{Calibrating the Cambridge e2v CCD201}
\label{sec:luckycam_2009_cals}
\begin{table}
\begin{center}

\begin{tabular}{|c|c|}
\hline 
\hline 
\multicolumn{2}{|c|}{LuckyCam 2009 detector properties} \\
\hline
\hline
 CCD type &  E2V technologies CCD201-20 \\
 Pixel format &  $1024\times1024$ active pixels \\
 Active image area & $13.3\times13.3\,\textrm{mm}$ \\
 Pixel size &  $13\times13\upmu\textrm{m}$ \\
 Frame rate & 21Hz \\
 Read noise & Typically 15 ADU \\
 Internally generated signal &  Equivalent to 0.05 photo-electrons per pixel per frame \\
 CICIR signal &  Equivalent to 0.04 photo-electrons per pixel per frame\\
\hline 
\hline
\end{tabular}
\caption[Properties of the July 2009 lucky imaging detectors]{Properties of detectors used in the the July 2009 lucky imaging camera. 
Full details of the E2V CCD201-20 can be found in the relevant data sheet \citep{e2v2005}. `Internally generated signal' refers to the combined effects of dark current and transfer CIC, see Section~\ref{sec:internal_signal}.
 See Section~\ref{sec:CIC} for an explanation of transfer CIC and CICIR.}
\label{tab:luckycam_props}
\end{center}
\end{table}

While the previous sections describe general techniques applicable to EMCCD data, the rest of this chapter details the specific characteristics of the detectors used in the \corr{July 2009 observing run at the Nordic Optical Telescope}, and the techniques used to calibrate these detectors as accurately as possible. A summary of the detector properties is provided in table~\ref{tab:luckycam_props}.

\subsection{Spatial gradient in bias pedestal}
\label{sec:bias_pedestal}

\begin{figure}[htp]
\centering
\subfigure[2D map of bias pedestal]{
	\includegraphics[width=0.3\textwidth]{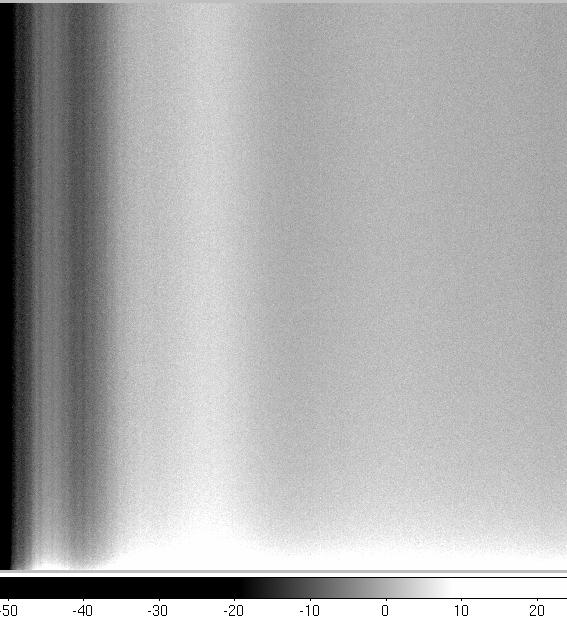}
	\label{subfig:per_pixel_bias}
}
\subfigure[Column based estimation]{
	\includegraphics[width=0.3\textwidth]{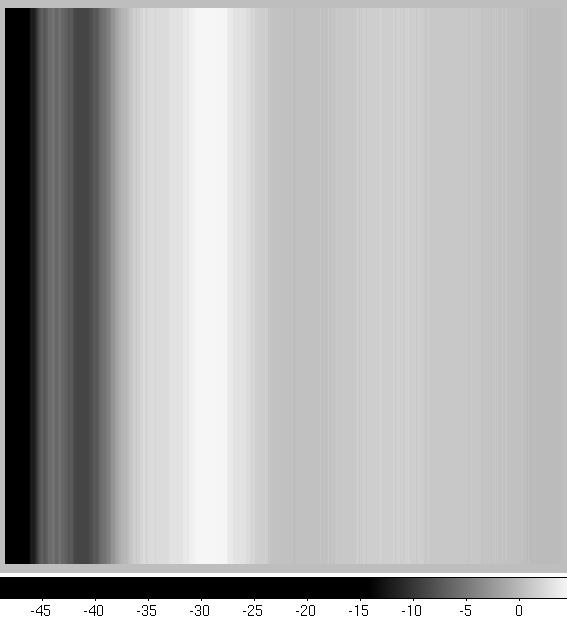}
	 \label{subfig:col_bias}
}
\subfigure[Row based estimation]{
	\includegraphics[width=0.3\textwidth]{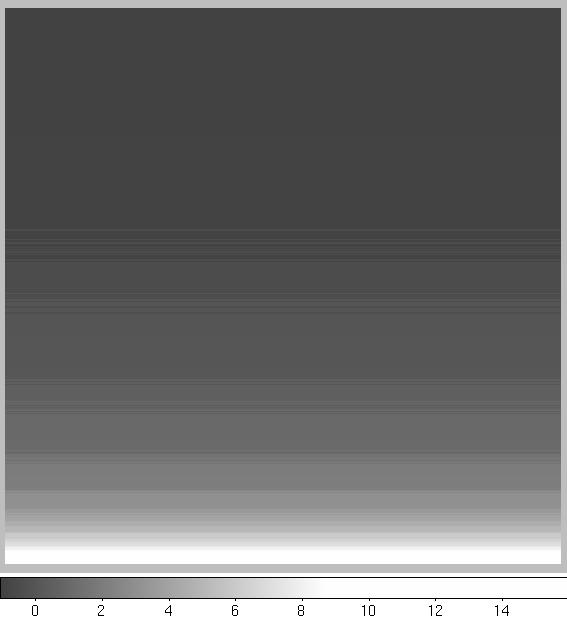}
	 \label{subfig:row_bias}
}

\caption[Calibrating the bias pedestal gradient]{The bias pedestal has a spatial gradient, which can be measured by estimating the bias pedestal for every pixel from a set of bias frames, \subref{subfig:per_pixel_bias}. Alternatively, the vertical and horizontal gradients may be measured separately, \subref{subfig:col_bias}, \subref{subfig:row_bias}. See text for details.}
\label{fig:bias_gradient}
\end{figure}

\begin{figure}[htp]
\begin{center}
 \includegraphics[width=0.6\textwidth]{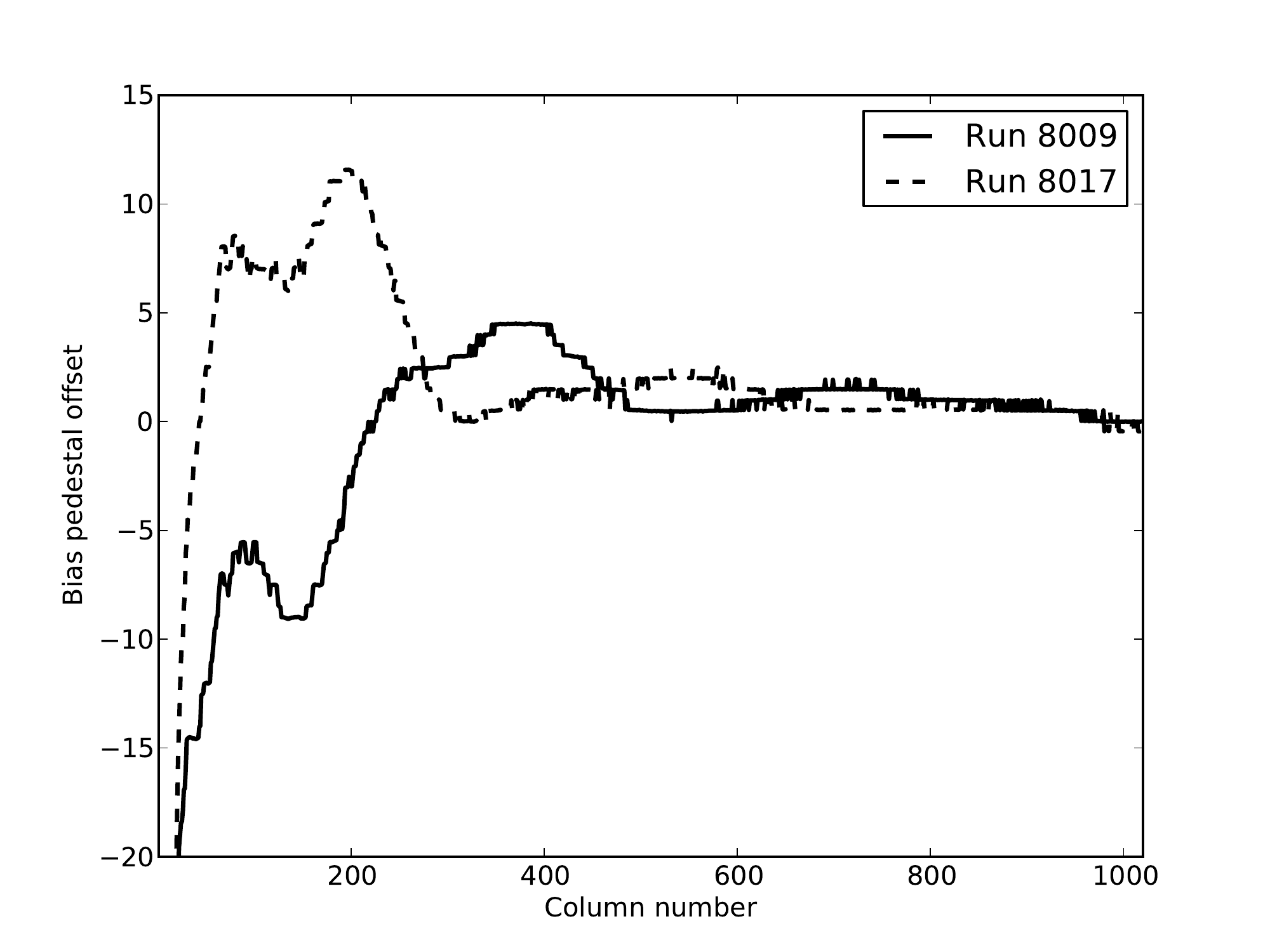}
\caption[Variation in horizontal bias gradient]{The horizontal (column to column) bias pedestal gradient varies between runs as the gain settings are altered --- run 8017 was taken with EM gain set to twice that of run 8009. Significance of the bias pedestal variation in terms of equivalent photo-electron signal will vary with gain, but typically 10 ADU is the equivalent signal level to 0.1 photo-electrons.}
\label{fig:bias_grad_variation}
\end{center}
\end{figure}


Conventionally, the bias pedestal in CCD images is estimated using a handful of bias frames. However, due to the high sensitivity of the EMCCD to \corr{low levels of internally generated signal such as CIC}, the bias pedestal must instead be estimated using the histogram analysis techniques described in Section~\ref{sec:bias_algos}. Whilst the bias estimation algorithm is insensitive to modest illumination levels, the best estimate will be made from a series of \corr{exposures taken with the camera shutter closed.} 

Creating a full two dimensional bias \corr{map} from a series of \corr{bias} frames requires recording a histogram for every pixel, for several thousand frames. I implemented routines to perform this analysis. Unfortunately, the bias map varies from run to run, most likely varying as the EM gain is adjusted (Figure~\ref{fig:bias_grad_variation}). As a result, a full 2d bias calibration can only be achieved by recording bias frames for every EM gain setting to be used. Alternatively, the variation across the bias map may be largely broken up into horizontal and vertical components (Figure~\ref{fig:bias_gradient}). While the horizontal (column to column) gradient varies with EM gain setting, the vertical (row to row) component appears stable. This allows us to gather column histograms for each \emph{on-sky} observation, which in most fields will largely contain pixels at low light levels, meaning we can determine the horizontal bias gradient component for each observation, then add it to the vertical component 
calibrated using a single set of \corr{closed shutter} bias frames.
\footnote{Note: when observing say, globular cluster fields, saturation of the column histograms by photon-events can and does become an issue. Signal from bright stars will dominate over bias pedestal variation, so if accurate photometry and PSF fitting is required then the observer should revert to taking \corr{closed shutter} bias frames between observations periodically. It may be possible to use pre-set levels of EM gain and build up a library of bias maps, \emph{if} the column to column bias pattern is not strongly dependent on other factors such as temperature.} 
%


\subsection{Bias pedestal drift}
\label{sec:bias_drift}
\begin{figure}[htp]
\begin{center}
 \includegraphics[width=0.5\textwidth]{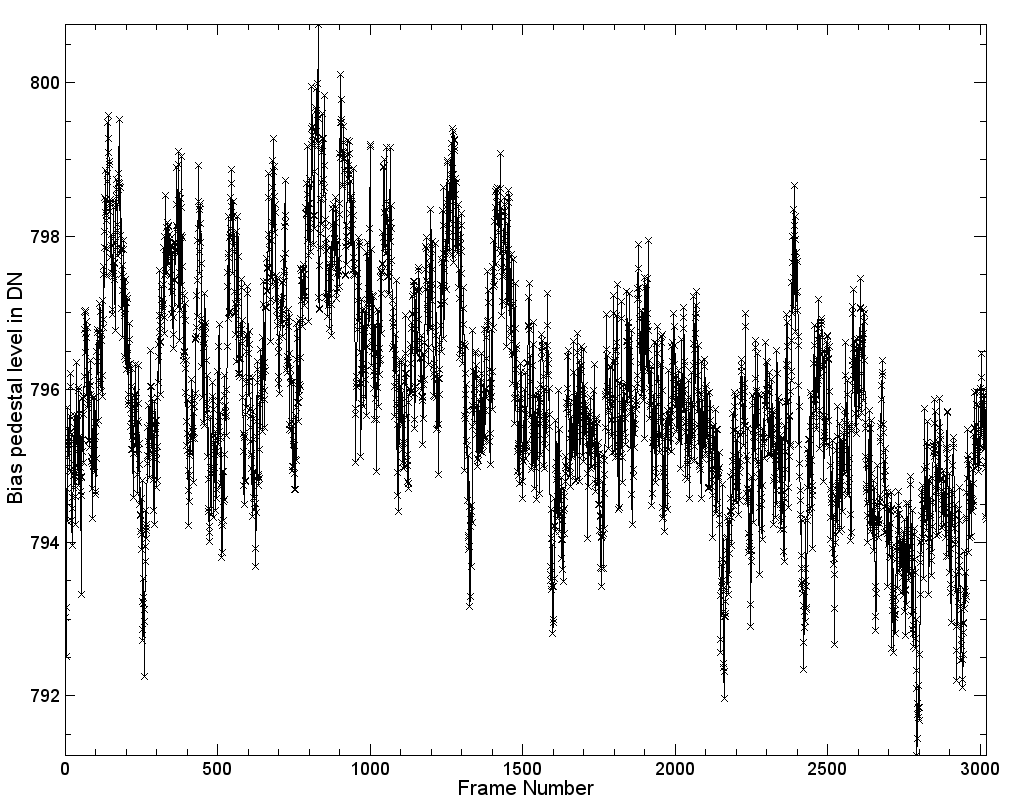}
\caption[Bias drift]{The bias pedestal also varies from frame to frame, termed `drift'.}
\label{fig:bias_drift}
\end{center}
\end{figure}
The bias pedestal also varies temporally, i.e. on a \emph{frame by frame} basis --- generally this is known as bias drift. The drift range is typically of the same order as the readout \corr{noise} (fig.~\ref{fig:bias_drift}), and so it is desirable to correct for this. Fortunately the spatial gradient appears stable on these timescales, with drift affecting the bias pedestal uniformly across the frame. Therefore it is possible to correct for the bias drift by analysing pixel histograms gathered from each frame, as long as a faintly illuminated region may be selected.

\subsection{Internally generated signal levels}
\label{sec:internal_signal}
\begin{figure}[htp]
\begin{center}
 \includegraphics[width=0.5\textwidth]{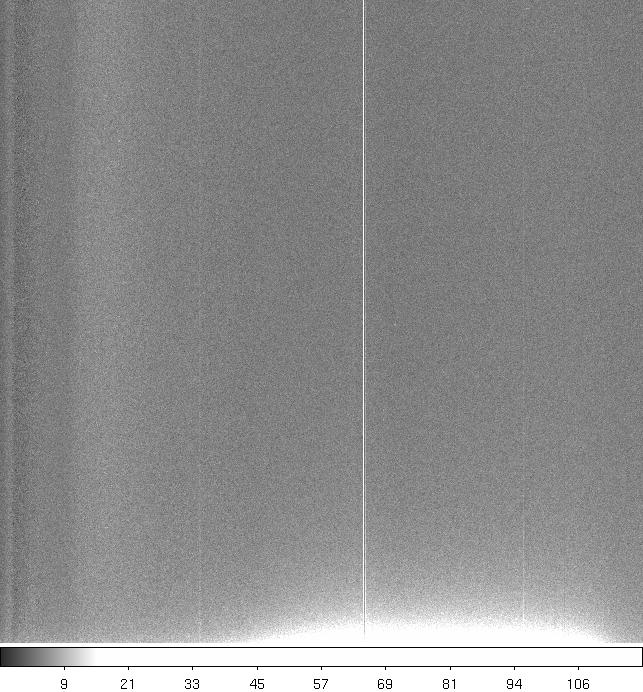}
\caption[Average of bias frames showing internally generated signal]{A \emph{mean} image composed from many bias frames, after subtraction of the bias \corr{map (which was generated via histogram methods)}. See text for details.}
\label{fig:dark_frame_avg}
\end{center}
\end{figure}

When processing bias frames, a significant and non-uniform signal is evident after subtraction of the (histogram calibrated) bias map, illustrated in Figure~\ref{fig:dark_frame_avg}. There are several possible sources of this residual signal:
\begin{itemize}
\item Dark current, electrons randomly generated due to thermal emissions
\item Clock induced charge as described in Section~\ref{sec:CIC}. 
\item Light leakage through the camera baffles.
\item Amplifier glow as described in \cite{Tulloch2004}.
\end{itemize}
Amplifier glow seems unlikely, as the resultant signal would be localised in one corner of the detector, near to the amplifier component location. 
Uniform light leakage would not be expected at the levels observed, leaving dark current and CIC effects.

While CICIR levels may be estimated by histogram fitting, the only way to distinguish transfer CIC contributions from dark current is to obtain closed shutter calibration frames with different exposure times --- dark current signal levels should increase with exposure time, while CIC levels remain constant. Hence the relative contributions may be inferred. Closed shutter datasets of differing exposure time were not available for calibrating the summer 2009 data. Interestingly, the internally generated signal level increases towards the bottom of the frame, suggesting some common causation factor along with the rise in bias pedestal. At any rate, the source of the internal signal is unimportant with regards to general data reduction as the resulting data reduction process is the same whatever the source \corr{--- we simply wish to subtract any signal not due to photons from the sky}.

\subsection{Clock induced charge levels}
\label{sec:serial_CIC_cal}
\begin{figure}[htp]
\begin{center}
 \includegraphics[width=0.8\textwidth]{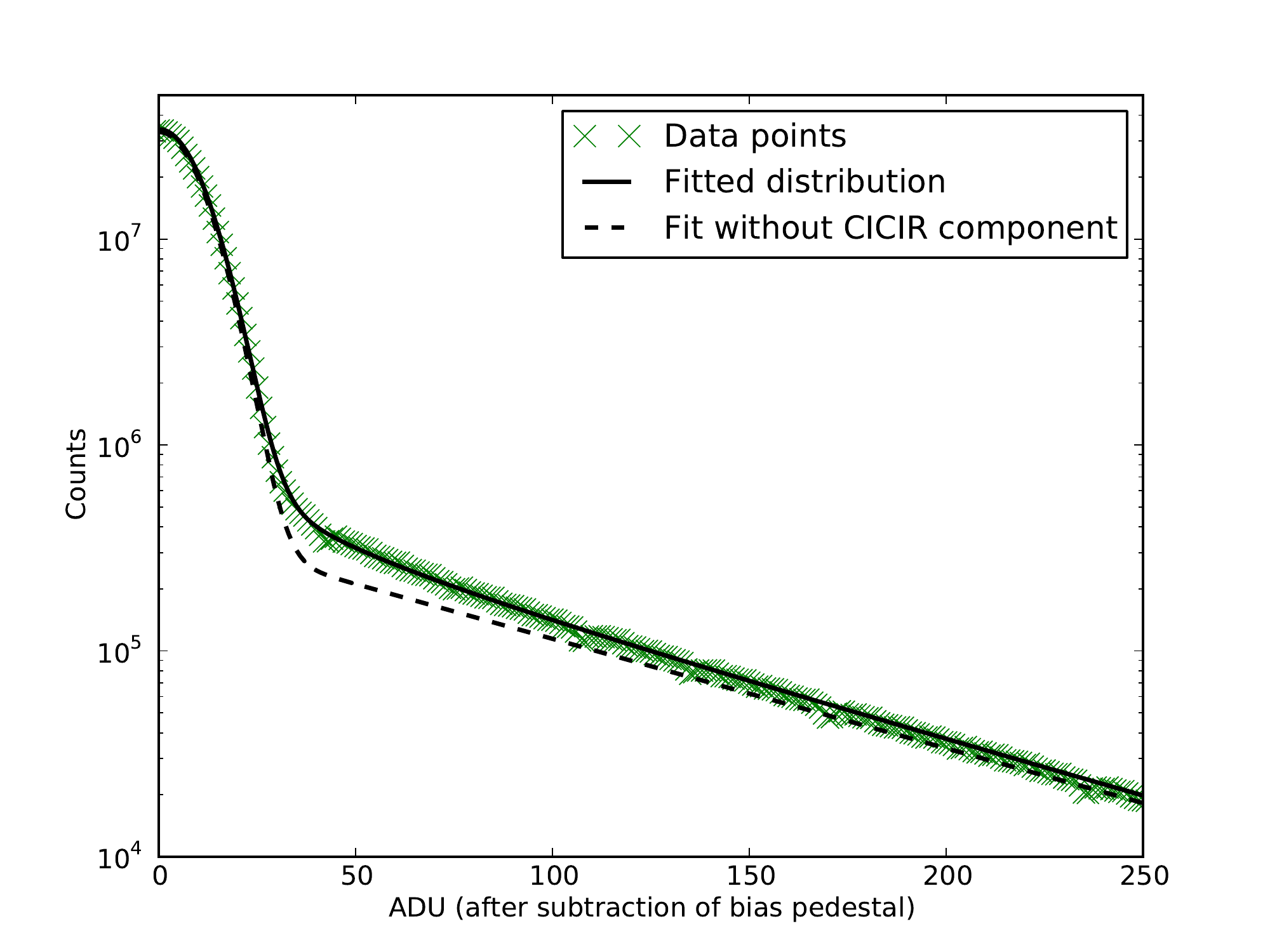}
\caption[Fitting the CICIR component]{Adding a CICIR component to the EMCCD output distribution model allows estimation of CICIR levels and provides a better fit to real data.}
\label{fig:CICIR_fit}
\end{center}
\end{figure}

Transfer CIC is indistinguishable from photo-electron signal, and so is calibrated as part of the internally generated signal (Section~\ref{sec:internal_signal}). However, using the multi-component model illustrated in Figure~\ref{fig:hist_model_components}, CICIR levels can be estimated. This is important not only for characterising and fine tuning EMCCDs, but also when determining thresholds for low light levels (cf. chapter~\ref{chap:thresholding}). Figure~\ref{fig:CICIR_fit} illustrates just such a fit to real data. 

\subsection{Flat fielding and gain uniformity}
Careful flat fielding of CCDs is essential for accurate photometry. \cite{Law2007} reported variations \corr{of up to 10\%} in EM gain across frames from the 2005 camera. To work around this variation the recommended method was to perform calibration with histograms from the same columns as any photometry target objects. To determine whether photometry across many objects in wide fields of view is feasible I investigated if this variation was still present in the 2009 camera.

\begin{figure}[htp]
\begin{center}
 \includegraphics[width=0.8\textwidth]{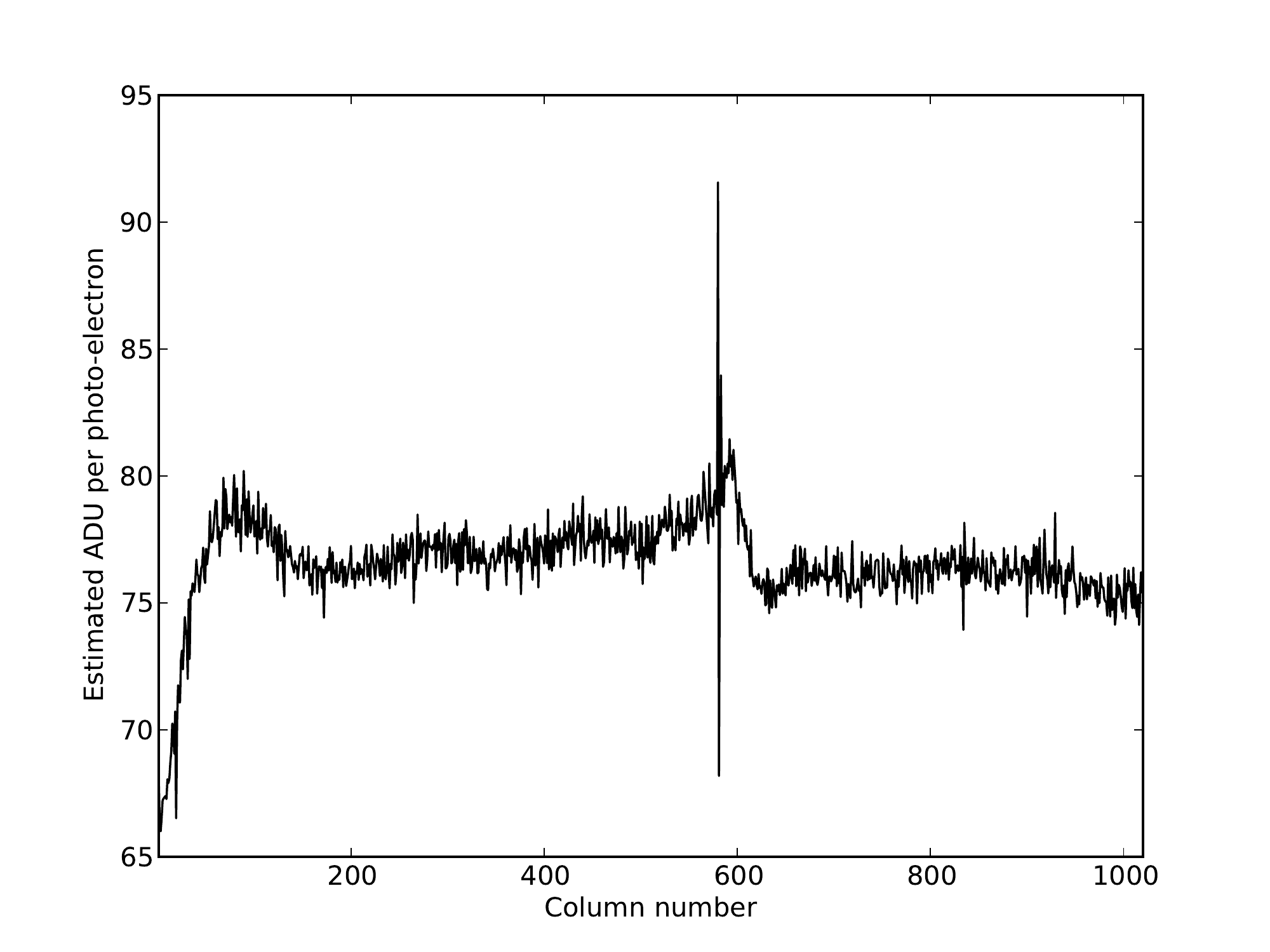}
\caption[EM gain variation]{There is a small variation in EM gain across the detector. See text for details.}
\label{fig:gain_variation}
\end{center}
\end{figure}

Investigation of inter-pixel gain variations is difficult, since the estimates of EM gain are inaccurate when using small pixel samples (see Figure~\ref{fig:gain_convergence}). Inspection of the \corr{mean internal signal levels in closed shutter calibration frames after subtraction of the bias map} suggested there may be some horizontal (column to column) gradient in the EM gain. I was able to implement a per-column estimation of EM gain levels by analysing the pixel histograms. Figure~\ref{fig:gain_variation} shows an example plot of estimated EM gain across the detector. The gain appears appears uniform to around the 5\% level over most of the detector (excepting some spikes due to bad columns which are rejected in reduced images), although there was greater deviation in columns 1 through 200. As a result, reasonably accurate photometry can only be extracted from the region right of column 200 in the 2009 datasets. If such variation remains in future detector configurations it should be possible to 
incorporate a normalizing `gain map' to correct for such effects, but calibration requires long series of \corr{closed shutter (or very weakly illuminated)} calibration frames. Such long period calibration datasets were not available for the 2009 data.

Flat fielding for other non-uniformities such as relative pixel area and lens obscuration is a relatively simple process of taking twilight sky flats, as with a conventional CCD, and provision for such flat fields is made in the reduction pipeline. \corr{The variations in both EM gain and pixel sensitivity can be disentangled in two ways. Firstly, the EM gain can be switched off and the conventional readout mode utilised when taking flat fields, to calibrate the pixel sensitivity alone. Alternatively the EM gain can be estimated first from sky flats using histogram analysis methods, then taken into account when estimating pixel sensitivity.}

\subsection{Bad pixels and pixel weighting}
\label{sec:bad_pixels}
It is common for CCDs to exhibit a few `bad pixels,' which consistently produce high or low values without much response to actual light levels. In the 2009 camera, each CCD contained at least a few columns of such pixels, easily identifiable by eye (e.g. the bright column in Figure~\ref{fig:dark_frame_avg}). The locations of such columns are stored in a configuration file for the pipeline, which then zeroes the pixels in average images, and also zeroes their allocated weight for the drizzle algorithm (see chapter~\ref{chap:data_reduction}). 

\section{Results}
\begin{figure}[htp]
\centering
\subfigure[Uncalibrated average image]{
	\includegraphics[width=0.3\textwidth]{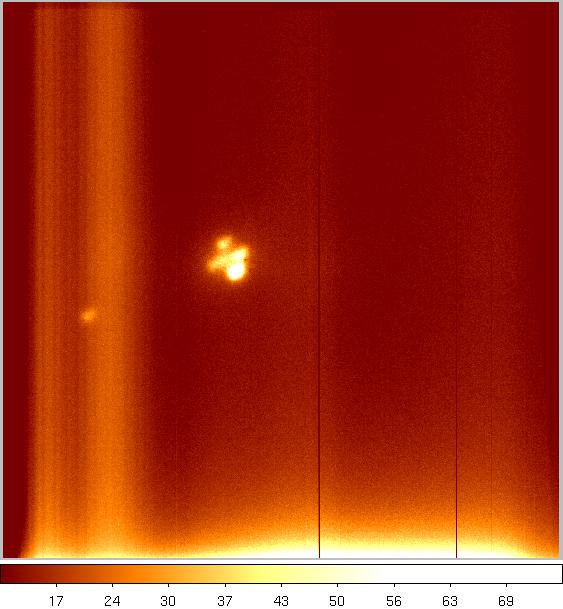}
	\label{subfig:EC_uncal}
}
\subfigure[Debiased average image]{
	\includegraphics[width=0.3\textwidth]{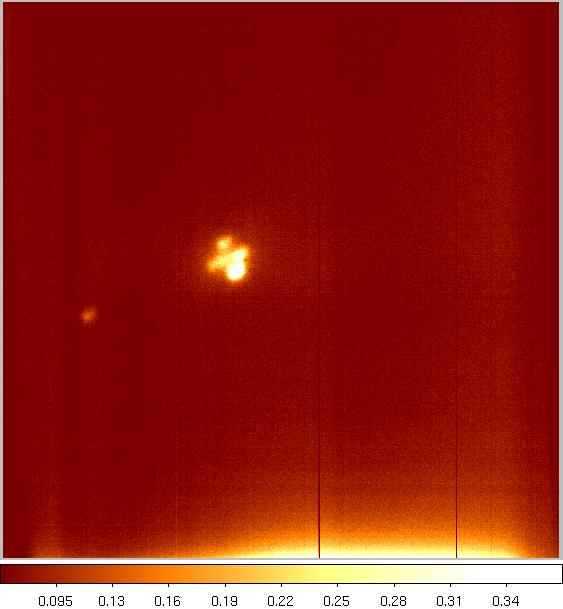}
	 \label{subfig:EC_db}
}
\subfigure[Fully calibrated and drizzled image]{
	\includegraphics[width=0.3\textwidth]{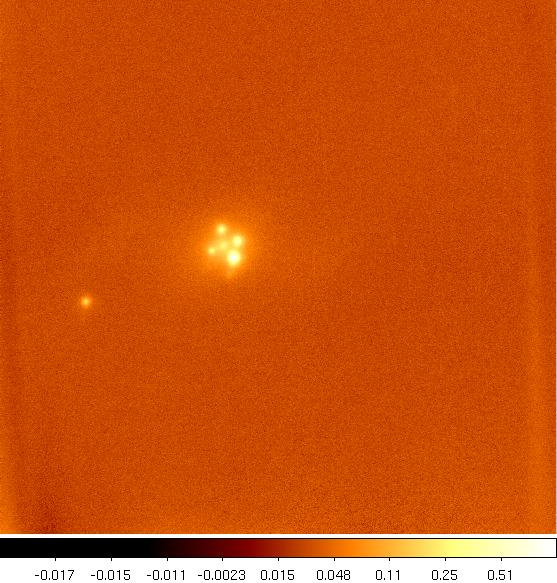}
	 \label{subfig:EC_driz}
}

\caption[Improvement in image quality due to calibration]{Improvement in image quality due to calibration.}
\label{fig:EC_cal_example}
\end{figure}
Figure~\ref{fig:EC_cal_example} illustrates the improvement in background uniformity due to accurate calibration. Bias pedestal variation typically has a range of levels equivalent to a signal level of 0.1 photo-electrons, evident in \ref{subfig:EC_uncal}. Dark current affects the bottom of the frame worst, where the additional signal reaches 0.2 photo-electrons, evident in \ref{subfig:EC_db}. Bad pixel columns are also noticeable (their pixel values have been zeroed). Figure~\ref{subfig:EC_driz} shows a fully calibrated, drizzled image. The RMS background variation over most of the image is of the order 0.005 photo-electrons, though some slight non-uniformities are still present near the edges at the level of 0.02 photo-electrons.

\chapter{Lucky imaging of faint sources: Thresholding techniques}
\label{chap:thresholding}

The stochastic nature of the electron multiplication gain in EMCCDs introduces an extra source of variance, so that compared to the ideal case of purely photon shot noise an EMCCD at high light levels will experience signal variation larger by a factor of $\sqrt{2}$ - this is termed the ``excess noise factor'' \citep{Basden2003}. At low light levels (where the mean photon flux per pixel per frame is much less than 1) we can be reasonably confident that any given pixel only receives 0 or 1 photo-electrons per exposure. In this case we can exploit the fact that, at high electron multiplication gain, most photon events result in ADU values many times the readout noise RMS above the background, and reduce the excess noise factor by employing a pixel thresholding scheme. 

The concepts behind this thresholding approach rely heavily upon the models developed in chapter~\ref{chap:EMCCD_calibration}, likewise the details of implementation rely upon careful calibration, and so the chapters are closely linked. \corr{Where the text would otherwise be unclear, I shall refer to EMCCD observations which do not employ thresholding as `linear mode' observations.}

\section{Summary and comparison with literature}
\cite{Basden2003} and \cite{Lantz2008} have explored the possibility of applying thresholding schemes across a range of light levels via either multiple thresholds or Bayesian estimators. However there are a number of practical issues with these schemes. 

Firstly, the scheme of \citeauthor{Basden2003} does not account for CIC effects, which would need to be taken into account and thresholds changed accordingly if a similar scheme were to be applied. The scheme of \cite{Lantz2008} requires the mean light level on every pixel to be constant so that an optimum Bayesian estimator can be chosen. In lucky imaging this is not the case as the atmospheric tip-tilt shifts the sources around the CCD from frame to frame. The suggested work-around is thresholding on frames which have already been recentred using the guide star. This is not entirely infeasible but still runs into the problems of guide star registration error and anisoplanatic effects --- so it seems likely that the technique could only be practically applied close to a bright guide star, and would require substantial computational effort.

Accordingly, I have focused upon developing and implementing photon counting methods which employ a single threshold to process data at low light levels. Thresholding is at any rate most needed at low light levels where the detector noise would otherwise dominate over the photon shot and stochastic multiplication noise. 

Photon counting is very simple in concept --- if a pixel has a value above a threshold it is recorded as a 1, below the threshold it is recorded as 0 --- but obtaining the best possible signal to noise and maintaining a linear signal response requires some care. This first aim boils down to choosing the optimum threshold level, given a set of known detector characteristics and a particular light level.

The question of optimum threshold choice for photon counting was partially treated in section 2 of \cite{Lantz2008}, but in that paper the authors focused on optimizing the exposure time, and hence light level, alongside the threshold to achieve a minimum ``misfit,'' i.e. to ensure absolute photon count estimations were as close as possible to their true values.

A more general approach is to assume that true photon counts can be estimated from the thresholded values, and focus on achieving optimum signal to noise ratio (SNR) in the thresholded estimate. To do this we require a model for estimating the SNR, which is non-trivial given the stochastic detector behaviour.

I presented a basic SNR equation for photon-counting with EMCCDS at low light levels in \cite{Staley2010a}. This result was independently reproduced in \cite{Tulloch2011}, but with more care taken over correction for signal loss due to photon coincidence (2 or more photons in one pixel), including calculation of the correction factor required to correct for coincidence losses. In the interests of consistency I shall adopt the notations of \citeauthor{Tulloch2011} herein.

In lucky imaging data there is always a range of light levels in a field of view, and a truly optimal thresholding scheme would require a variation of threshold for each pixel, depending upon the light level of that pixel. However, as previously mentioned, atmospheric effects make it difficult to estimate the mean light level on any given pixel a priori; additionally multiple thresholds would require a complex data reduction process. As a simple and practical alternative I explore the effectiveness of applying a single threshold over a range of light levels.

Finally I present results of applying the thresholding techniques to real lucky imaging data. To my knowledge results from application of the techniques to real data have not been published elsewhere \citep[except in][]{Staley2010a}.

\section{Signal to noise ratio for conventional and electron multiplying CCDs}
Given an ideal detector, the only noise would be the photon-shot noise of the Poisson distribution, and so the SNR of a single pixel in an ideal detector is simply:
\begin{equation}
 SNR = \frac{M}{\sqrt{M + K}} 
\end{equation}
where M is the mean signal in photo-electrons per pixel from the target, and K is the background sky flux, in the same units.

Of course, detectors are imperfect. If we take into account the various sources of noise described in chapter~\ref{chap:EMCCD_calibration}, then in a conventional CCD with read noise of standard deviation $\sigma_N$, dark current of mean signal $D$, transfer CIC
\footnote{See Section~\ref{sec:CIC} for an explanation of transfer CIC and CICIR.}
of mean signal $\eta_c$ and CICIR of mean signal $\nu_c$
\footnote{We denote the frequency with which CICIR events occur as $B_c$, but since they undergo a lower mean gain than photo-electron events their signal contribution is $\nu_c = B_c / ln(g_A)$. See Section~\ref{sec:CIC} for details.}
 then the SNR equation of a conventional CCD becomes the rather unwieldy:
\begin{equation}
 SNR = \frac{M}{\sqrt{M + K + D + \eta_c + \nu_c + \sigma_N^2}}\,. 
\end{equation}
Typically, in a conventional CCD run at slow frame rates the CIC contributions are negligible compared to the other factors. 
With a dark sky and a well cooled CCD, the limiting noise contribution becomes the readout noise, $\sigma_N$, which is typically of the order of a few photo-electrons and will increase when run at high frame rates. From this we can see that at signal levels any lower than 10 photo-electrons per pixel per frame the readout noise will dominate over the signal.

For an EMCCD, the output for a fixed input of $n$ photo-electrons (i.e. without photon-shot noise) to the electron multiplication serial register obeys the distribution described in Section~\ref{sec:pixel_PDFs}, with mean $ng_A$ and variance $ng_A^2$. This stochastic gain variance is added in quadrature to the shot noise, producing a factor of 2 in the SNR denominator. If we assume the dark current $D$ is negligible (it is at any rate indistinguishable from transfer CIC in effect), then the SNR equation for EMCCD observations \corr{taken in linear mode} with readout noise $\sigma_{EM}$ becomes, after dividing through by $g_A$:
\begin{equation}
 SNR = \frac{M}{\sqrt{2(M + K + \eta_c + \nu_c) + (\sigma_{EM}/g_A)^2}}\,.
\label{eq:SNR_EMCCD_linear}
\end{equation}
 From this equation we can deduce a couple of points. First, at high light levels the photon count dominates over the other terms and performance is a factor of $\approx\sqrt{2} $ worse than a conventional CCD. Second, at low light levels the dominant noise term will depend largely on the relative magnitudes of the readout noise and the CIC event frequencies. However, as is clear in Figure~\ref{fig:hist_model_components}, the readout, CICIR and signal components have considerably different probability distributions, and so employing a thresholding scheme helps to distinguish between them.

\section{Thresholded signal to noise equation}
Since thresholding reduces the digitized signal to a binary one, we may assume that the frequency with which pixel values cross the threshold is determined by a Poisson distribution, with the mean frequency determined by the fraction of the original pixel value distributions which lie above the threshold. If we denote the CICIR event frequency $B_C$ and the threshold `pass fractions' \corr{--- that is, the fraction of the component PDFs (c.f. Figure~}\ref{fig:hist_model_components}) \corr{which lie above a given threshold} ---  as $F_M$, $F_B$, and $F_R$ for the photo-electron, CICIR and readout distributions respectively, then the thresholded signal to noise equation becomes:
\begin{equation}
 SNR = \frac{F_M M}{\sqrt{(F_M(M + K + \eta_c) + F_{B}B_{C} + F_{R}(1 - (M+K+\eta_c))}}\,.
\end{equation}
Generally the optimum threshold is several times the readout noise standard deviation, and so $F_{R}\approx0$. Neglecting this term, we may also account for the loss of signal induced by coincidence losses \citep{Tulloch2011}, which is of order $ (M+K+ \eta_c)^2$; the equation then becomes:
\begin{equation}
 SNR = \frac{F_M M}{\sqrt{exp[(F_M(M + K + \eta_c) + F_{B}B_{C}] - 1}}\,.
\label{eq:thresh_snr}
\end{equation}

\section{Choosing the best detector mode for an observation}
\label{sec:three_regimes}
\begin{figure}[ht]
\begin{center}
 \includegraphics[width=0.8\textwidth]{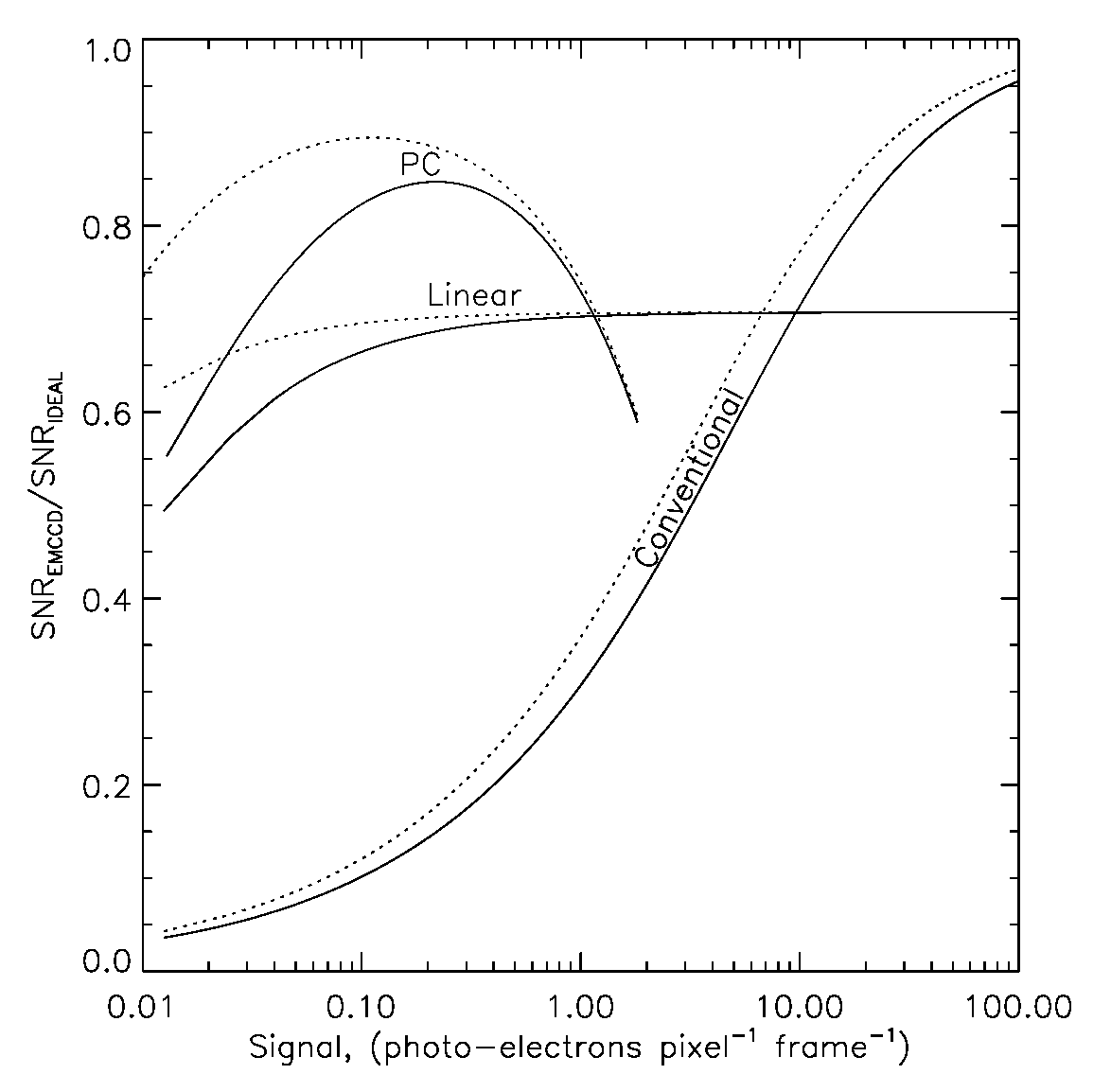}
\caption[Comparison of SNR for different CCD observing modes]{
A comparison of SNR for different CCD observing modes --- conventional CCD readout, linear mode EMCCD, and thresholded or photon-counting EMCCD (labelled PC on plot). When planning fast imaging observations the signal level should be estimated and the observing mode chosen accordingly to achieve best possible SNR. Reproduced from~\cite{Tulloch2011}. Solid lines depict the SNR corresponding to the camera tested by \citeauthor{Tulloch2011}, while the dotted lines represent results for a camera with reduced levels of clock induced charge and readout noise.
}
\label{fig:three_regimes}
\end{center}
\end{figure}
\corrbox{Having derived SNR equations for data from conventional CCDs, linear mode EMCCDs, and thresholded EMCCDs, we may consider their relative merits across a range of light levels. A comparative plot is shown in Figure~\ref{fig:three_regimes}. We may summarise this comparison by dividing it into three regimes. At high light levels (greater than $\sim10$ photo-electrons per pixel per frame) the excess noise factor introduced by the stochastic electron multiplication gain means that conventional mode CCD observations still give best signal to noise. At light levels between about 1 and 10 photo-electrons per pixel per frame, linear mode EMCCD observations win out due to the effectively suppressed read noise, while photon-counting is impossible due to coincidence losses. At still lower light levels photon-counting using thresholding methods gives the best result. In particular, at light levels of around 0.1 photo-electrons per pixel per frame, photon-counting EMCCD observations are getting close to achieving 
the maximum theoretically possible SNR (ignoring quantum efficiency effects), but are currently limited by noise from clock induced charge.
See\cite[][]{Tulloch2011} for further detail.}

\section{Threshold optimization}
\corrbox{Considering equation~\ref{eq:thresh_snr}, it is clear that the thresholded signal to noise ratio will be partially determined by the choice of threshold, which in turn determines the `pass fractions' of the various contributing signals. I now consider the matter of choosing an optimal threshold to maximize SNR.}

Given a chosen threshold level, the pass fractions for the photo-electron event pixel distribution and readout noise distributions may be estimated with simple analytical expressions; however, the pass fraction for the CICIR is calculated numerically from the rather complex distribution model. In practice the photo-electron pass fraction is estimated numerically too, since this makes it trivial to account for the contribution from pixels which receive 2 photo-electrons. Once this is achieved, it is simple to implement a numerical optimization routine. 

Conceptually, the trade off is between higher noise from readout and CICIR at low threshold levels, and higher relative shot noise levels when the signal pass fraction is small at high threshold levels. This trade off evidently depends upon the light level. However, since we employ thresholding techniques to improve the light level estimate accuracy, we do not expect to have an accurate estimate a priori. As such, it is interesting to investigate the effect of using a fixed threshold across a range of light levels. 

\begin{figure}[htp]
\begin{center}
 \includegraphics[width=1.0\textwidth]{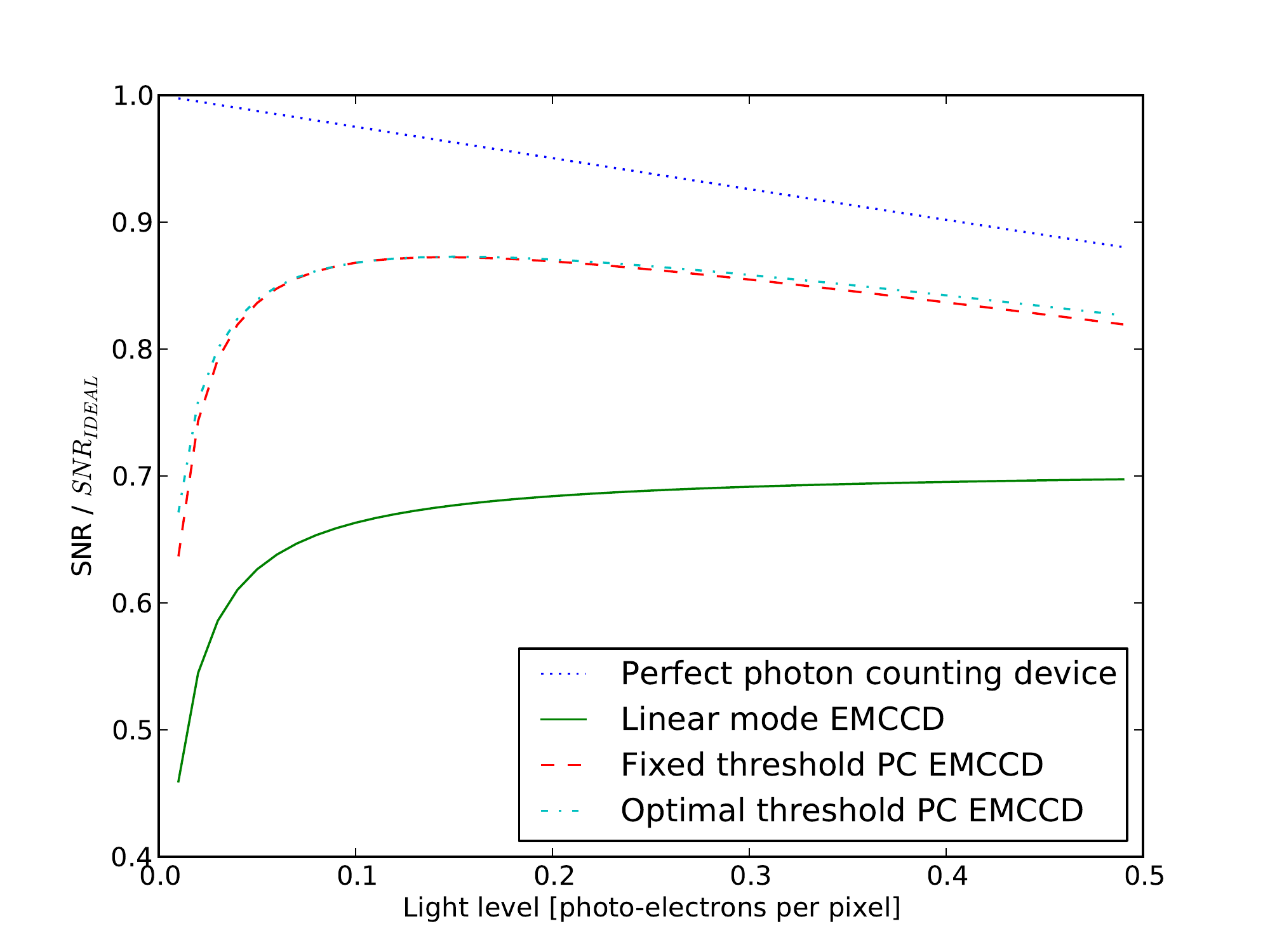}
\caption[SNR comparisons]{Comparison of signal-to-noise ratios for linear mode observations, thresholding with a (varying) optimal threshold, and thresholding with a fixed threshold (optimized for 0.1 photo-electrons signal level).  
Dotted line shows SNR for an ideal binary photon-counting device, which suffers loss of signal due to coincidence losses. 

The SNR for a fixed threshold is very nearly as good as the SNR for a threshold varied to be optimal across the range of light levels plotted.

The EMCCD model used for the linear and thresholded plots has an EM gain of ten times the readout noise. CICIR (c.f. Section~\ref{sec:CIC}) frequency $B_C$ was modelled at 0.05 CICIR events per pixel per frame. Sky background $K$ and transfer CIC $\eta_c$ are both set to zero ---  non-zero values of these uniformly lower the effective SNR of both thresholded and linear EMCCD models, but do not change the shape of the graph or alter the effect of using a fixed threshold. }
\label{fig:thresh_comparison}
\end{center}
\end{figure}

\begin{figure}[htp]
\begin{center}
 \includegraphics[width=0.8\textwidth]{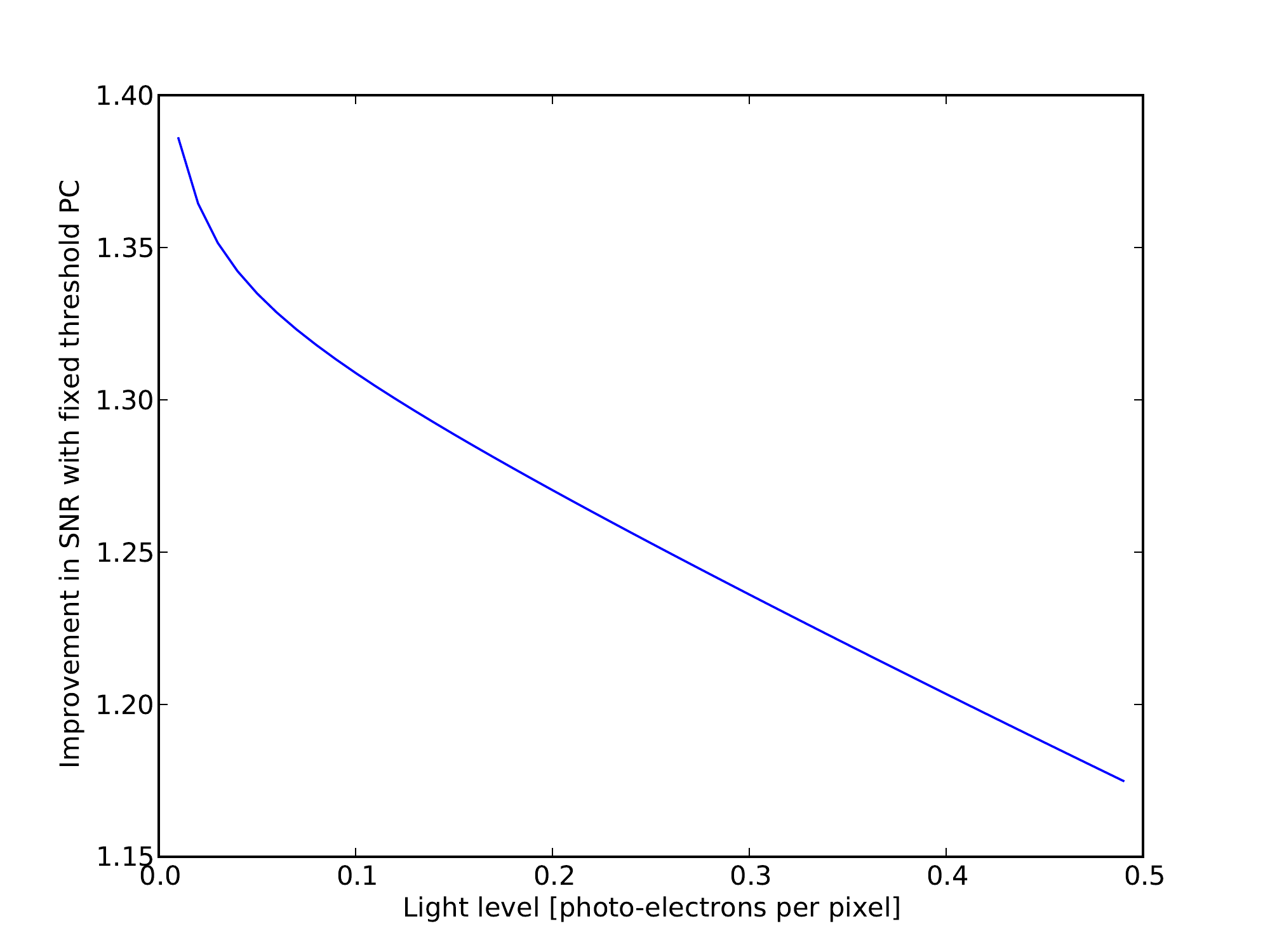}
\caption[SNR improvement with thresholding]{SNR improvement predicted by applying a fixed threshold, compared to linear mode detection with an EMCCD.}
\label{fig:thresh_improvement}
\end{center}
\end{figure}

Fortunately, as long as the threshold is chosen to exclude almost all readout noise and the highest levels of CICIR, the SNR equation is fairly insensitive to further variation of the threshold. As a result, a fixed threshold may be used effectively across a reasonably wide range of illumination levels, as illustrated in Figure~\ref{fig:thresh_comparison}. Figure~\ref{fig:thresh_improvement} displays the corresponding improvement predicted  in SNR at low light levels when using a fixed threshold.

\section{Combining thresholded and linear data}
\label{sec:combining_thresholded_data}
To linearise the photometry of thresholded images, and allow for direct comparison with linear mode images, we must correct for the effects of the thresholding process. 

To calculate the correction factor for loss of multiple photo-electron pixel events, we consider the mean number of pixels with photo-electron counts of 1 or greater, $n$, for a mean illumination level of $N$ photo-electrons. From the Poisson distribution we have
\begin{equation}
 n = 1 - e^{-N}\,,
\end{equation}
therefore
\begin{equation}
 N = -ln(1-n) \quad [\approx n + \frac{1}{2}n^2, \quad n<<1]\,.
\end{equation}

We must also consider the reduction in signal level due to the threshold level, and the contribution of CICIR events, which will be different for thresholded and linear reductions. If we consider pixel values of $X_{lin}$ and $X_{thresh}$ in the reduced linear and thresholded images respectively, then the signal estimates may then be calculated as:
\begin{equation}
 N_{lin} = \frac{ X_{lin}}{g_A } - \frac{B_c}{ln(g_A)}
\end{equation}
and
\begin{equation}
 N_{thresh} = -ln(1 - n), \textrm{ where } n = \frac{X_{thresh} - B_c }{F_M}\,.
\end{equation}

Once these corrections have been applied, it is possible to combine linear mode and thresholded images created from the same dataset. The linear mode image may be used to estimate the light level of any given pixel, and any pixels with light levels below a chosen threshold, e.g. 0.25 photons per frame, are replaced with the pixel values from the thresholded image. Such an approach achieves high dynamic range coupled with the SNR improvements of thresholding, but requires extra care when undertaking further analysis. Since the noise level estimates will be different for thresholded and linear images, if they are combined a separate ``pixel flag'' image must be maintained so that any analysis algorithms can employ the correct noise estimator on a per pixel basis.

\section{Results of applying thresholding techniques to real data}
\label{sec:threshold_real_data}
\begin{figure}[htp]
\begin{center}
 \includegraphics[width=1\textwidth]{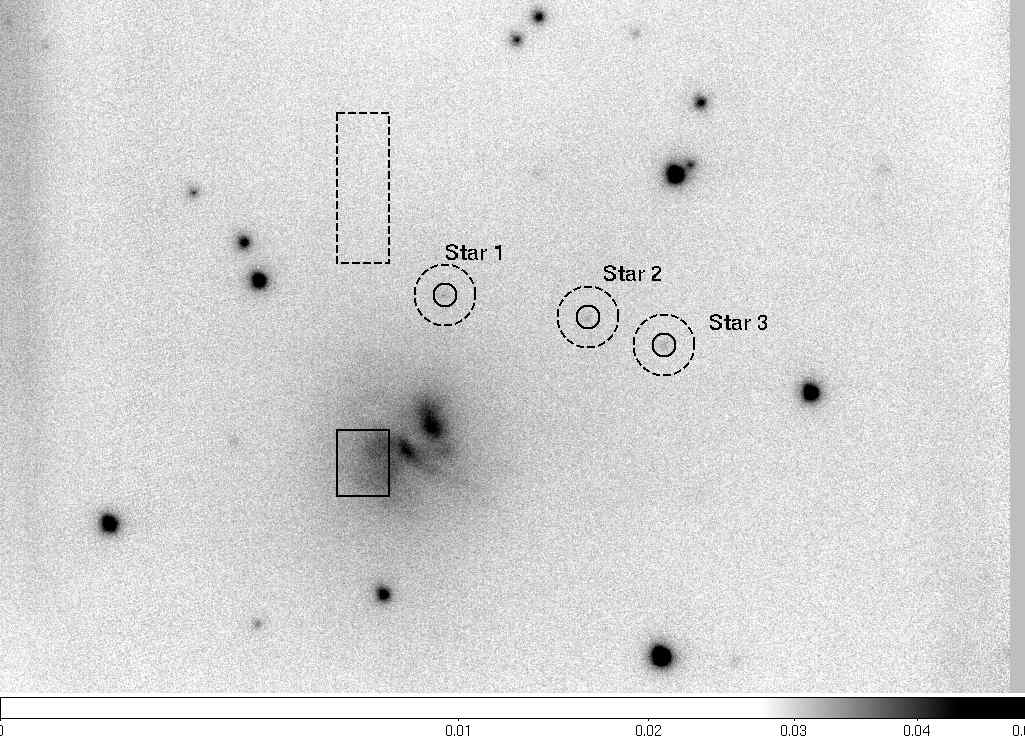}
\caption[SNR test image]{The image used to compare the SNR for analogue mode and photon counting mode reduction techniques. This is a fully reduced image of the radio galaxy 3C405 composed from a 50 percent selection drawn from $\sim$75000 short exposures taken over 1 hour. North is up and East to the left. The field of view is approximately 30x30 arcseconds in size.

The colour scale is a square root stretch from 0.0 to 0.05, in units of photo-electrons per pixel per exposure .

The dashed regions were used to estimate the background sky flux and noise variance levels, while the solid line regions are photometric apertures used to estimate signal in the images. More details can be found in the text.}
\label{fig:SNR_test_image}
\end{center}
\end{figure}

To assess the impact of applying photon counting techniques to real data, I undertook an analysis of a sample dataset --- 1 hour of short exposures of the radio galaxy 3C405. These data were taken under excellent seeing conditions of $\sim$0.4 arcseconds seeing FWHM, during the July 2009 Nordic Optical Telescope observing run \corr{detailed in sections}~\ref{sec:luckycam_2009} \corr{and} \ref{sec:luckycam_2009_cals}. The guide star is about 20 arcseconds from the centre of the field of view \corr{and was imaged on a different detector in the mosaic field of view to 3C405. This meant it was possible to use a high EM gain setting for observing 3C405, without it resulting in detector saturation}. FWHM at this radius from the guide star was estimated at $\sim$0.25 arcseconds.%
\footnote{%
\corrbox{See Section~\ref{sec:hires_wide_field} for details of how the FWHM varies across the field}.
}%
This observation was chosen as it is an unusually long timespan dataset, and as a result the faint limit is largely due to detector noise.

The dataset was reduced in the usual manner and final images were produced using a 50 percent selection of the dataset, applying both the standard linear mode reduction and photon counting thresholding technique. I then compared the SNR for both a resolved region of extended emission in the galaxy profile (solid line box in Figure~\ref{fig:SNR_test_image}) and for 3 faint point sources (solid-line circles in the figure). For the region of extended emission we estimated background level and variance using a remote region which was judged to only contain sky pixels (dashed box in the figure), while for the point sources we used an annulus around the photometric aperture  (dashed circles denote outer radius, inner radius was set at photometric aperture boundary). The signal to noise ratio was then estimated using the formula:

\begin{equation}
 SNR = \frac{ F - (N_{pix}\times b) }{ \sqrt{ F + N_{pix}\times \sigma^2  } }
\end{equation}
where $b$ and $\sigma$ are the mean value and variance of the relevant background pixels, $F$ is the sum flux of the pixels in the photometric aperture, and $N_{pix}$ is the number of pixels in the photometric aperture.

Numbering the stars 1 through 3 from left to right, the results are:
\begin{center}
  \begin{tabular}{| l || r | r | r| }
    \hline
    Region & Analogue SNR & Photon Counting SNR & SNR Increase\\ \hline \hline
    Galaxy region & 239 & 391 & 63.5\% \\ \hline
    Star 1 & 4.55 & 4.28 & -6\% \\ \hline
    Star 2 & 2.44 & 3.46 & 42\% \\ \hline
    Star 3 & 5.12 & 6.24 & 22\% \\ \hline
    \hline
  \end{tabular}
\end{center}

The decrease in SNR for star 1 is due to a relatively increased variance in the annulus used to determine the local background --- while not immediately apparent in the image, inspection at high contrast shows that star 1 is in a region of relatively steep decline of the faint wings of the galaxy. As such the background variability is largely due to real signal variation which would need to be modelled and subtracted. Stars 2 and 3 also suffer from this, but to a lesser extent such that it is not the dominant source of background variation. This is, of course, a quick and crude comparative test - more optimal SNR ratios could be achieved through PSF fitting. We note that the \emph{peak} pixel in the region of star 2 has a photon-flux per frame of only 0.005 above the sky flux level - i.e one excess photon in 200 short exposures, on average.

Thresholding techniques are clearly useful at these low light levels, and have potential to further the limits of lucky imaging, especially as the detectors continue to improve and produce lower CIC levels.
\chapter{Optimising and predicting the image formation process}
\label{chap:frame_registration}
In this chapter I consider the extrinsic variation in the short exposures recorded for lucky imaging --- variations in the point spread function (PSF) caused by the fluctuations in atmospheric seeing conditions due to the turbulent processes described in Section~\ref{sec:atmos_effects} --- and how we may deal with them. 
To place my work in a meaningful context I first recall some salient points of sampling theory, and describe the drizzle algorithm used to combine multiple exposures. I then summarise the characteristics of the PSF in short exposures obtained at ground based telescopes. With this grounding in the image formation processes, the frame analysis and registration algorithms are compared and optimal methods determined. Finally, error budgets are explored with the dual aims of predicting image quality and enabling multiple guide star frame-registration techniques.

\section{Background --- Sampling theory and image combination}
\label{sec:image_formation_background}
When designing a camera for astronomy, pixel size is a crucial parameter, due to the competing requirements of wide field of view, high signal-to-noise ratio, and good sampling of the PSF. The relevant pixel measurement in this discussion is subtended angular width, the width of the section of sky from which light is focused upon one pixel. 

\begin{figure}[htp]
\begin{center}
 \includegraphics[width=0.8\textwidth]{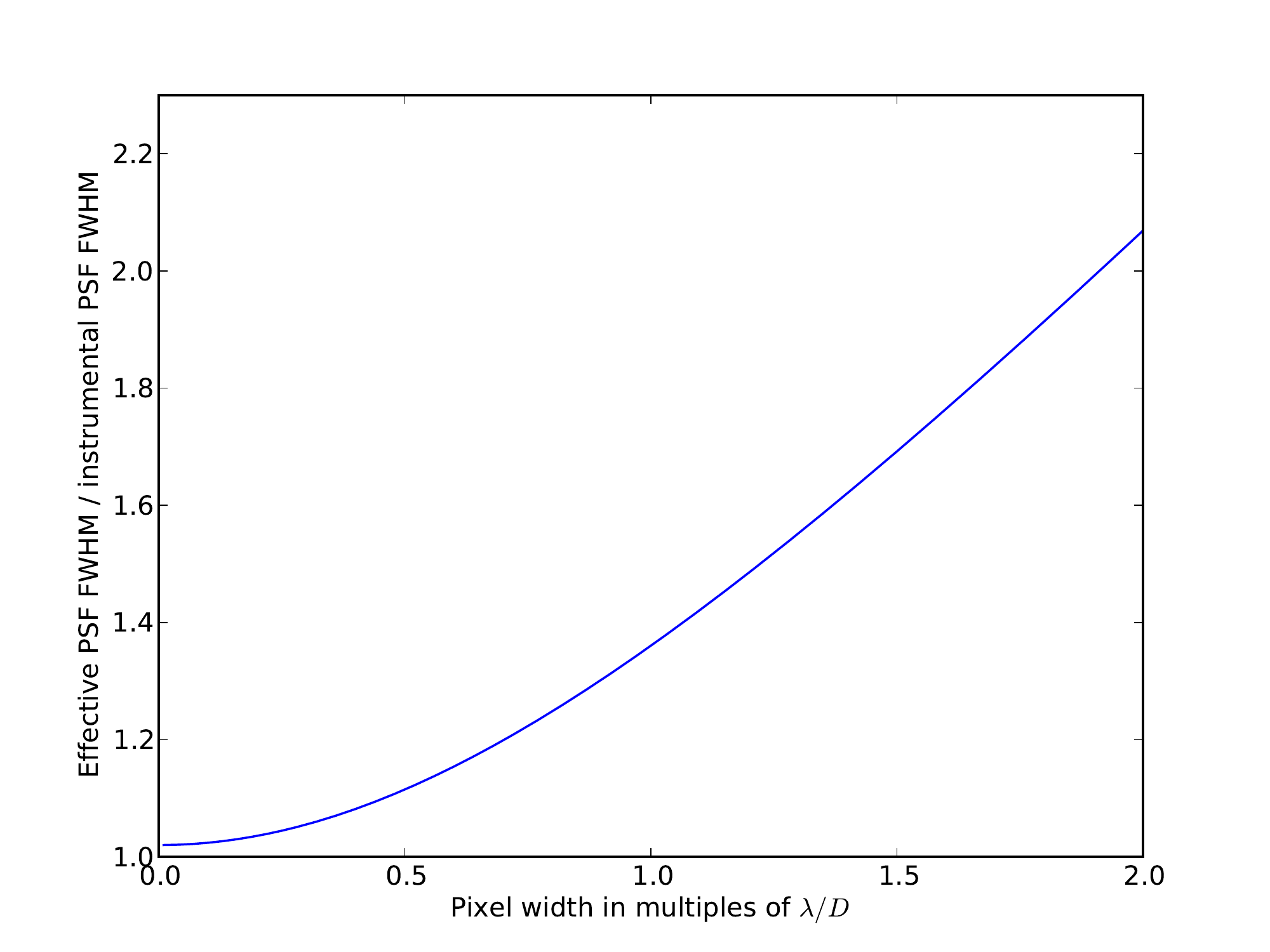}
\caption[Effective PSF]{Due to finite pixel width, the effective PSF has a full width at half maximum (FWHM) always greater than the instrumental FWHM. (Instrumental Airy disc PSF and pixel response modelled assuming Gaussian approximations.)}
\label{fig:ePSF}
\end{center}
\end{figure}

A pixel of large angular width gathers more photons and so has a better signal level. However, the Nyquist-Shannon theory \citep{Shannon1998} tells us that to properly encode a signal we require a sampling rate of twice the maximum frequency present. In the case of a plane wavefront entering perfect telescope optics, the corresponding critical pixel angular width is often taken to be $\lambda/2D$, where $\lambda$ is the wavelength and D the telescope primary aperture diameter.
However, there is a further complexity in that the Nyquist-Shannon theory refers to instantaneous or point sampling, whereas a pixel integrates a signal over a finite area. This is most easily analysed using the formalism of an ``effective PSF'' \citep{Anderson2000}; the PSF resulting from the convolution of the actual light intensity distribution (called the instrumental PSF) with the pixel response function. We can then apply point sampling theory to this effective PSF. This formalism highlights the fact that to obtain the highest instrumental resolution we actually need sampling of the PSF with pixels of angular width smaller than $\lambda/2D$ (fig.~\ref{fig:ePSF}), although realistically this would be impractical for most cases. 

\begin{figure}[htp]
\begin{center}
 \includegraphics[width=0.8\textwidth]{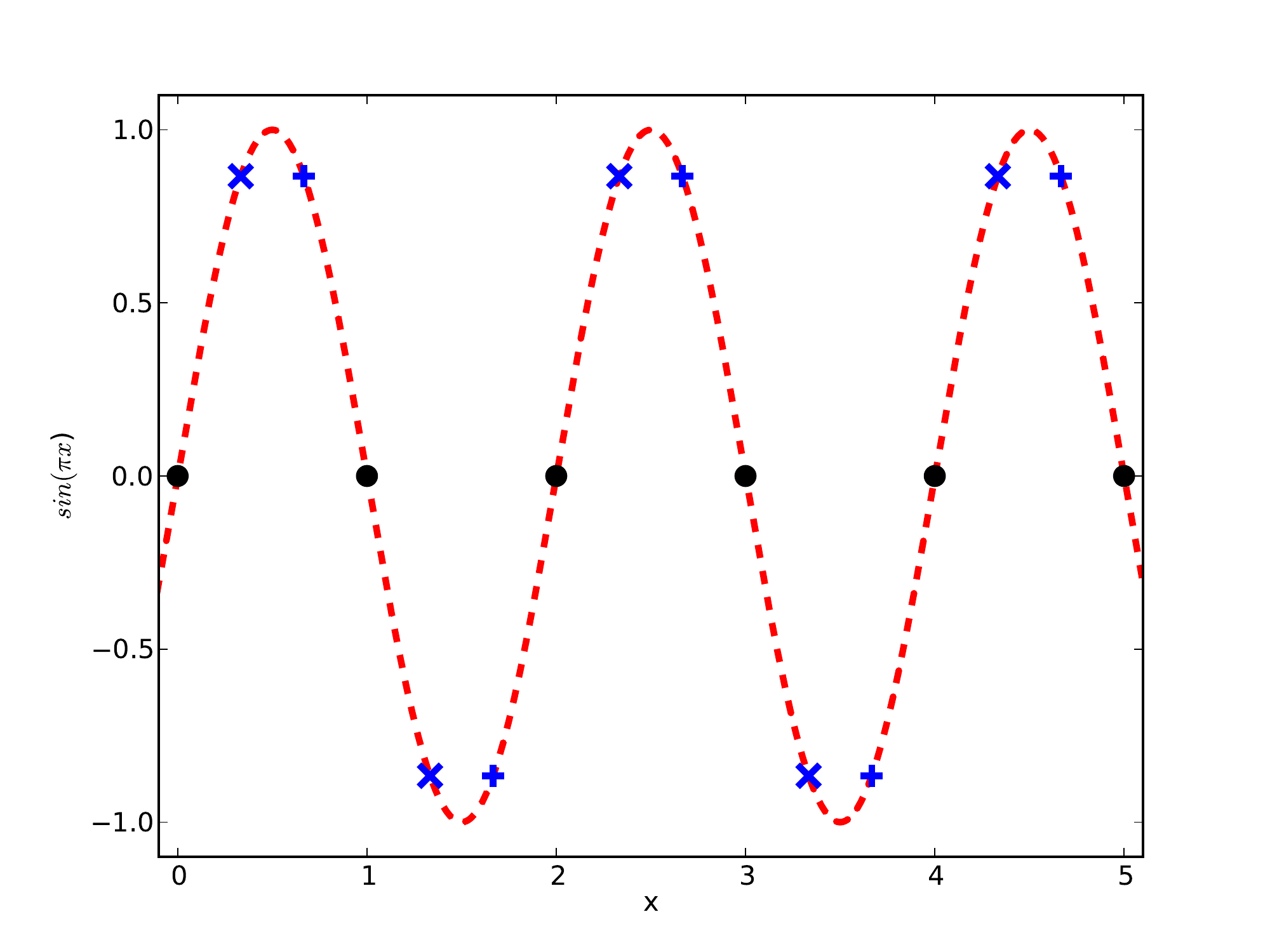}
\caption[Sampling rates]{An insufficient sampling rate results in an incorrect measurement of the underlying signal (black dots). Additional sampling points (x's and +'s) can be obtained by increasing the sampling rate, or sequentially sampling at different phase offsets.}
\label{fig:sampling_rate}
\end{center}
\end{figure}

To obtain a higher sampling rate of a signal, we can do one of two things --- sample at smaller intervals, or if the signal is stable we may sample multiple times at different phase offsets (fig.~\ref{fig:sampling_rate} ). In optical astronomy these correspond to pixels subtending a smaller angle on sky, or multiple pointings at sub-pixel angular offsets. The optimal choice of pixel size will depend upon many factors such as desired angular width of the field of view, readout noise and other detector characteristics, data transfer rates, dynamic range and saturation levels, and so on. As a result, it is often useful to observe a field of view with multiple pointings.

\begin{figure}[htp]
\begin{center}
 \includegraphics[width=1.0\textwidth]{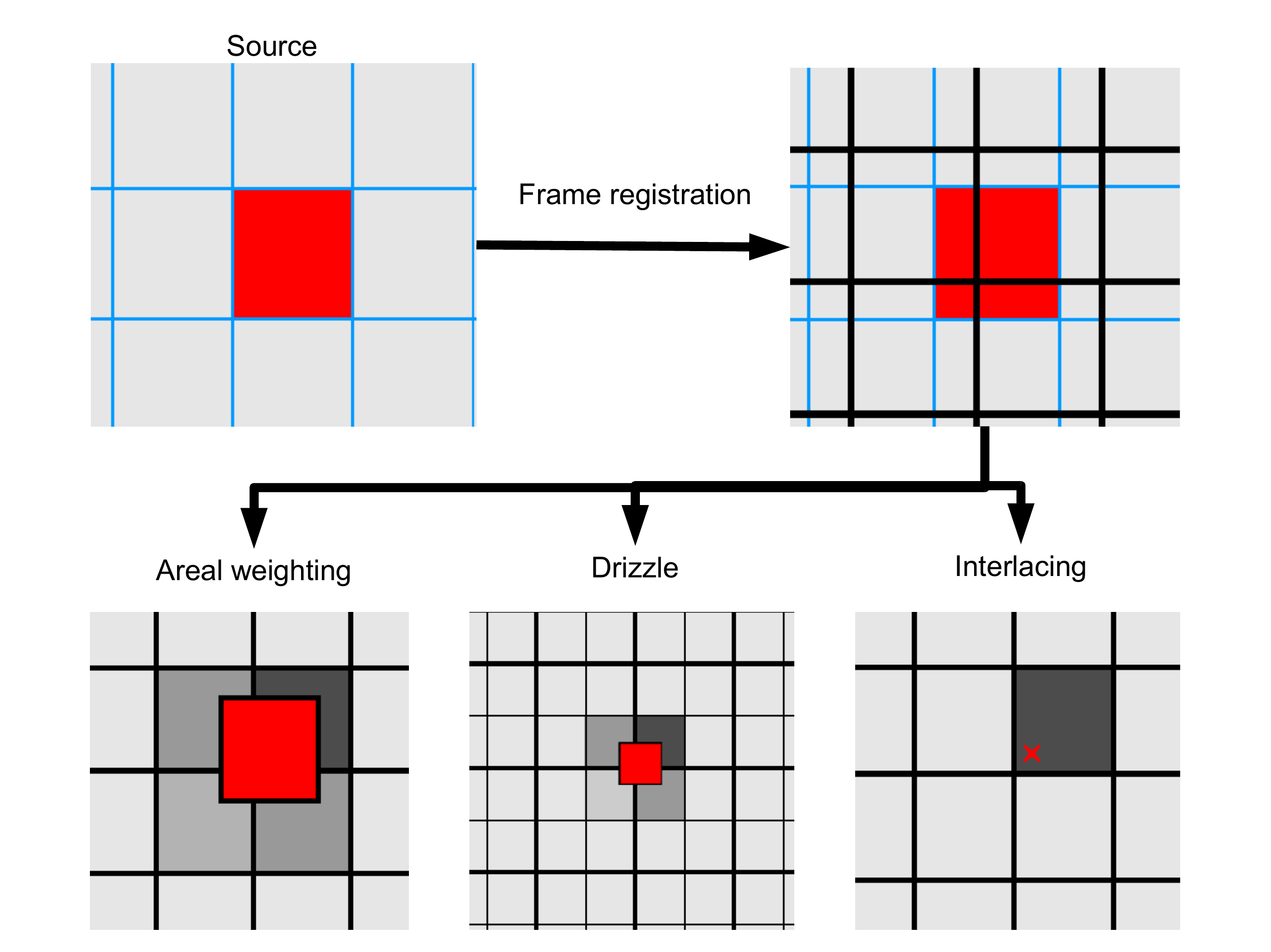}
\caption[Drizzle schematic]{Schematic comparison of different image combination techniques. 
Upper half: Some method is required to determine the relative position of each input frame, ideally with sub-pixel accuracy (this is known as frame registration). 

Lower half: Consider a single pixel from the input image (highlighted in red). How do we represent the input signal?

Lower right: Interlacing assigns all signal from each input pixel to a single output pixel based on its central location. This works well for multiple input images at regular and accurate spacings, but is equivalent to poor accuracy frame registration if positionings are random. 

Lower left: Area overlap weighting deals well with random positioning, but results in further convolution and loss of resolution --- a single pixel may effectively be blurred to twice its original size if it happens to lie across an output pixel boundary.

Lower centre: Drizzle algorithm \citep{Fruchter2002} essentially uses area based weighting, but allows for adjustment of the input and output pixel sizes, resulting in a compromise solution.

A simple linear translation is portrayed, but the drizzle algorithm may be used to apply many sorts of transformations and deformations, e.g. correction for the HST WFPC2 field distortion \citep{Casertano2001}.
}
\label{fig:drizzle_schema}
\end{center}
\end{figure}

If multiple pointings are used, then the individual exposures recorded are usually combined to produce an average image, in order to reduce data storage requirements, ease data analysis procedures, and produce datasets that are easily inspected visually. We desire an image combination algorithm which minimises further convolutions with the pixel response function and resulting loss of resolution, while at the same time providing methods for bad pixel data rejection. The drizzle algorithm \citep{Fruchter2002} provides parameters that may be varied towards one or the other of these aims, as illustrated in  Figure~\ref{fig:drizzle_schema}. 

\section{Background: Speckle patterns and speckle imaging}
We can calculate the theoretical (instrumental) PSF for a telescope of circular or annular aperture with relative ease. It is described by the Airy function, known as the Airy disc, and has an angular width proportional to the ratio of the observed wavelength and the aperture size, $\lambda / D$. However, as all astronomers know, the PSF observed using a conventional long exposure camera behind a medium or large ground-based telescope (of say, a metre plus aperture size) is very different to the Airy disc. Instead of a compact, sharply peaked profile we observe a wide, shallow PSF known as the seeing disc, well modelled using the Moffat function \citep{Trujillo2001}. 

The increased width of the seeing PSF causes loss of information. If we maintain a pixel size suited to sampling of an Airy disc then the source flux will be spread across many pixels and so the signal-to-noise ratio will be much lower, due to readout noise. If we increase the angular pixel width then we lose resolution due to convolution with the pixel response function. Even in the ideal case of a well sampled PSF at high signal level, the wider PSF blurs away much of the information present at high spatial frequencies. This introduces degeneracy in signal from multiple close sources, so that for example a binary or triple star system may not be distinguished. 

\begin{figure}[htp]
\begin{center}

\subfigure[Airy disc]{	
	\includegraphics[width=0.3\textwidth]{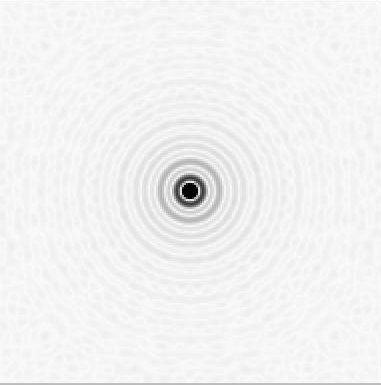}
	\label{subfig:airy}
}
\subfigure[Instantaneous speckle pattern]{
		\includegraphics[width=0.3\textwidth]{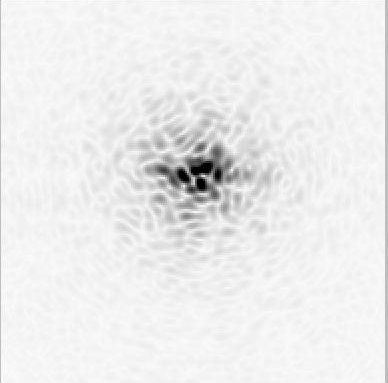}
	\label{subfig:speckle}
}
\subfigure[Seeing disc]{
	\includegraphics[width=0.3\textwidth]{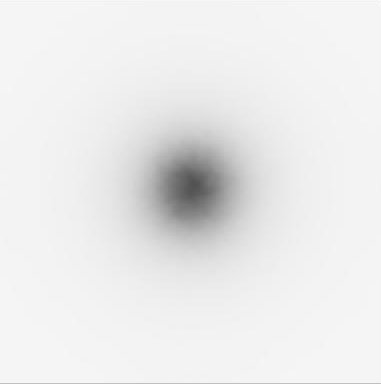}
	\label{subfig:seeing}
}
\\
\subfigure{	
	\includegraphics[width=0.3\textwidth]{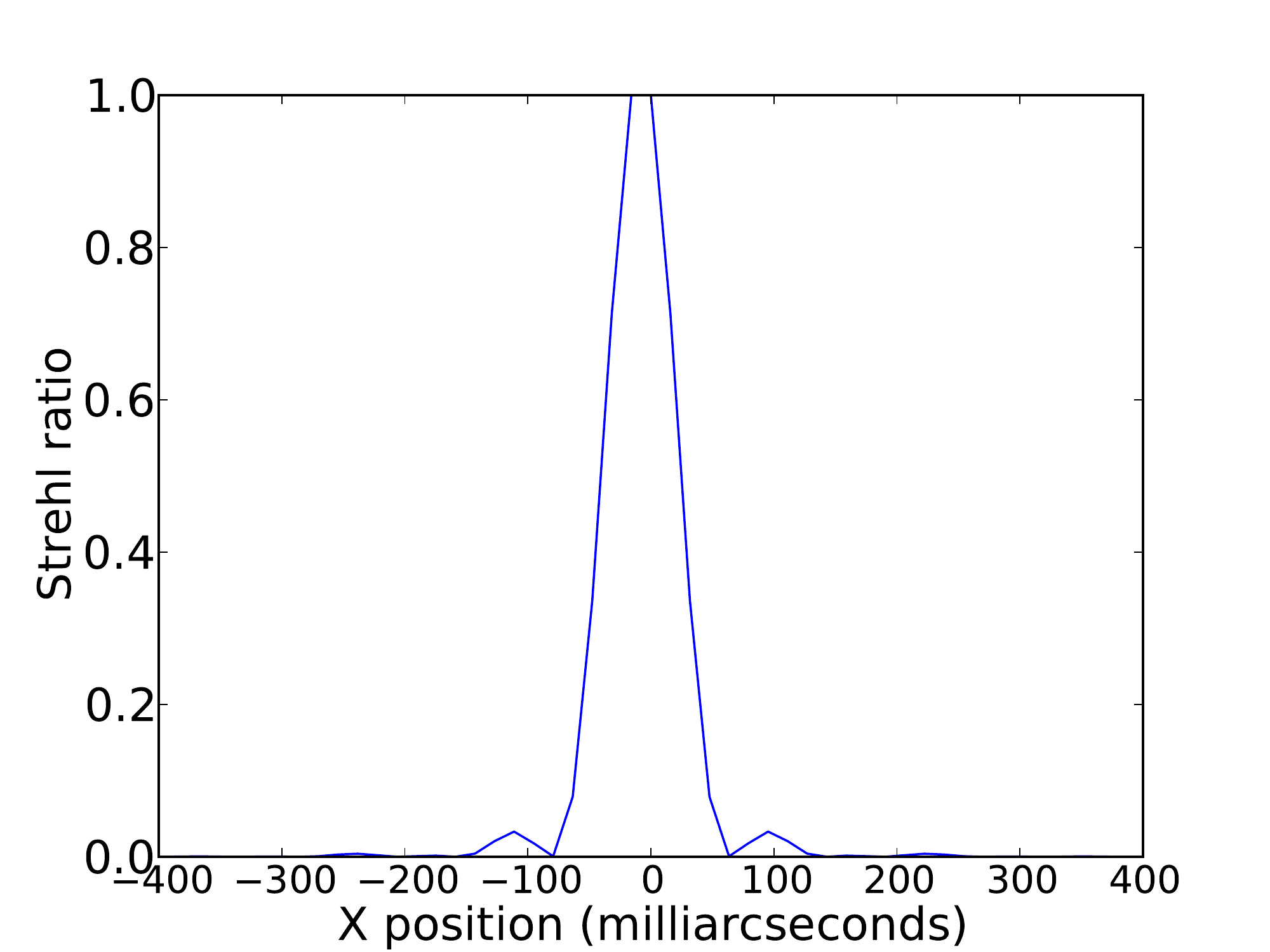}
	\label{subfig:airy_x}
}
\subfigure{
		\includegraphics[width=0.3\textwidth]{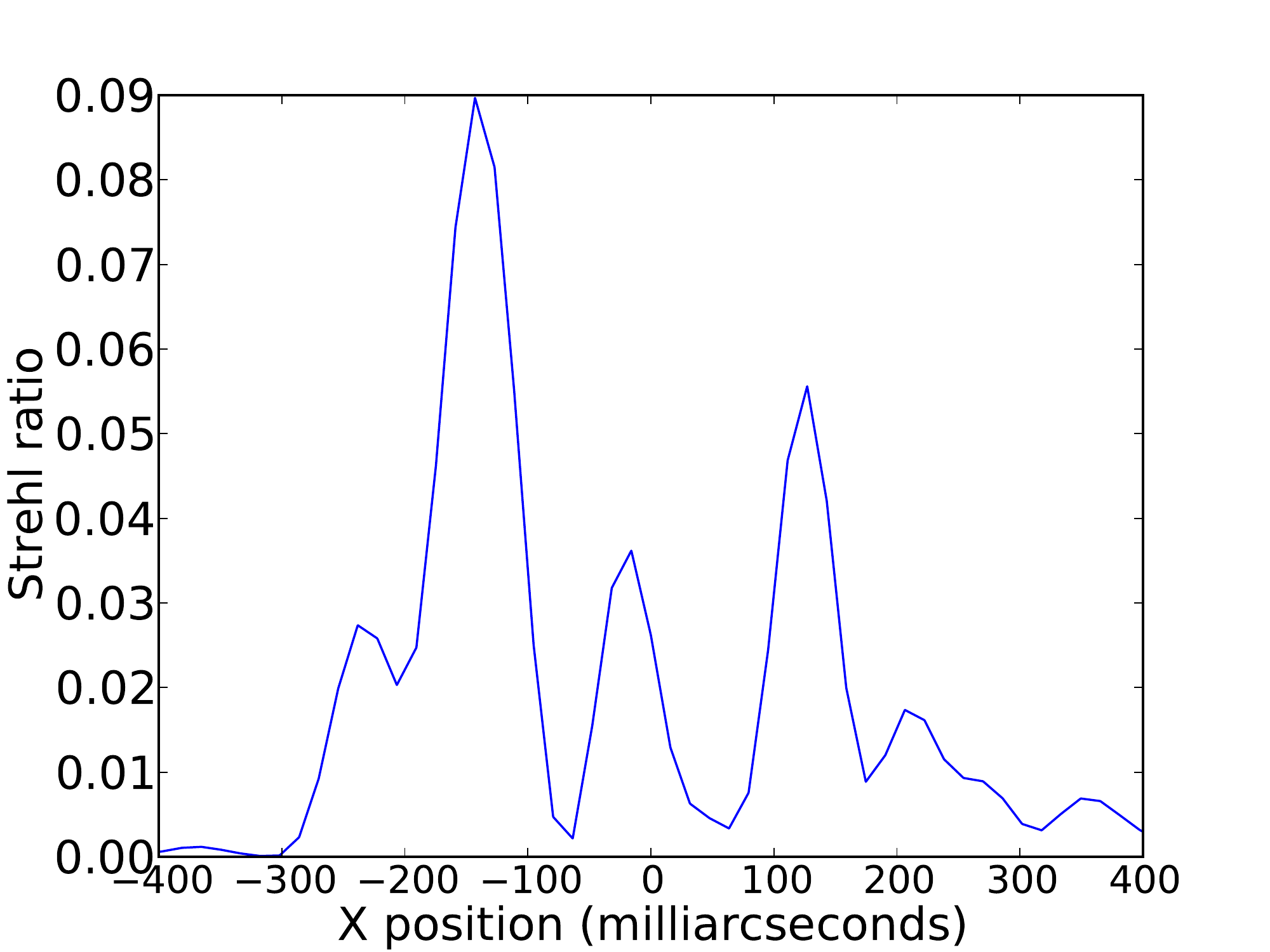}
	\label{subfig:speckle_x}
}
\subfigure{
	\includegraphics[width=0.3\textwidth]{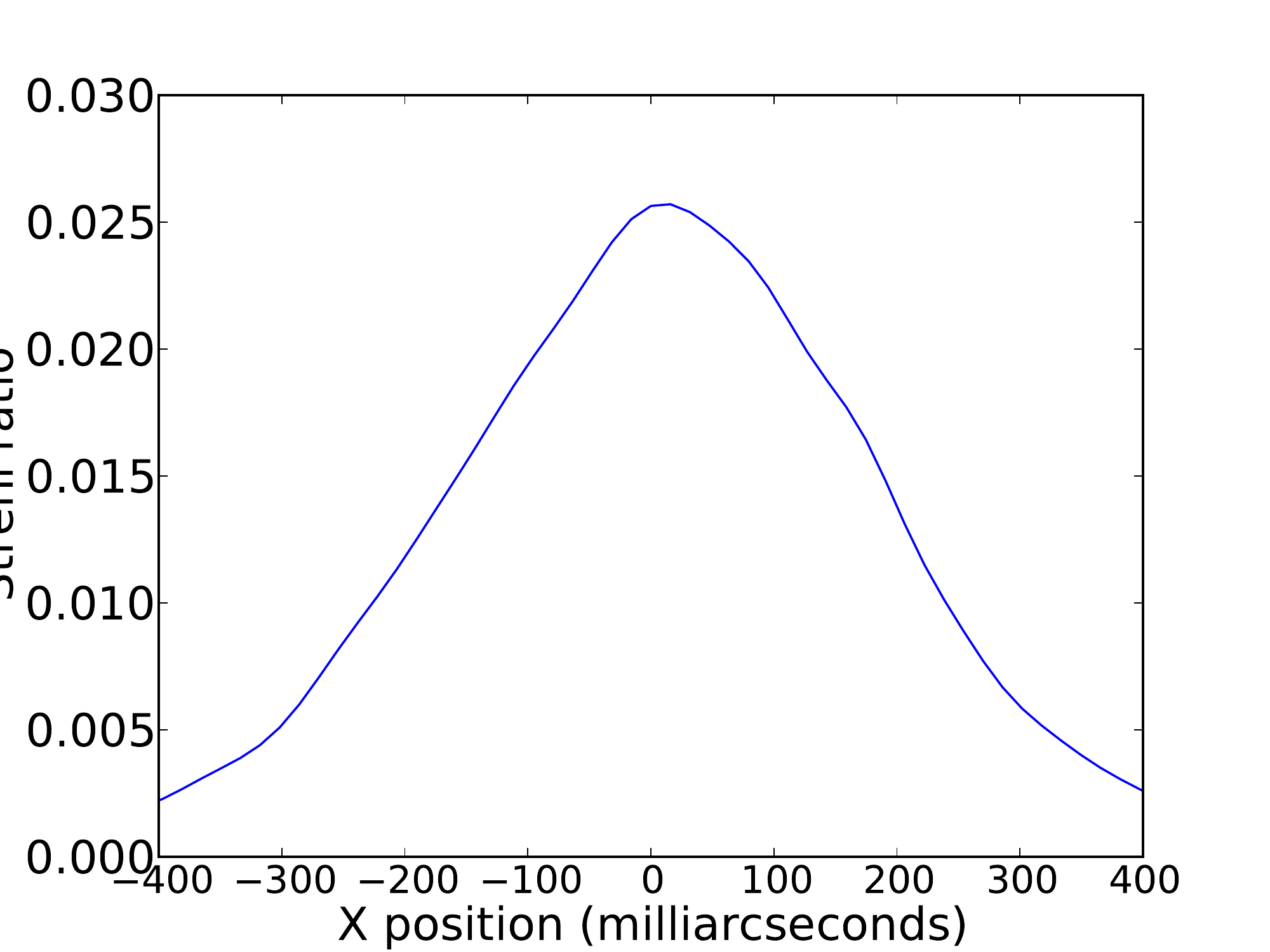}
	\label{subfig:seeing_x}
}
\caption[Simulated PSF examples]{
	Simulated intensity functions to illustrate the PSF obtained \subref{subfig:airy}: with no atmospheric turbulence, \subref{subfig:speckle} with a very short exposure in the presence of atmospheric turbulence, and \subref{subfig:seeing}: with a long exposure in the presence of atmospheric turbulence. Plots below depict values across a cross-section for each of the PSFs. The angular width measurements correspond to results for a 2.5 metre diameter telescope in 0.5 arcsecond seeing, as observed at a wavelength of 770nm.
}
\label{fig:sample_PSFs}
\end{center}
\end{figure}

The reason for this loss of resolution is perturbations in the phase of the incoming starlight, caused by the turbulent atmospheric processes described in Section~\ref{sec:atmos_effects}. Due to the power law scaling of these effects, the phase disturbance across a small aperture is too small to cause significant deviation of the PSF from the Airy model. As the telescope aperture size increases, for a fixed observation wavelength the phase perturbations increase according to the power law until, at phase disturbances of around 1 radian, the PSF becomes noticeably distorted. For larger apertures still, the aperture phase may be considered in terms of multiple coherent regions, within which the phase difference is around 1 radian. The size of such a coherent region is equivalent to the Fried coherence length, denoted $r_0$ \citep{Fried1965}, and is dependent upon the wavelength being observed according to the formula:

\begin{equation}
r_0 = \left [ 0.423 \, k^2 \, \sec \zeta \int_{\mathrm{Path}} C_n^2(z) \, dz \right ]^{-3/5}.
\label{eq:r0}
\end{equation}
where $k = 2\pi / \lambda$ is the wavenumber, $\zeta$ is the angle of the telescope from the zenith, and $C_n^2$ is the ``structure constant,'' denoting the strength of the turbulence optical effects at each altitude \citep{Hardy1998}. It is worth noting that $r_0$ is dependent upon wavelength, varying as 
$r_0\sim  \lambda^{6/5}$.
 This is a useful quantity, since we may discuss severity of turbulence for any given site and telescope in terms of the dimensionless ratio $D/r_0$, where $D$ is the diameter of the telescope primary aperture. At values of $D/r_0 > 1 $ the multiple coherent phase regions typically result in an instantaneous PSF of multiple local maxima, which we may consider as a superposition of distorted copies of the Airy disc --- these are known as ``speckles'' or collectively as speckle patterns. The speckle patterns evolve as the atmospheric phase disturbances change, and their long-term average light intensity distribution is the seeing disc PSF observed in long exposure images (see Figure~\ref{fig:sample_PSFs} for an illustration). Note that the compact nature of the individual speckles represents high frequency spatial resolution, lost during the long exposure averaging process as the speckle patterns evolve and shift position in the focal plane.

It has been recognised for some time that it is possible to retrieve some of the high frequency spatial information present in speckle patterns, through speckle interferometry methods \citep{Labeyrie1970}. Exposures are recorded on a timescale short enough to give a reasonable approximation to the instantaneous PSF (or colloquially ``to freeze the seeing''), and then analysed post-exposure to retrieve information present at high resolution, e.g. parameters of binary star systems.

Lucky imaging is a  different approach to analysis of similar datasets. By selectively combining the short exposures we may exploit the stochastic nature of the turbulence \citep[as noted in][]{Fried1978}, combining data from moments when the RMS phase disturbance across the telescope aperture is smaller than average. Ideally we select short exposures in which the PSF closely resembles an Airy disc. However, even under poorer seeing conditions, or when using all recorded short exposures, the resolution is improved compared to the seeing disc --- the data reduction process preserves some of the high frequency spatial resolution lost in a long exposure.

\section{Frame registration overview}
The speed at which the atmospheric phase disturbances and corresponding speckle patterns evolve can be characterised using a defined `atmospheric coherence time' \citep{Scaddan1978,Lopez1992}. \corr{The relevant formulation for speckle imaging, which is likely most relevant to lucky imaging, is } \citep{Roddier1982a,Aime1986}: 
\begin{equation}
\tau_{speckle} \propto \frac{r_0}{\Delta v}
\end{equation}
\corr{where $\Delta v$ is the mean velocity dispersion for the turbulent layers, and is typically of the order 10ms} \citep{Lopez1993}. To preserve the high spatial resolution information present in the instantaneous PSF, the length of the short exposures must be of the same order as this atmospheric coherence time. As a result, to gain useful information about the instantaneous PSF we must observe a reasonably bright source, in order that a sufficient number of photons are detected from that source within every atmospheric coherence time. The difference between a naive and an optimal analysis algorithm applied to the short exposure data will determine the faint limit, i.e. the faintest source which the technique may be usefully applied to, and also how close we get to the diffraction resolution potentially obtainable under favourable observing conditions. Clearly, this is a crucial aspect of the data analysis to get right.

For lucky imaging, frame registration and estimation of image quality may be equated to the problem of locating and estimating the peak value of the short exposure PSF. By aligning the images upon the point of brightest intensity (i.e. the brightest speckle), these add coherently to produce a narrow core, while secondary speckles add incoherently, converging to an axisymmetric, wider PSF component of lower intensity, or `halo,' over many exposures.
In the seeing regime of $ D/r_0 \sim 7$ or less, a small but significant portion of frames are reasonably close to the diffraction-limited instrumental PSF (see table~\ref{tab:fried_selection_table}). If signal levels are high enough then potentially the diffraction-limited frames may even be aligned with sub-pixel precision, such that the atmospheric motion is exploited in the manner of multiple pointings to improve PSF sampling. 

Since the guide star registration is such a critical step in determining the lucky imaging system performance, I decided to investigate the registration algorithms to see if they could be improved further, particularly with the aim of extending the guide-star faint limit. 

\section{Registration methods}
\label{sec:reg_methods}
The simplest method of estimating the PSF peak is simply to use the brightest pixel, but this is a flawed method --- even when considering pixellation effects in the absence of noise, it only locates the PSF peak to the nearest pixel, and results in underestimation of the PSF peak value depending on sub-pixel location. \cite{Tubbs2003} implemented a Fourier resampling and filtering algorithm to overcome these problems, but this is computationally expensive and introduces artefacts. \cite{Law2007} implemented a combination of bicubic resampling and a cross-correlation algorithm, using the core of an Airy PSF as a reference model. If we consider the short exposures closely resembling an Airy PSF, cross correlation with an Airy reference is a reasonable approximation to the maximum likelihood estimator of position, due to the well known result that cross correlation is equivalent to the maximum likelihood position estimator for a pixellated signal in the presence of uniform noise (see e.g. the appendix of \cite{
Gratadour2005} for a derivation).

We can do even better, by dropping the false assumption of uniform noise. As detailed in chapter~\ref{chap:thresholding}, the largest source of signal variance at all but the lowest light levels is photon shot noise, increased by a factor of $\sqrt{2}$ due to the stochastic electron multiplication gain. \corr{Hence, a more optimal reference image can be obtained by weighting using this simplified noise estimate proportional to $\sqrt{N}$, where N is the mean light level in photo-electrons at each pixel --- or equivalently by replacing each pixel value of the reference by its square root (referred to hereafter as `normalising the reference'). Note that \emph{uniform} scaling of the reference has no effect on the estimation of best matching position.}

If sub pixel accuracy is desired, then we must also perform some kind of interpolation. A simple and efficient method is to calculate the cross-correlation at the original pixel scale, and then fit a parabola in the X and Y directions about the maximum correlation value to determine the correlation peak, as suggested in \cite{Poyneer2003}. However, this method is sub-optimal for a critically-sampled or under-sampled image, since the signal is only ever compared to a reference generated at a particular sub-pixel offset, artificially lowering the cross correlation value in some instances. In a high signal-to-noise regime (i.e. when using a bright guide star) it therefore makes sense to either attempt cross-correlation with a reference of varied sub-pixel shift \citep{Gratadour2005}, or to interpolate the signal data prior to cross-correlation at each interpolated point.

\subsection{Implementation details}
Since lucky imaging deals with many short exposures per second, and we may wish to perform frame registration using multiple stars, it is desirable that any frame registration routine be implemented so as to achieve the best computational speed possible. To calculate the cross-correlation we have a choice of computational methods. We can either calculate the convolution at each pixel index $(m,n)$ in the spatial domain:
\begin{equation}
 C[m,n]:=s[m,n]*r[i,j]= \sum_i \sum_j s[m,n]r[m+i,n+j]
\end{equation}
where $i,j$ are defined over a small range of negative and positive offsets about the central pixel of the reference image; 
or we may multiply their discrete Fourier transforms \corr{(DFT)} pixel-wise and then transform back:
\begin{equation}
C:=DFT^{-1}\Big(	 DFT( s ) \times DFT( r)         	\Big)\,.
\end{equation}
Both methods have pros and cons. 

In computational terms, spatial convolution is very expensive for large references, scaling as $O(N_r^2N_s^2)$ where $N_r^2$, $N_s^2$ are the pixel widths of the reference and signal images respectively, and so Fourier convolution ($O(N_slogN_s)$) is favoured for `scene' cross-correlation \corr{wherein a large reference image is used} \cite[see e.g. ][]{Poyneer2003}. However, for the purposes of lucky imaging I favour spatial convolution, for the following reasons:
\begin{itemize}
 \item For a critically sampled Airy PSF the reference will always be small, since beyond a radius of a few pixels the signal becomes negligible compared to noise.  Therefore, typically $O(N_r^2)\sim10$.
 \item It allows for calculating the cross-correlation value only at input pixels above a certain threshold, e.g. 50\% of the input maximum pixel value. If the cross-correlation region only covers the guide star and background sky without further bright sources, the number of calculations is drastically reduced.
 \item Spatial convolution enables the masking of bad pixels, useful if the guide star happens to be observed near to bad detector columns (cf. Section~\ref{sec:bad_pixels}).
\item Even if the PSF is oversampled, photon and detector noise will introduce signal variation at the single pixel level. Therefore, the image data must be embedded in an array of twice the size before performing the DFT to prevent aliasing effects due to this high-frequency pixel noise, which further increases DFT computational cost \citep{Thomas2006}.
\end{itemize}
   
I implemented spatial cross-correlation routines for use in the pipeline described in Section~\ref{sec:pipeline}. The routines accept a reference image class which keeps track of subtleties such as the nominal sub-pixel position of the reference central location. I also implemented subroutines to generate a reference image using any axisymmetric function defined by the user. Alternatively the reference image may be externally generated and loaded into the pipeline program. The reference image may be generated either at the pixel scale of the original data, or at finer pixel scales for cross-correlation with data that has been interpolated. I implemented a bicubic resampling algorithm via convolution with a smoothing kernel \citep{Park1983}, as previously used by \cite{Law2007}, the algorithm has been implemented anew and optimized for high computational speed. The interpolation algorithm of \citeauthor{Park1983} has also been found to be effective by \cite{Diolaiti2000}. Optionally, parabolic fitting of the 
cross-correlation maxima may also be applied. 


\section{Simulation methods used for testing}
\label{sec:reg_sims}
In order to test various frame registration methods under controlled conditions I used the packages of code described in chapter~\ref{chap:lucky_AO} to perform end-to-end Monte Carlo simulation of a lucky imaging system representative of the system used at the Nordic Optical Telescope in summer 2009, with a telescope primary aperture diameter of 2.5m and central obscuration diameter of 0.5m. The code was used to produce time-evolving short exposure images, sampled at 
%
four times
the Nyquist rate and at 0.05 second intervals. A seeing width of 0.5 arcseconds (as observed at 500nm, which is the standard reference wavelength for measurements of seeing) was chosen as representative of good, but not exceptional, observing conditions for La Palma in the summer months. The atmospheric turbulence layers were simulated according to the model detailed in table~\ref{tab:atmos_model_layers}. An observation wavelength of 770nm was simulated, representative of the central wavelength of the SDSS \textit{i'} band filter.

The generated focal-plane images were normalised such that the peak pixel value of each short exposure represented the Strehl ratio of that frame.
After using the peak pixel to obtain a precise estimate of the Strehl and PSF peak location in each frame (hereafter referred to as the ``true'' values), the images were rebinned to represent pixel angular widths corresponding to those used in the real lucky imaging observations, specifically pixel scales of $\sim32$ and $\sim 96$ milliarcseconds per pixel. These correspond to Nyquist sampling at the observation wavelength, and under-sampling by a factor of 3, respectively. 

Next, the stochastic processes of photon arrival and detector response were simulated to produce realistic images. Various source photon flux levels were simulated. Transfer CIC events were simulated at a signal level equivalent to 0.05 photo-electrons per pixel per frame and CICIR events were simulated at an occurrence rate of 0.04 per pixel per frame, representative of calibrated levels from the real detector (cf. Section~\ref{sec:internal_signal}). The background sky flux level was assumed to be negligible in comparison (equivalent to dark time observing). 

The realistic datasets were then processed by the lucky imaging pipeline described in chapter~\ref{chap:data_reduction}. While reduction steps involving calibration frames are not applied to simulated data, all other aspects of the pipeline reduction process are identical to real data.

\section{Testing registration algorithms}
\subsection{Results from comparison via simulation}
\begin{figure}[htp]
\begin{center}

\subfigure[Typical standard deviation in Monte Carlo realisations]{	
	\includegraphics[width=0.45\textwidth]{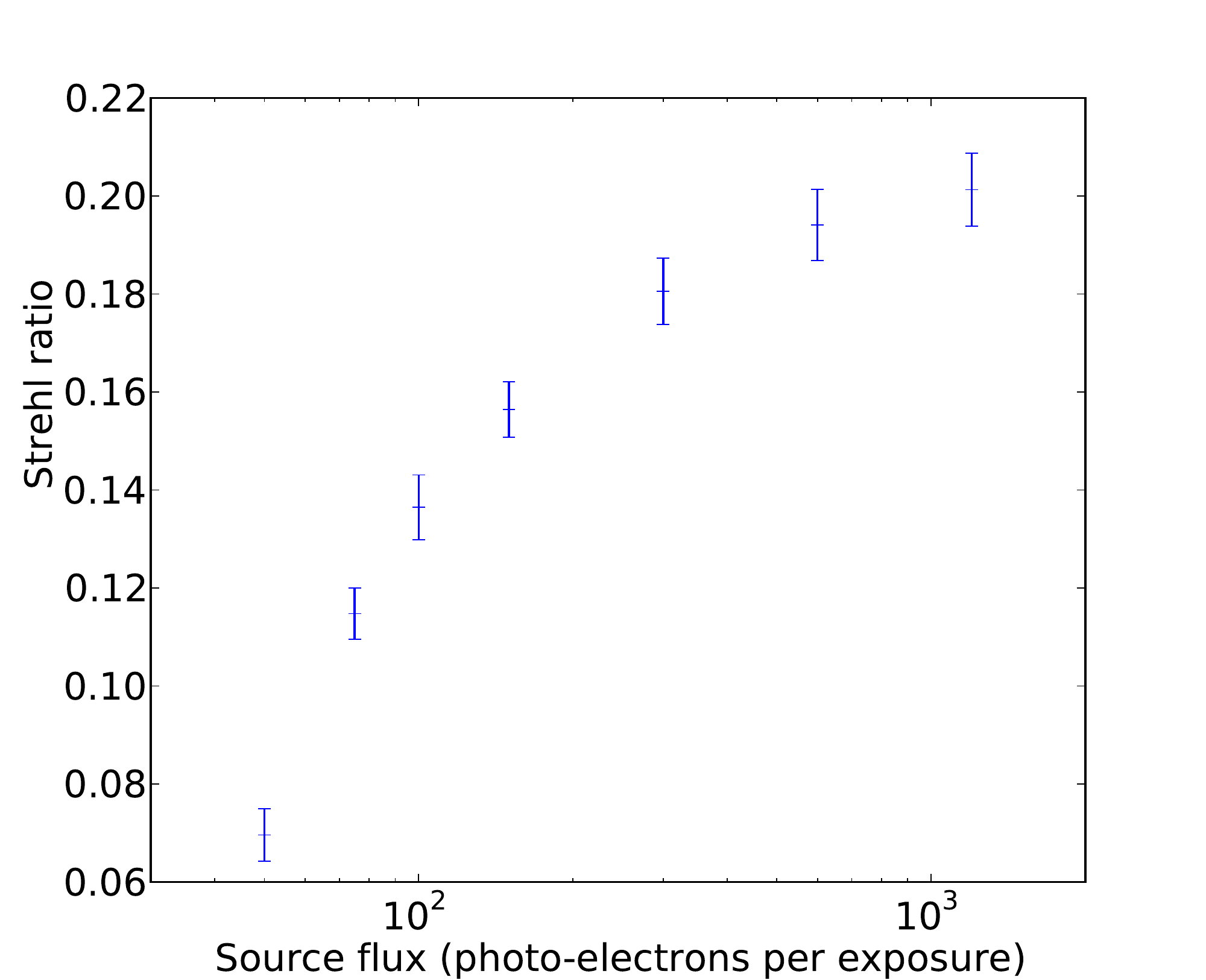}
	\label{subfig:sim_std_dev}
}
\subfigure[Histogram of Strehl ratios at source flux of 150 photo-electrons per frame]{
	\includegraphics[width=0.45\textwidth]{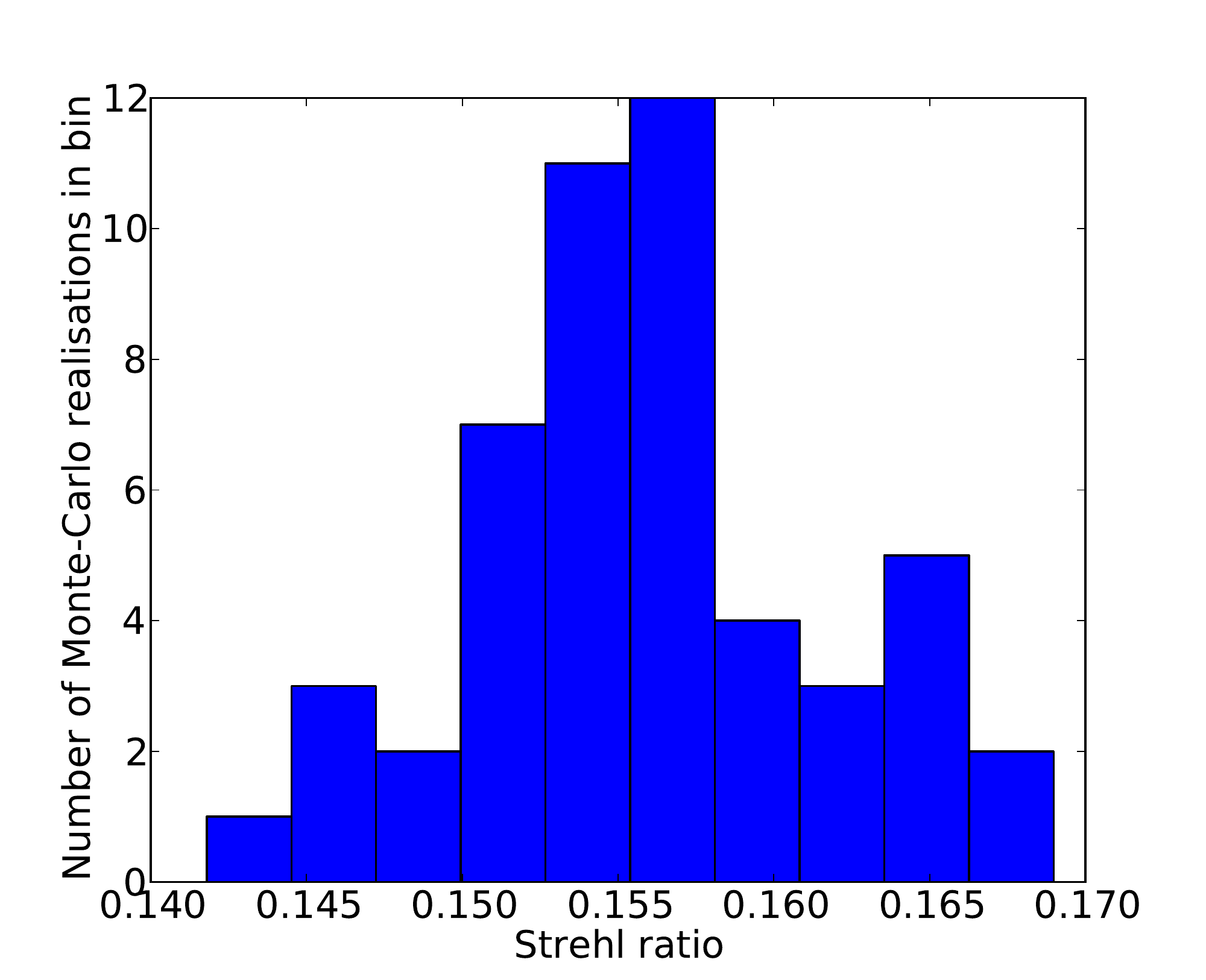}
	\label{subfig:sim_err_hist}
}
\caption[Typical range in simulated results]{
	\subref{subfig:sim_std_dev}: A plot depicting Strehl ratios obtained at 10\% frame selection from images at Nyquist sampling pixel widths, using an Airy cross-correlation template with interpolation to 4x the original resolution. Error bars depict standard deviation in the results from a range of Monte-Carlo realisations (midpoints mark mean value). \\
	\subref{subfig:sim_err_hist}: A histogram displaying the range of results for the point at source flux of 150 photo-electrons per frame in \subref{subfig:sim_std_dev}.
}
\label{fig:simulation_range}
\end{center}
\end{figure}

The chain of simulated data generation and processing steps was controlled via scripts written in Python, to allow for feasible variation, processing and analysis of many different parameters. ten atmospheric simulations of 1800 frames each (90 seconds worth of data) were generated. For each simulated detector pixel width and source light level, these datasets were processed with 5 different Monte-Carlo realisations to obtain a good sampling of the photon and detector noise processes. \corr{The simulated exposures were then processed} using the pipeline, and the output image qualities estimated using the image analysis libraries I developed for the pipeline described in chapter~\ref{chap:data_reduction}. \corr{A typical distribution of the simulated results is plotted in figure}~\ref{fig:simulation_range}. \corr{The simulated Strehl ratios for any given parameter set can be seen to have reasonably small standard deviation and range compared to differences between parameter sets, and appear to cluster around 
a single mode. Hence error bars are suppressed in later plots for clarity.}

Initial investigations with simulated data tested the relative performance of different cross-correlation algorithm implementations. Extensive investigation of the various parameter combinations were undertaken. The results may be summarised as follows:
\subsubsection{Interpolation method}
\begin{figure}[htp]
\begin{center}
 \includegraphics[width=0.8\textwidth]{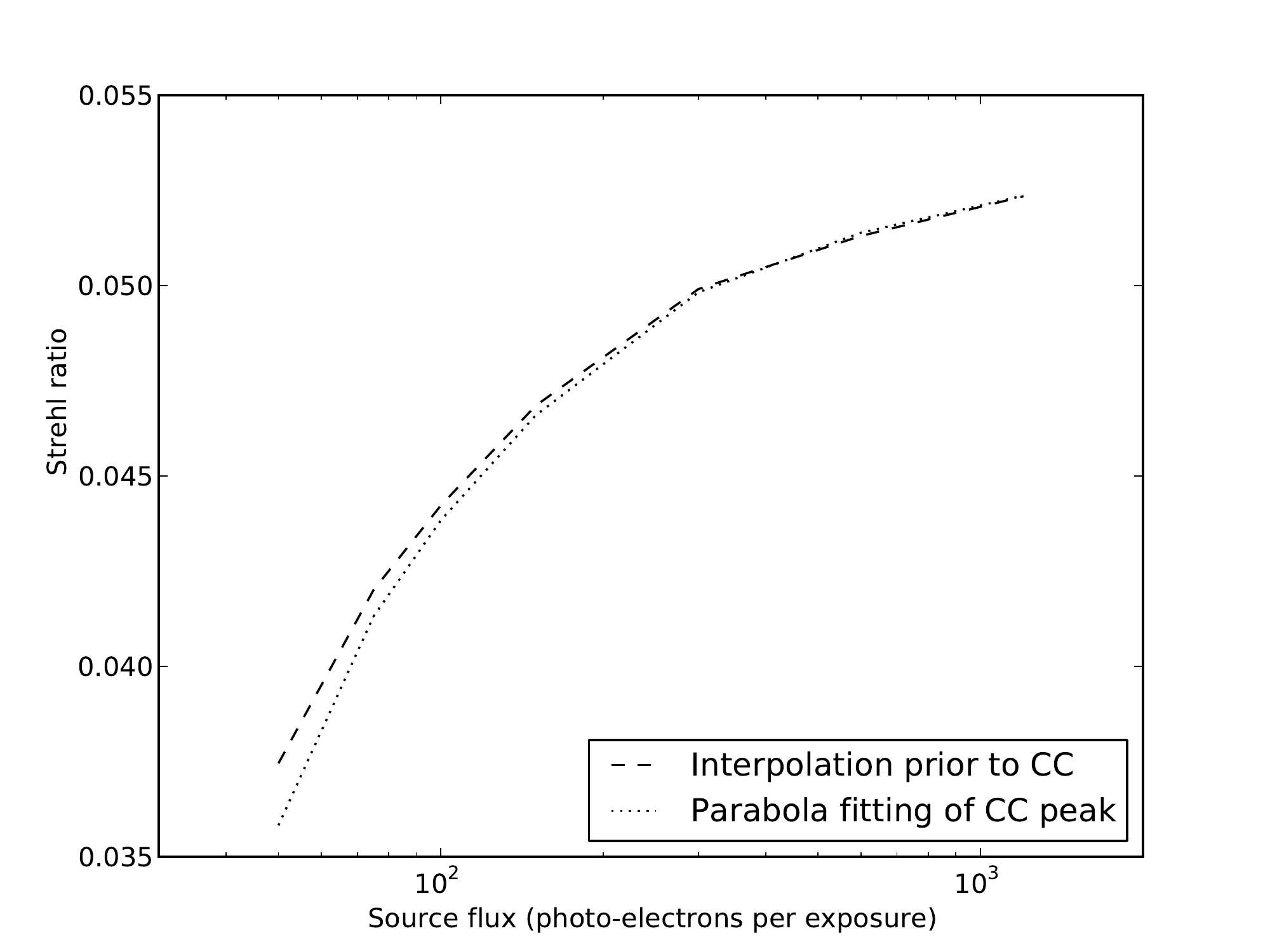}
\caption[Comparison of interpolation methods]{Comparison of interpolation methods employed for the cross-correlation (CC) algorithm. Plot depicts Strehl ratios obtained at 50\% frame selection from under-sampled images (at pixel widths 3 times Nyquist sampling width). The difference in resulting Strehl ratio is small to negligible. Note that a combined approach of interpolation followed by parabola fitting was tested, again with negligible improvement. Further details in text. }
\label{fig:interpolation_comparison}
\end{center}
\end{figure}

The simulations indicate that, so long as one of the described interpolation methods is employed --- e.g. interpolate data and then cross-correlate at interpolated positions, or cross-correlate data and then fit a parabola about the maximum --- the differences between them are small (Figure~\ref{fig:interpolation_comparison}). At high and intermediate flux levels the difference in resulting Strehl ratio is negligible. This null result is of some use, since the interpolation process and cross-correlation at the resulting higher pixel resolution incur a significant penalty in computational speed. On workstations with less powerful CPUs which are not performance limited by the data input/output, switching to parabola fitting may result in a useful speed boost. Interestingly, at the lowest light levels (source flux of 50 photo-electrons per frame) the interpolation process does provide a small but non-negligible benefit (relative increase of 10\% in Strehl ratio, although in absolute terms this is only $\sim0.5\%
$ at most), presumably due to the smoothing effect of the interpolation process.

\subsubsection{Cross-correlation reference}
\begin{figure}[htp]
\begin{center}
 \includegraphics[width=0.8\textwidth]{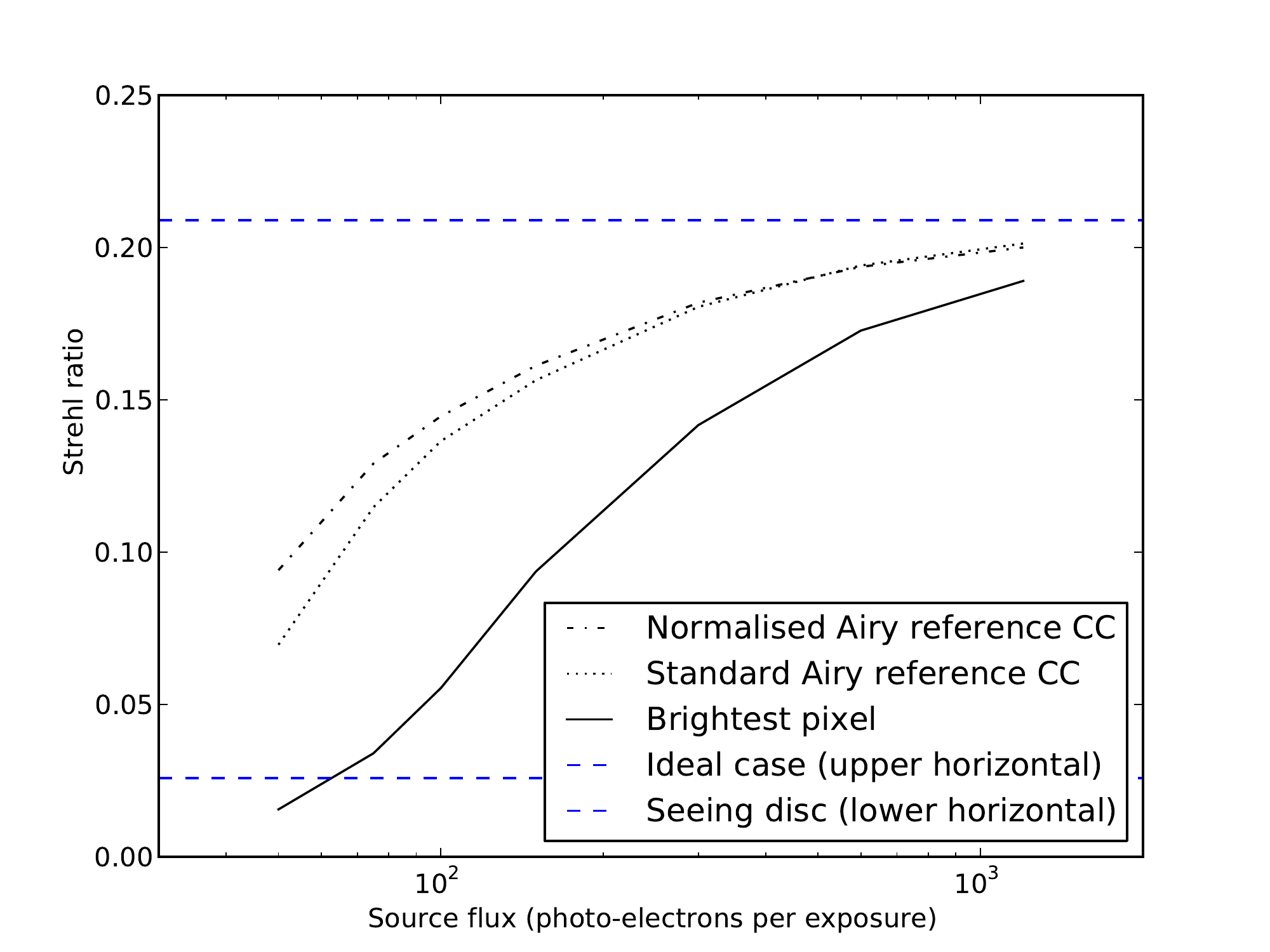}
\caption[Effect of normalising CC reference]{Normalising the cross-correlation reference to account for non-uniform noise significantly improves Strehl ratio at low light levels, particularly if the PSF is Nyquist sampled as depicted here (and hence has a lower signal level in each pixel). At a source flux of 50 photo-electrons per frame Strehl ratio is increased from 7\% to 9.4\%, a relative improvement of about one third. Plots are shown for Strehl ratios obtained at 10\% frame selection. The Strehl ratios obtained using a simple `brightest pixel' registration technique are plotted for comparison. Also plotted (upper horizontal) is the Strehl ratio obtained from 10\% frame selection after analysing the intensity images \corr{(i.e. those which represent the true speckle pattern, prior to pixellation and addition of detector noise)}, representing the ideal case of a very bright source. The lower horizontal dashed line represents the Strehl ratio obtained from a simple average of the frames to obtain the long 
exposure average (seeing disc).}
\label{fig:reference_comparison}
\end{center}
\end{figure}

The most interesting results came from variation of the cross-correlation reference. Normalising the reference to account for a non-uniform noise distribution as described in Section~\ref{sec:reg_methods} significantly improves performance at low light levels, as seen in Figure~\ref{fig:reference_comparison}.

\subsection{Comparison using real data}
\begin{figure}[htp]
\begin{center}

\subfigure[Left: Image reduced from data registered using Airy cross-correlation reference. Right: Data registered using normalised CC reference. Images are shown at same colour scale and stretch.]{	
	\includegraphics[width=0.8\textwidth]{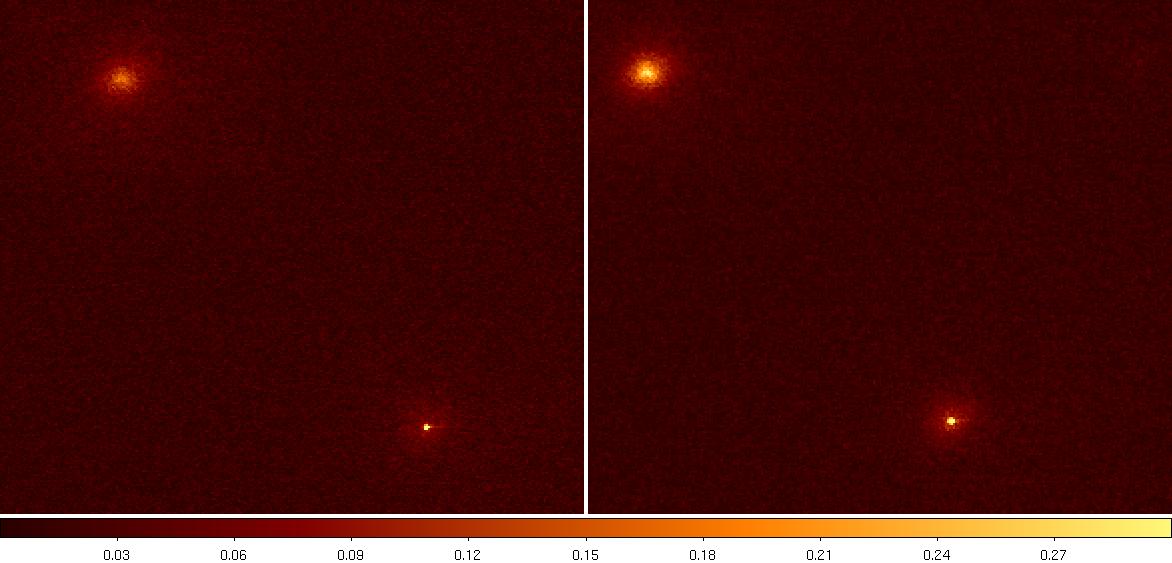}
	\label{subfig:CC_real_img}
}
\\
\subfigure[Cross sections of the upper left star from the images above]{
	\includegraphics[width=0.6\textwidth]{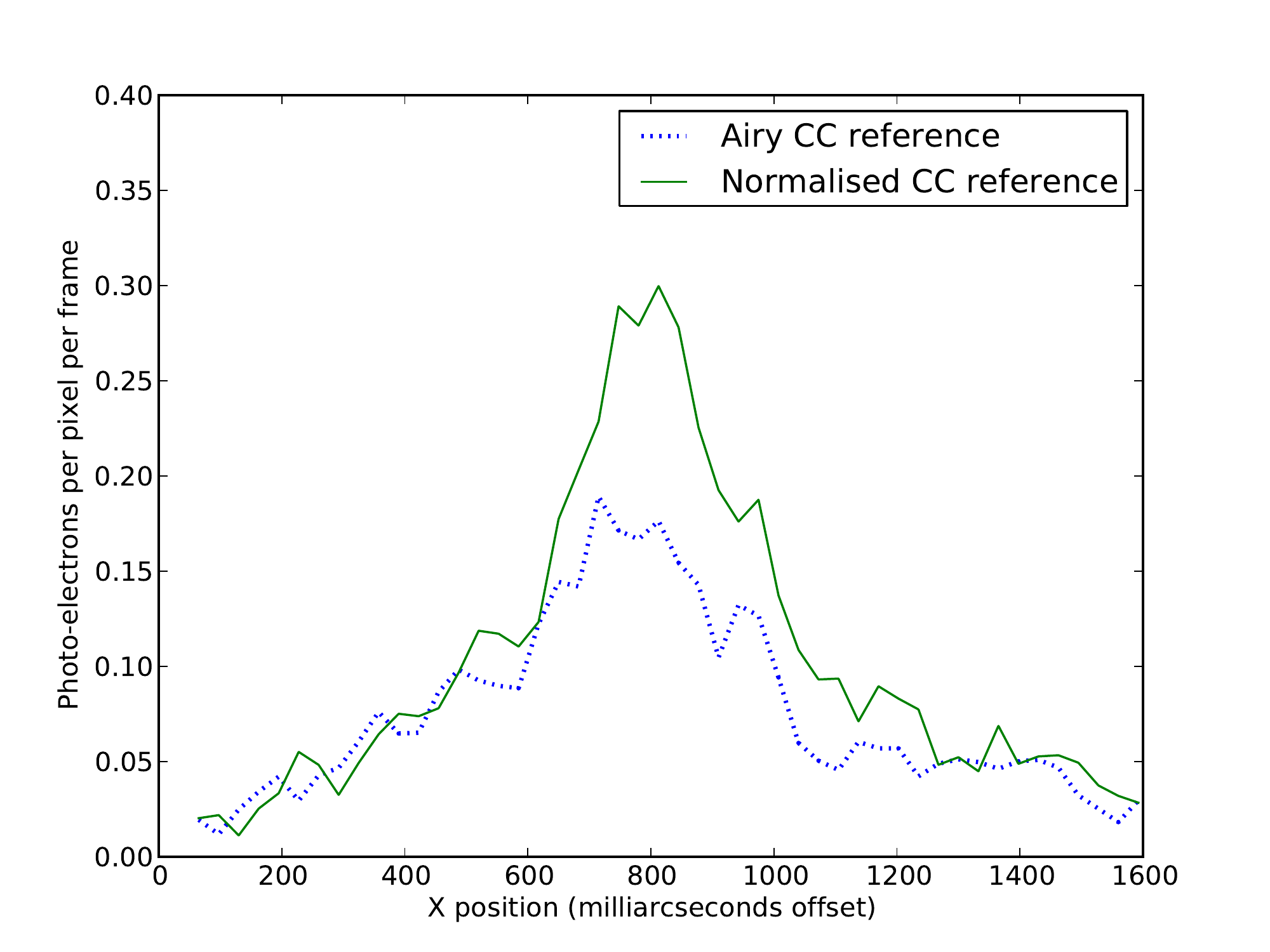}	
}
\caption[Improvement in Strehl ratio in real data through use of normalised CC reference]{
Improvement in Strehl ratio in real data through use of the normalised cross-correlation reference. Top image depicts a cropped field of view from observations of a binary star candidate. Seeing width was around 0.8 arcseconds and the pixel width was 32.5 milliarcseconds. The lower right source in the field of view has a flux of around 50 photo-electrons per frame (\textit{i'} band apparent magnitude $\approx$ 18 ) and is utilised as the guide star. 
The plots depict cross sections from the star in the upper left of the images, about 7.5 arcseconds from the guide star. Use of the normalised cross-correlation reference results in a FWHM of around 0.4 arcseconds, and a Strehl ratio improvement of around 70\% compared with the Airy cross-correlation reference.
}
\label{fig:faint_gs_example}
\end{center}
\end{figure}

The trends observed in simulated results were verified using real data. Various fields were reduced using the standard and normalised cross-correlation references, and for faint guide stars use of the normalised reference does indeed result in improved image quality. Figure~\ref{fig:faint_gs_example} illustrates one particularly nice example, where a Strehl ratio improvement of around 70\% is observed.

\section{Image formation processes}
In this chapter thus far, I have only considered the Strehl ratio in reduced images as a metric of the frame registration and image quality estimation process. This is often the single most useful number when summarising image quality, and gives an easy point of comparison for different frame registration algorithms.
However, having settled upon an optimum registration algorithm, it is of interest to explore the detailed error budgets applicable under different observing conditions. If we can characterise and predict the errors for any particular set of observing conditions then we may predict the image quality of the lucky imaging results, and adjust our observing strategy accordingly. If this could be done reliably it would be invaluable in both designing and undertaking large scale lucky imaging observing campaigns. 

First however, we require a model of the image formation process. If we neglect the direct effects of pixellation, then the image formation process for lucky imaging can be broken down as follows. First, I consider the average PSF resulting from perfect alignment of many instantaneous PSF speckle patterns upon their position of maximum intensity (brightest speckle). The resultant Strehl ratio will simply be the mean of the Strehl ratios present in each instantaneous PSF. Fortunately, a model for the instantaneous Strehl ratio probability distribution function exists. Second, I consider the Strehl misestimation, since this affects our frame selection process. Finally I consider the deleterious effect of frame registration positional error.

\subsection{Instantaneous Strehl ratio probability distribution function}
 The instantaneous Strehl ratio probability distribution function (PDF) is equivalent (after multiplication by some constant normalisation factor) to the PDF of the peak intensity in the focal plane, which has been investigated under the term ``speckle statistics.'' \cite{Fitzgerald2006} give an excellent summary of the speckle statistics PDF originally derived in \cite{Goodman1975} in the context of laser speckles, and later applied in the context of astronomical adaptive optics by \cite{Canales1999a}. The intensity PDF in the context of uncorrected or partially-corrected seeing is well modelled by a modified Rician (MR) distribution: 
\begin{equation}
 P_{MR}(I) = \frac{1}{I_{s}} exp \left( \frac{-I + I_{c}}{I_{s}} \right) I_{0}  \left(  2 \frac{ \sqrt{II_c}}{I_s}  \right) 
\label{eq:mod_rician}
\end{equation}
where $I$ is the intensity, and $I_{c}$,$I_{s}$ represent the coherent and speckle wavefront amplitude components respectively. It is possible to derive values for the parameters of the function purely from knowledge of the atmospheric seeing conditions and the number of Zernike modes corrected by any adaptive optics present, however this assumes perfect correction of the Zernike modes \citep{Canales1999a}. Since correction at higher Zernike modes in particular tends to be far from perfect, the actual intensity PDF is often better described by a semi-empirical distribution obtained by fitting the parameters $I_{c}$, $I_{s}$ to observed data. \citeauthor{Fitzgerald2006} present just such an experimental verification of the model PDF for \corr{short exposures obtained on-sky with partially corrected adaptive optics}.

\cite{Gladysz2010b} point out that the modified Rician model is not valid in the PSF core of high order, high Strehl-ratio adaptive optics systems, as made evident by the fact that it has positive skew for all values of the parameters, whereas the intensity distribution in a high Strehl system has negative skew. However, it is an effective model of both conventional lucky imaging and hybrid lucky imaging adaptive optics systems --- in high Strehl ratio regimes lucky imaging produces rapidly diminishing returns and so is unlikely to be applied to such systems.

\begin{figure}[htp]
\begin{center}
 \includegraphics[width=0.7\textwidth]{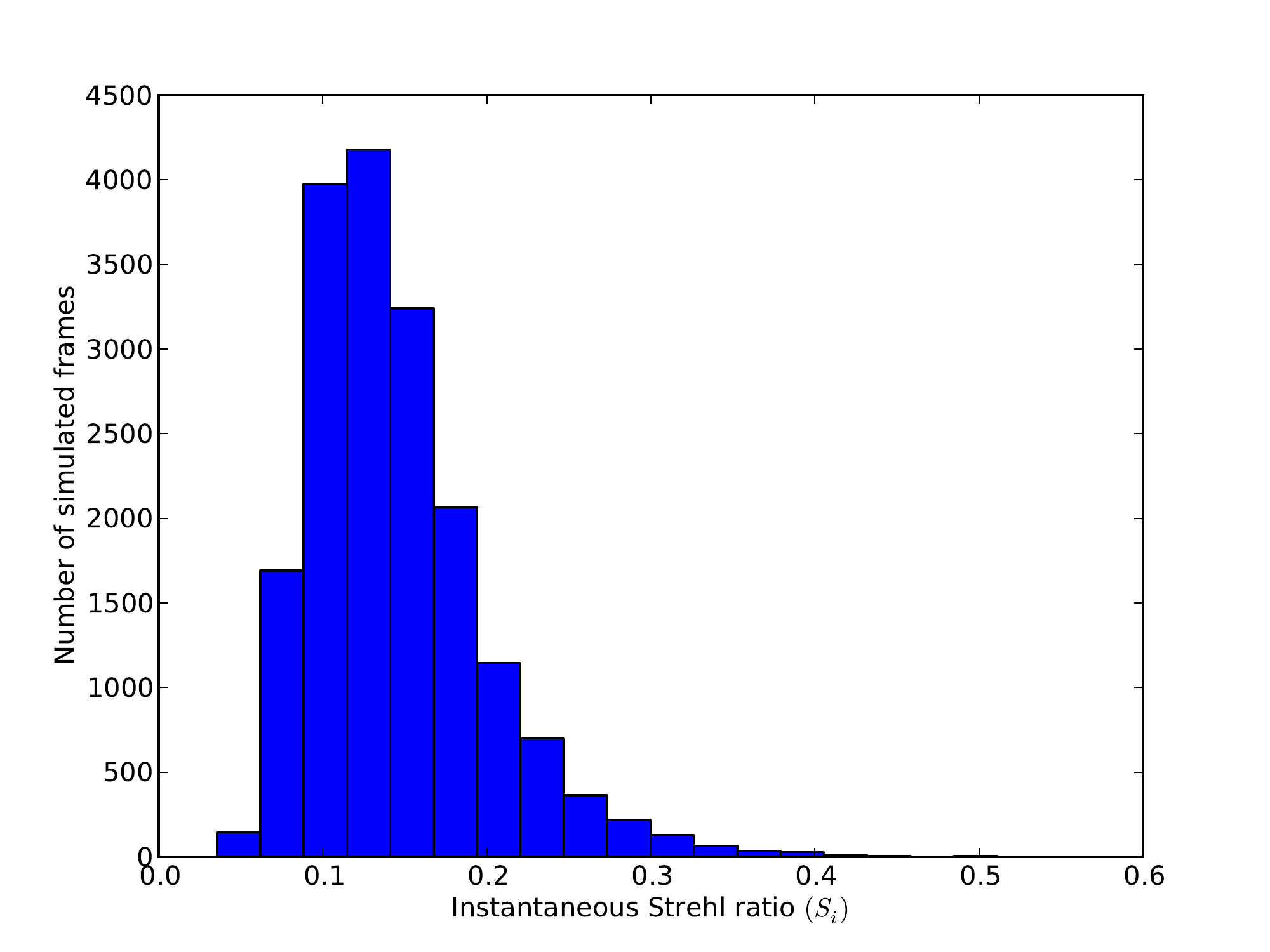}
\caption[Strehl ratio histogram]{Histogram of the Strehl ratios (true values, not estimated) observed in the 18000 simulated short exposures.
}
\label{fig:strehl_distribution}
\end{center}
\end{figure}

Note that when using the modified Rician intensity PDF to estimate the performance of conventional lucky imaging, it is appropriate to assume a phase correction of the first 3 Zernike modes (piston and tip-tilt), since we are assuming perfect re-centring upon the brightest speckle. Figure~\ref{fig:strehl_distribution} illustrates the Strehl ratio histogram observed in the simulated data.

\subsection{Strehl estimation error}
With a model for the instantaneous Strehl PDF, we may begin to estimate the performance of lucky imaging under favourable conditions, by considering the mean Strehl ratio attained by selecting a certain portion of frames with the highest Strehl ratios, e.g. the mean Strehl ratio of the top 10\% of the PDF. However, applying this in practice requires a reliable estimate of the Strehl ratio in each short exposure, at least via a proxy measure such as the cross-correlation maxima. Errors in the estimation process will lower the Strehl ratio in the final image if frame selection is applied. A reliable estimator of Strehl ratio in short exposure images is also of interest when testing models of the atmospheric turbulence and the speckle statistics that result, since we may begin to investigate the short-timescale properties of the models and real data.

\begin{figure}[htp]
\begin{center}
\subfigure[Noisy data, source flux of 150 photo-electrons per frame.]{
	\includegraphics[width=0.3\textwidth]{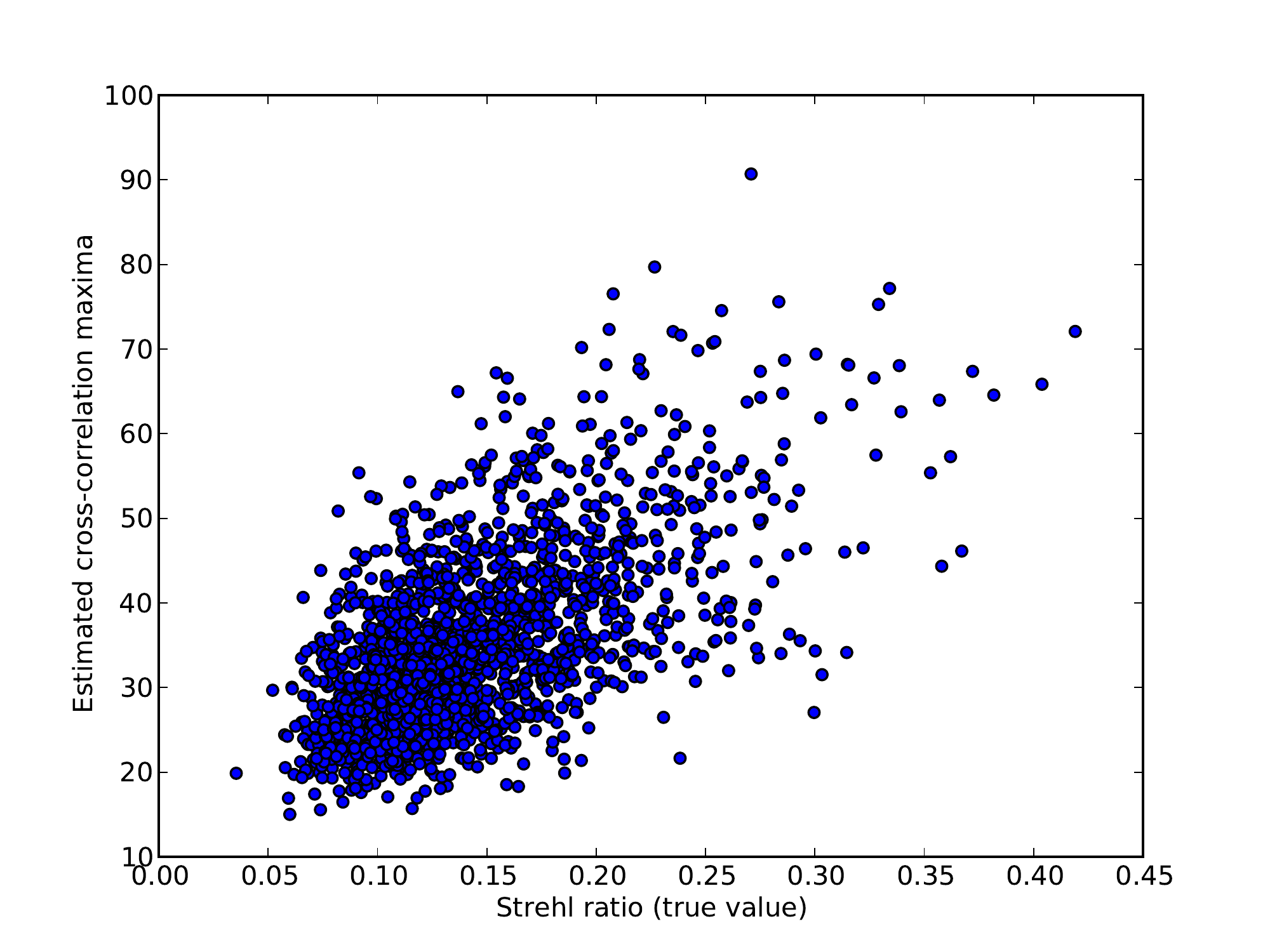}
	\label{subfig:150}
}
\subfigure[Noisy data, source flux of 600 photo-electrons per frame.]{
	\includegraphics[width=0.3\textwidth]{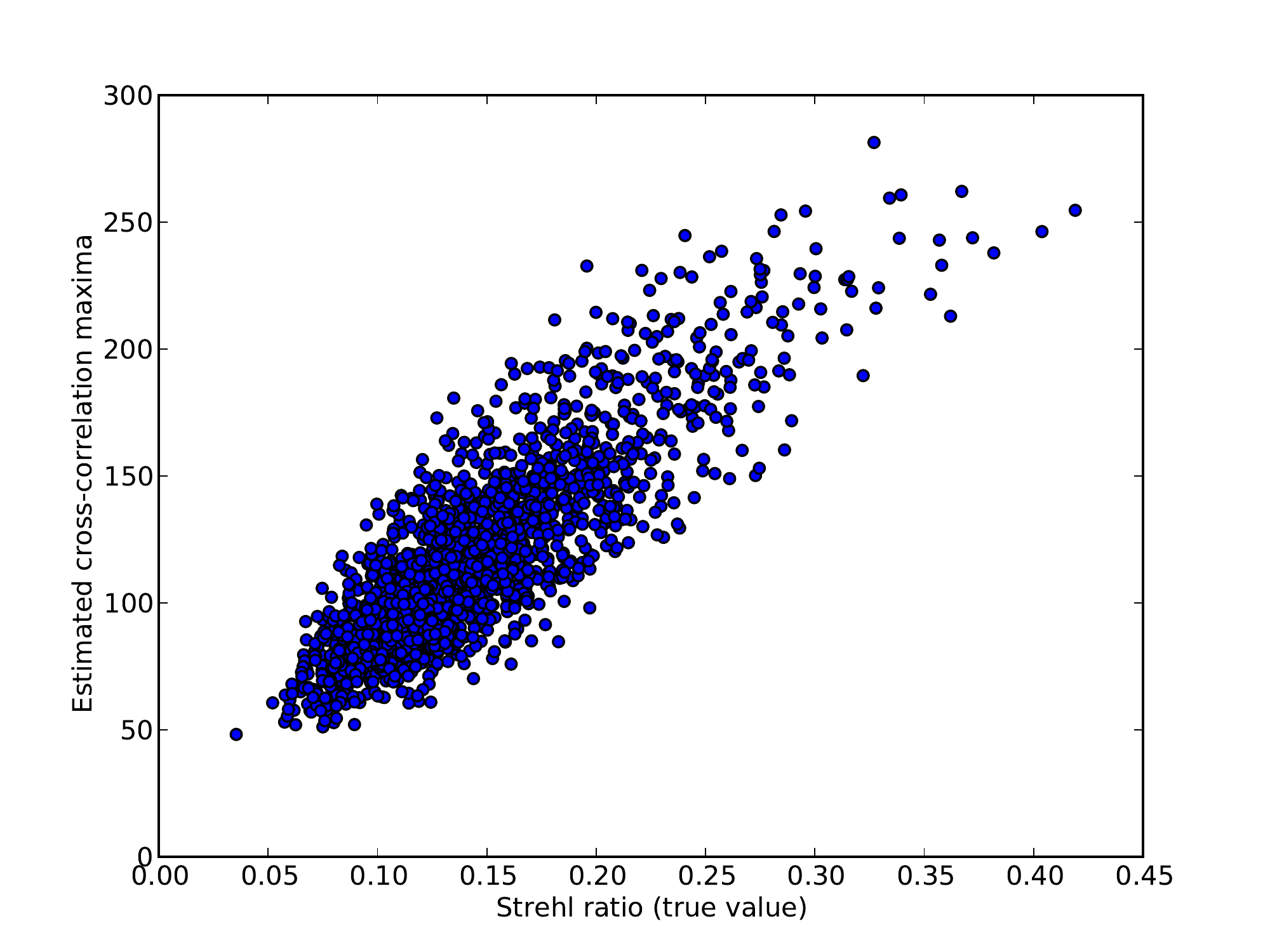}
	\label{subfig:600}
}
\subfigure[Noiseless (pixellated only) data.]{
	\includegraphics[width=0.3\textwidth]{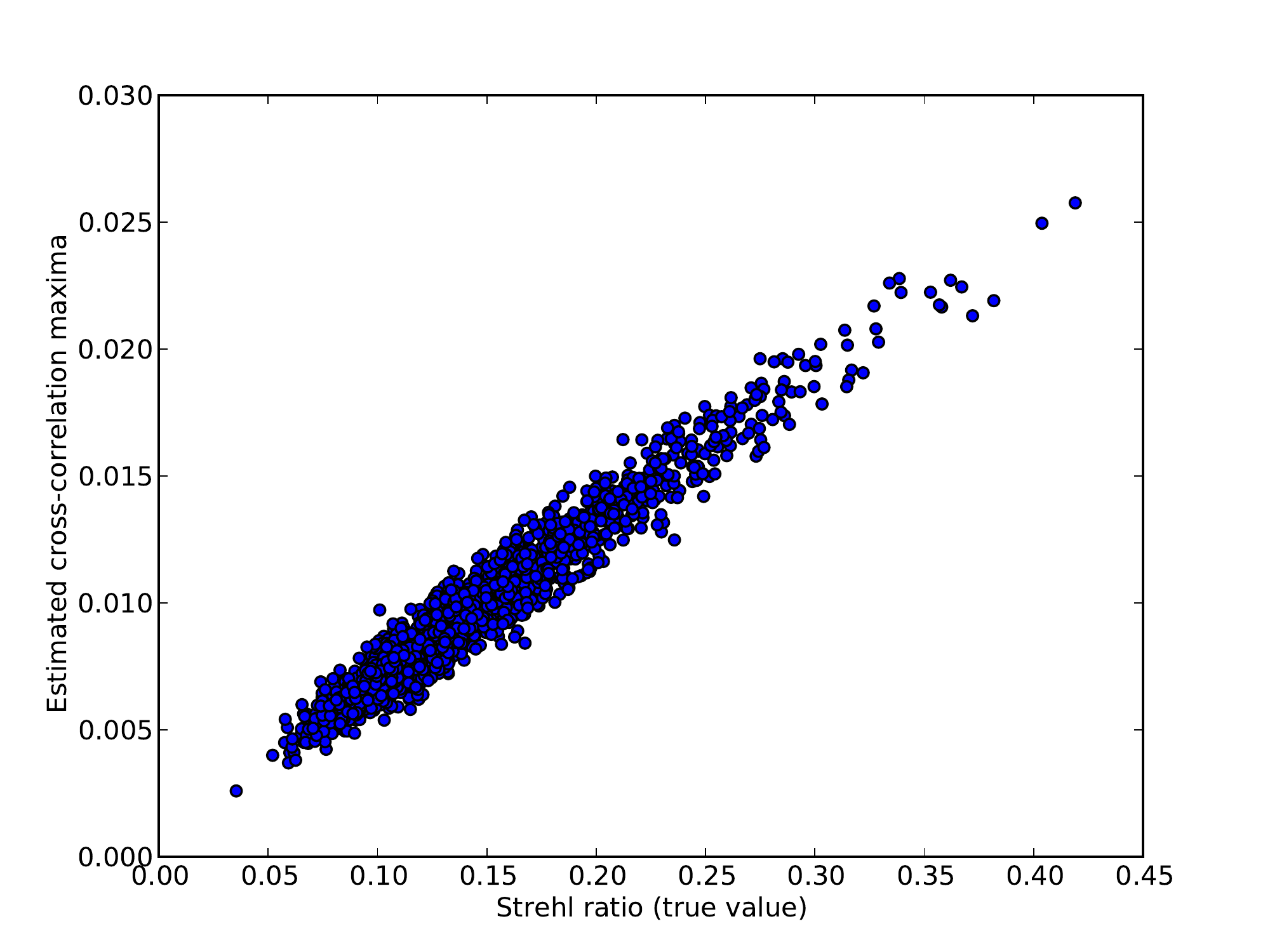}
	\label{subfig:clean}
}
\caption[Scatter plots of Strehl ratio estimates]{Example scatter plots of Strehl ratio estimates against the true Strehl ratio in each frame, for simulated datasets with source flux of 150 photons, 600 photons, and noiseless (but still pixellated) data. Pearson correlation coefficients for these particular datasets are 0.62, 0.86, and 0.97 respectively. There are 1800 simulated short exposures in each dataset.
}
\label{fig:strehl_est_scatter}
\end{center}
\end{figure}

\begin{figure}[htp]
\begin{center}
 \includegraphics[width=0.7\textwidth]{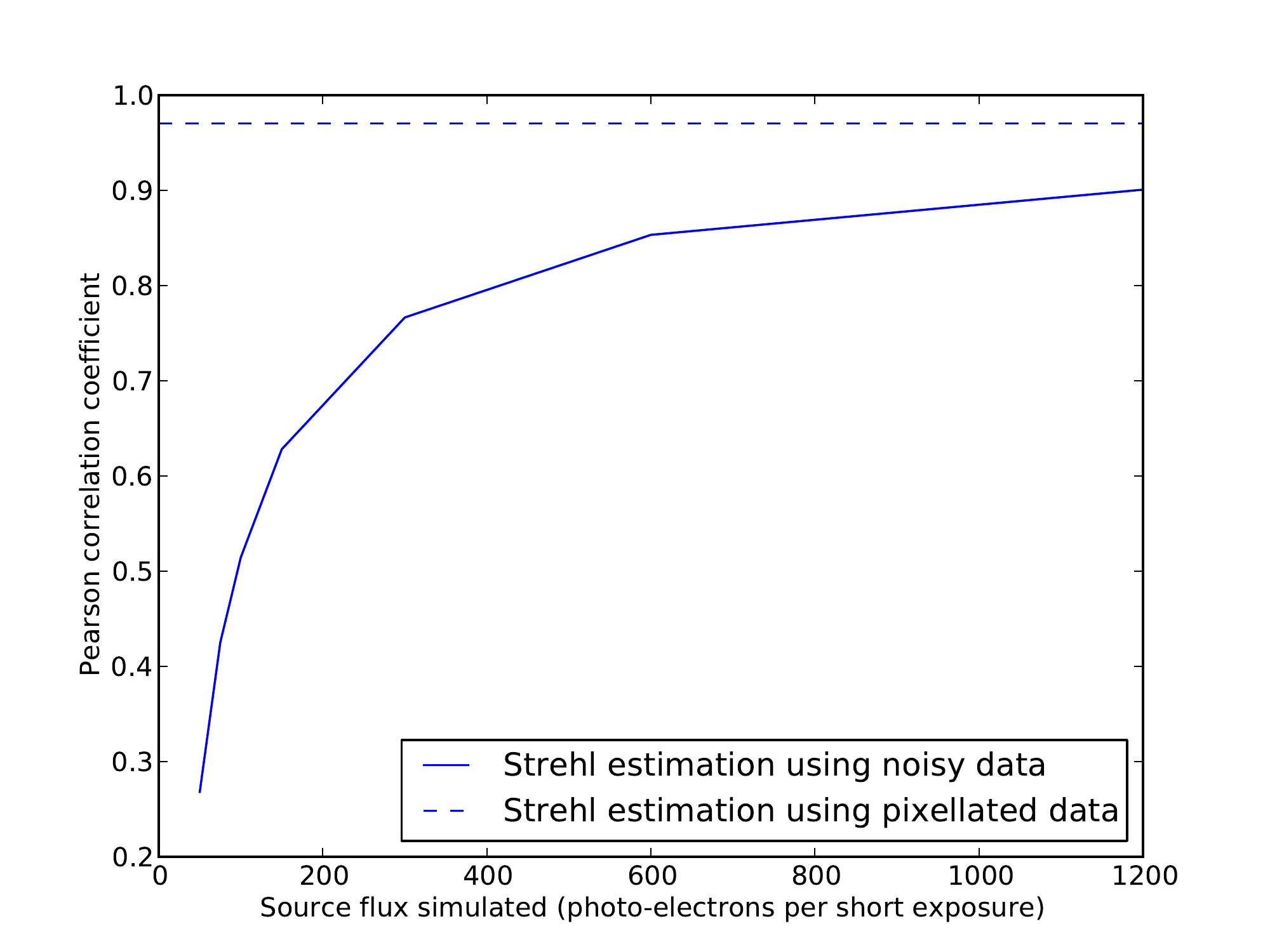}
\caption[Correlation values between estimated and true Strehl ratio]{
Correlation between estimated and true Strehl ratio, when estimated using the cross-correlation algorithms, at various signal levels. Correlation values are derived from a collation of estimated and true Strehl ratios in all simulated datasets --- some 90000 frames in total --- for each simulated source flux level. Also plotted is the correlation resulting from application of the cross-correlation algorithm to the noiseless (pixellated only) data, representing ideal conditions of excellent signal-to-noise ratio. The results suggest that, at least for this particular set of simulated conditions, attempting to perform frame selection using any guide source with a signal level lower than say, 100 photo-electrons per frame is likely to produce diminishing returns compared to simply utilising 100\% of the frames.
}
\label{fig:strehl_est_correlation}
\end{center}
\end{figure}

\begin{figure}[htp]
\begin{center}
 \includegraphics[width=0.7\textwidth]{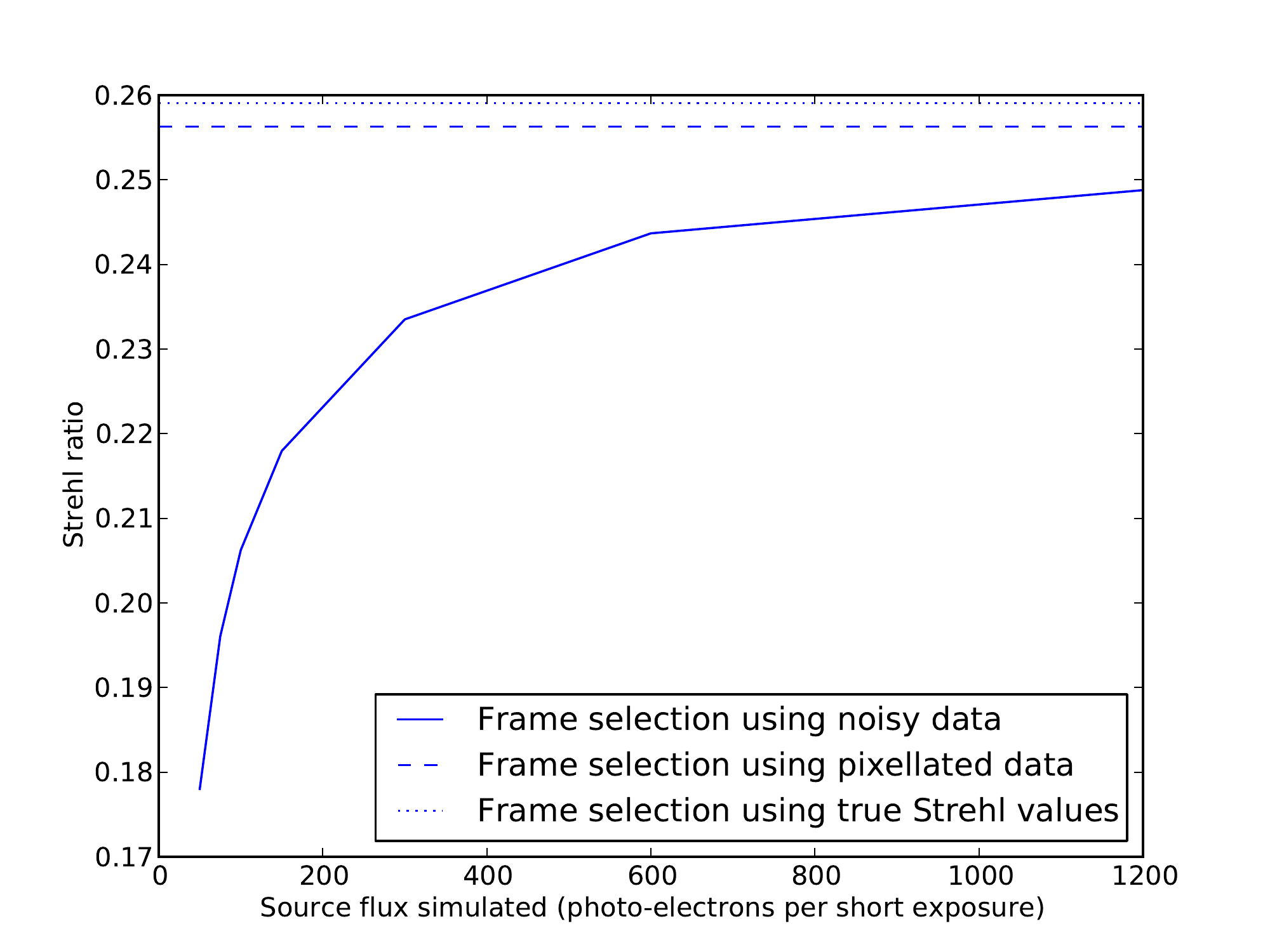}
\caption[Mean Strehl ratio of selected 10\%]{\corr{Mean \emph{true} Strehl ratio of the top 10\% of frames, selected according to their \emph{estimated} Strehl. This partially accounts for the reduction in image quality when fainter guide stars are used. The Strehl ratio in the reduced combined image is lower still, due to position estimation error in the frame alignment process.}
}
\label{fig:actual_selected_strehl}
\end{center}
\end{figure}

An investigation of Strehl estimation errors in real data is deferred to future work. For the time being I present results from the simulations described in Section~\ref{sec:reg_sims} as an illustration of the typical errors that might be expected. Figures~\ref{fig:strehl_est_scatter} and~\ref{fig:strehl_est_correlation} give some idea of the scatter in estimated versus true Strehl ratio, while Figure~\ref{fig:actual_selected_strehl} plots the resultant mean Strehl ratios resulting from the 10\% selection cut-offs based on the cross-correlation quality estimates at various source photon fluxes.

\begin{figure}[htp]
\begin{center}
 \includegraphics[width=0.7\textwidth]{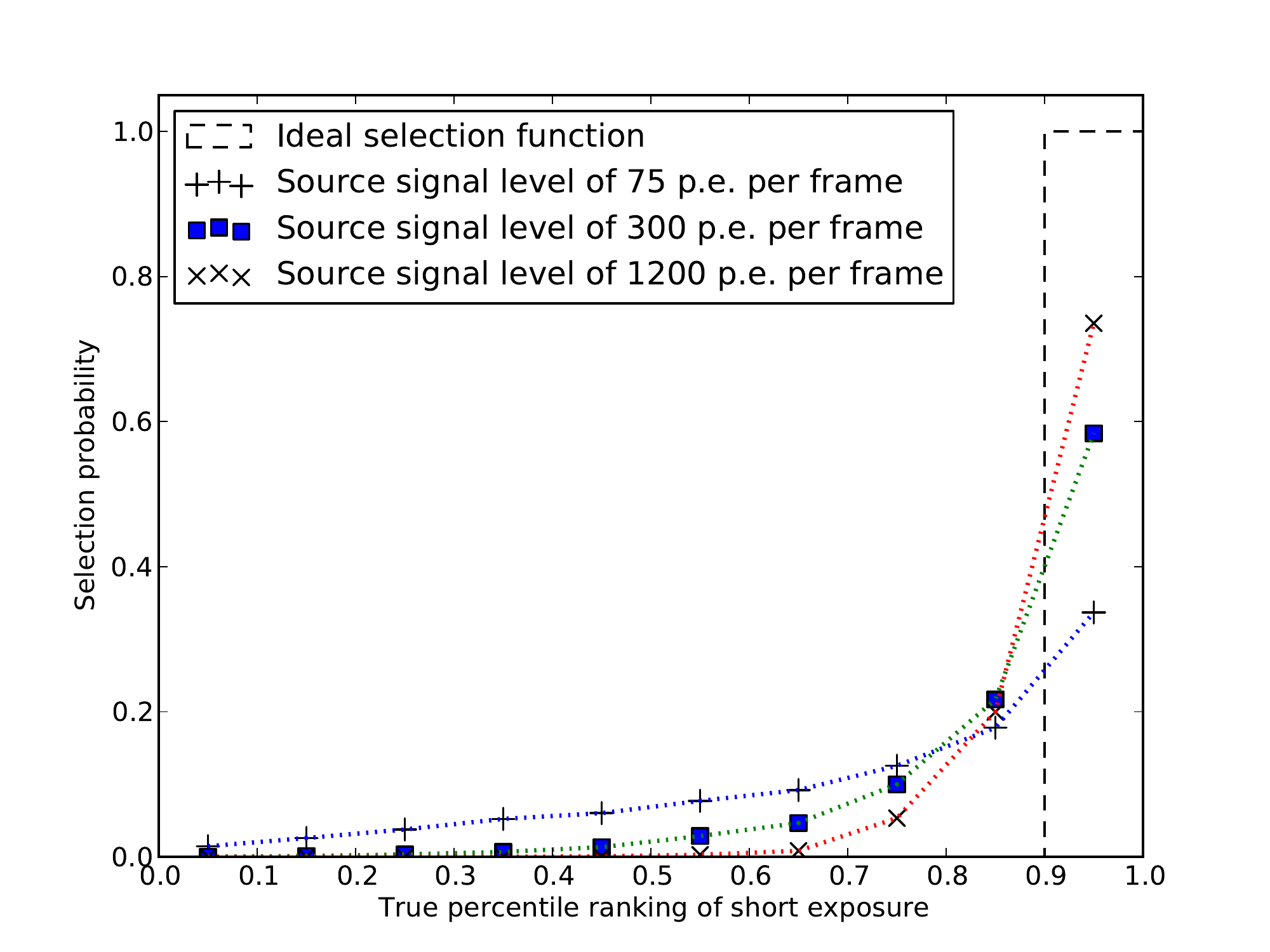}
\caption[Frame selection probability functions]{Selection probability functions (circa equation \ref{eq:selection_func_ideal}) empirically determined from the simulated data, for a desired selection of the best 10\% of short exposures. \corr{Plots are for guide stars of different brightness, with fluxes stated in photo-electrons (p.e.) per frame.} The ideal selection function is a step function (dashed line in plot), \corr{but the noisy Strehl estimation process causes erroneous selection of frames of lower Strehl ratio.}
}
\label{fig:selection_functions}
\end{center}
\end{figure}

We may characterise the effect of Strehl estimation error upon the image formation process by defining a selection probability function, which I shall denote $\chi(S_i)$, where $S_i$ is the true Strehl ratio in a short exposure. This represents the probability that a short exposure with true Strehl ratio $S_i$ will be selected and drizzled to produce the final reduced image. For example, we may wish to construct a reduced image using the best 10\% of the short exposures --- i.e. using all images where $S_i > S_{i 90}$, $S_{i 90}$ denoting the 90th percentile short exposure Strehl ratio. Then the ideal selection function is simply a step function:
\begin{equation}
\chi_{ideal}(S_i) = \begin{cases} 0, & S_i < S_{i90}, \\ 1, & S_i \ge S_{i90}, \end{cases} 
\label{eq:selection_func_ideal}
\end{equation}
However, due to Strehl estimation error the transition from a probability of 0 to 1 is much broader for the selection function resulting from quality estimation with real data. As can be seen in Figure~\ref{fig:selection_functions}, at very low source signal levels the frame selection is not very much better than a random selection.

\subsection{Frame registration positional error}
Once we have determined the Strehl ratio of the short exposures which will be used to produce the reduced image, all that's left to be modelled is the alignment and combination process. Neglecting the direct effects of pixellation, we may consider the reduced image resulting from registration, alignment and summing of short exposures. 

Designating the focal-plane co-ordinates with the two-vector $\mathbf{x}$, and the intensity distribution for each short exposure (speckle pattern) as $I(\mathbf{x} ; \quad t)$, with peak intensity at $\mathbf{x_p}(t)$ then we may consider the sum image resulting from perfect alignment of the short exposures, which I shall call the `zero-error' sum PSF with light intensity in the focal plane described by the function $I_0(\mathbf{x})$:
\begin{equation}
 I_0(\mathbf{x}) = \sum_t I(\mathbf{x} - \mathbf{x_p}(t) ; \quad t)\,.
\end{equation}

The effects of pixellation, photon shot noise and detector noise will cause error in the frame registration process, represented by the two-vector $\boldsymbol\epsilon$ , such that the estimated position of peak intensity is 
\begin{equation}
 \mathbf{x_e}(t) = \mathbf{x_p}(t) + \boldsymbol\epsilon(t)
\end{equation}
resulting in the mean image with registration error described by the function
\begin{equation}
 I_e(\mathbf{x}) = \sum_t I(\mathbf{x} - \mathbf{x_p}(t) -  \boldsymbol\epsilon(t);\quad t)\,.
\end{equation}

Now, if we consider the two-dimensional error probability distribution function $E$, such that 
\begin{equation}
P\big(\epsilon = \mathbf{x}\big) = E(\mathbf{x})
\end{equation}

then over many thousands of frames, the summing of images with error offsets $\boldsymbol\epsilon$ determined by the PDF $E$ is equivalent to convolution by the error distribution, and so we get the result:
\begin{equation}
 I_e(\mathbf{x}) =  I_0(\mathbf{x}) \ast E(\mathbf{x})
\end{equation}
where $\ast$ represents spatial convolution. 

If $E$ may be reasonably approximated as a Gaussian distribution, then knowledge of the standard deviation $\sigma_E$ leads to an extremely simple, if somewhat crude, analytical estimate of the Strehl ratio in the final reduced image. We expect the core of the zero-error sum PSF to be essentially an Airy disc core, of full width at half maximum (FWHM) $\approx \lambda  / D $. This may be approximated by a Gaussian of the same FWHM, with parameter $\sigma_{Airy} \approx \frac{\lambda}{2.3548D}$.
\footnote{(2.3548 is approximately the ratio between the FWHM of a Gaussian and its parameter $\sigma$.)}
The PSF resulting from convolution with the error distribution may then be approximated by a Gaussian of parameter 
\begin{equation}
\sigma = \sqrt{ \sigma_{Airy}^2 + \sigma_E^2}
\end{equation}
with corresponding reduction in Strehl ratio, which I shall denote $R$:
\begin{equation}
 R=\frac{I_e(\mathbf{0})}{I_0(\mathbf{0})} = \frac{\sigma_{Airy}^2}{\sigma_{Airy}^2 + \sigma_E^2}\,.
\label{eq:strehl_reduction}
\end{equation}
This obviously neglects contributions from the halo of the PSF, which will mitigate the effect somewhat at low Strehl ratios.

The error distribution is complex, depending upon many parameters, not least of which is the Strehl ratio of any given short exposure. Ideally we require a formula to estimate the typical error without undertaking time-consuming Monte-Carlo simulations. 
I attempted a breakdown of the error variance into various contributing factors, along the lines of the analyses presented in \cite{Thomas2006}, in the hopes of producing a useful approximation. As explained below, the assumptions used are somewhat flawed and the simulations show the estimates to be poor, but I present the mathematical treatment nonetheless as a starting point for further investigations.

\begin{figure}[htp]
\begin{center}

\subfigure[Sample scatter plot for a single simulation run]{	
	\includegraphics[width=0.45\textwidth]{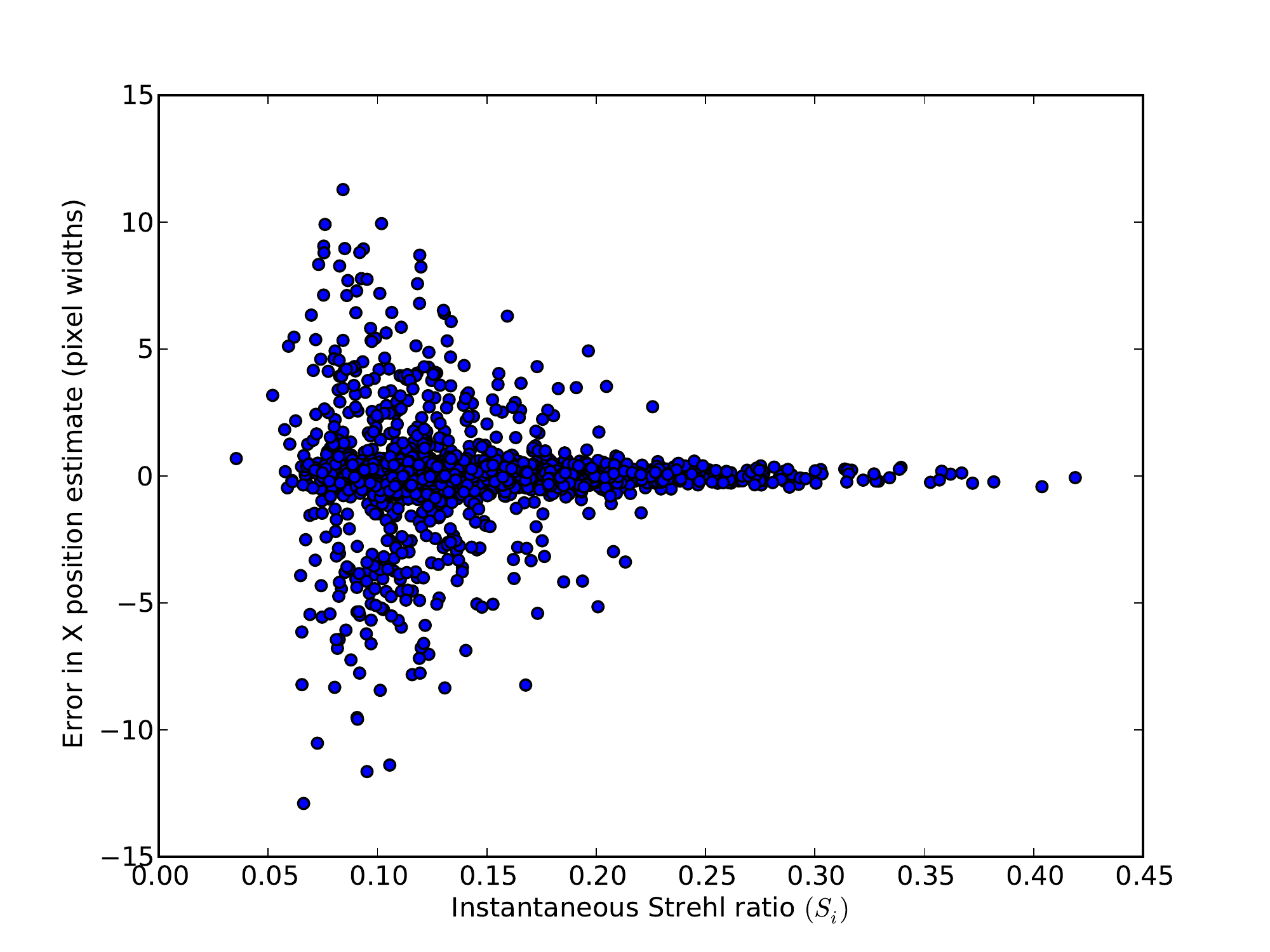}
	\label{subfig:scatter}
}
\subfigure[Histograms of x error at high and low Strehl ratios]{
	\includegraphics[width=0.45\textwidth]{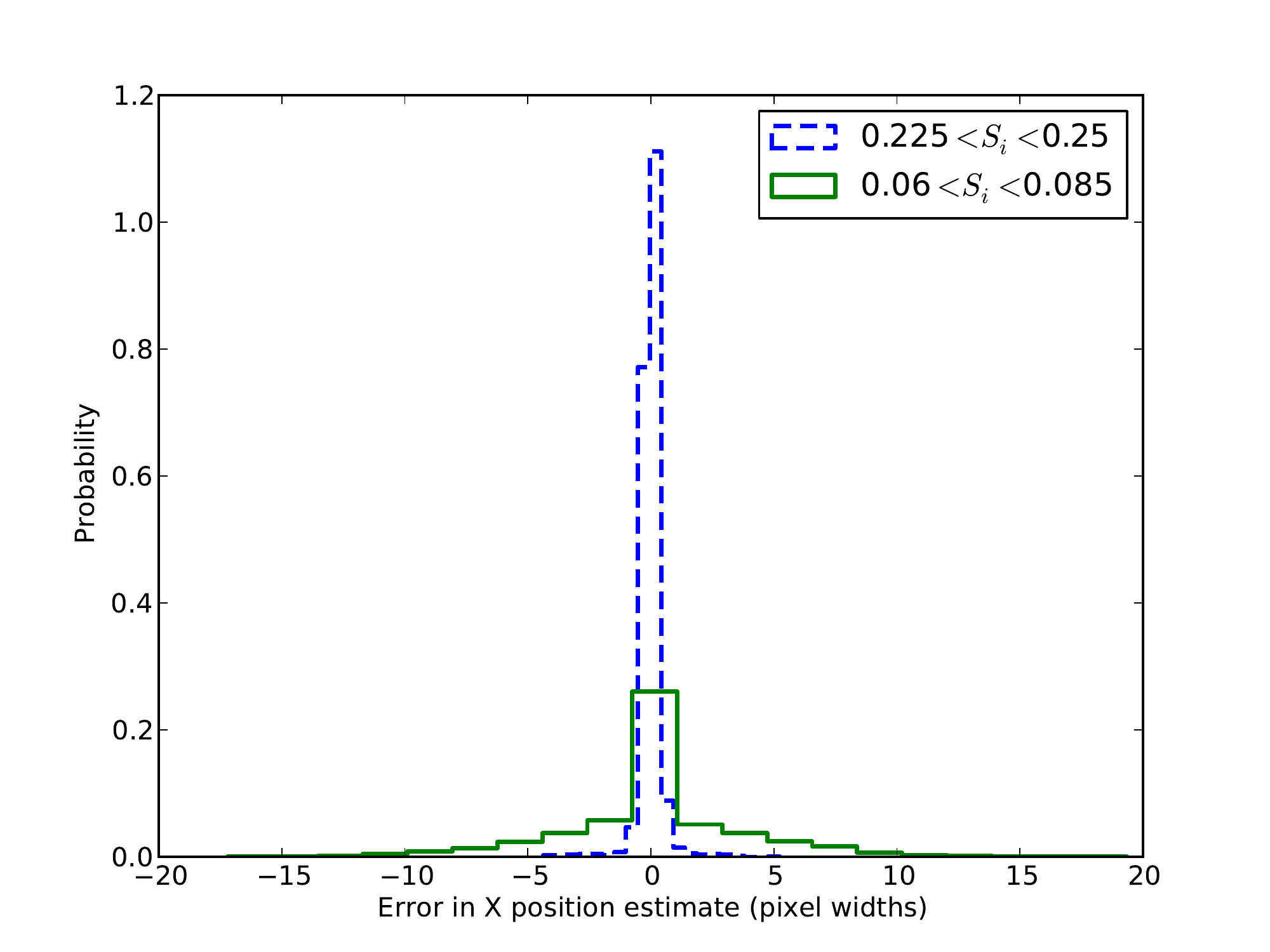}
	\label{subfig:histograms}
}
\caption[Sample plots of x position estimation error]{
	Sample plots of x position estimation error, for a simulated source signal level of 600 photo-electrons per frame. Plot \subref{subfig:scatter} is a scatter plot of one simulation run, containing 1800 data points.
	Plot \subref{subfig:histograms} is a histogram depicting the range of error values for short-exposures in two Strehl bins, at Strehl ratios around 0.07 and 0.25. The histogram is composed using the whole dataset of 90000 simulated short exposures.
}
\label{fig:x_position_estimation_error_samples}
\end{center}
\end{figure}

In order to apply analytic estimates, I first estimated the flux present in the bright speckle against which the reference image is cross-correlated. Assuming that the brightest speckle is reasonably similar in shape to an Airy disc, we can make the approximation that the proportion of total source flux concentrated in the brightest speckle is proportional to the Strehl ratio of the short exposure image. Then, the proportion of flux within the first minima of an Airy disc is around 80\% of the total Airy disc flux (the exact proportion depends on the aperture obscuration ratio, \corr{i.e. the ratio between the diameters of the inner and outer radii of the telescope's primary aperture annulus}). Denoting the mean flux level $F$,
we then have 
\begin{equation}
F_{core} = 0.8 \times F_{total} \times  SR 
\end{equation}
where $SR$ denotes the short exposure Strehl ratio. Then, for a pixel width corresponding to Nyquist sampling of the Airy disc, we may estimate the cross-correlation positional error variance in the x-direction due to shot noise after detection by the EMCCD as \citep{Thomas2006}
\begin{equation}
\sigma^2_{x,phot} \approx \frac{2\sigma_{Airy}^2}{ F_{core} }\,,
\end{equation}
where an additional factor of 2 has been introduced to account for the additional variance due to stochastic signal multiplication in the EMCCD multiplication register (cf. Section~\ref{sec:pixel_PDFs}), which is equivalent to a photon flux reduced by a factor of $\sqrt{2}$. \corr{(Note I have chosen the x-axis without loss of generality as a convenient co-ordinate system for representing positional error along any given axis.)}

We must then consider error due to light in the halo, since this will dominate over readout noise if the typical pixel illumination due to halo flux is greater than around 0.1 photo-electrons per frame. \citeauthor{Thomas2006} give an estimation of cross-correlation error variance due to a uniform readout noise. We may utilise this to give a crude approximation of error-variance due to halo flux, though I note that the approximation is a poor one at low light levels as the halo flux noise will of course be Poissonian and not Gaussian in nature. With this in mind, we may derive a first order approximation to the halo intensity in the vicinity of the PSF core by approximating it as a Gaussian with FWHM equivalent to the seeing width, which I denote here as $\omega$, measured in multiples of pixel width. For a sum halo flux 
\begin{equation}
 F_{halo} = F_{total} - F_{core}
\end{equation}
we then get a halo light level per pixel per frame, $h$, in the vicinity of the core of 
\begin{equation}
 h = \frac{ F_{halo} }{2\pi (\frac{\omega}{2.3548})^2}
\end{equation}
which we may then insert into the appendix result of \citeauthor{Thomas2006} \corr{(equation C10)} to estimate the positional error variance in the X-direction due to background flux from the halo:
\begin{equation}
\sigma^2_{x,halo} \approx \frac{8 \delta^2 h^2}{ F_{core}^2 }
\end{equation}
where an extra factor of 2 has been inserted again to account for added variance due to stochastic multiplication, and $\delta$ is the FWHM of the reference auto-correlation --- $\delta$ has an angular width of $\approx 1.44 \lambda / D$ in this case.

\begin{figure}[htp]
\begin{center}
 \includegraphics[width=0.7\textwidth]{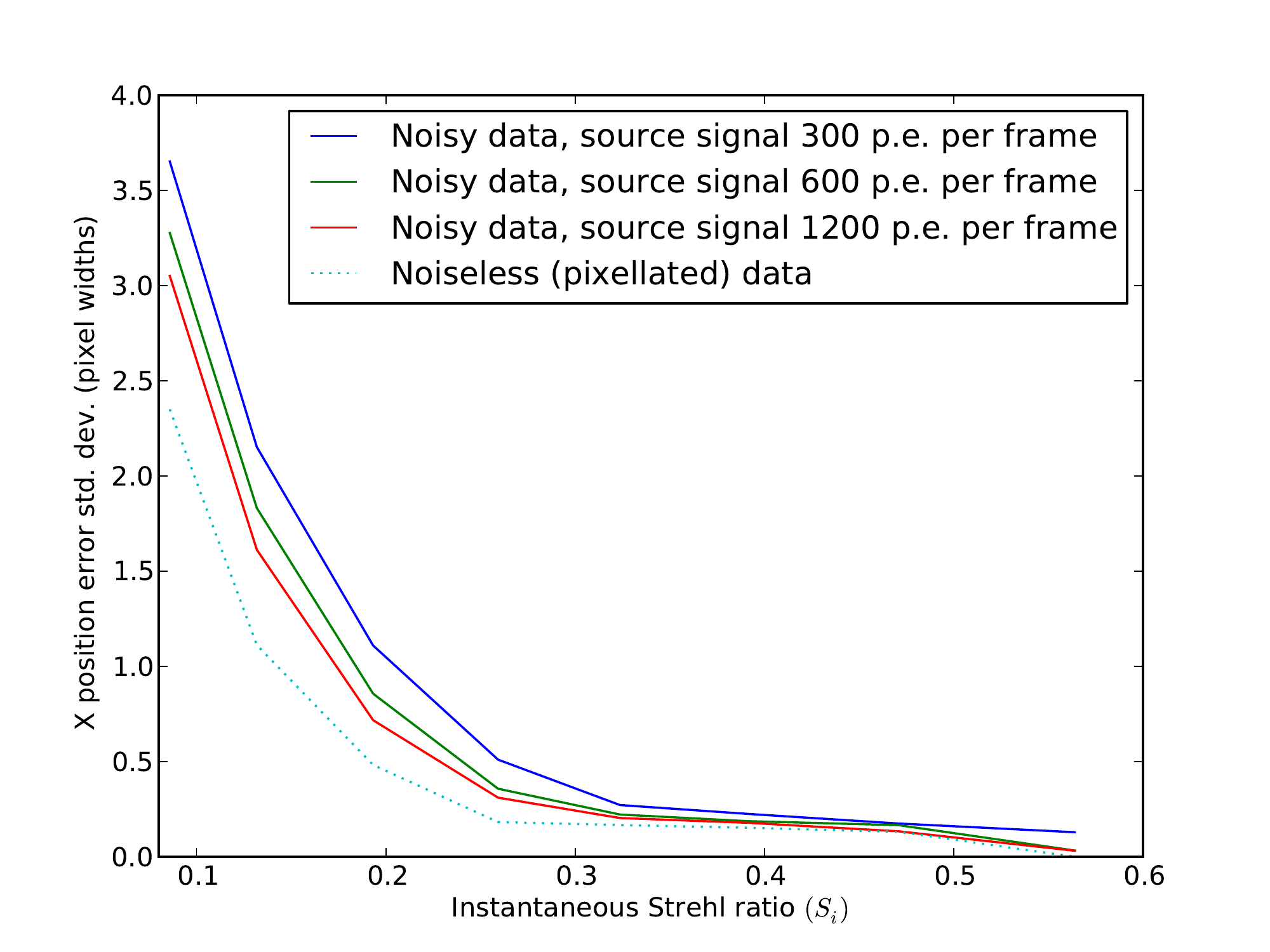}
\caption[Variation in position estimation error standard deviation with Strehl ratio]{
Variation in the standard deviation of the position estimation error in the X-axis, plotted against Strehl of the short exposures. To obtain a good sample size, samples were aggregated by Strehl ratio into 8 bins of equal width across the range of simulated Strehl ratio values. Dotted line depicts errors obtained from applying the frame registration algorithm to noiseless, pixellated data, and therefore gives some indication of the error purely due to spatially correlated background signal noise (speckle noise). Solid lines denote, from top to bottom, source signal levels of 300, 600, and 1200 photo-electrons per short exposure.
}
\label{fig:posn_err_against_strehl}
\end{center}
\end{figure}

\begin{figure}[htp]
\begin{center}
 \includegraphics[width=0.7\textwidth]{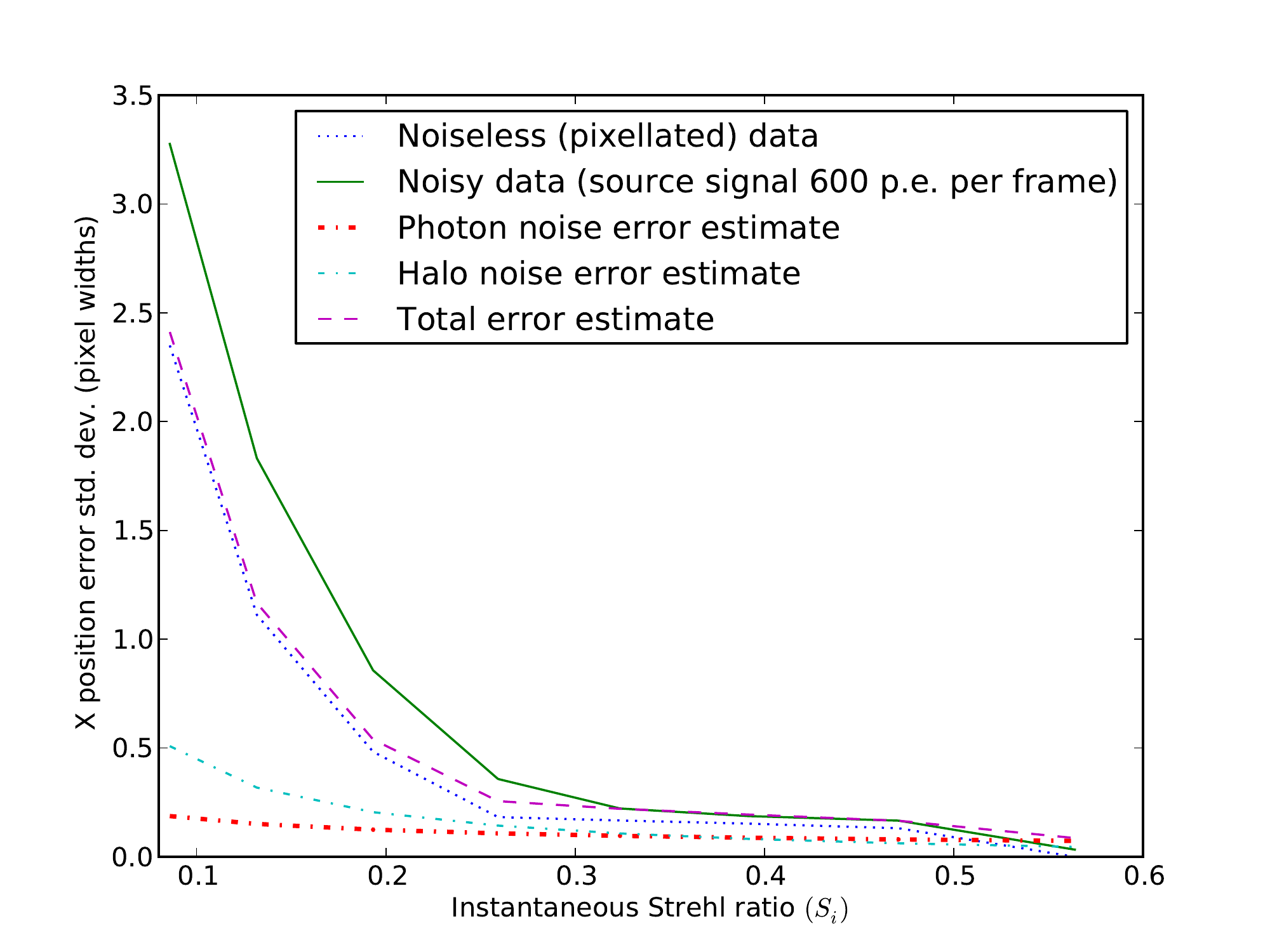}
\caption[Comparison between analytically predicted and simulated positional estimation error]{A comparison between the predicted and simulated position estimation errors. Simulated results are plotted for a source signal level of 600 photo-electrons per frame, and for pixellated intensity data (representing error purely due to speckle noise). Error variance components due to \corr{source shot noise (labelled photon noise in plot)} and halo background noise are also plotted, together with a semi-empirical ``Total error estimate'' derived by adding the variance predicted due to shot noise and halo noise, to the variance due to speckle noise determined from the pixellated simulation. 
}
\label{fig:posn_err_predictions}
\end{center}
\end{figure}

Figure~\ref{fig:posn_err_against_strehl} plots the standard deviation of the error distance in the X-axis, against true Strehl of the short exposures, for various simulated source signal levels. Figure~\ref{fig:posn_err_predictions} compares this with the analytical error variance estimates derived above, which clearly underestimate the simulated error. This discrepancy is to be expected if we take into consideration the fact that, unlike readout noise, halo noise is not in fact independent between pixels, but highly spatially correlated (hence the observed speckle patterns). Unfortunately, to the best of my knowledge an analytical formula for estimation of registration error in the presence of correlated background noise is not available in the literature. Furthermore, even when I take into account an empirical estimate of the position error variance due to speckle, the combined error estimate is still lower than the simulated error for much of the range of instantaneous Strehl ratios. It is interesting to 
note that the disparity becomes particularly clear around the instantaneous Strehl ratio of 0.3, which is about the level where secondary speckles become clearly noticeable in short exposures.

\section{Image formation model}
Having considering the processes affecting the formation of the reduced image, we may now consider a model that draws them together. I propose the following formalism for a model predicting Strehl ratio in reduced lucky images. I propose that the Strehl ratio in the final, reduced image, $S_f$ may be estimated as:
\begin{equation}
S_{f} = \frac{ \int_{S_{i}} \!R \, S_i \, \chi \, P_{MR}\, \mathrm{d}S_i }{\int_{S_{i}} \!\chi \,P_{MR}\, \mathrm{d}S_i}
\end{equation}
where $S_i$ is the Strehl ratio in a short exposure, $R$ is the reduction in Strehl ratio due to position estimation error, $\chi$ is the short exposure selection probability function and $P_{MR}$ is the modified Rician distribution of instantaneous Strehl ratios (this assumes the short exposure time is less than the atmospheric coherence time). Validity of the model largely depends upon whether the effects of positional estimation error may be well approximated in this fashion. Verification of the model is deferred to future work.

\chapter{Scientific applications of lucky imaging}
\label{chap:science}
To date, the applications of lucky imaging have been limited in scope. The choice of useful observation targets has been restricted by the narrow field of view entailed by good sampling with the older $512\times512$ pixel EMCCD detectors, and the difficulties of accurate detector calibration required to observe faint and extended sources. These factors, together with the generally developmental nature of the technique, have meant that conservative choice of observing targets was essential in order to achieve scientific results. As such, productive observing programs have largely revolved around high-resolution stellar binarity surveys, typically looking at very-low-mass stellar populations, for example \cite{Law2005,Law2008,Lodieu2009,Bergfors2010}. Such programs benefit from the short target acquisition time overheads associated with lucky imaging, when compared to adaptive optics systems, for comparable spatial resolution. 

I begin this chapter with further work on stellar binarity, this time in the context of planetary transit host candidates. This is a relevant topic for current research, given the current plethora of transit surveys. Well quantified, preferably high contrast, companion detection limits are a necessity, as explained in Section~\ref{sec:transit_science}.

I present analysis algorithms and application of a matched filter companion detection technique, which produces quantified detection limits with a challenging dataset, and results in several companion detections at fainter contrasts than previously achieved with lucky imaging.

I then explore some of the possibilities for more general application of lucky imaging techniques, given the improvements in field of view and detector calibration achieved with the summer 2009 Cambridge LuckyCam observing run. The possibility of high temporal-resolution photometry with EMCCDs is explored in a preliminary fashion, and recommendations for further investigations made. A case is made for lucky imaging as a viable alternative to Hubble space telescope observations for high resolution imaging with an excellent faint limit. Furthermore, the possibilities of lucky imaging coupled with adaptive optics systems are considered, specifically in the context of probing binarity distributions at previously unobtainable levels of resolution in the crowded central regions of globular clusters.

\section{Binarity of planetary transit hosts}
\subsection{Planetary transit surveys and the need for follow-up observations}
\label{sec:transit_science}
\begin{figure}[htp]
\begin{center}
 \includegraphics[width=0.8\textwidth]{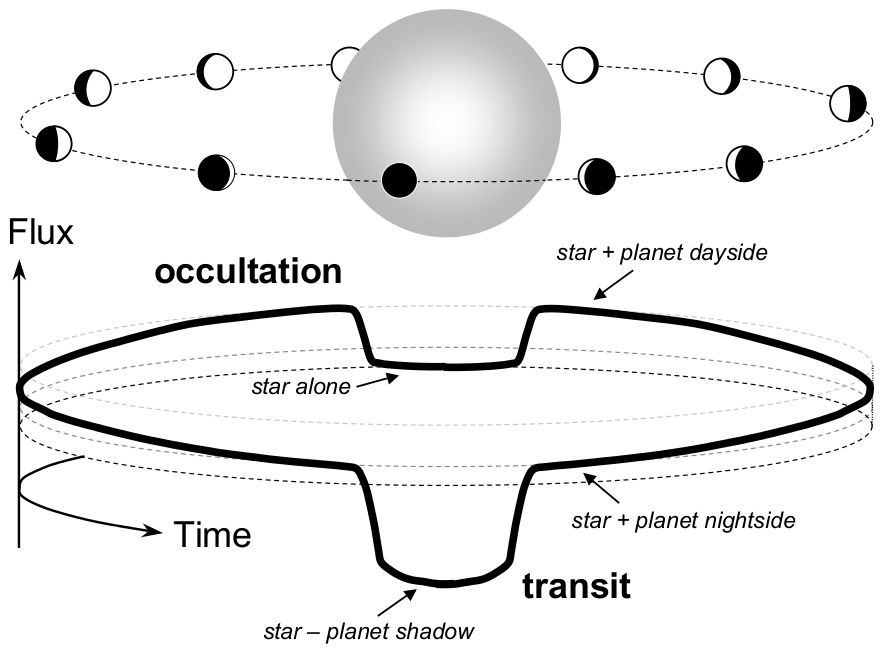}
\caption[Illustration of transits and occultations]{
Illustration of transits and occultations. 
When monitoring a planet host system we observe the combined flux of the star and planet. 
During a transit, the flux drops because the planet blocks a fraction of the starlight. 
Then the flux rises as the planet’s dayside comes into view. 
The flux drops again when the planet is occulted by the star.
Reproduced from \cite{Winn2010}.
}
\label{fig:transit}
\end{center}
\end{figure}
\corrbox{
Since the discovery of the first planet around a Sun-like star in the mid-nineties \citep{Marcy1995}, exoplanet science has grown to become a major sub-field of astronomy, and was listed as a priority in the most recent Astronomy and Astrophysics decadal survey \citep{Decadal2010}. 
While radial-velocity surveys have produced the majority of confirmed exoplanets to date, transit surveys have an important role to play. 
The \textit{a priori} chance of discovering any particular exoplanet system through transit detection is much smaller than for the radial-velocity method, due to two reasons: 
first, transit surveys are sensitive only to those planetary systems which are by chance aligned such that their orbit crosses our line of sight (see Figure~\ref{fig:transit}), and second, we must be lucky enough to observe the system during transit.
However, while radial velocity surveys require large amounts of time on large telescopes to perform spectroscopic measurements of host star candidates, transit surveys can monitor many thousands of stars simultaneously through use of instruments with wide fields of view. 
Consequently, in recent years ground-based transit surveys have detected exoplanets in numbers rivalling the radial-velocity method \citep{Winn2010}, and the Kepler space mission has detected over 2000 planetary-host candidates since its launch in 2009 \cite{Borucki2010,Batalha2012}. Transits also give us information that is complementary to the mass estimates obtained via radial-velocity measurements --- for transiting systems we may derive the planetary radius, orbital inclination, and even atmospheric composition (though the latter does of course require spectroscopic follow-up).

Unfortunately, since identification of planetary candidates is based primarily on detecting shallow transits wherein the detected flux drops by a few percent, we often observe similar phenomena which contaminate the candidate set. 
The two primary contaminants are binary star systems with grazing eclipses, and binary systems close to an unresolved third star, both of which result in a shallow drop in the observed lightcurve \citep{Charbonneau2004}. 
As a result some ground-based surveys have a false/real candidate ratio of ten to one, and so candidate verification becomes a major problem \citep{Winn2010}. 
To conclusively confirm or reject a planetary-host candidate found via transit surveys requires spectroscopic follow up, but making the required observations requires a considerable amount of allocated time on a large telescope. 
Techniques which can detect a subset of the false-positives and so winnow down the remaining candidates are thus highly desirable. 
Lucky imaging can perform admirably in this role by providing a low-cost, highly efficient observing tool to produce excellent constraints on the presence, or lack, of close secondary stellar sources in the vicinity of planetary-transit candidates.
}

As part of the 2009 observing run, we observed a number of planetary host candidates identified by transit detection programs such as WASP \citep{Pollacco2006} and HAT \citep{Bakos2004} with the aim of determining their binarity. Placing constraints on the possible binary companions of planetary hosts is important for verifying the significance of transit data, as an undetected secondary star may add variability to a photometric  signal that can be misinterpreted as a planetary transit. The binary fraction is also an interesting statistic, giving insight into phenomena that govern planetary system formation such as Kozai migration \citep{Takeda2009}. 

Observations were made during the period 18th-22nd July 2009 at the 2.56 metre Nordic Optical Telescope on La Palma with the Cambridge LuckyCam visitor instrument. Seeing ranged from $\sim0.6''$ to $\sim1.65''$ as measured by the differential image motion monitor (DIMM). All observations were made in SDSS \textit{i'} band, using a plate scale of 32.4 milliarcseconds per pixel, providing good sampling of the point spread function (PSF). The camera frame rate was 21 frames per second using full chip readout ($\sim1\textrm{k}\times1\textrm{k}$ pixels ). Table~\ref{tab:obs} lists the observations made.

\def \ObsTableScale {0.85}
\begin{table}

%

 \scalebox{\ObsTableScale}{
	\begin{tabular}{|lllcccc|}
    \hline

		\begin{minipage}{ 1cm }%
		\begin{center}
		\vspace{1ex}
		
		\textbf{ Run ID } 
		
		\end{center}
		\end{minipage}

		&

		\begin{minipage}{ 2cm }%
		\begin{center}
		\vspace{1ex}
		
		\textbf{ Target } 
		
		\end{center}
		\end{minipage}

		&

		\begin{minipage}{ 1.5cm }%
		\begin{center}
		\vspace{1ex}
		
		\textbf{ Date } 
		
		\end{center}
		\end{minipage}

		&

		\begin{minipage}{ 3cm }%
		\begin{center}
		\vspace{1ex}
		
		\textbf{ Companions detected within 6.5 as } 
		
		\end{center}
		\end{minipage}

		&

		\begin{minipage}{ 3cm }%
		\begin{center}
		\vspace{1ex}
		
		\textbf{ Exposure time (s) } 
		
		\end{center}
		\end{minipage}

		&

		\begin{minipage}{ 3cm }%
		\begin{center}
		\vspace{1ex}
		
		\textbf{ Seeing @ 500nm (as) } 
		
		\end{center}
		\end{minipage}

		&

		\begin{minipage}{ 3.5cm }%
		\begin{center}
		\vspace{1ex}
		
		\textbf{ Background detection limit \\ ($\Delta$i' mag) } 
		
		\end{center}
		\end{minipage}

	\\[3.6ex]
    \hline

            3012
             
               &

            TrES-1
             
               &

            18/07/09
             
               &

            2
             
               &

            370
             
               &

            0.96
             
               &

            8.9

        \\

            3013
             
               &

            TrES-2
             
               &

            18/07/09
             
               &

            1
             
               &

            336
             
               &

            0.91
             
               &

            8.7

        \\

            3019
             
               &

            HD209458
             
               &

            18/07/09
             
               &

            0
             
               &

            480
             
               &

            1.05
             
               &

            9.5

        \\

            4001
             
               &

            Hat-P-2
             
               &

            18/07/09
             
               &

            0
             
               &

            384
             
               &

            1.09
             
               &

            9

        \\

            5025
             
               &

            HAT-P-8
             
               &

            20/07/09
             
               &

            0
             
               &

            437
             
               &

            1.64
             
               &

            8.7

        \\

            6006
             
               &

            TrES-4
             
               &

            20/07/09
             
               &

            1
             
               &

            480
             
               &

            1.23
             
               &

            8.8

        \\

            6008
             
               &

            HAT-P-5
             
               &

            20/07/09
             
               &

            1
             
               &

            432
             
               &

            1.11
             
               &

            8.5

        \\

            6010
             
               &

            WASP-3
             
               &

            21/07/09
             
               &

            0
             
               &

            317
             
               &

            1.21
             
               &

            8.7

        \\

            6011
             
               &

            CoRoT-2
             
               &

            21/07/09
             
               &

            1
             
               &

            288
             
               &

            1.49
             
               &

            6.8

        \\

            6012
             
               &

            CoRoT-3
             
               &

            21/07/09
             
               &

            1
             
               &

            248
             
               &

            1.57
             
               &

            5.8

        \\

            6013
             
               &

            HAT-P-7
             
               &

            21/07/09
             
               &

            1
             
               &

            248
             
               &

            1.2
             
               &

            8.7

        \\

            7007
             
               &

            XO-1
             
               &

            22/07/09
             
               &

            0
             
               &

            480
             
               &

            0.87
             
               &

            9

        \\

            7012
             
               &

            WASP-3
             
               &

            22/07/09
             
               &

            0
             
               &

            240
             
               &

            0.71
             
               &

            9.4

        \\

            7022
             
               &

            HAT-P-11
             
               &

            22/07/09
             
               &

            0
             
               &

            288
             
               &

            0.7
             
               &

            9.4

        \\

            7036
             
               &

            WaSP-10
             
               &

            22/07/09
             
               &

            0
             
               &

            240
             
               &

            0.81
             
               &

            8.4

        \\

            7037
             
               &

            HAT-P-1
             
               &

            22/07/09
             
               &

            0
             
               &

            259
             
               &

            0.71
             
               &

            6.3

        \\

            7038
             
               &

            HAT-P-6
             
               &

            22/07/09
             
               &

            1
             
               &

            240
             
               &

            0.73
             
               &

            9.5

        \\
    
    \hline
    \end{tabular}
}

    \caption{Planetary host candidates observed.}
    \label{tab:obs}
\end{table}

The data were reduced using the LuckyCam pipeline as described in Section~\ref{sec:pipeline}. 


\subsection{General techniques for detecting faint secondary sources}
The technique most widely applied when attempting to identify faint or crowded point sources in astronomical images is that of PSF fitting and subtraction. A step crucial to this process is the choice and evaluation of PSF model, which may be derived semi-analytically, empirically, or by some combined analytical model fit with empirical corrections --- see \cite{Dolphin2000, Diolaiti2000, Stetson1987} for examples.

Ideally, for complex PSFs such as those produced by adaptive optics systems, a fully empirical PSF model is created by analysing a number of bright, isolated calibrator star images, close to the time of observation of the binary candidate to ensure similar atmospheric and instrumental conditions. However, unless the field of view containing the binary candidate happens to also contain suitable bright calibrator stars simultaneously this is very costly with regard to observing time, as it requires separate calibration observations for every target observation. Also, it may not be practical to use separate calibration observations if the system parameters are evolving on timescales similar to observation timespans, for example when quasi-static speckles \citep{Gladysz2008} are present \corr{due to imperfect correction in an adaptive optics system}. 
In contrast, conventional long exposure observations produce a PSF which may be expected to conform to a Moffat profile \citep{Trujillo2001}. As such it is largely parametrised by a single number, the seeing width, which is relatively easy to extract and use as a model parameter, although obviously the data is inherently lower resolution. 

The case of lucky imaging is somewhere between the two described above. 
The data contains information at higher resolution than conventional long exposures, and we should be able to calibrate the PSF with relative ease compared to the complexities of adaptive optics observations. 
We expect the PSF \corr{in the reduced images} to be axisymmetric, however the radial profile is non-trivial as the PSF consists of a narrow core surrounded by a wide halo (\corr{see cross-section in figure}~\ref{fig:WASP_resids_x_section}). 
At any rate, for this dataset, a fully empirical PSF model was not an option, due to small non-axisymmetric components that changed between targets, as can be seen from the residual images in Figure~\ref{fig:WASP_resids}. 

One approach I considered was fitting an axisymmetric multi-component analytic model representing the core and the halo components. Just such a multi-component fitting routine is described in chapter 5 of \cite{Law2007}, with a ``sliding box'' method used for faint companion detection, \corr{whereby a local estimate of noise levels is used to isolate particularly bright regions as candidates for PSF fitting}. Unfortunately the code was not available for testing and comparison. Such a model requires 4 or more parameters, and preliminary work suggested that attempting to fit such a model using a single image which evidently has unmodelled residuals would be unreliable. Instead, I settled on the simple and robust semi-empirical method described below, which also provides a pixel variance estimate as a function of radius from the primary star.

\subsection{Primary star PSF modelling and subtraction}
\label{sec:psf_model_gen}
To create a PSF model for subtraction of the primary star flux, I wrote a program to create an axisymmetric, semi-empirical model. \corr{For this data set, all the candidate exoplanet host stars are bright, and so may be used to calibrate the PSF model with reasonable accuracy. The PSF model may then later be used for fitting companion star candidates at any sub-pixel position.}
First, the central location of the PSF was determined to sub-pixel precision by fitting a Gaussian profile to 9 pixels around the peak pixel. The pixel values around this nominal centre were then collected into bins by radius (I refer to these as ``annulus bins'') to get a median value and standard deviation at approximately one pixel width radius intervals. A continuous radial profile can then be produced --- the values of the Gaussian core fit are used at radii within 1.5 pixels, and the annulus bin median values are interpolated to provide model values at larger radii.

\begin{figure}[htp]
\begin{center}
 \includegraphics[width=1.0\textwidth]{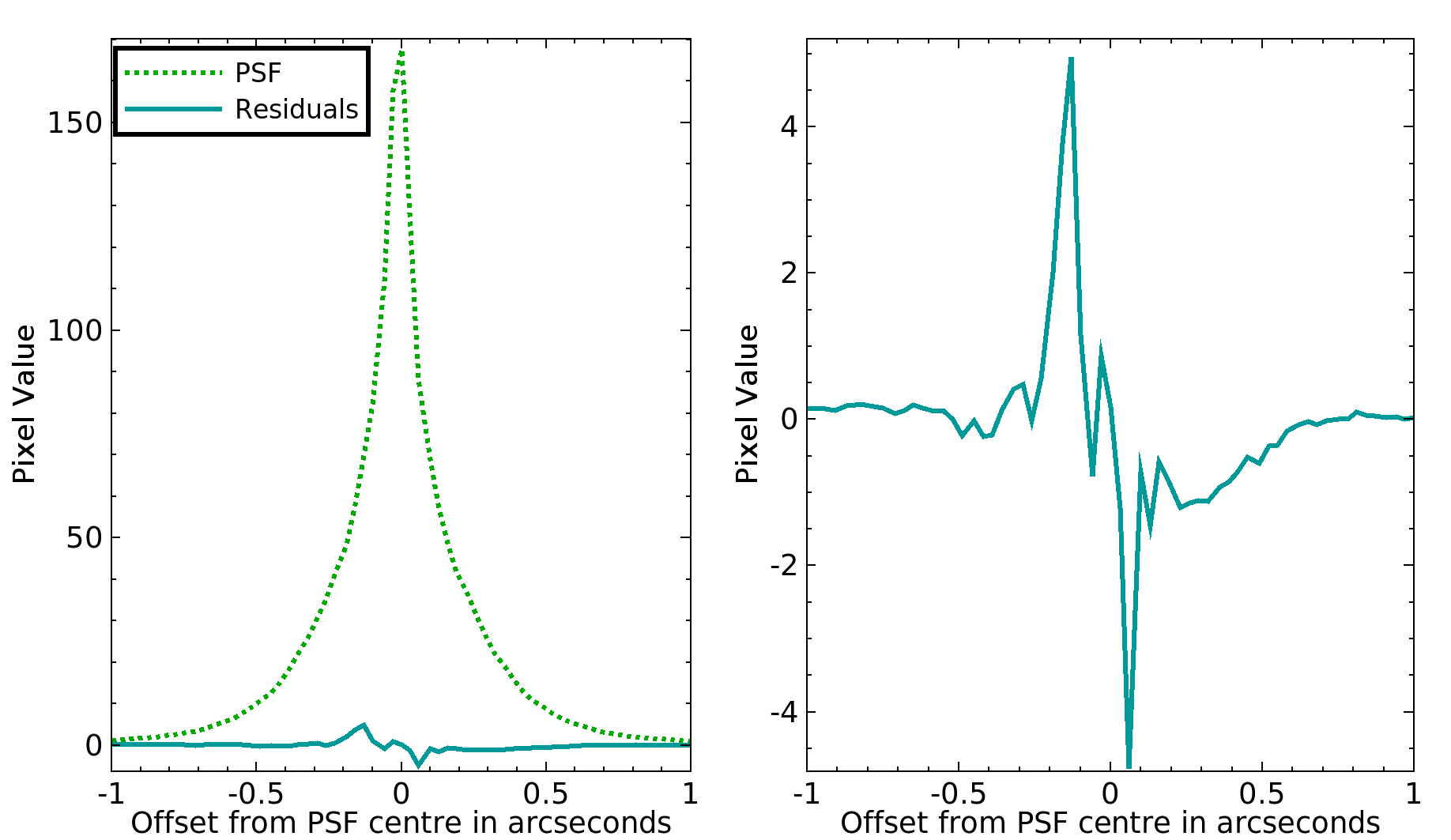}
\caption[PSF residuals cross section]{Left: A cross-section plot comparing the pixel values of HAT-P-1 before and after subtraction. Right: The residuals replotted on a larger scale. The residuals image is displayed in Figure~\ref{fig:WASP_resids}.
}
\label{fig:WASP_resids_x_section}
\end{center}
\end{figure}

\begin{figure}[htp]
\begin{center}
 \includegraphics[width=1.0\textwidth]{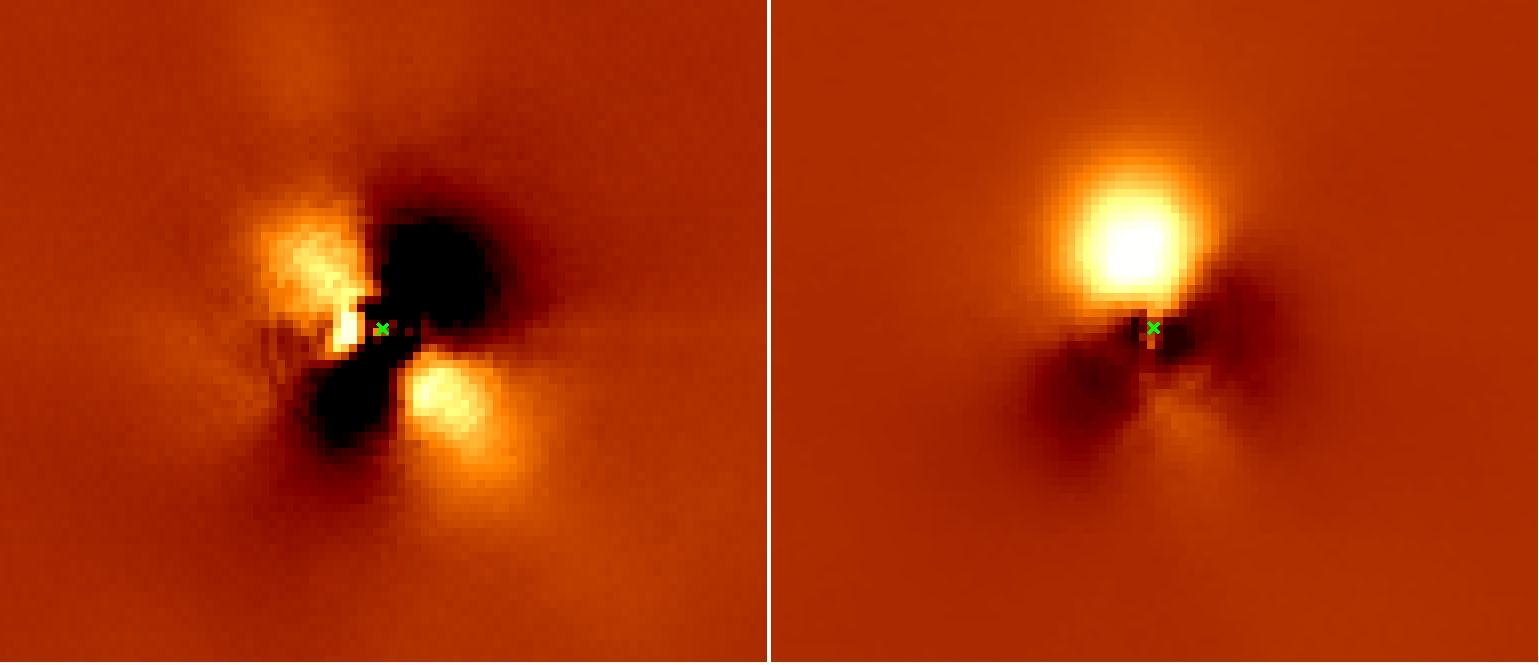}
\caption[PSF residuals ]{Residuals images resulting from subtracting an axisymmetric PSF model from observations of HAT-P-1 (left) and HAT-P-6 (right). \corr{The axisymmetric PSF model was generated as described in section}~\ref{sec:psf_model_gen}. \corr{As can be seen, the non-axisymmetric residuals vary between observations, and so refinement of the PSF model was impossible.}
}
\label{fig:WASP_resids}
\end{center}
\end{figure}

Figures~\ref{fig:WASP_resids_x_section} and \ref{fig:WASP_resids} display results from fitting and then subtracting the primary star PSF in this manner. Unfortunately the images were affected by some degree of aberration which was not obvious during the observing run. As a result, subtraction of the axisymmetric model produces much larger residuals than would be expected simply due to shot and detector noise. As Figure~\ref{fig:WASP_resids} shows the non-axisymmetric PSF components vary from target to target, and so a full empirical calibration to further remove these residuals from the raw data is not possible --- without a stable PSF, any empirical corrections might attribute a faint companion source to PSF aberration effects. \corr{The source of the non-axisymmetric PSF components is unclear --- it may have been due to problems with the atmospheric dispersion corrector optics, or imperfect focussing. Unfortunately the problems did not become apparent until the PSF analysis was undertaken, after the 
observing run was completed. If such problems cannot be rectified in future observing runs, calibrator stars may be observed to improve the PSF modelling and subtraction process.}

\subsection{Applying a matched filter}
To enable automated detection of companion stars, and ameliorate the variance introduced by the non-axisymmetric PSF components, I applied a matched filter to the residual images, using the cross-correlation routines described in chapter~\ref{chap:frame_registration}. This was done by using the PSF model to generate a small reference image for cross-correlation with the residuals image. Before cross-correlation, the reference image pixel values are rescaled, such that the mean pixel value of the reference is zero. This has the effect that convolving the reference with a uniform or slowly varying background section of an image will produce pixel values that are close to zero, while copies of the PSF will produce high pixel values (this is essentially the same algorithm employed by the DAOphot FIND routine \citep{Stetson1987}). Obviously, this algorithm relies on the assumption that the PSF of companion stars will be a good match to the modelled guide star PSF. The assumption is valid for this dataset for the 
following reasons. Firstly, the search region of interest is within the typical isoplanatic patch for lucky imaging \citep{Tubbs2002}. Secondly the guide stars are all relatively bright and therefore should not suffer from the central bright pixel effect that becomes an issue when guiding at low photon rates \corr{due to the frame registration process locking on to pixels which are bright simply due to shot noise} \citep{Christou1991, Law2007}.

One parameter to be determined is the size of the reference image to use in the filtering process. This partly depends on the proximity to the primary star --- at larger radii where the primary star residuals are small and detector noise dominates, a wider reference image is more effective. Close to the primary star a narrower reference image is best. After thorough experimentation I settled on a 15x15 pixel reference at large radii, and an 11x11 pixel reference for the region close to the primary star. I also experimented with using different selection cut-off levels for the initial lucky imaging reduction process, but it appears that for the relatively short observations in this dataset the best signal-to-noise is obtained from analysing images created using 100\% of the short exposures, despite the slight increase of FWHM and decrease in Strehl ratio. 

\begin{figure}[htp]
\begin{center}
 \includegraphics[width=1.0\textwidth]{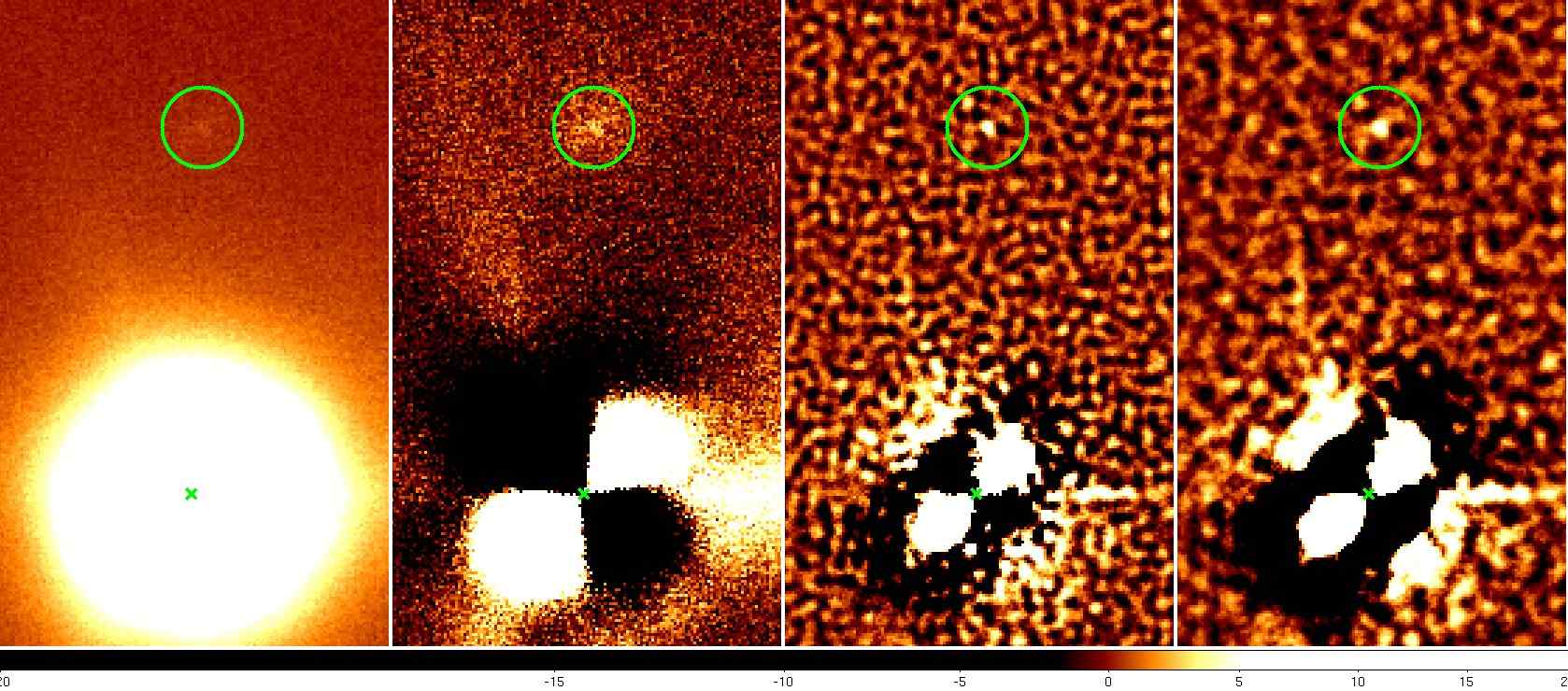}
\caption[Images from the analysis procedure.]{Images of HAT-P-5 from various stages in the analysis process. From left to right: The initial input image, the residuals after subtraction of the PSF model, residuals after application of the narrow filter, and residuals after application of the wide filter. The images have been normalized by their background noise level estimated using the standard deviation of pixel values in a user defined background region. The central location of the primary is marked with a small x, and a faint secondary is circled (note that in these images, north is left and west is up). Separation between the primary and secondary is 4.25 arcseconds.}
\label{fig:filtering}
\end{center}
\end{figure}

Figure~\ref{fig:filtering}  illustrates the results of the filtering process. The procedure is effective in reducing the noise levels and partially suppressing the non-axisymmetric features.

\subsection{Automated companion candidate detection}
Finally, the filtered images were analysed for companion candidates at a user specified threshold level. At a threshold of 5 sigma the companion candidates are almost entirely automatically identified \corr{in agreement with visual inspection}, while at 4 sigma a few false candidates appear along with fainter real companions. For this dataset it was practical to visually confirm or reject all detections at the 4 sigma level. Close to the primary star, the noise estimates are based on the standard deviation in annulus bins (reanalysed using the filtered image), while far from the primary a simple background standard deviation noise estimate is used (this is obtained by calculating the standard deviation of pixels in a user defined background region). The switch-over radius is chosen as the radius at which the standard deviation in the annulus bins equals the background noise estimate. Typically this is at a radius of around 90 pixels ($\sim$3 arcseconds).

Certain criteria must be satisfied for a pixel to be marked as a companion candidate: 
\begin{itemize}
 
\item	First, the pixel value must be above a user specified multiple of the noise estimate (i.e., above the sigma threshold). The signal-to-noise ratio (SNR) is stored for reference.
	
\item	Second, the pixel must be a local maximum - to prevent small noise spikes becoming candidates this is assessed by computing whether the pixel has a higher value than all the pixels in a surrounding 9x9 pixel box.

\item Third, the region about the pixel must also have non-negligible signal-to-noise in the unfiltered residuals image. The flux is measured in a small aperture about the candidate pixel in the unfiltered image \corr{(measurement in the filtered image would underestimate the flux)}, and the aperture SNR calculated using equation~\ref{eq:SNR}. If the unfiltered aperture SNR is below 0.5 the candidate is rejected.

\item Finally, an extra caveat is applied to filter out false detections caused by a detector artefact --- charge transfer inefficiency, which causes a faint trail of raised pixel values to the right of very bright pixels. All candidates that are within one row of the primary star centre, and to the right, are automatically rejected as companion candidates, but saved for further visual inspection.
\end{itemize}

\subsection{Companion candidate analysis}
 After detection in the filtered image, analysis of companions is performed on the unfiltered residuals image. Each companion candidate is fitted with a Gaussian to determine the central location, then flux is measured in a circular aperture of diameter 6 times the primary target star FWHM (this is measured from the reduced image, and is much smaller than the seeing disc FWHM).
 The flux of the primary star is measured in a similar manner, and the magnitude difference calculated from the ratio. The SNR of the companion in the unfiltered image is calculated taking into account background and photon shot noise using the formula:
 \begin{equation}
 SNR = \frac{ F - N_{pix}b }{ \sqrt{ N_{pix}\sigma^2 + F } }
 \label{eq:SNR}
\end{equation}
where $b$ and $\sigma$ are the mean value and variance of the relevant background pixels, $F$ is the sum flux of the pixels in the photometric aperture, and $N_{pix}$ is the number of pixels in the photometric aperture.

The algorithm was implemented in C++ using the image processing routines described in chapter~\ref{chap:data_reduction}. The program is run twice, once to initially identify candidates, and then a second time with the confirmed or rejected candidates input so that any companions close to the primary target star may be masked during creation of the PSF model.

\subsection{Results}

Companion detections are detailed in table~\ref{tab:companions}. Targets were often observed slightly off-centre on the CCD detector to achieve better positioning of the mosaic field of view, \corr{in order to visually confirm the target being observed using the positions of nearby stars}. The target observed closest to the CCD edge has an unbroken observation area of radius 6.5 arcseconds. As a result, to give a uniform dataset we only list detections within 6.5 arcseconds in table~\ref{tab:companions}.

\begin{figure}
\begin{center}
 \includegraphics[width=1\textwidth]{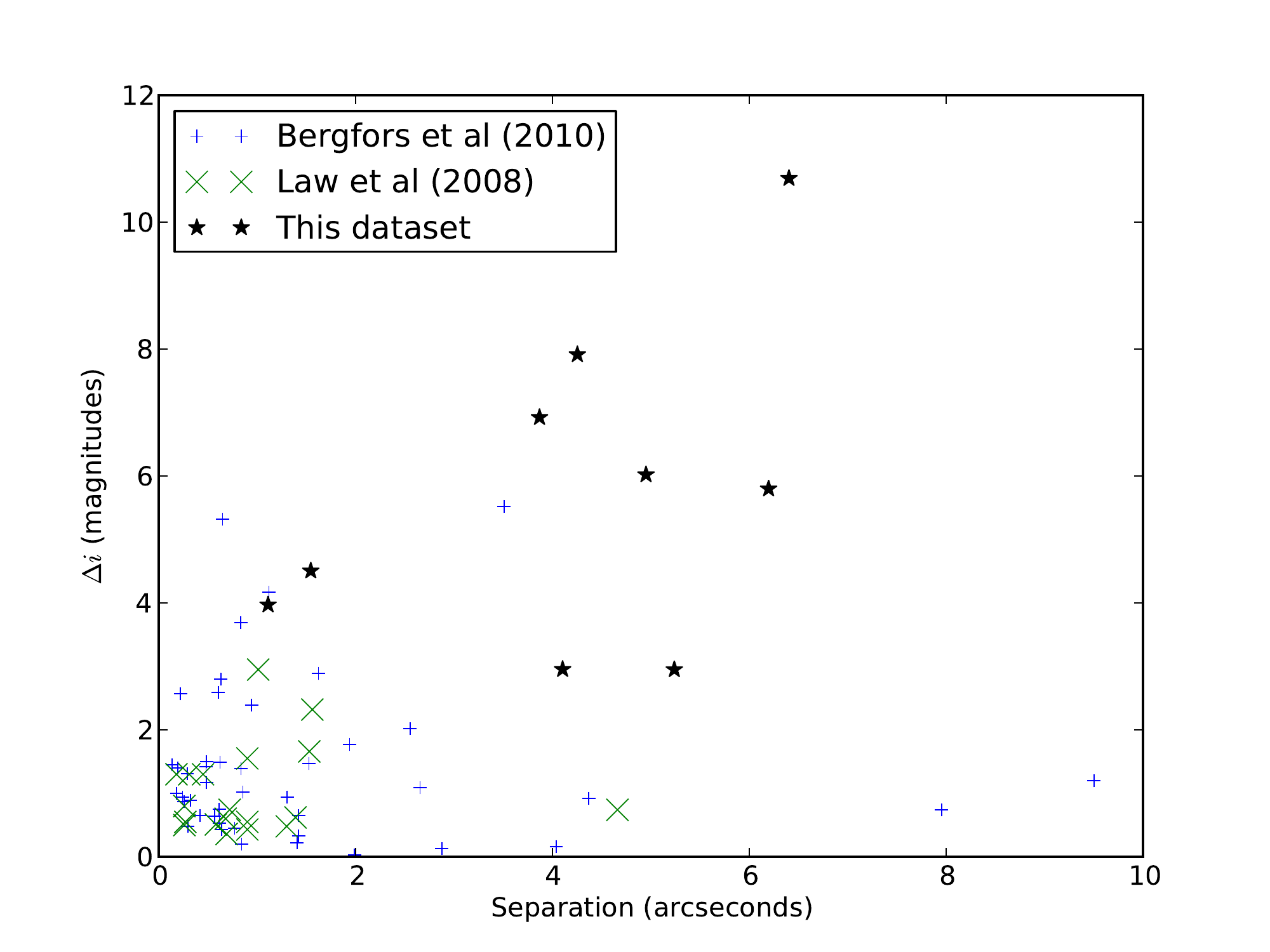}
\caption[Binary detections]{Binary detections: a plot of source separation against contrast for detected binaries. Also plotted are values from the literature for \cite{Law2008} and \cite{Bergfors2010}, \corr{which both employed lucky imaging techniques}. The new detections clearly expand the proven range of detection in terms of contrast ratio between binary components.}
\label{fig:binary_detections}
\end{center}
\end{figure}

8 new companions are detected. These detections significantly expand the range of binary contrast ratios for which lucky imaging detections exist, as clearly shown in Figure~\ref{fig:binary_detections}.

\def \CompanionTableScale {0.75}
\begin{table}
    \centering

%

 \scalebox{\CompanionTableScale}{
	\begin{tabular}{|l|c|c|c|c|c|c|c|}
	\hline
	 \textbf{Target} & \multicolumn{7}{c|}{\textbf{Companion Parameters}} \\
	\hline

		\begin{minipage}{ 2cm }%
		\vspace{1ex}
		\begin{center}
		
		\textbf{ Identifier } 
		
		\end{center}
		\end{minipage}

		&

		\begin{minipage}{ 2.5cm }%
		\vspace{1ex}
		\begin{center}
		
		\textbf{ Separation angle (as) } 
		
		\end{center}
		\end{minipage}

		&

		\begin{minipage}{ 2.5cm }%
		\vspace{1ex}
		\begin{center}
		
		\textbf{ Distance (pc) } 
		
		\end{center}
		\end{minipage}

		&

		\begin{minipage}{ 2.5cm }%
		\vspace{1ex}
		\begin{center}
		
		\textbf{ Separation (au) } 
		
		\end{center}
		\end{minipage}

		&

		\begin{minipage}{ 2.5cm }%
		\vspace{1ex}
		\begin{center}
		
		\textbf{ Parallactic Angle } 
		
		\end{center}
		\end{minipage}

		&

		\begin{minipage}{ 2.5cm }%
		\vspace{1ex}
		\begin{center}
		
		\textbf{ $\Delta$ i' (mag) } 
		
		\end{center}
		\end{minipage}

		&

		\begin{minipage}{ 2.5cm }%
		\vspace{1ex}
		\begin{center}
		
		\textbf{ Aperture SNR } 
		
		\end{center}
		\end{minipage}

		&

		\begin{minipage}{ 2.5cm }%
		\vspace{1ex}
		\begin{center}
		
		\textbf{ Filtered SNR } 
		
		\end{center}
		\end{minipage}

	\\[3ex]
    \hline
		& & & & & \\[-2ex]

				CoRoT-2

               &

					4.1 \raisebox{0.1ex}{ \small{$\pm0.03$ } }

               &

					300 \raisebox{0.1ex}{ \small{$\pm100$ } }

               &

					1230 \raisebox{0.1ex}{ \small{$\pm410$ } }

               &

					208.41 \raisebox{0.1ex}{ \small{$\pm0.4$ } }

               &

					$2.95$ ${ }_{ -0.03 }^{ + 0.03  }$

               &

				40.3

               &

				28.3

        \\[1ex]

				CoRoT-3

               &

					5.24 \raisebox{0.1ex}{ \small{$\pm0.03$ } }

               &

					680 \raisebox{0.1ex}{ \small{$\pm160$ } }

               &

					3563 \raisebox{0.1ex}{ \small{$\pm839$ } }

               &

					173.99 \raisebox{0.1ex}{ \small{$\pm0.39$ } }

               &

					$2.95$ ${ }_{ -0.11 }^{ + 0.12  }$

               &

				9.3

               &

				6.1

        \\[1ex]

				HAT-P-5

               &

					4.25 \raisebox{0.1ex}{ \small{$\pm0.03$ } }

               &

					340 \raisebox{0.1ex}{ \small{$\pm30$ } }

               &

					1445 \raisebox{0.1ex}{ \small{$\pm128$ } }

               &

					268.47 \raisebox{0.1ex}{ \small{$\pm0.44$ } }

               &

					$7.91$ ${ }_{ -0.46 }^{ + 0.82  }$

               &

				1.9

               &

				6.2

        \\[1ex]

				HAT-P-6

               &

					6.4 \raisebox{0.1ex}{ \small{$\pm0.03$ } }

               &

					200 \raisebox{0.1ex}{ \small{$\pm20$ } }

               &

					1280 \raisebox{0.1ex}{ \small{$\pm128$ } }

               &

					39.85 \raisebox{0.1ex}{ \small{$\pm0.22$ } }

               &

					$10.69$ ${ }_{ -1.32 }^{ + 14.31  }$

               &

				0.4

               &

				3.5

        \\[1ex]

				HAT-P-7

               &

					3.87 \raisebox{0.1ex}{ \small{$\pm0.03$ } }

               &

					320 \raisebox{0.1ex}{ \small{$\pm40$ } }

               &

					1238 \raisebox{0.1ex}{ \small{$\pm155$ } }

               &

					90.41 \raisebox{0.1ex}{ \small{$\pm0.49$ } }

               &

					$6.92$ ${ }_{ -0.11 }^{ + 0.12  }$

               &

				9.7

               &

				9.7

        \\[1ex]

				Tres-1

               &

					4.95 \raisebox{0.1ex}{ \small{$\pm0.03$ } }

               &

					157 \raisebox{0.1ex}{ \small{$\pm6$ } }

               &

					777 \raisebox{0.1ex}{ \small{$\pm30$ } }

               &

					149.61 \raisebox{0.1ex}{ \small{$\pm0.51$ } }

               &

					$6.02$ ${ }_{ -0.08 }^{ + 0.08  }$

               &

				13.9

               &

				41.1

        \\[1ex]

				TRes-1

               &

					6.2 \raisebox{0.1ex}{ \small{$\pm0.03$ } }

               &

					157 \raisebox{0.1ex}{ \small{$\pm6$ } }

               &

					973 \raisebox{0.1ex}{ \small{$\pm37$ } }

               &

					47.36 \raisebox{0.1ex}{ \small{$\pm0.22$ } }

               &

					$5.8$ ${ }_{ -0.06 }^{ + 0.07  }$

               &

				17.09

               &

				50.2

        \\[1ex]

				TRes-2

               &

					1.11 \raisebox{0.1ex}{ \small{$\pm0.03$ } }

               &

					220 \raisebox{0.1ex}{ \small{$\pm10$ } }

               &

					244 \raisebox{0.1ex}{ \small{$\pm13$ } }

               &

					136.78 \raisebox{0.1ex}{ \small{$\pm2.36$ } }

               &

					$3.97$ ${ }_{ -0.01 }^{ + 0.01  }$

               &

				86.1

               &

				102.3

        \\[1ex]

				TRes-4

               &

					1.54 \raisebox{0.1ex}{ \small{$\pm0.03$ } }

               &

					479 \raisebox{0.1ex}{ \small{$\pm26$ } }

               &

					738 \raisebox{0.1ex}{ \small{$\pm43$ } }

               &

					1.16 \raisebox{0.1ex}{ \small{$\pm1.2$ } }

               &

					$4.5$ ${ }_{ -0.02 }^{ + 0.02  }$

               &

				51.6

               &

				64

        \\[1ex]
    
    \hline
    \end{tabular}
} 

    \caption{
Companions detected within 6.5 arcseconds of target. Some targets were independently observed using the Astralux lucky imaging camera \citep{Hormuth2008a}, with results presented in \cite{Daemgen2009}. Of these, we independently confirm companions for TrES-2 and TrES-4. HAT-5 and TrES-1 were also observed by \cite{Daemgen2009}, but companions were not detected --- the new detections in this dataset are due to deeper faint companion detection limits. 
\newline *The companion to HAT-P-6 is a visual detection. The residual images are presented in Figure~\ref{fig:HATp6_resids}.
}
    \label{tab:companions}
\end{table}

\begin{figure}
\begin{center}

\subfigure[Left: Original image. Right: Image after subtraction of primary and application of matched filter.]{
	\includegraphics[width=0.7\textwidth]{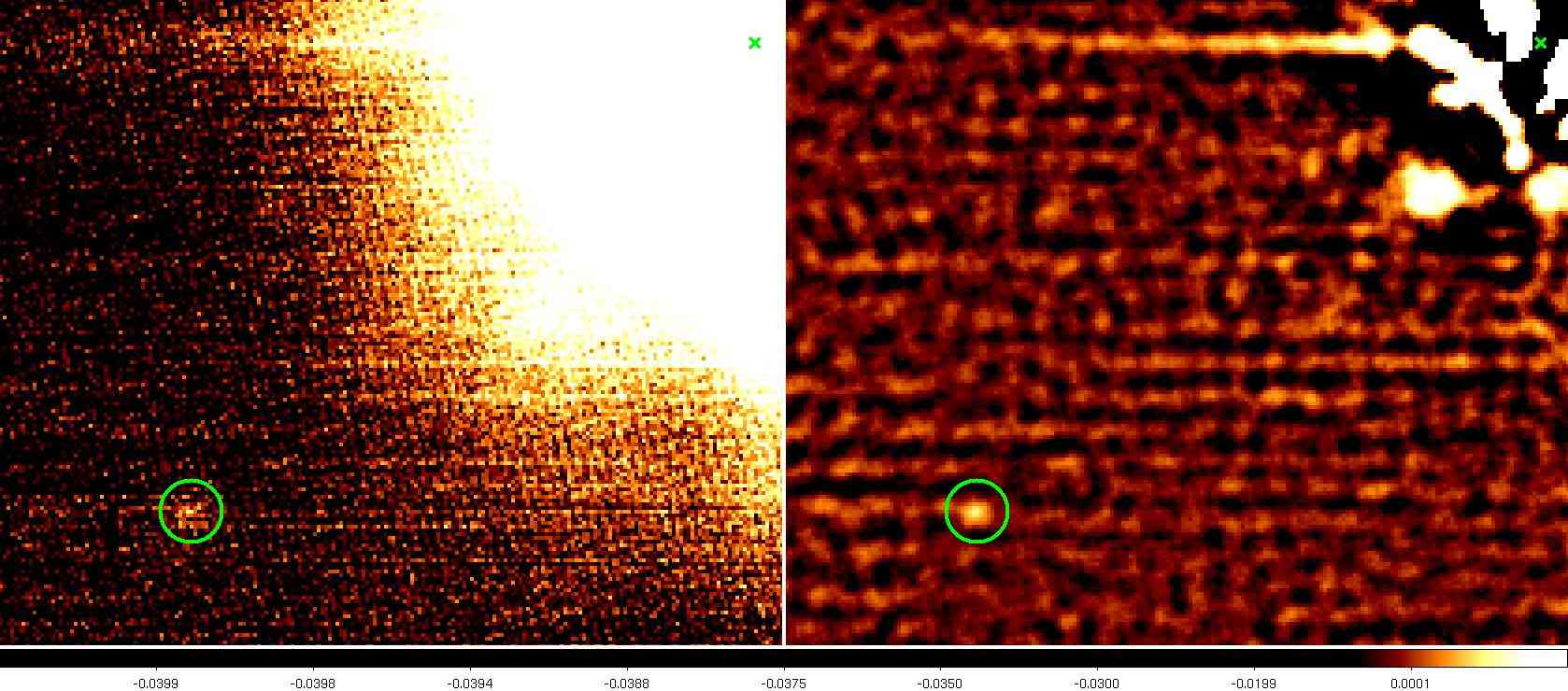}
}
\\
\subfigure[Cross section from original image.]{
	\includegraphics[width=0.45\textwidth]{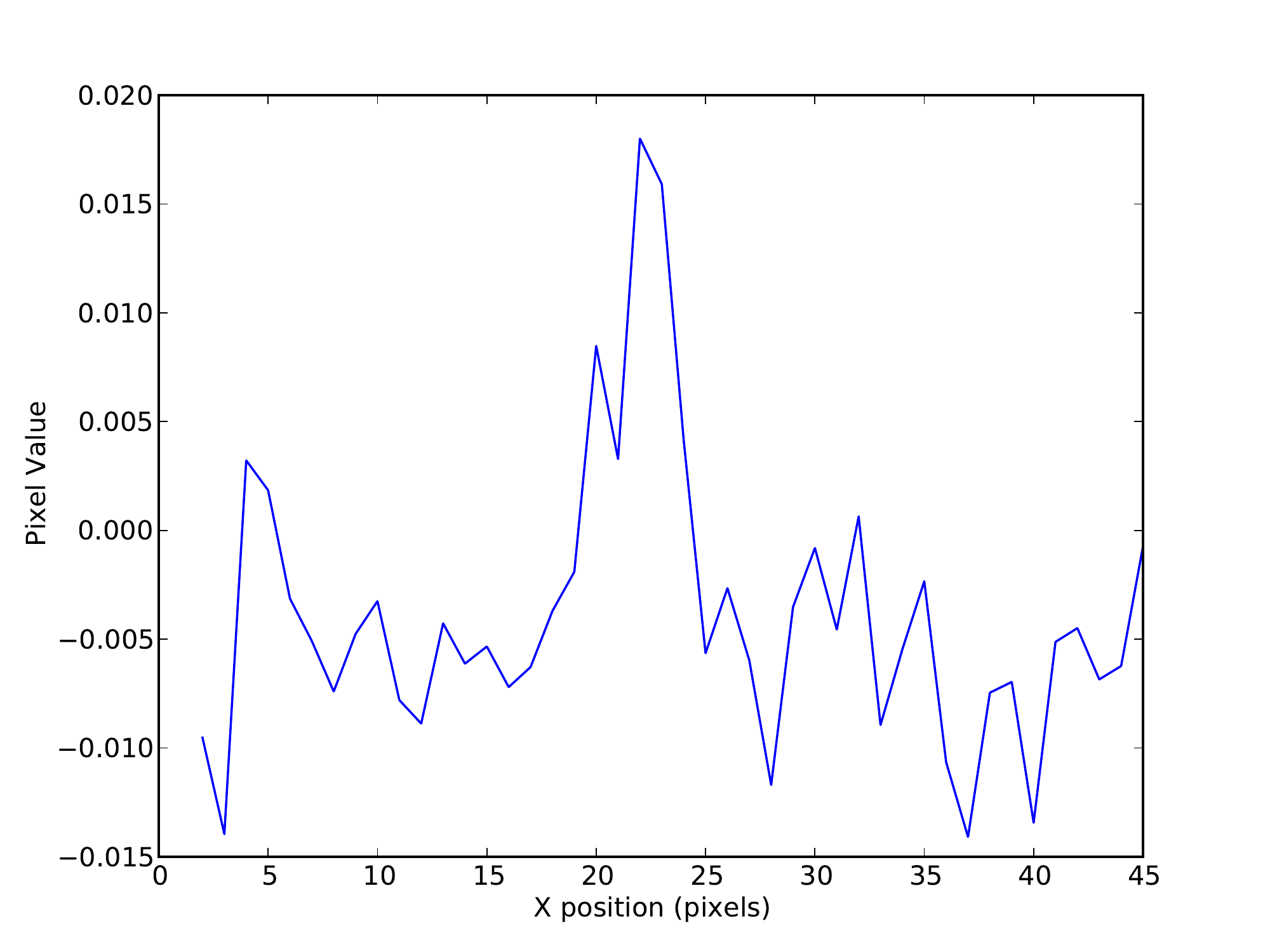}
}
\subfigure[Cross section from processed image.]{
	\includegraphics[width=0.45\textwidth]{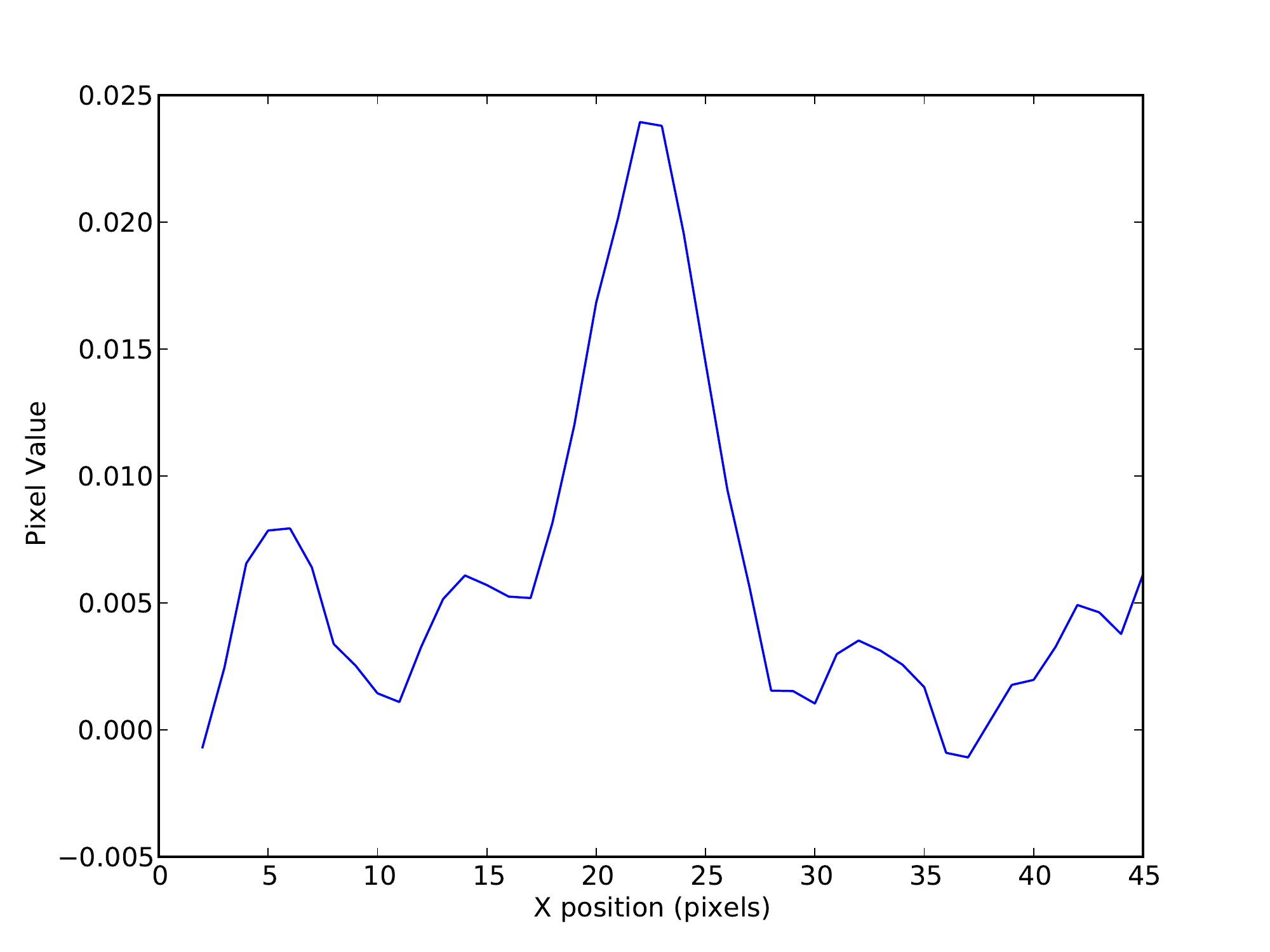}
}
\caption[HAT-P-6 residual images]{Visual detection of a faint companion to HAT-P-6. The faint companion is circled in the images. Images are $\sim$7 arcseconds across. Plots are shown of cross-sections across the circled regions to give an impression of the signal-to-noise ratio.}
\label{fig:HATp6_resids}
\end{center}
\end{figure}

\begin{figure}[htp]
\begin{center}
\subfigure{
	\includegraphics[scale=0.7]{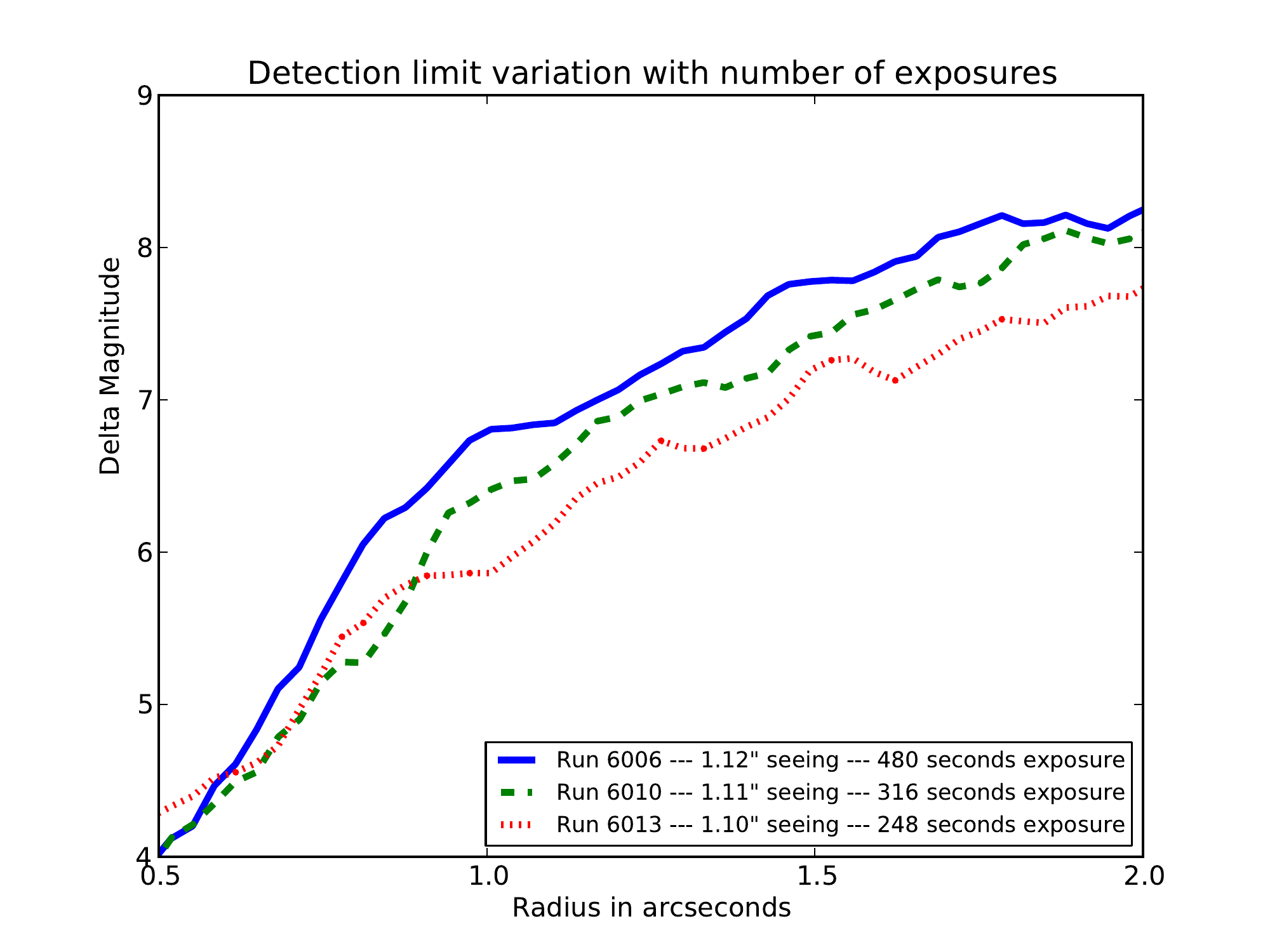}
	\label{subfig:detection_limits_time}
}
\\
\subfigure{
	\includegraphics[scale=0.69]{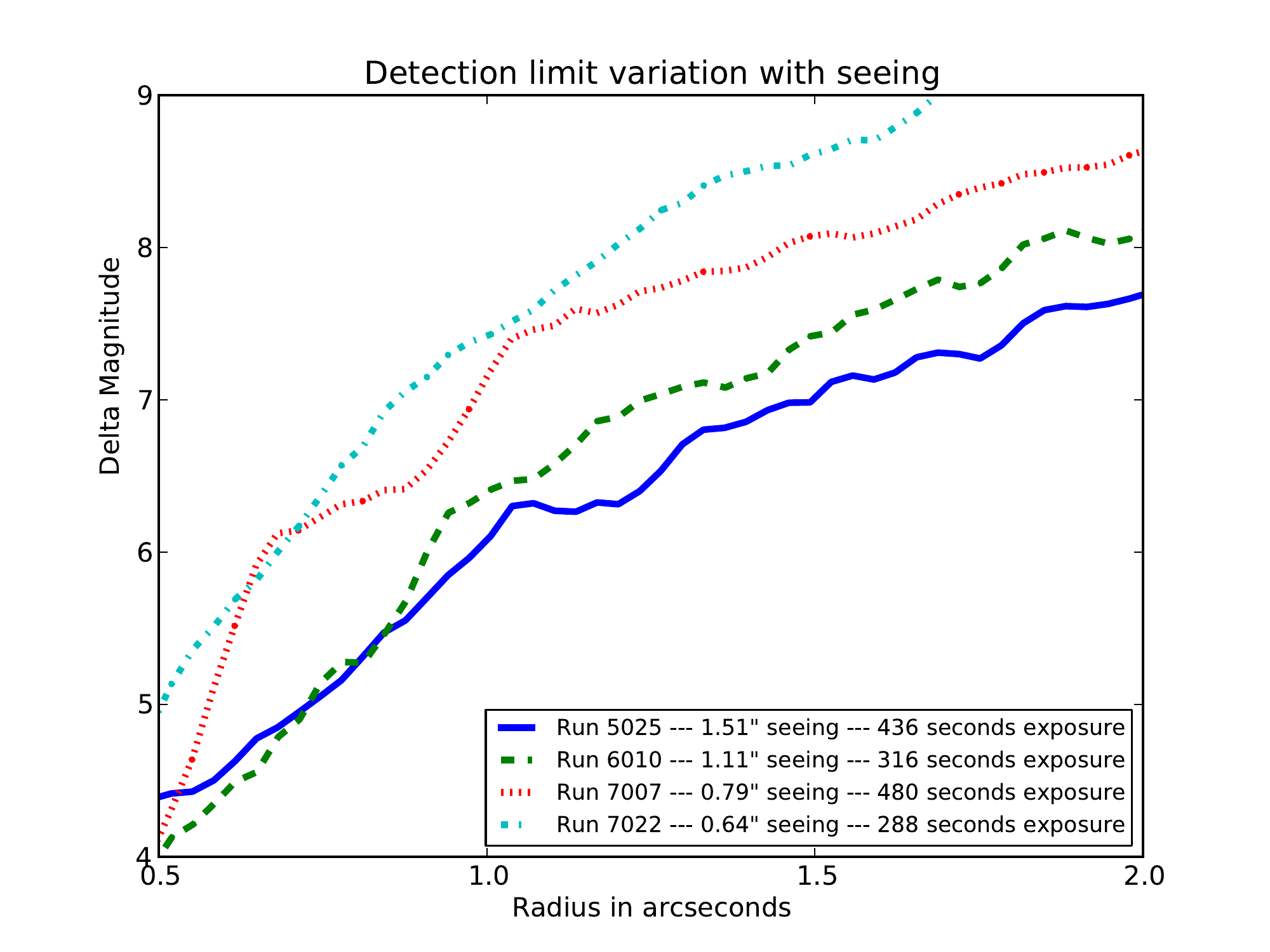}
	\label{subfig:detection_limits_seeing}
}
\caption[Binary detection limits]{Faint binary companion detection limits}
\label{fig:binary_detection_limits}
\end{center}
\end{figure}

Detection limits at a SNR threshold of 4$\sigma$ are plotted in Figure~\ref{fig:binary_detection_limits}. As one would expect, the limits improve quite rapidly as the seeing improves, and also improve gradually as more exposures are gathered. Beyond these general trends it is difficult to draw any more quantitative conclusions from this relatively heterogeneous dataset --- as the plots show the limits are somewhat noisy, as one would expect from exposure times on the order of 5 minutes. However, what is clear is that even with slightly aberrated images good binary detection limits can be achieved on short timescales with lucky imaging, and excellent detection limits are achievable in good seeing conditions.


\section{High temporal resolution photometry with EMCCDs}
\label{sec:fast_photometry}
Observations at high temporal resolution are rare in optical astronomy, leaving potential for a range of investigations if such data were available. Notably, the `Ultracam' group have undertaken a very successful science program with a low noise, high speed camera based on frame transfer CCD technology \citep[see for example][]{Dhillon2007a}.

\cite{Law2006} gave a proof of concept demonstration that lucky imaging EMCCD cameras could be used for observation of rapidly varying optical sources. Observations of the Crab pulsar were taken with a 100Hz frame rate, clearly displaying the 33 millisecond variation pattern. The larger format cameras available for the 2009 observing run provide further opportunities to exploit the high temporal resolution of lucky imaging data. The CCDs can run in a 1000x200 pixel format with a frame rate of approximately 100Hz. The wider format now available (relative to the previous detector size of $512\times512$ pixels) gives a better capability for positioning multiple bright sources in the field of view, enabling relative photometry at these high frame rates. Potentially, the very low effective read noise levels of EMCCD cameras may improve the faint limit of such techniques. However, there is a trade-off at intermediate signal levels due to increased noise from the stochastic multiplication process. 

A small number of observations were recorded during the 2009 observing run with two aims. Firstly, detecting high speed optical variability in the x-ray source Cygnus X-1 as has been previously detected in other sources \citep{Durant2011}, and secondly providing some data on the accuracy of the high-speed photometry achievable with LuckyCam.

\subsection{Cygnus X-1}
\corrbox{
Cygnus X-1 is the best known example of a high-mass X-ray binary, consisting of a black hole of mass $\sim15M_{\odot}$ in a tight orbit about a giant star of mass $\sim19 M_{\odot}$ \citep{Orosz2011}. 
X-ray binaries are an intensely studied class of objects with a rich set of phenomena arising from the interacting extremes of strong gravity and dense matter \citep[see e.g.][]{Remillard2006}. Insight into these systems has been gained primarily by studying the highly variable emission across a range of wavelengths. The emission results from accreting matter in clumpy stellar winds releasing gravitational potential energy during infall \citep[see e.g.][and references therein]{Oskinova2012}. 
Flux variation from these sources can be extremely rapid --- the dynamical timescales resulting from accretion onto a compact object such as a black hole or neutron star are on the order of milliseconds \citep{Klis2000}, and observations of X-ray binaries have recorded variations over a wide range of timescales, right down to the predicted millisecond regime \citep{Westphal1968,Motch1982,Meekins1984}. However, while rapid variation has been detected in both X-ray and optical, \emph{simultaneous} observations in both bands with high-temporal resolution are rare \citep{Durant2011}. The July 2009 observing run gave us a chance to test LuckyCam to see if it would be suitable for performing such observations.
}
\begin{figure}
\begin{center}

\subfigure[Reduced image]{
	\includegraphics[width=1\textwidth]{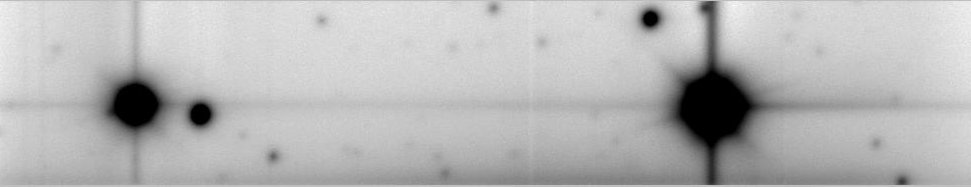}
	\label{subfig:drizzled}
}
\\
\subfigure[Single frame of data]{
	\includegraphics[width=1\textwidth]{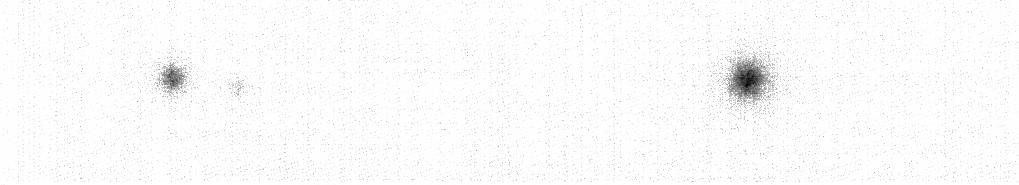}
	\label{subfig:single_frame}
}
 
\caption[Cygnus X-1]{
\subref{subfig:drizzled}: A reduced image created from the high-frame rate observations of Cygnus X-1. Cygnus X-1 is the bright source to the right of the image. Width of the field of view shown is around 90 arcseconds, North is left and West is up.
\subref{subfig:single_frame}: A single short exposure from the dataset.
}
\label{fig:cygnus_x_1_fov}
\end{center}
\end{figure}

As part of the 2009 observing run, observations of Cygnus X-1 were recorded in the 1024x200 pixel frame-size mode. Around 700 000 short exposures were recorded over two observations, over a total observing time of around 2 hours. The lens resulting in a pixel size of 95 milliarcseconds was employed, since high angular resolution is not a priority here, and larger pixel widths result in better signal levels.
Figure~\ref{fig:cygnus_x_1_fov} depicts the field of view observed for the CCD containing the target.

\subsection{Estimating a lower bound to signal variance}
When considering the likely accuracy of high temporal resolution photometry, we may derive a first estimate of the expected signal variance due to photon shot noise and detector noise. Obviously this neglects variance in the photometry due to PSF variations induced by atmospheric turbulence, and thus forms a lower bound. Recalling equation~\ref{eq:SNR_EMCCD_linear}, we may first consider what are likely to be dominant sources of variance for this bright target. Total signal from Cygnus X-1 was around 4000 photo-electrons per frame. Seeing width was estimated at around 15 pixels (1.42 arcseconds). 
While every short exposure will of course have a different PSF, we may proceed to analyse the per pixel signal-to-noise ratio equation by simply assuming a uniform illumination within an aperture of the same diameter as the seeing FWHM, such that the mean signal per pixel is 
\begin{equation}
 M = \frac{4000  }{\pi (7.5)^2} \approx 20 \textrm{ photo-electrons per pixel}
\end{equation}
To avoid detector saturation, Cygnus X-1 was observed at an EM gain (c.f. section \ref{sec:pixel_PDFs}) $g_A=10.5$ ADU / photo-electron. Readout noise had an estimated standard deviation of 12.2 ADU, equivalent to 1.15 photo-electrons signal level. Background signal, including sky flux and CIC signal, was around 0.3 photo-electrons per pixel per frame. Clearly then, for such a bright target the major contributor to signal variance will be the shot noise, as might be expected. Including the additional variance factor due to stochastic multiplication, the per frame SNR estimate is then simply
\begin{equation}
 SNR = \frac{ M }{\sqrt{2M}} \approx 45
\label{eq:fast_phot_SNR}
\end{equation}
or equivalently, for the Cygnus X-1 data we expect a minimum standard deviation in the per-frame source signal of around 0.02 times the mean source signal, due to photon-shot noise and detector effects alone. 

Of course, the real signal variance will unavoidably be larger. Atmospheric effects cause variation in the PSF so that some of the source flux may fall outside the photometric aperture, the aperture may be misplaced due to a position estimation error, and there may be intrinsic variation in atmospheric transparency. Minimising this additional variation requires judicious choice and application of data reduction techniques.

\corrbox{In hindsight, the observations of Cygnus X-1 would have achieved a signal-to-noise ratio better by a factor of $\sqrt2$, if the detectors had been switched to the conventional readout, rather than utilising the electron multiplication register. This is obvious when considering the `three regimes' as outlined in Section~\ref{sec:three_regimes}, but unfortunately was not considered during the observing run.%
\footnote{In defence of the LuckyCam observing team, the observing run was fairly intense, with 24 hours to assemble the camera shortly after arrival, and residual software glitches requiring some last minute bug fixing. Future instrumentation scientists pay heed and plan your observations carefully!}
}

\subsection{Fast photometry data reduction techniques}
\begin{figure}
\begin{center}
 \includegraphics[width=1\textwidth]{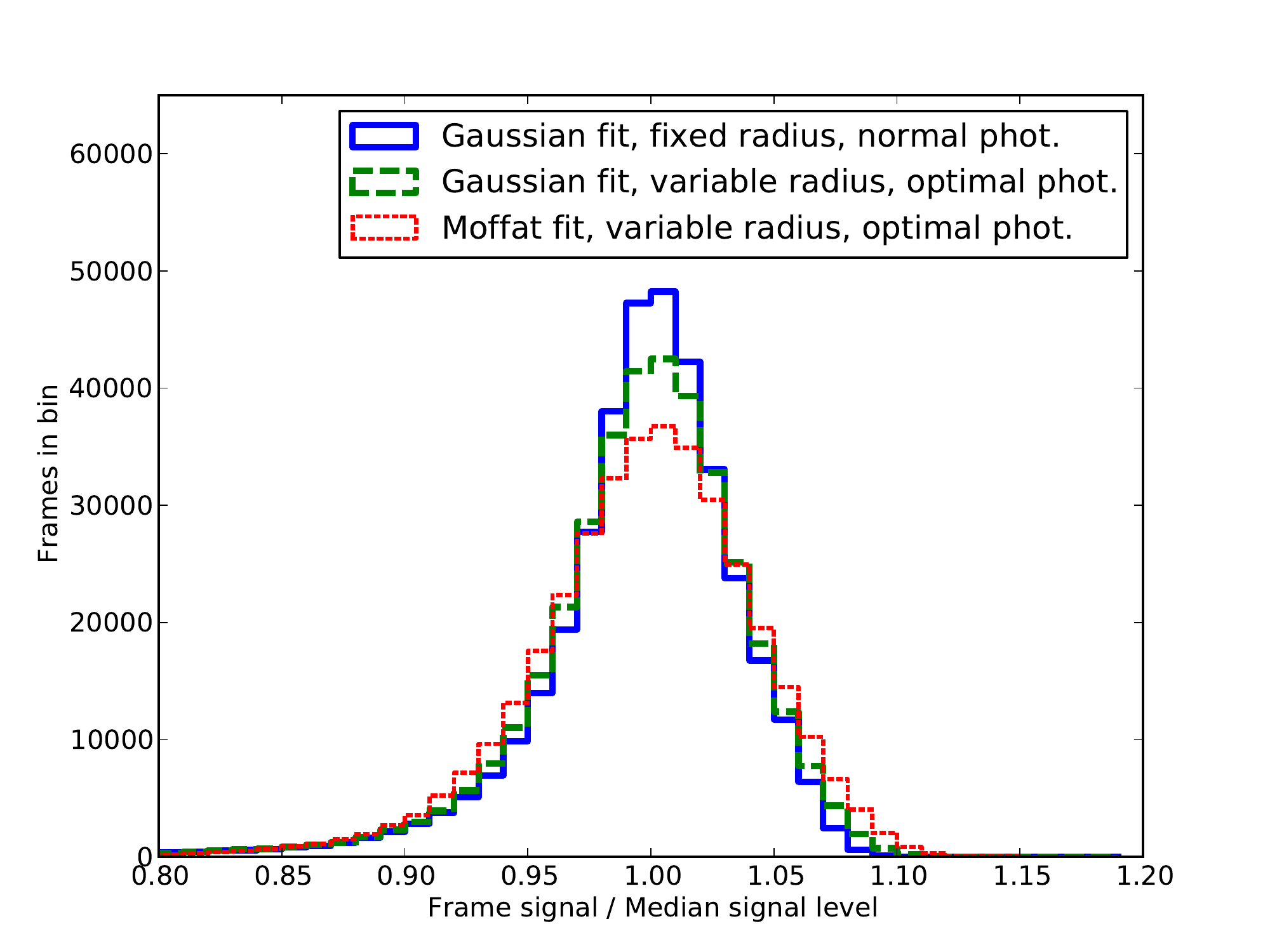}
\caption[Cygnus X-1 photometry methods]{A comparison of the different photometry extraction methods available in the Ultracam reduction package, as applied to the Cygnus X-1 data. Photometric counts obtained using different methods were normalised by the median of each dataset, then plotted in a histogram. A larger variance results in a wider histogram. The legend refers to various parameters of the extraction process; whether the PSF in each frame is fitted using a Gaussian or Moffat profile, whether the photometric aperture has a fixed or variable radius, and whether optimal or regular photometry is employed. For this dataset, employing a simple photometric aperture of fixed radius gives the least variance. Further details in text.
}
\label{fig:ultracam_method_hists}
\end{center}
\end{figure}

I developed two methods for reducing the Cygnus X-1 dataset. The first was to make some alterations to the lucky imaging pipeline described in Section~\ref{sec:pipeline}, to implement a simple aperture photometry algorithm. First, the user selects the target sources by selecting regions using a simple average image. For each source, a circular aperture is defined about the local maximum pixel. The brightest source is then utilised as a guide star to adjust the aperture positions in each frame --- a thresholded centroiding algorithm is used to locate the weighted centre of the source, and the position compared to the position measured in the average frame. The apertures are then shifted accordingly. This approach is robust, since the centroiding algorithm is very reliable, and fast, as the pipeline multi-threading capabilities may be employed. A reasonable processing speed is useful since an hour's worth of data is around 150 gigabytes when uncompressed. However, simple aperture photometry has disadvantages. 
Over the course of an hour's observation we expect the PSF width to vary significantly, and as a result the proportion of flux within the defined aperture will vary (unless we define a very large aperture, which will give unnecessarily poor signal-to-noise for most frames).

With these considerations in mind I employed the Ultracam data reduction program as described in \cite{Dhillon2007a}. This involved some modification of source code to allow conversion from lucky imaging FITS format to the Ultracam `.ucm' format, and a somewhat laborious process of decompressing all the individual frames before storing to disk. However, it enabled use of the well developed Ultracam pipeline feature set, which includes optimal photometry \citep[photometry using annular apertures weighted according to SNR, see][]{Naylor1998}, and a choice of position estimation and frame by frame PSF width estimation via fitting of a Moffat or Gaussian profile. 

\begin{figure}
\begin{center}
\subfigure{
	\includegraphics[width=0.3\textwidth]{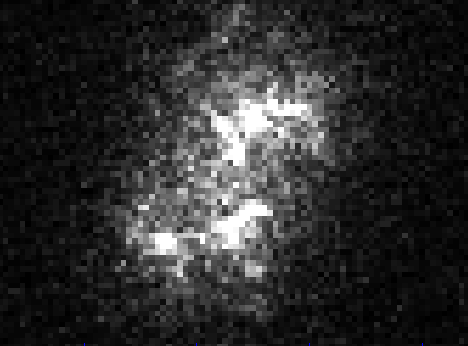}
}
\subfigure{
	\includegraphics[width=0.3\textwidth]{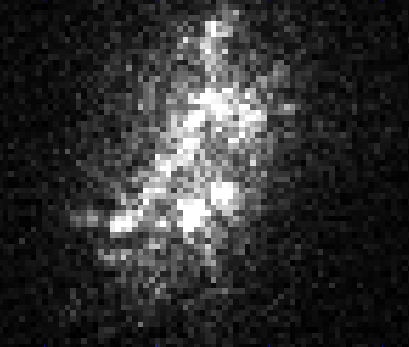}
}
\includegraphics[width=0.3\textwidth]{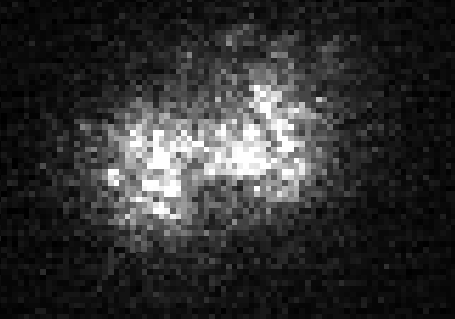}
\caption[Frames of data rejected by the Ultracam pipeline]{
Close-ups of the PSF from frames of data from the Cygnus X-1 observation, rejected by the Ultracam pipeline due to inability to locate a PSF position with a Gaussian profile fit.
}
\label{fig:ultracam_bad_frames}
\end{center}
\end{figure}

Experimentation with the Ultracam pipeline found that for this dataset simple aperture photometry, using a fixed radius aperture, was in fact the method which produced least variance in the recorded photometric signal, as illustrated by the histograms in Figure~\ref{fig:ultracam_method_hists}. With a bright source such as Cygnus X-1 the benefits of sophisticated photometric techniques are outweighed by the variation introduced by aperture adjustment with a varying PSF, and a fixed, large aperture size is more reliable. Indeed, some of the frames were assigned error codes by the Ultracam pipeline and rejected, since no fit could be made to the PSF to determine position. These generally occur when the PSF clearly has multiple bright speckles, as depicted in Figure~\ref{fig:ultracam_bad_frames}, which presumably cause the fitting routine to not converge.%
\footnote{\corr{From conversations with Vik Dhillon, I have since learned this is usually preventable with appropriate adjustment of the Ultracam pipeline centroiding 
settings.}
}

\begin{figure}
\begin{center}
 \includegraphics[width=1\textwidth]{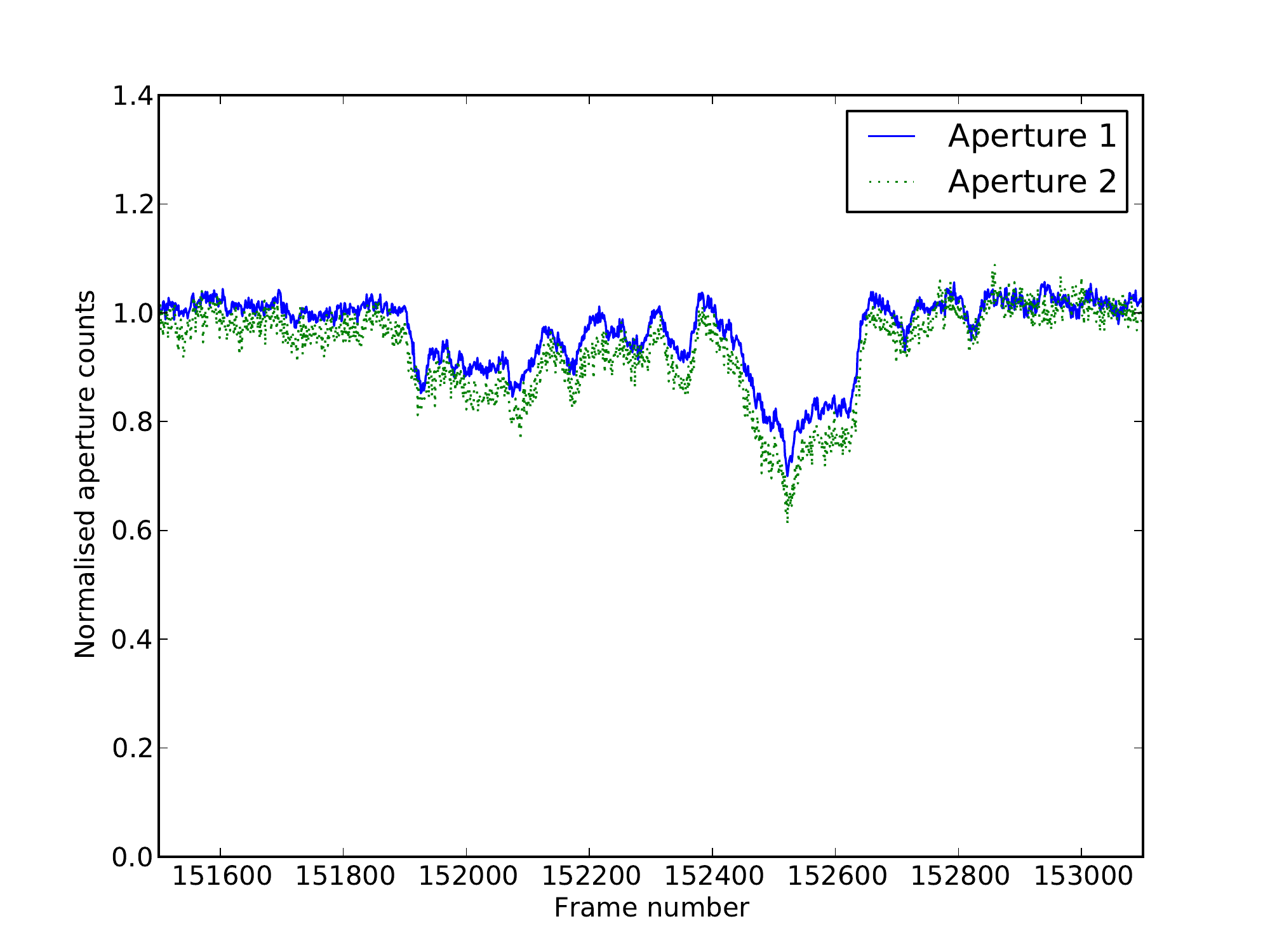}
\caption[Atmospheric effects on fast photometry]{
A plot depicting normalised photometric counts from the Cygnus X-1 data, for a selected time interval illustrating a drop in photometric counts due to atmospheric effects. Apertures 1 and 2 correspond the the sources on the right and left respectively in Figure~\ref{fig:cygnus_x_1_fov}.
}
\label{fig:atmos_drop}
\end{center}
\end{figure}

\subsection{Results}

\begin{figure}
\begin{center}
\subfigure[Aperture 1]{
	\includegraphics[width=0.45\textwidth]{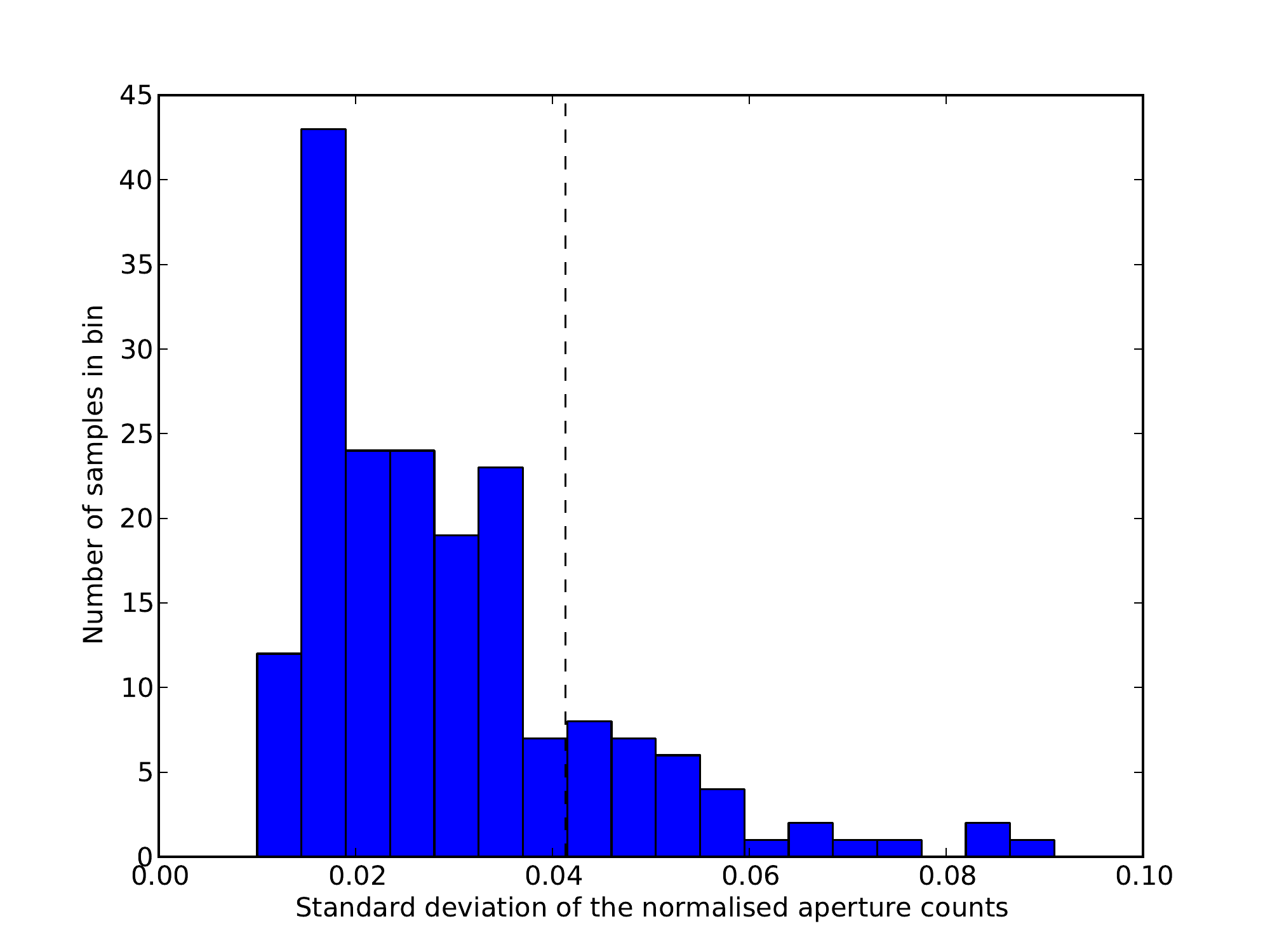}
}
\subfigure[Aperture 2]{
	\includegraphics[width=0.45\textwidth]{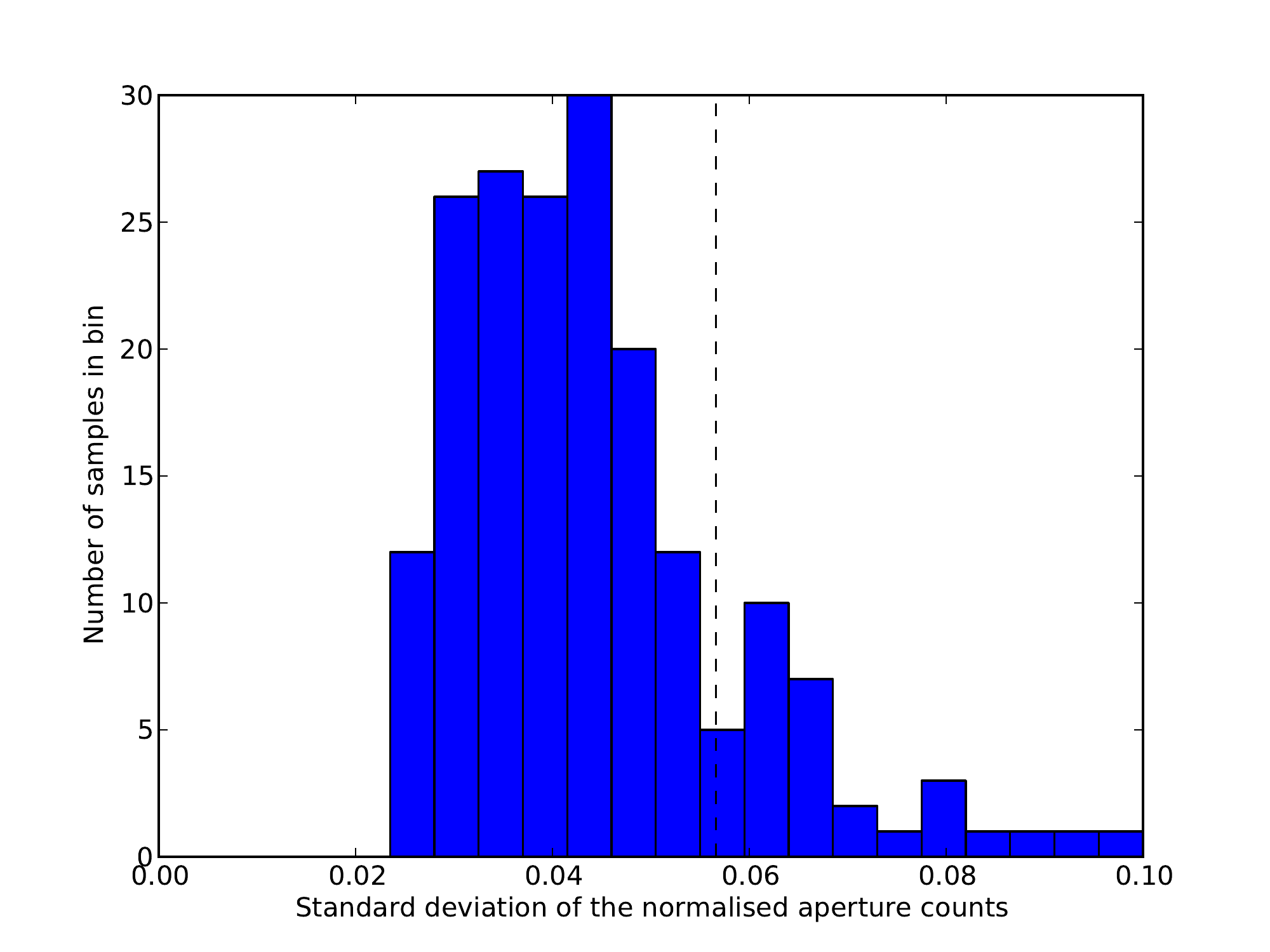}
}
\\
\subfigure[Ratio of normalised aperture counts]{
	\includegraphics[width=0.5\textwidth]{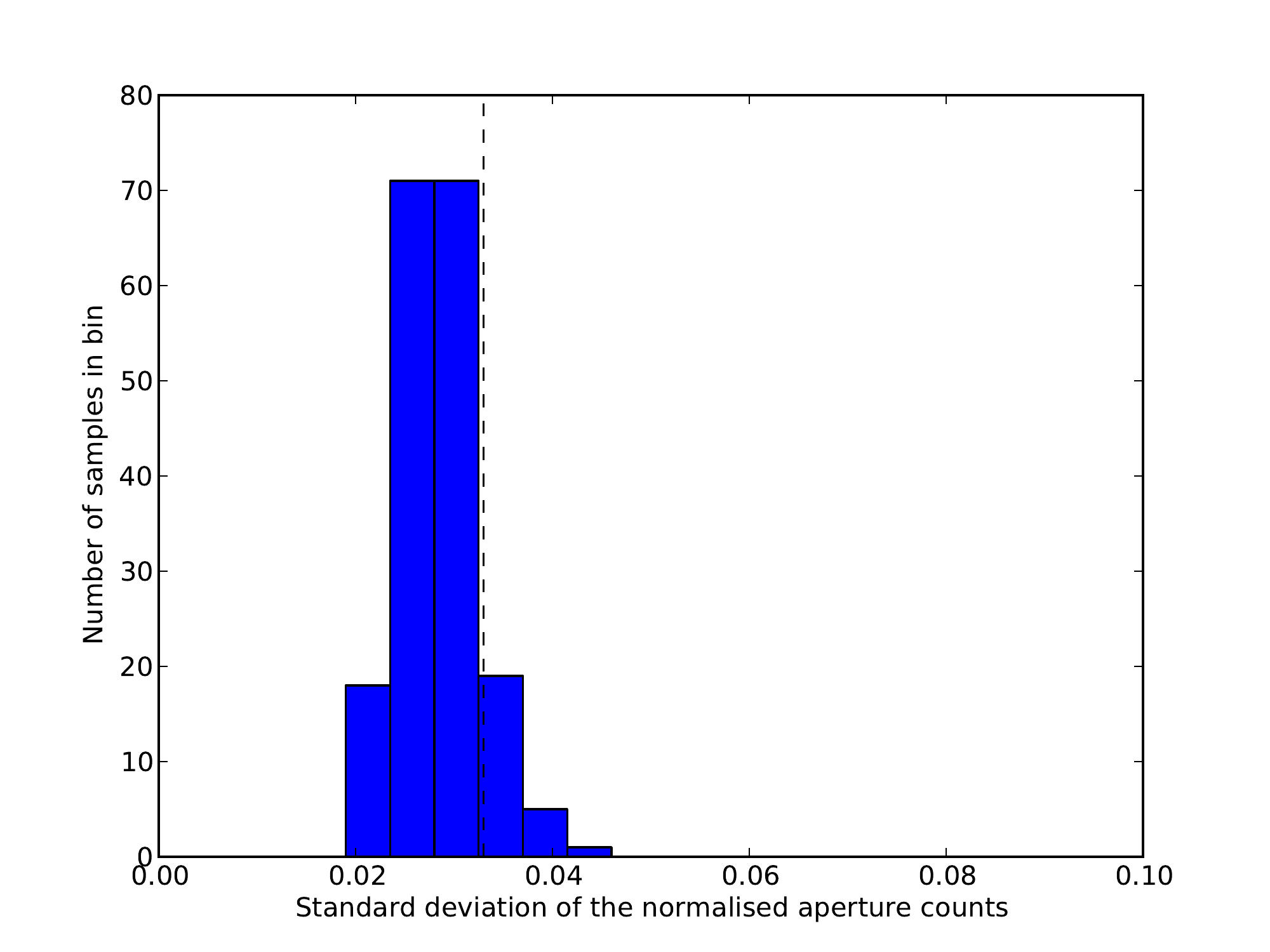}
	\label{fig:fast_phot_ratio_subsamples}
}
\caption[Standard deviation in sub-samples of the fast photometry]{
Standard deviation in sub-samples of the normalised Cygnus X-1 photometry datasets. Sub-samples were created by simply splitting up the full dataset into bins of 2000 frames, corresponding to around 20 seconds worth of data per bin. Dashed vertical lines denote the standard deviation in each full dataset.
}
\label{fig:fast_phot_subsamples}
\end{center}
\end{figure}

Regarding optical variability in Cygnus X-1, a preliminary analysis of the extracted photometric time series was undertaken by Poshak Gandi \cite[who has some experience in this area, see][]{Durant2011}. \corr{Power spectrum analysis} revealed no indication of an intrinsic signal. Some coherent variations at slightly longer timescales was present in the data due to the atmospheric variability, which was considerable --- drops in photometry of up to 30\% were encountered, as illustrated in Figure~\ref{fig:atmos_drop}. Regarding the general feasibility of high temporal resolution with EMCCDs, the poor atmospheric conditions again rather limit any conclusions that may be drawn from the data. However, I did consider the variation in the data during intervals shorter than the full observing period, since we expect at least short periods of relative atmospheric stability. Figure~\ref{fig:fast_phot_subsamples} depicts the histograms of standard deviations in sub-samples of the data, created by simply binning the 
full dataset into sections of 2000 frames, corresponding to around 20 seconds of observations per sample. Normalised aperture counts derived by dividing the raw counts by the median aperture count were used for the analysis. Note that the histogram peaks at a standard deviation in the normalised counts of around 0.02, which is the value predicted by the simple estimate of variance due to photon shot noise and detector noise in equation~\ref{eq:fast_phot_SNR}. Similarly, the number of sample standard deviations for the second aperture peak at around 0.04 --- this doubling of standard deviation is to be expected if photon shot noise is the dominant factor, since the source of aperture 2 has a flux level around 1/4 of Cygnus X-1. This gives some supporting evidence that when atmospheric conditions are stable, the noise levels in high temporal resolution photometry performed with EMCCDs conform to the noise level predicted using simple signal-to-noise considerations --- at least in the context of bright sources. 

Of course, there will always be some degree of atmospheric variation, which is why we hope to employ relative photometry. Figure~\ref{fig:fast_phot_ratio_subsamples} displays a similar histogram of sample standard deviations, created using a time-series of the ratio between the normalised photometric counts for apertures 1 and 2. Much of the large range variation due to atmospheric effects is removed as a result. 

\subsection{Future work}
This preliminary study gives some evidence that EMCCDs are a viable detector for high temporal-resolution photometry, but perhaps more importantly it has raised a number of issues for further consideration when attempting future observations. 

First and foremost is the matter of observation timing. When collating multi-wavelength data from multiple observatories, an accurate time stamp to accompany the data is crucial. For the summer 2009 LuckyCam set-up, absolute time-stamping of short exposures was a secondary consideration, since the primary concern was to ensure that the multiple EMCCDs in the camera mosaic were simply synchronised. As such, absolute timings relied upon an uncalibrated PC clock stamp, with an undetermined error due to delay in the data acquisition pipeline. As such, an absolute timing accuracy of around 0.5 seconds is probably the best that could be expected. If future observing campaigns are to focus on high temporal resolution work, then an added time-stamping facility, possibly along the lines of the GPS module described in \cite{Hormuth2008}, will be a key consideration.

Second is the consideration of data reduction software. If large numbers of observations are to be recorded it may be worthwhile to implement the per-frame photometric algorithms using the multi-threaded lucky imaging pipeline as a basis, or to adapt the Ultracam pipeline to a multi-threaded form.

\corr{Finally, care must be taken to ensure observations are made with the detector in the appropriate mode (conventional or electron-multiplication readout), depending upon which of the `three regimes' the observation falls under} (sec.~\ref{sec:three_regimes}). On a more interesting note we may also consider faint targets, where a bright companion may be used for photometric comparison and guiding of the photometric aperture position, while photon-thresholding techniques are employed to enhance the signal-to-noise ratio of the faint variable source. The mosaic nature of the camera will enable this sort of observation, since a high EM gain may be set on one EMCCD and conventional readout employed on another, so that both the bright and faint sources are observed with an optimal detector set-up. If the investigator were so inclined, a numerical study of signal-to-noise ratio in such observations could be undertaken using the simulation tools I describe in chapter~\ref{chap:lucky_AO}.


\section{General high resolution imaging in the visible}
\label{sec:general_science}
With a wide field of view and a sensitive detector, lucky imaging techniques should now be applicable to many more targets, allowing a wide variety of science applications. In this section I give some supporting evidence for the claims that lucky imaging can be applied as a general observing tool.

\subsection{Faint limits}
The faint limit of lucky imaging observations has not previously been explored. If the detector readout noise can be controlled to sufficiently low levels, then lucky imaging should have very good faint limits, compared to other ground-based optical astronomy techniques. The narrower PSF and associated increase in encircled energy at small radii give the technique a significant advantage when considering the case where sky background flux is the dominant noise source. Since the observations are at visible wavelengths, thermal background is less of a problem compared to adaptive optics observations in the infrared. A mosaic field with independently variable electron multiplication gain gives practically unlimited dynamic range, since a bright guide star can be observed on one CCD at low or zero gain without saturation, while a faint target is observed on a nearby sub-field at very high gain. When thresholding techniques are employed the detector noise can be lowered even further.

The best example of faint source detection is the observation of 3C405 discussed in Section~\ref{sec:threshold_real_data}. A source of estimated magnitude $m_i\sim22.5$ was observed with signal-to-noise ratio (SNR) of 6 in approximately 1 hour of observation. For comparison, I generated some SNR estimates using the signal-to-noise calculator 
\footnote{http://www.not.iac.es/observing/forms/signal/v2.2/index.php}
for ALFOSC, the Andalucia Faint Object Spectrograph and Camera, which is a permanent instrument on the NOT. Modelling a set of twelve 5 minute exposures with seeing width of 0.4 arcseconds (matched to our observed seeing) under dark time observing conditions, the SNR for 1 hour of observing time on a 22.5 magnitude source in the \textit{I} band is estimated at around 50. Note that this represents a pixel sampling of 0.19 arcseconds per pixel, rather than the 0.032 arcseconds sampling used for the lucky imaging observations. 
Since our observation is largely limited by detector noise, increasing the pixel scale to match could result in a 34 fold reduction in our noise levels. 
Observing with a pixel size of 0.1 arcseconds would have likewise reduced detector noise levels by around a factor of 9, \corr{giving SNR of around 50} while still benefiting from significant resolution improvements over seeing- limited observations \corr{(considering a reduction process involving 100\% selection of the short exposures, though this is effectively tip-tilt correction rather than lucky imaging)}. 
The CCD configuration used for our observation experienced a significant level of noise \corr{due to clock induced charge} (c.f. Section~\ref{sec:CIC}) and it seems very likely that further tuning of the camera electronics will reduce noise levels significantly.

\subsection{High resolution across a wide field of view}
\label{sec:hires_wide_field}
\begin{figure}[htp]
\begin{center}
 \includegraphics[width=1\textwidth]{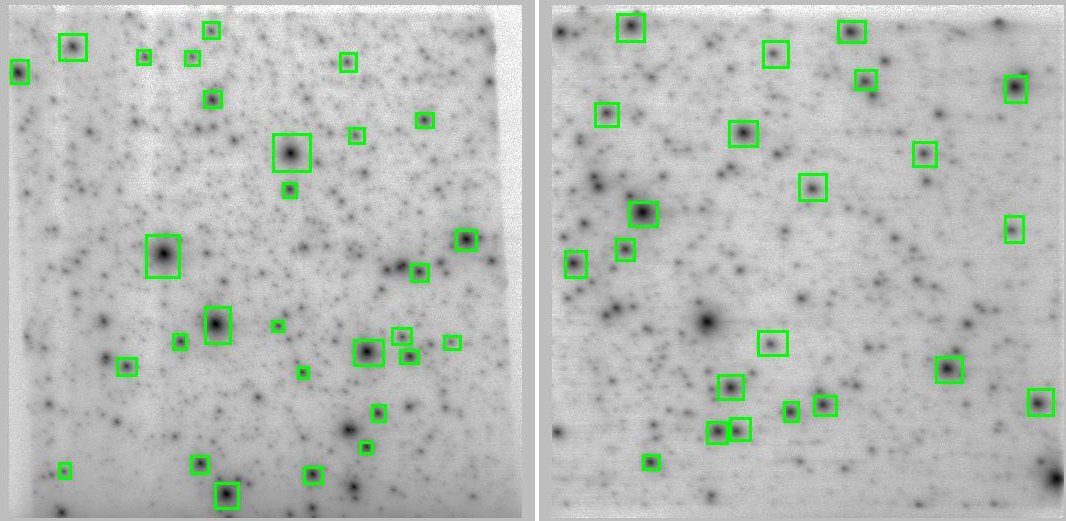}
\caption[M13 FWHM analysis targets]{Drizzled images from two CCDs, produced using a 10\% selection from observations of M13. Further details in text. Green boxes mark manually selected targets for FWHM measurement. Results are plotted in Figure~\ref{fig:M13_FWHM_results}
}
\label{fig:M13_FWHM_targets}
\end{center}
\end{figure}

\begin{figure}[htp]
\begin{center}
 \includegraphics[width=1\textwidth]{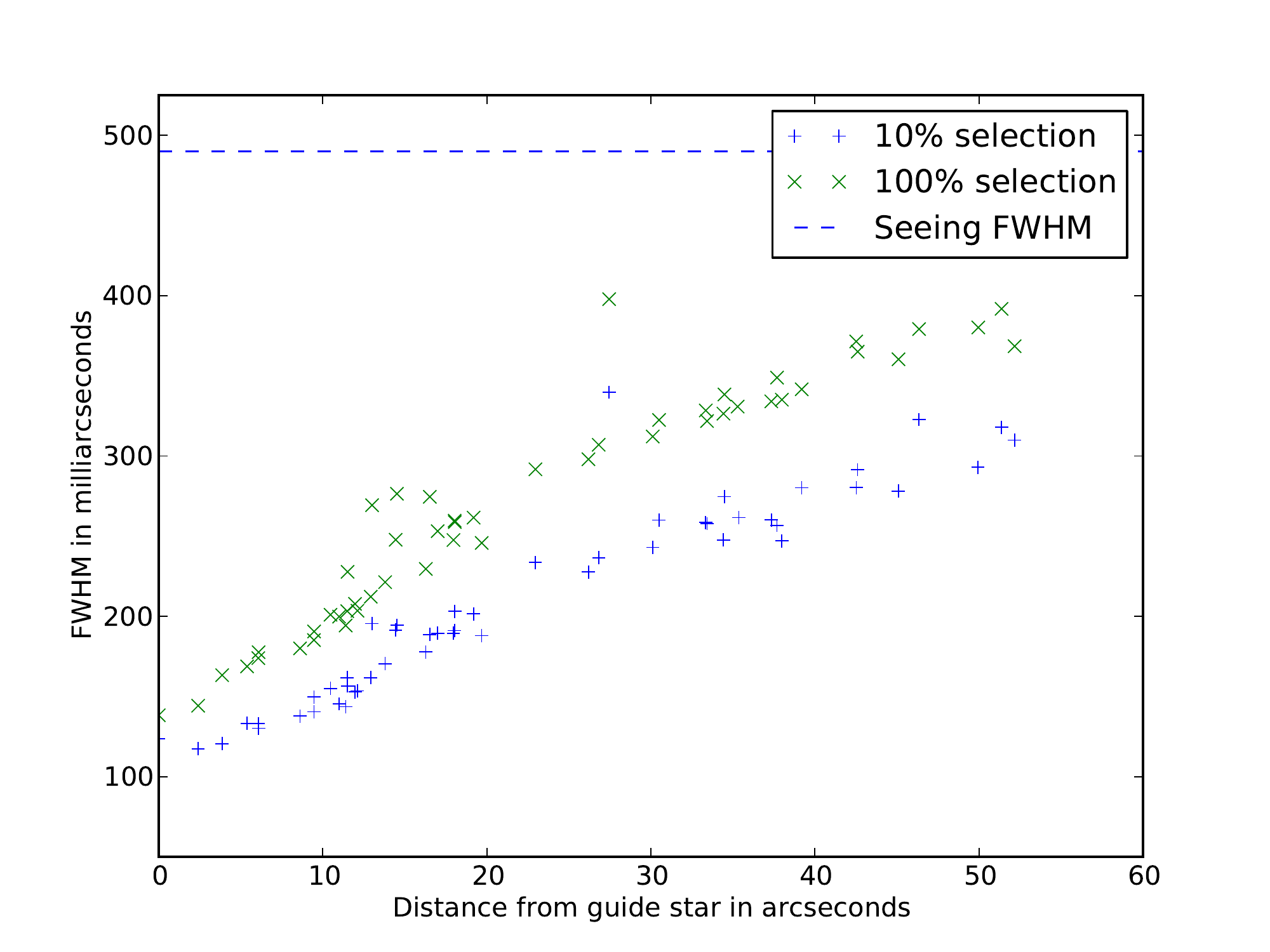}
\caption[FWHM across a wide field of view]{A plot of FWHM variation with separation from the guide star. FWHM improvement by more than a factor of 2 is obtained right out to radii of 30 arcseconds with 10\% selection.}
\label{fig:M13_FWHM_results}
\end{center}
\end{figure}

Previous work has suggested that lucky imaging observations have a wide isoplanatic angle, typically in the region of 15--30 arcseconds  \citep{Tubbs2003,Law2007}. The availability of a wide, well sampled field of view offers the opportunity to verify this estimate and extend investigations to wider angles.

However, accurate measurements of Strehl ratio across a wide field of view are hard to determine. 
Ideally a crowded field should be utilised, to give a good number of sources, but accurate photometry in such fields is difficult --- since the lucky imaging PSF varies slightly across the field, conventional PSF fitting routines will give poor estimates of photometry, and hence poor Strehl ratio estimates. 
However, globular cluster data does allow for measurement of the full width at half maximum (FWHM) across a wide angle, \corr{since this does not rely on accurate characterisation of the faint halo away from the PSF core}. 
To my knowledge investigation of this particular pair of variables (FWHM vs. axis offset) has not been previously published; instead only plots of Strehl vs. axis offset are available in the literature \citep{Tubbs2003,Law2007}.

Figure~\ref{fig:M13_FWHM_targets} displays a pair of drizzled images from different CCDs, produced using a 10\% selection drawn from 6000 short exposures (around 5 minutes total observing time) of the M13 globular cluster. Pixel angular width was 32.5 milliarcseconds, resulting in good sampling of the PSF. Seeing width was estimated at 0.49 arcseconds at the observation wavelength, which was SDSS \textit{i'} band. The boxes overlaid in the image mark manually-selected stars chosen for FWHM analysis. The criteria were that the star should be the brightest source in the local region of the image, without nearby companions of comparable brightness that would significantly affect FWHM measurements.

Figure~\ref{fig:M13_FWHM_results} plots the FWHM measurements resulting from analysis of stellar PSFs over 2 CCDs of the full mosaic, for images resulting from drizzling 10\% and 100\% frame selections. The FWHM measurements were obtained using custom analysis routines written using the lucky imaging libraries (chapter~\ref{chap:data_reduction}). The plots clearly display an improvement in FWHM by more than a factor of 2 even beyond radii of 30 arcseconds, when a 10\% frame selection is employed.


\section{Science with lucky imaging-enhanced adaptive optics: Probing the binary star distribution in globular clusters}
\label{sec:lucky_AO_science}

\begin{figure}[htp]
\begin{center}
\subfigure[A comparison of data from HST WFPC2 (left) and LAMP (right). ]{
	\includegraphics[width=0.6\textwidth]{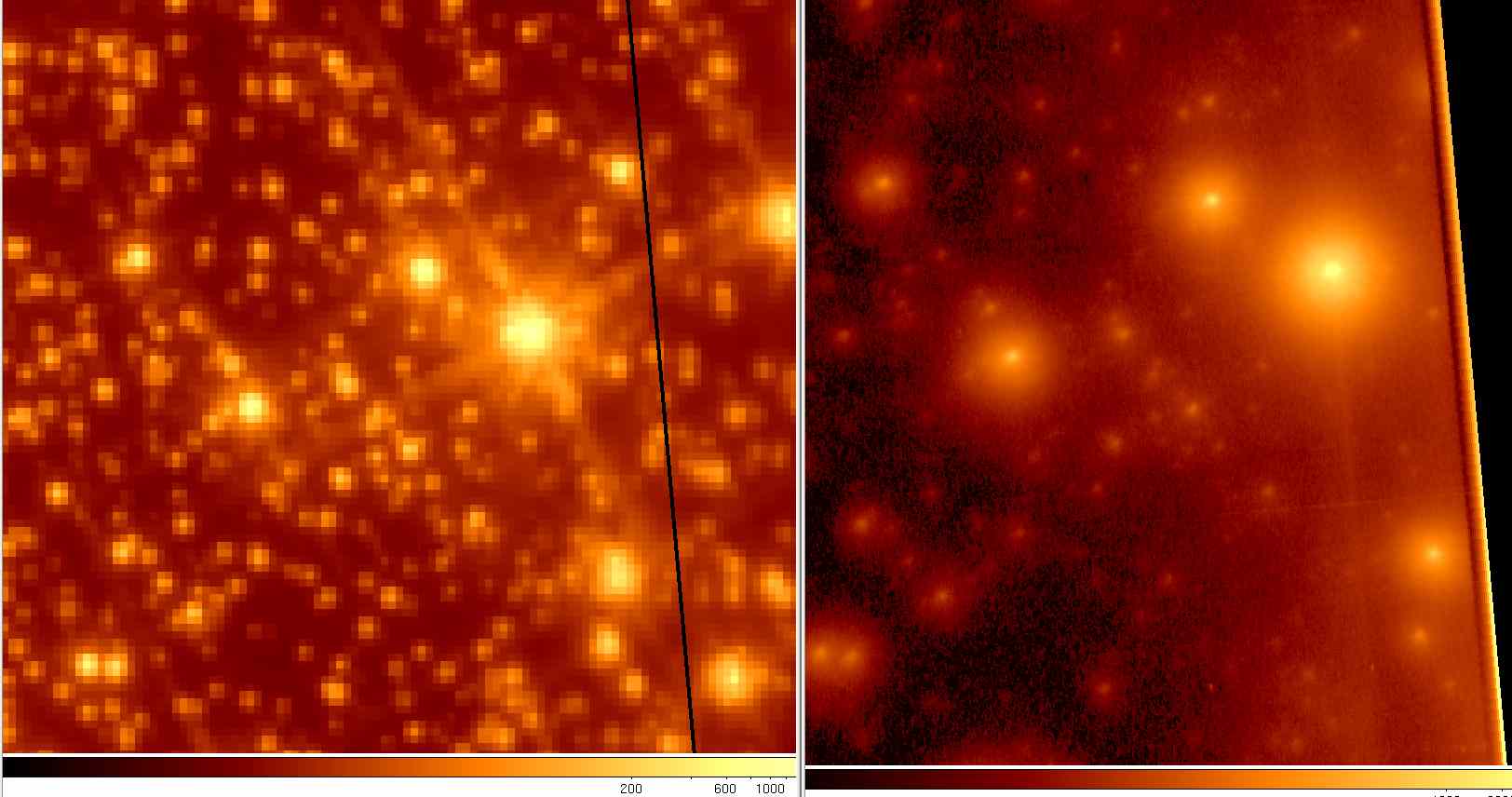}
}
\\
\subfigure[Close up from central region]{
	\includegraphics[width=0.6\textwidth]{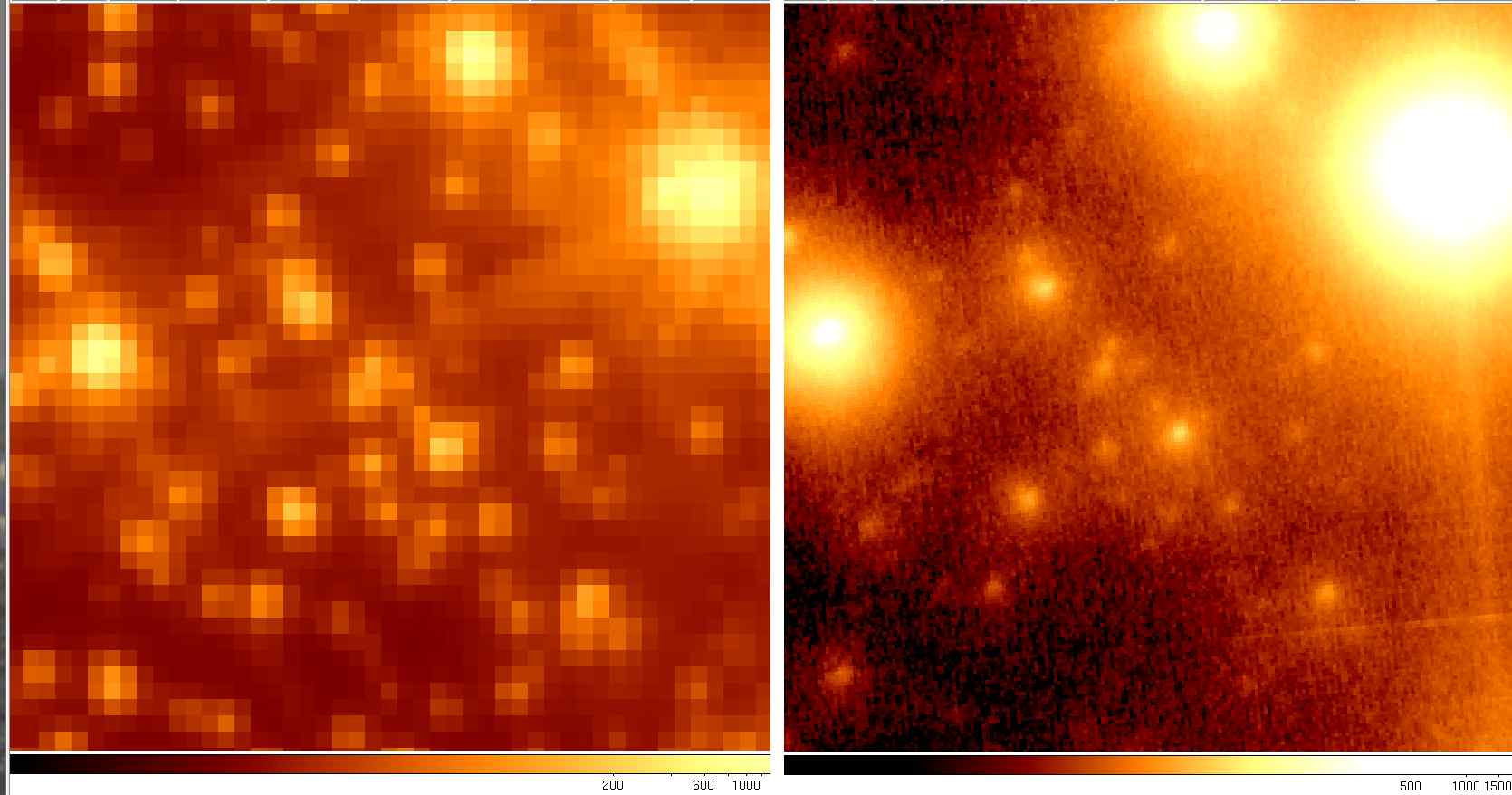}
}
\caption[Comparison of LAMP and HST data]{
A comparison of `Lucky At Mount Palomar' (LAMP) and archival HST WFPC2 images of M13. LAMP observations were taken behind the PALMAO adaptive optics system on the 5 meter Hale telescope. The LAMP dataset was obtained in under 5 minutes of time on-sky. The HST image results from $\sim$1.5 hours of observations (archival data retrieved from the Hubble Legacy Archive, dataset proposal ID is 8278). Close inspection of the images reveals distinct sources in the LAMP image which are unresolved in the HST image \corr{(dead centre in close-up images)}.
}
\label{fig:LAMP_HST_comparison}
\end{center}
\end{figure}

\begin{figure}[htp]
\begin{center}
 \includegraphics[width=1\textwidth]{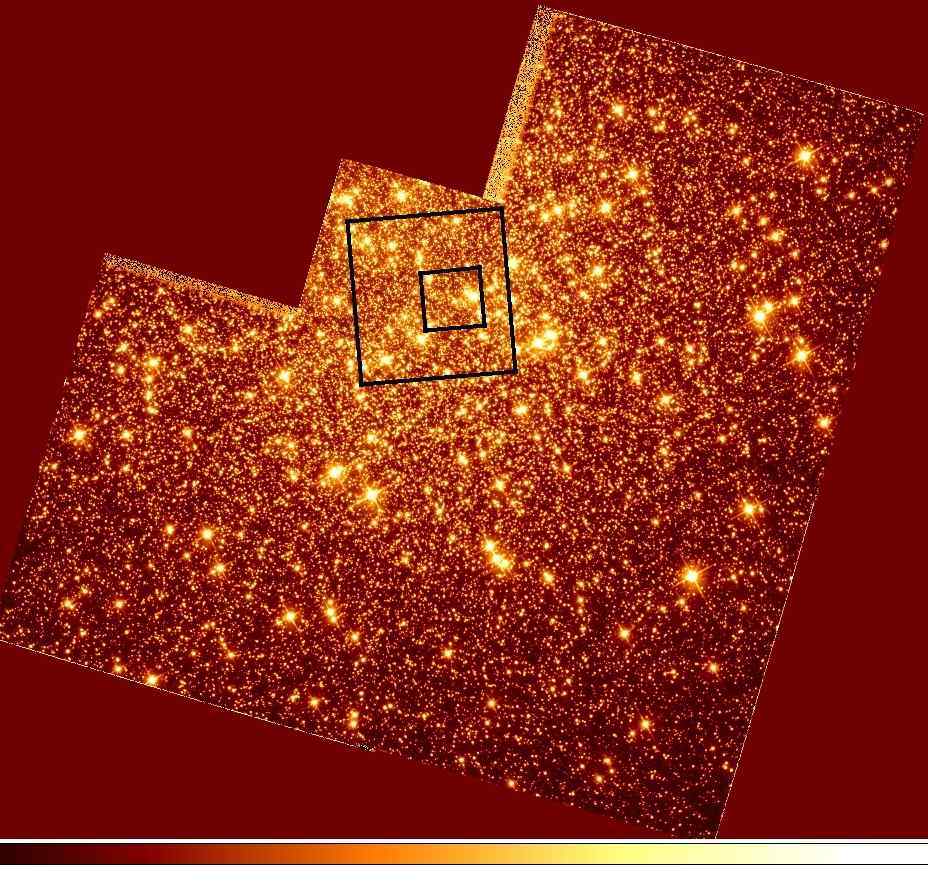}
\caption[LAMP footprint]{HST archival image of the globular cluster Messier 13 (NGC6205). The 2 black squares represent the `LAMP' fields of view, with magnification set to give 20mas and 60mas pixel angular widths. Note that the current mosaic EMCCD camera has a field of view area 16 times that of the EMCCD camera used for LAMP in 2007, given the same pixel angular width.}
\label{fig:LAMP_fov}
\end{center}
\end{figure}

As discussed in chapter~\ref{chap:lucky_AO}, lucky imaging techniques may be employed behind adaptive optics systems, enabling high-Strehl imaging at shorter wavelengths, and hence at higher resolutions, than previously achievable. 
This has already been tried using the (soon to be upgraded) adaptive optics system on the 5 metre Hale telescope at the Mount Palomar observatory \citep{Law2009}, resulting in the highest resolution images obtained using direct imaging, as illustrated in Figure~\ref{fig:LAMP_HST_comparison}. \corr{Figure}~\ref{fig:LAMP_fov} \corr{shows a footprint of the fields of view for the camera used in the tests at Mount Palomar, overlaid on an image from the Hubble Space Telescope for comparison of field sizes.} Use of the technique with the latest multi-EMCCD mosaic camera opens up the intriguing possibility of very high resolution imaging across a field of view several arcminutes across, as opposed to tens of arcseconds. 
 
For a Nyquist sampled PSF in \textit{i'} band on the Hale telescope, the current configuration would correspond to a field angular width of around $64 \times16$ arcseconds, although a redesign is in development for a $32\times32$ arcsecond mosaic configuration. One set of attractive targets for such an instrument would be globular clusters, since they provide a multitude of bright sources which may be used for guiding, and present fields of view with stellar densities impossible to resolve with current imaging techniques.

\subsection{Globular clusters}
Globular clusters are some of the oldest surviving structures in the universe \citep{Krauss2003}, providing a rich source of information about the early galactic conditions under which they first formed, and the evolutionary processes that have determined their behaviour since. 
\corrbox{
Their densely populated inner regions give rise to relatively large numbers of exotic stellar objects such as blue stragglers \citep{Bailyn1995,Knigge2009}, millisecond pulsars \citep{Davies1998,Ivanova2005}, and X-ray binaries \citep{Hut1991}. 
Imaging at greatly increased resolution might allow more accurate observations of many targets in these dense regions, and one could imagine very-high resolution multi-colour photometry surveys as an extremely interesting project. 
On a wider scale, we might gain insight into the dynamical history of the globular cluster as a whole by examining the binary fraction, and it is this avenue of investigation we consider in detail here. 
}

In the central, densely populated regions of globular clusters, the binary fraction is key to the dynamical evolution and lifetime of the globular cluster \citep{Hut1992}. In a process analogous to a single star burning nuclear fuel to support the core, globular clusters undergo a continual process of kinetic energy transfer to delay the onset of core collapse. Binary systems provide the fuel, transferring energy to the globular system through close proximity gravitational interactions. These interactions result in dissolution of wide binaries and tighter binary coupling of close pairs. By probing the binary distribution in these regions we gain an insight into these dynamical processes, which produce many interesting objects such as blue stragglers and cataclysmic variables.

In contrast, beyond the half mass radius the binary fraction is thought to remain fairly stable throughout the lifetime of the globular cluster \citep{Hurley2007}. Determining this primordial binary distribution gives a probe of the earliest star formation processes.

Presently, a number of globular cluster populations have been investigated for evidence of the binary fraction. 
\corr{The traditional method of testing for binarity is to take spectroscopic observations and look for radial velocity perturbations, but taking such an approach for a globular cluster would be infeasible due to the observing time required. Another technique is undertake monitoring observations which look for source dimming due to binary transits, but these will only observe binaries whose orbit is viewed edge on, and again requires a large amount of observing time}.
The investigative method of choice for globular clusters is photometric colour measurement. 
On a colour magnitude diagram, a binary system of equal mass components appears twice as bright as a single star of the same spectral type, and so binary fraction can be inferred. 
However, this technique is inherently insensitive to binaries of unequal mass. 
Also, with current facilities it is only applicable to the closest globular clusters, and open clusters, where typical angular separation between sources is sufficient such that crowding does not prevent individual photometric measurements. 
The resulting estimates of binary frequency vary considerably. 

\begin{figure}[htp]
\begin{center}
 \includegraphics[width=1\textwidth]{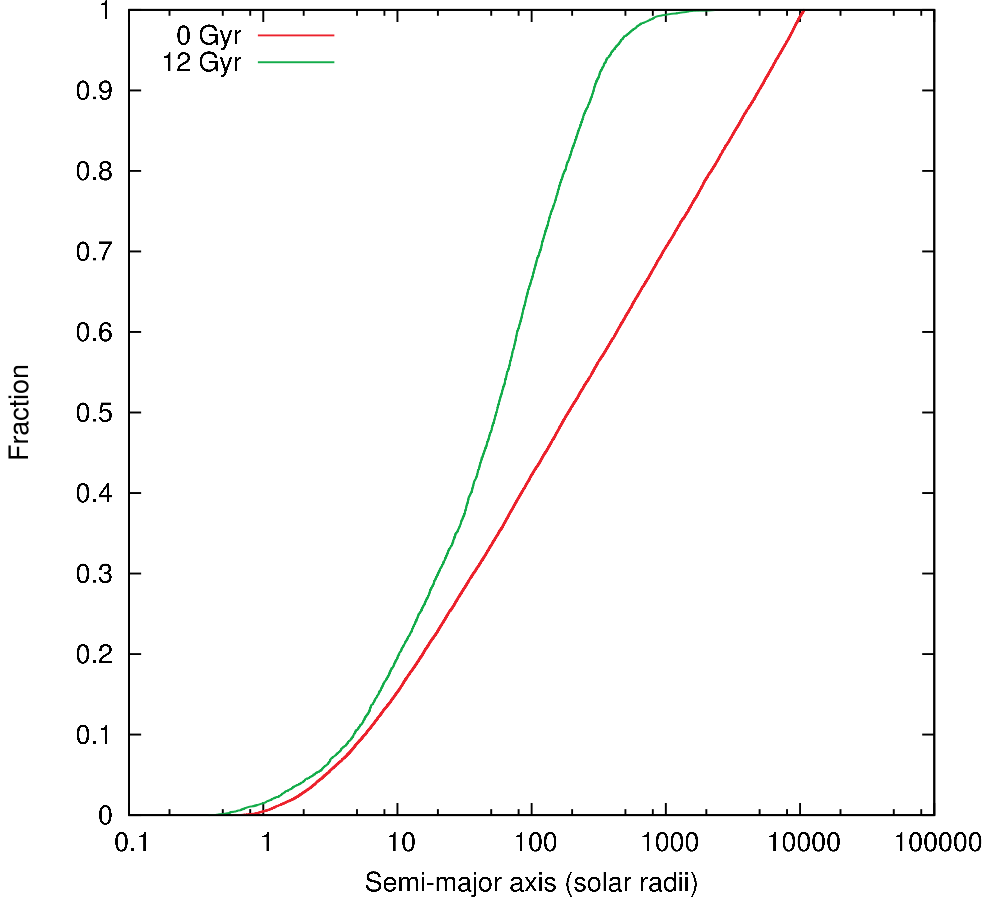}
\caption[GC Binary separation distribution]{Cumulative distribution function of the semi-major axes of binaries in a Monte-Carlo simulation of the globular cluster M4, at 0  and 12Gyrs age (initial and current states). Reproduced from \cite{Heggie2008}.}
\label{fig:GC_binary_sep_dist}
\end{center}
\end{figure}

The high spatial resolution provided by hybrid lucky imaging adaptive optics systems at visible wavelengths makes it possible to begin probing the binary distribution of close globular clusters through \emph{direct} binary detection, i.e. resolving the separate components of \corr{some binaries which would otherwise remain unresolved}. 
We should be able to make binarity estimates based on single epoch data using the spatial distributions of sources. Detailed Monte Carlo simulations of individual clusters are now available (e.g. \cite{Heggie2008, Giersz2009}) which provide precise observational hypotheses which may be explored with such data. For example, it should be possible to probe the globular cluster Messier 4, at a distance of 1.72kpc \citep{Richer2004}, down to a spatial resolution of around 60 astronomical units (AU), assuming a PSF FWHM of 35 milliarcseconds as obtainable on a 5m class telescope. Observations in \textit{i'} band on the VLT could reduce the resolvable distance to around 40AU. At these scales we should be able to place observational constraints upon the binary populations predicted by simulation (see Figure~\ref{fig:GC_binary_sep_dist}). I performed a preliminary investigation to assess the feasibility of such a study.

\subsection{Metric: star separations detected vs. random positioning model}
Ideally, high resolution globular cluster observations should be tested against full Monte-Carlo simulations, but this requires simulating and recording parameters which have not been focused upon in the past, \corr{such as typical `nearest neighbour' distances and other clustering metrics, for a projection of the star positions onto a plane representing an observational image}. For this preliminary investigation I resorted to using a simple random positioning model as a point of comparison to the observational data.
By comparing the spatial distribution of the detected sources to distribution expected from random positioning, we should be able to look for over-densities at typical binary separation distances. 

In order to analyse the data I first wrote a simple Monte-Carlo simulation to estimate the spatial distributions, assuming random positioning.%
\footnote{\corr{While globular clusters are clearly denser in their central regions, and so a random positioning simulation would obviously not be comparable, the field of view analysed covers only a small sub-region of the globular cluster.}} 
Given a detected number of sources and a particular detector layout, the simulation produces many thousands of source position datasets, so that an average distribution of, for example, nearest neighbour distances may be built up for comparison with the real data. To account for the fact that real data has a minimum distance between resolvable sources, any sources randomly positioned within this radius of a pre-existing simulated source are rejected, and the random positioning algorithm runs again to determine a new location. 

I then needed some metrics for comparing the simulated datasets with the observations. The first of these was a set of simple algorithms for producing ``Nth nearest neighbour'' distances, given a set of co-ordinates representing source detections. Written in C++, these run quickly for a few hundred sources. Histograms of neighbour distances can then be compared. I also experimented with algorithms for estimating local over-densities, but quickly came to the conclusion that the datasets analysed are too small for such tests to have much significance, with local estimates being easily influenced by artificial elements such as a bright saturated star.

\subsection{Probing the binary distribution of M13 with LAMP}
\begin{figure}[htp]
\begin{center}
\subfigure[Source detections]{
	\includegraphics[width=0.45\textwidth]{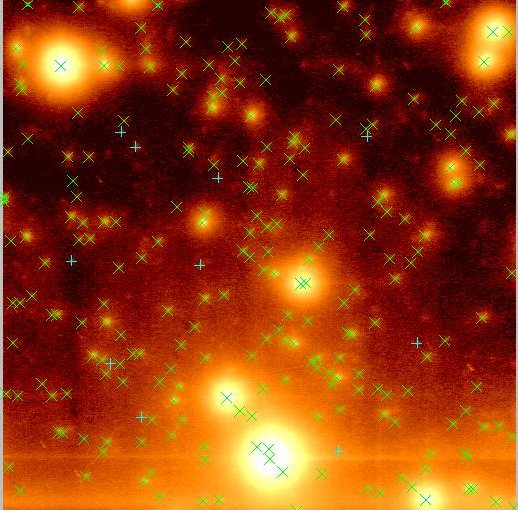}
}
\subfigure[``Nearest neighbour'' pairings]{
	\includegraphics[width=0.45\textwidth]{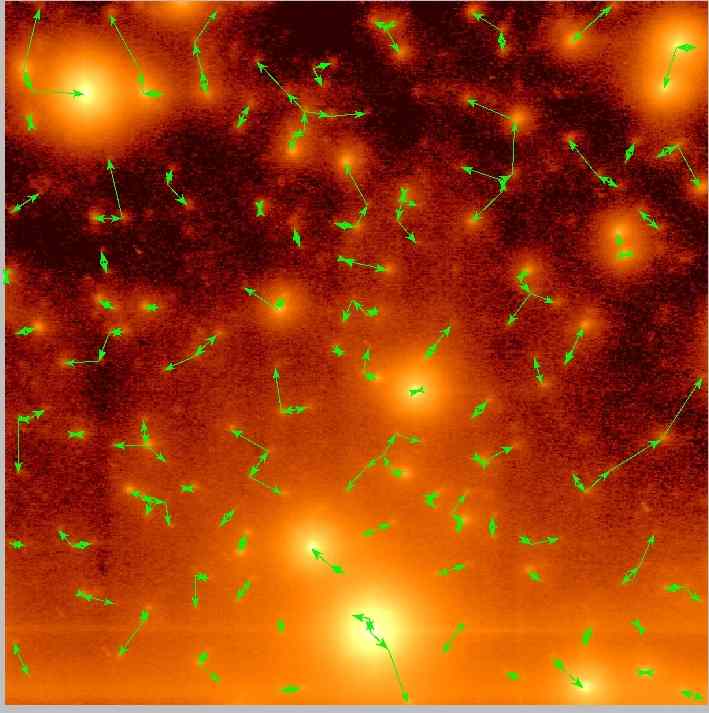}
}
\caption[Source extraction in the LAMP M13 data]{Left: sources extracted from the LAMP observation of M13 with a 20 mas pixel angular width. Image angular width is approximately 10 arcseconds. A 50\% selection cut-off was used to created the reduced image. 244 sources were extracted, after visual confirmation and removal of a few false sources due to detector artefacts. 
Right: The ``nearest neighbour'' pairings assigned by the clustering analysis routines described in text.}
\label{fig:LAMP_M13_sources}
\end{center}
\end{figure}

Only one pre-existing dataset of lucky adaptive optics observations was available for analysis, the `Lucky at Mount Palomar' (LAMP) run from 2007 \citep{Law2009}. Figure~\ref{fig:LAMP_M13_sources} gives an illustration of the source extraction and analysis. Many of the nearest globular clusters which would offer the smallest resolvable spatial scales have declinations well below the equator \citep{Harris1996, Harris2010}, and so were not viable targets for this observing run. However, observations were made of NGC6205 / M13, which has an estimated distance of 30.4 kpc from the sun. The smallest angular scale we can hope to resolve with the LAMP data corresponds to around 1000AU, and so if the binary distribution is dominated by close pairs we would expect few if any resolved binary detections. Despite this, analysis of the data is useful since it provides a practical test of applying source extraction techniques to crowded fields observed in this manner.

The M13 observations were analysed using Starfinder \citep{Diolaiti2000}. Of the currently available source extraction packages this is probably the best suited to lucky+AO observations, since it uses a number of user-identified sources to build a fully empirical model of the PSF. 

\begin{figure}[htp]
\begin{center}
 \includegraphics[width=1\textwidth]{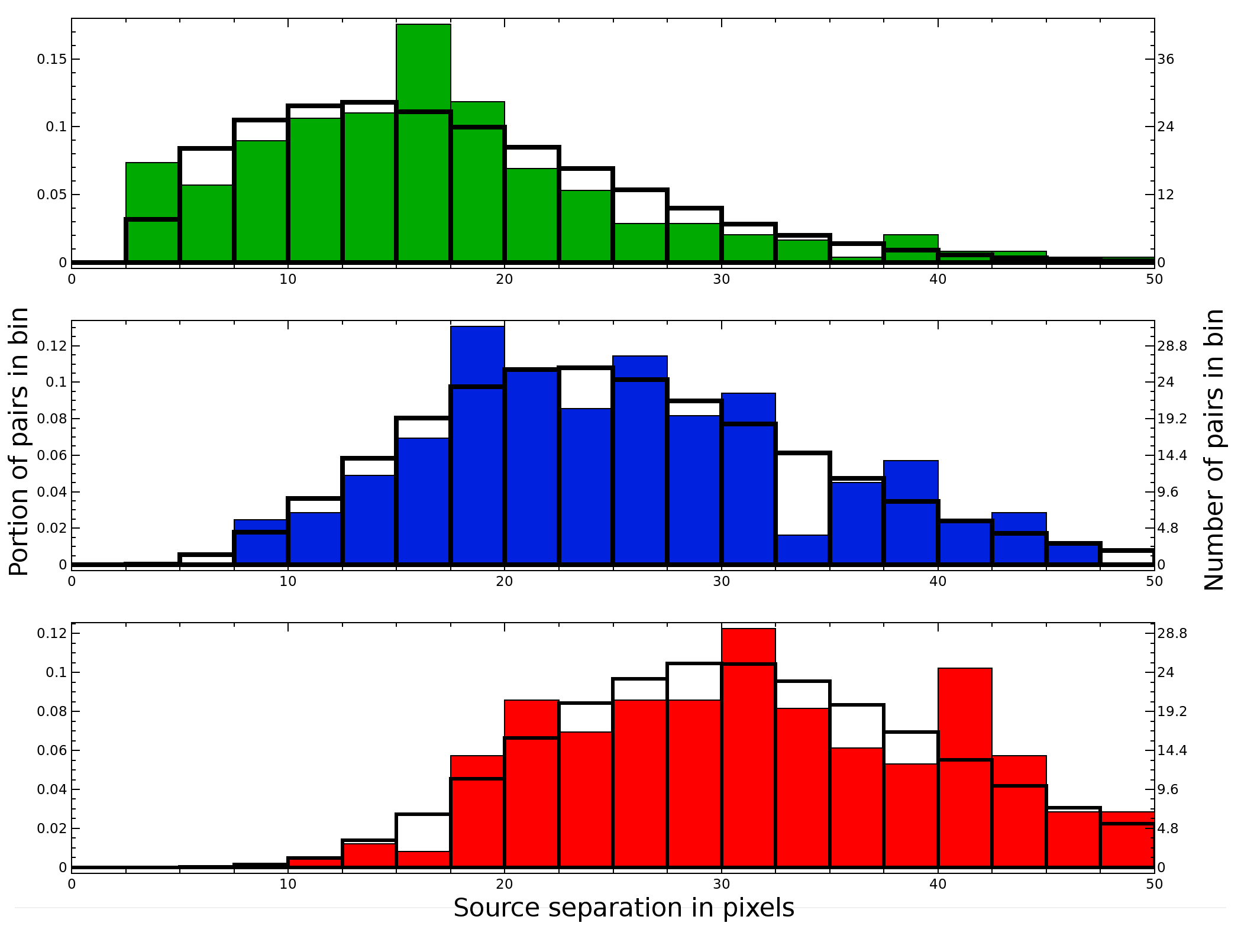}
\caption[Observed and simulated source separation histograms]{Observed (solid colour) and simulated (open heavy line) separation histograms for the LAMP observations of M13 at 20mas pixel scale. From top to bottom, 1st 2nd and 3rd nearest neighbour separation diagrams are displayed. 
}
\label{fig:LAMP_GC_source_sep}
\end{center}
\end{figure}

Figure~\ref{fig:LAMP_fov} shows the fields of view observed with LAMP at 2 pixel scales. The best source extraction was achieved with the smallest pixel scale, however the $512\times512$ pixel squared detector size does not allow for much sky coverage at this resolution. As a result, the number of detected sources is relatively small, around 250. Figure~\ref{fig:LAMP_GC_source_sep} shows the histograms resulting from analysis of this dataset. The observed data seem to conform fairly well to a random distribution, as expected at the large separation distances resolved here. There is an intriguing excess in the observed data for the number of sources detected at closest separations, but with small number statistics this may be a simple statistical anomaly. Sources were de-blended right down to separations of 3 pixels, which proves the data can provide source separation right down to the resolving limit.

\chapter{Modelling of lucky imaging systems}
\label{chap:lucky_AO}

Whether fine tuning data reduction techniques or designing an entirely new instrument, models and simulated data are often a useful tool. Provided that the real phenomena are adequately modelled, simulated data can be used to verify that reduction algorithms give accurate results, provide a range of datasets with carefully controlled parameters to investigate the interplay of external factors, and offer insight into potential performance under entirely new instrumental configurations.

Both lucky imaging and electron multiplying CCD observations are relatively new techniques in astronomy, and as such there are no standard simulation packages. During my PhD I developed a body of software which may be used to simulate many aspects of lucky imaging. Specifically, I developed two packages; one aimed at simulating atmospheric effects and optical propagation, and one focusing upon simulating the stochastic processes of photon arrival and detector effects, as detailed below.

A large portion of this chapter is spent giving a brief review of the models, data generation techniques and parameter choices required for such simulations. A good understanding of all of the above is vital if one wishes to produce computationally efficient, accurate simulations, and to my knowledge no review materials of this particular niche area exist in the literature (\cite{Roddier1981} and \cite{Hardy1998} provide excellent reviews of the models and subject matter, but do not give any information on simulation techniques). I also highlight recent advances both in modelling and sensing atmospheric turbulence, which make possible more comprehensive and better informed simulations, tailored to specific observing sites.

Once developed, the simulations were used for verification and improvement of various aspects of the data reduction process, as detailed in chapters~\ref{chap:EMCCD_calibration} and \ref{chap:frame_registration}. I also began developing simulations to look into possible configurations of hybrid lucky imaging adaptive optics systems, and preliminary results are presented at the end of this chapter.

\section{End-to-end Monte Carlo simulation of atmospheric effects and optical systems}
A number of methods and tools have been developed in order to simulate and predict the behaviour of ground based astronomical systems, usually aimed at adaptive optics or occasionally interferometry. These may be grouped into three main categories \citep{LeLouarn2010}:
\begin{itemize}
 \item Error budgets and analytic estimators: For well understood and well characterised sources of error or image degradation effects, there is sometimes a neat formula that estimates the average effect or outcome. For example, given a typical wavefront phase RMS error we may use the Marechal approximation to estimate the Strehl ratio \citep{Ross2009}:
\begin{equation}
\textrm{Strehl ratio} \approx e^{-\sigma^2}\,,
\end{equation}
where $\sigma$ is the phase RMS error.

Where error sources do not have analytic estimations, simulations or large empirical datasets may be used to generate an error budget table, so that the effect may then be incorporated into calculations. Where available, these methods often provide the quickest and easiest route to an answer, but they may neglect correlation between error sources, and are obviously unavailable when examining poorly understood phenomena. The simplicity may also be a drawback, if additional details are required that may not be obtained from these methods --- for example we may wish to analyse some metric of the PSF other than the Strehl ratio.

\item Semi-analytical Fourier space methods: If the investigator requires a detailed estimate of the long exposure PSF, numerical simulations may be obtained using Fourier space methods \citep[see e.g.][]{Jolissaint2006}. These methods are based upon calculations involving the ensemble average power spectrum of the atmospheric turbulence. By calculating the average filtering effect of any given system component upon the power spectrum, a long exposure PSF may be obtained. However, these methods are not well suited to examining the behaviour of the system over short timescales, and some phenomena such as the laser guide star `cone effect' described in Section~\ref{sec:ao_sky_coverage} are hard to model in this fashion.

\item End-to-end Monte Carlo methods: These produce the most accurate and comprehensive simulations, but are also orders of magnitude slower than the semi-analytical methods. Accurate simulation of lucky imaging requires use of end-to-end Monte Carlo (or simply MC, for brevity) simulations, since the method is very much dependent on short term variations (although I have attempted to determine approximate analytical estimators for the reduced image Strehl ratio, as covered in chapter~\ref{chap:frame_registration}). It is these methods upon which this chapter focuses.
\end{itemize}

\subsection{Atmospheric phase screens}
\subsubsection{Basic generation algorithms and application}
Recalling the model of optical propagation through atmospheric turbulence described in Section~\ref{sec:atmos_effects}, we can reproduce the relevant effects upon the wavefront by defining a ``phase screen.'' This is a two-dimensional scalar field (or equivalently, a three-dimensional surface) where the value at each position in the x-y plane represents the optical path difference incurred by the atmospheric variations in refractive index. For a given observational wavelength this may be converted into fractions of the wavelength, so that each value represents the phase perturbation in radians.

We may generate an array of values to represent discrete samples drawn from such a phase screen at regular spacings. We expect the values to conform to some random distribution, resulting from the chaotic turbulent mixing. In order to successfully model the real phenomena, we must ensure this distribution of the random values conforms to the structure function derived from the Kolmogorov turbulence model by Tatarski:
\begin{equation}
 D_\phi(r) = 6.88 {\left( \frac{r}{r_0} \right) }^{5/3}
\end{equation}
or equivalently, that the power spectrum of the random values conforms to the Kolmogorov $-11/3$ power law.

In truth, generation of \emph{truly} random values is a complex issue, since we expect computers to behave in a deterministic manner, but we may employ cunning generation algorithms to generate \emph{pseudo-random} numbers which appear random to the casual observer, and suffice for the purposes of simulation.

To generate an array of pseudo-random numbers distributed according to the desired power spectrum, we make use of the discrete Fourier transform. If we assume a Gaussian distribution to the phase disturbances, an array of complex pseudo-random numbers representing phase and intensity at different wavenumbers can be drawn from the Gaussian normal distribution and then modulated by the relative amplitude for the appropriate part of the power spectrum. Taking the discrete Fourier transform then produces a phase screen array which satisfies the Tatarski structure function (for most length scales at least --- to ensure conformity at the largest scales extra care must be taken, as described below).

The final stage is to produce an image of a star as seen through the turbulence. Simply applying an annular mask to the wave amplitude values produces sections of the phase screen that represent the phase of light entering the telescope pupil. Taking the Fourier transform of this subregion then simulates Fraunhofer diffraction to produce a focal plane image. \corr{To model time evolution,} the phase screen can be shifted past the mask at an appropriate speed to simulate the turbulent layer being blown past the telescope. This is known as the ``frozen flow'' model, \corr{ since it assumes that the turbulent timescales which produce intrinsic changes in the turbulent layer are much slower than the `crossing time', i.e. the time it takes for a single point in the turbulent layer to be blown past the telescope aperture. This simplifying assumption is common in atmospheric modelling for astronomy, and there is some experimental evidence it is valid} --- see \cite{Poyneer2009}.

The simulation as described thus far uses an over-simplified model, but does produce many of the qualitatively observed atmospheric effects such as speckle patterns and image motion due to tip-tilt. The biggest failing point is that only one turbulent layer is simulated, directly above the telescope (i.e. at ground level). Observations uniformly suggest atmospheric turbulence is usually concentrated in two or more turbulent layers (for example, \citealt{Caccia1987}), but in the past single layer simulations have been used for simplicity and to reduce the necessary computational power. By varying the $r_0$ parameter a single strongly turbulent layer represented by a phase screen of large phase perturbations can approximate two layers of weaker turbulence; but although the long term seeing PSF will be much the same, the high speed behaviour of short exposure speckles is markedly different. For a single frozen layer the speckles track uniformly across the focal plane in the direction of the phase screen's 
associated wind velocity, whereas a superposition of two or more layers causes more of a `random-walk' behaviour \citep[cf.][]{Roddier1982a}.

With the basic model set out, we may consider the further refinements and subtleties that are required to produce datasets of adequate realism. Note that the subject of the turbulence intensity vertical distribution (i.e. turbulent layer heights) is rather dependent upon choice of observing site, and so discussion of this aspect is delayed until Section~\ref{sec:NOT_model}.

\subsubsection{Outer Scales and subharmonic methods} 
Kolmogorov's $-11/3$ power law only applies over an intermediate inertial range, well removed from either the outer or inner scales. The von Karman spectrum shown in Figure~\ref{fig:spectrum} is the ubiquitous model used in astronomy to accommodate the deviations from the Kolmogorov power spectrum beyond the inner and outer scales.

Naturally, the question arises of exactly what those inner and outer scales are, and when they are significant in astronomy.
The inner scale generally goes unmentioned in astronomical literature. \cite{Roddier1981} asserts in passing that the inner scale for the velocity field, $l_0$ ranges from a few millimetres near the ground to 1cm near the tropopause, and that the scale for the temperature fluctuations will be of the same order. The universal assumption is that fluctuations at or below the inner scale are generally negligible for modelling wave propagation since the minor random phase fluctuations tend to cancel.

In contrast, the outer scale is significant because it places an upper limit on the wavefront difference across a large aperture. If the outer scale is larger than the aperture in question there could potentially be a continuous underlying tilt causing a large image shift in the focal plane, whereas a smaller outer scale should result in decorrelation and hence no tilt above a certain range except in rapidly shrinking statistical tails. This information allows instrumentation designers to estimate the maximum stroke required for adaptive optics deformable mirrors, interferometer fringe trackers, etc. 

In the context of lucky imaging, outer scale is mainly significant when considering targets off-axis, i.e. at a significant angular separation from the guide star. While the tip-tilt motion of the guide star is hopefully corrected by the re-alignment part of the data reduction process, the outer scale will still determine at what angular separation the atmospheric effects totally decorrelate and the tip-tilt correction becomes invalid.

It seems likely that the outer scales of the atmospheric turbulence that causes atmospheric seeing vary to some extent with weather, location, and altitude. However, there has been a sustained effort to constrain the numbers, and estimates seem to have now converged upon an outer scale for the phase fluctuations ranging from 10-- 30 metres. For a recent discussion of the best models for outer scale, see \cite{Maire2008}. It is important to note that this is the relevant number for designing astronomical instruments, but there are a variety of more general outer scales for atmospheric turbulence relating to the vertical and horizontal scales of boundary layers and their velocity fields, etc. For the simulations of this thesis an outer scale parameter of 30 metres was used.

\begin{figure}[htp]
\begin{center}
 \includegraphics[width=0.8\textwidth]{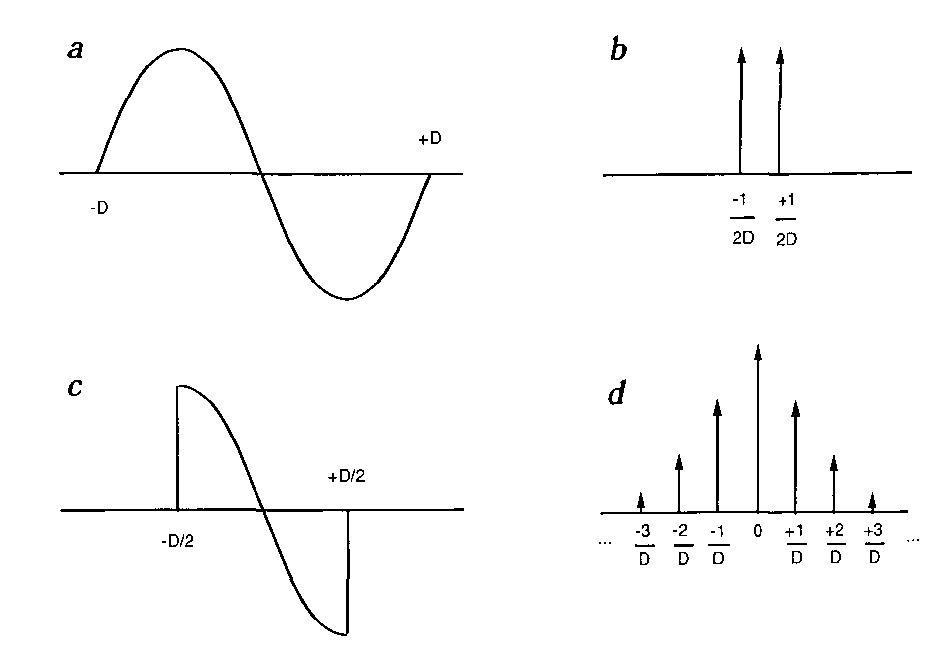}
\caption[Aliasing of non-periodic low frequencies]{
A schematic illustrating the effect of applying a discrete Fourier transform to a non-periodic signal, reproduced from \cite{Lane1992}. This effect results in generation of phase screens with an incorrect structure function, unless corrective measures are employed.\newline
(a): Consider a sine wave signal which just completes one period within a sampling window of length 2D. \newline
(b): Performing a discrete Fourier transform upon samples of the signal results in a double delta function response in the power spectrum, representing a single frequency propagating in the positive and negative directions. \newline
(c),(d): Now consider the same signal, but sampled only through a window of length D. The signal is non-periodic in this window (it is a ``subharmonic''), and aliasing occurs in the discrete Fourier transform. The power originating at frequency $1/2D$ is artificially represented by power at higher frequencies, an artefact called ``spectral leakage.''
}
\label{fig:lane_dft}
\end{center}
\end{figure}

\begin{figure}[htp]
\begin{center}
 \includegraphics[width=0.5\textwidth]{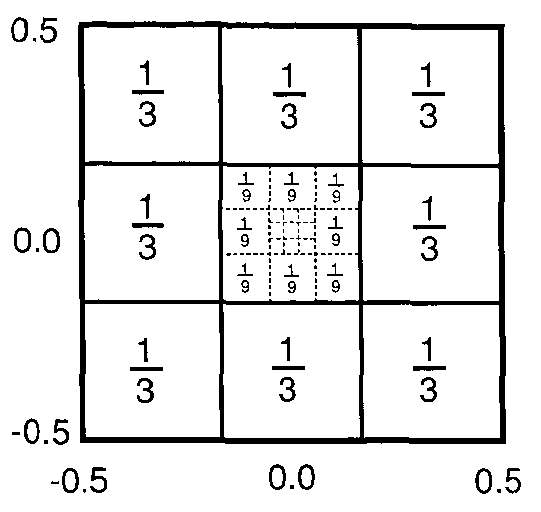}
\caption[Subharmonic sampling points in the Fourier domain]{
A schematic illustrating the the subharmonic frequencies sampled in Fourier space in order to generate a phase screen with the correct structure function. The fractional values correspond to widths of a sampling interval in the standard Fourier sampling.
Reproduced from \cite{Lane1992}.
}
\label{fig:lane_subgrid}
\end{center}
\end{figure}

\begin{figure}[htp]
\begin{center}

\subfigure[Kolmogorov model (infinite outer scale)]{	
	\includegraphics[scale=0.5]{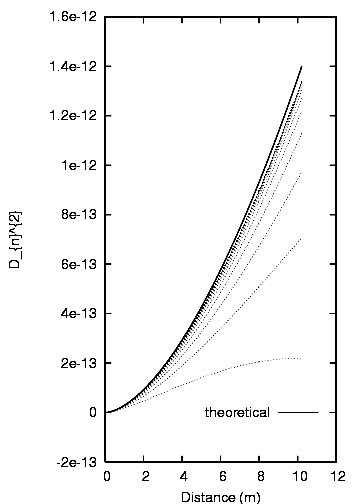}
}
\subfigure[von Karman model (30 metre outer scale)]{
	\includegraphics[scale=0.5]{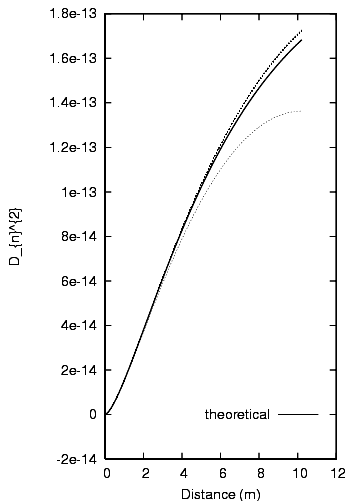}
}
\caption[Improvement in structure function with subharmonics]{
Plots depicting improvement in structure function with subharmonics, reproduced from validation tests of Arroyo \citep{Britton2004}. The solid line represents the theoretical structure function. The various dotted lines represent various levels of subharmonic correction, from zero to fifteen corrected modes. The lowest dotted line represents zero subharmonic modes corrected, i.e. the standard technique, which evidently does not reproduce the theoretical structure function well for either the Kolmogorov or von Karman model. The simulations were performed by generating the structure function corresponding to a phase screen with integrated turbulence profile $C_{n}^{2}$ of $10^{-14}$. Phase screens of size 20 metres were generated, with sampling of 2 cm.
 
}
\label{fig:subharmonic_structure}
\end{center}
\end{figure}

Having chosen an outer scale, care must be taken to ensure that the corresponding structure function is accurately represented by the simulated data. The method described above utilises normalisation of the random data using the desired power spectrum in Fourier space, followed by a discrete Fourier transform to get the spatial phase screen. However, if we consider the inverse process, it is clear that a Fourier-space representation will not accurately represent the lowest frequency components at long spatial scales (see Figure~\ref{fig:lane_dft}). Fortunately there are methods for dealing with this problem. If we wish to simulate a finite outer scale, we may simply ensure that the phase screens generated are always significantly larger than the outer scale, at the cost of increasing the computational requirements (especially memory). However, if we wish to test the Kolmogorov spectrum with infinite outer scale, or simply avoid excessive array sizes, then an alternative method is available.

\begin{figure}[htp]
\begin{center}

\subfigure[No subharmonic correction]{	
	\includegraphics[width=0.6\textwidth]{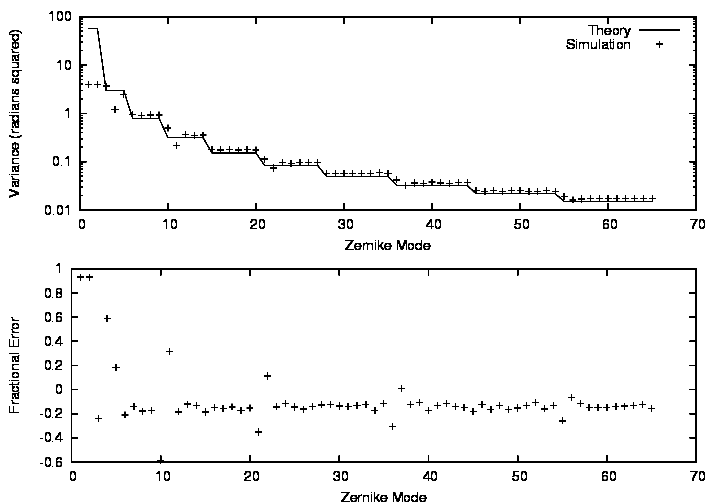}
}
\\
\subfigure[Subharmonics added using the generalized method implemented in Arroyo]{
	\includegraphics[width=0.6\textwidth]{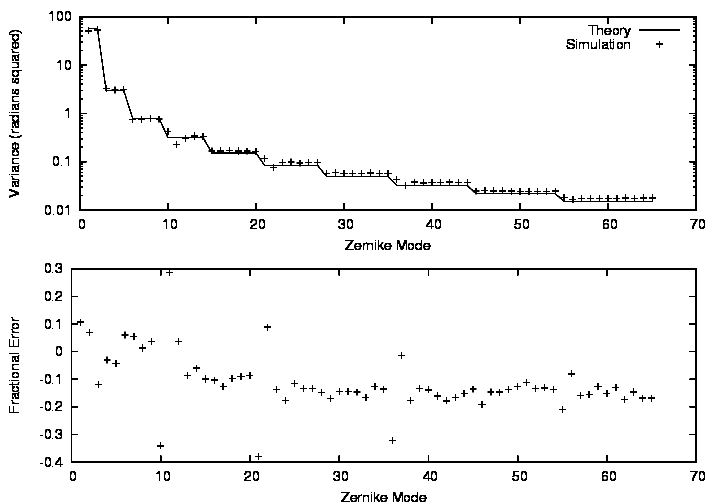}
}
\caption[Improvement in Zernike polynomial statistics with subharmonics]{
Plots depicting improvement in Zernike polynomial statistics with subharmonics, reproduced from validation tests of Arroyo \citep{Britton2004}.
The simulation was performed on a 2 metre aperture, with 1 cm pixel sampling of the phase screen. The Fried parameter was chosen as 20 cm at 1 micron. A Kolmogorov (infinite outer scale) model was assumed and variances were computed using 5000 random screens. The variances shown in the plots are the phase variances computed for electromagnetic radiation with a wavelength of 1 micron. The fractional errors are computed as the difference between the theoretical and simulated variance, divided by the theoretical variance. }
\label{fig:subharmonic_zernikes}
\end{center}
\end{figure}

\cite{Lane1992} proposed the `subharmonics' method of phase screen correction, later refined by \cite{Johansson1994}. First, a phase screen is generated using the standard Fourier transform method described above. Then additional random samples are generated and normalised by the power spectrum positions on a subgrid in Fourier space representing low frequencies not sampled in the standard phase screen (Figure~\ref{fig:lane_subgrid}). For each sample the corresponding waveforms in real space are calculated and the values added, pixel by pixel, to the standard phase screen. While fairly computationally intensive, this requires only the memory allocated to the standard phase screen, and is still faster than generating an oversized phase screen.
Figures~\ref{fig:subharmonic_structure} and \ref{fig:subharmonic_zernikes} display the resulting improvement in phase screen statistics when employing the generalized subharmonic methods implemented in the Arroyo simulation package (described in Section~\ref{sec:MC_packages}).
The subharmonics clearly produce a more accurate simulation of the tip-tilt Zernike modes, which result in image motion in the focal plane.

A second consideration with regard to generating phase screens with the correct statistics is that of correctly representing the smaller scales. This is simply a case of computing enough samples to give a sufficiently small spatial sampling interval. To my knowledge there is no guidance on this matter in the literature, but Francois Rigaut (author of \corr{the `YAO' AO simulation package}, see Section~\ref{sec:MC_packages}) recommends a sampling rate producing at least two and preferably three sample intervals per $r_0$.
\footnote{Source: http://frigaut.github.com/yao/manual.html}
 This seems sensible considering that phase disturbances on scales smaller than this are of order 1 radian or less, and are therefore likely to have negligible effect.

\subsubsection{Intermittency}
\label{sec:intermittency}
\begin{figure}[htp]
\begin{center}
 \includegraphics[width=0.5\textwidth]{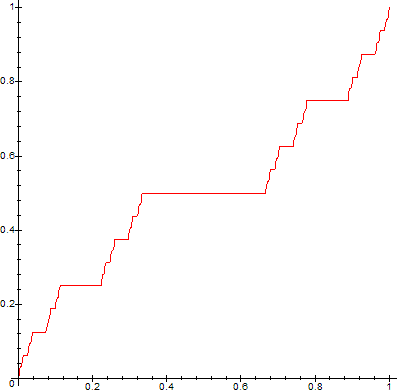}
\caption[The Devil's Staircase]{A graphical approximation to the `Devil's Staircase,' which serves as an example of intermittency (albeit a rather abstract one). This function is constant almost everywhere, but still rises from 0 to 1.}
\label{fig:staircase}
\end{center}
\end{figure}
When performing simulations where the focus is upon phenomena at short timescales, it is worth noting that verifying the statistical averages of the power law tells us nothing about short timescale evolution within the turbulence. Long exposure images smear out the rapidly varying speckle statistics. Balloon measurements should be sensitive to very rapid variations, but are almost universally focused on gathering large scale information about the structure and distribution of turbulence \cite[for example][]{Azouit2005}. One of the first major criticisms of Kolmogorov's 1941 universal  theory of homogeneous turbulence was an observation by Landau, that at the outer scale turbulence is often observed to be intermittent. 

\begin{figure}[htp]

 \includegraphics[width=\textwidth]{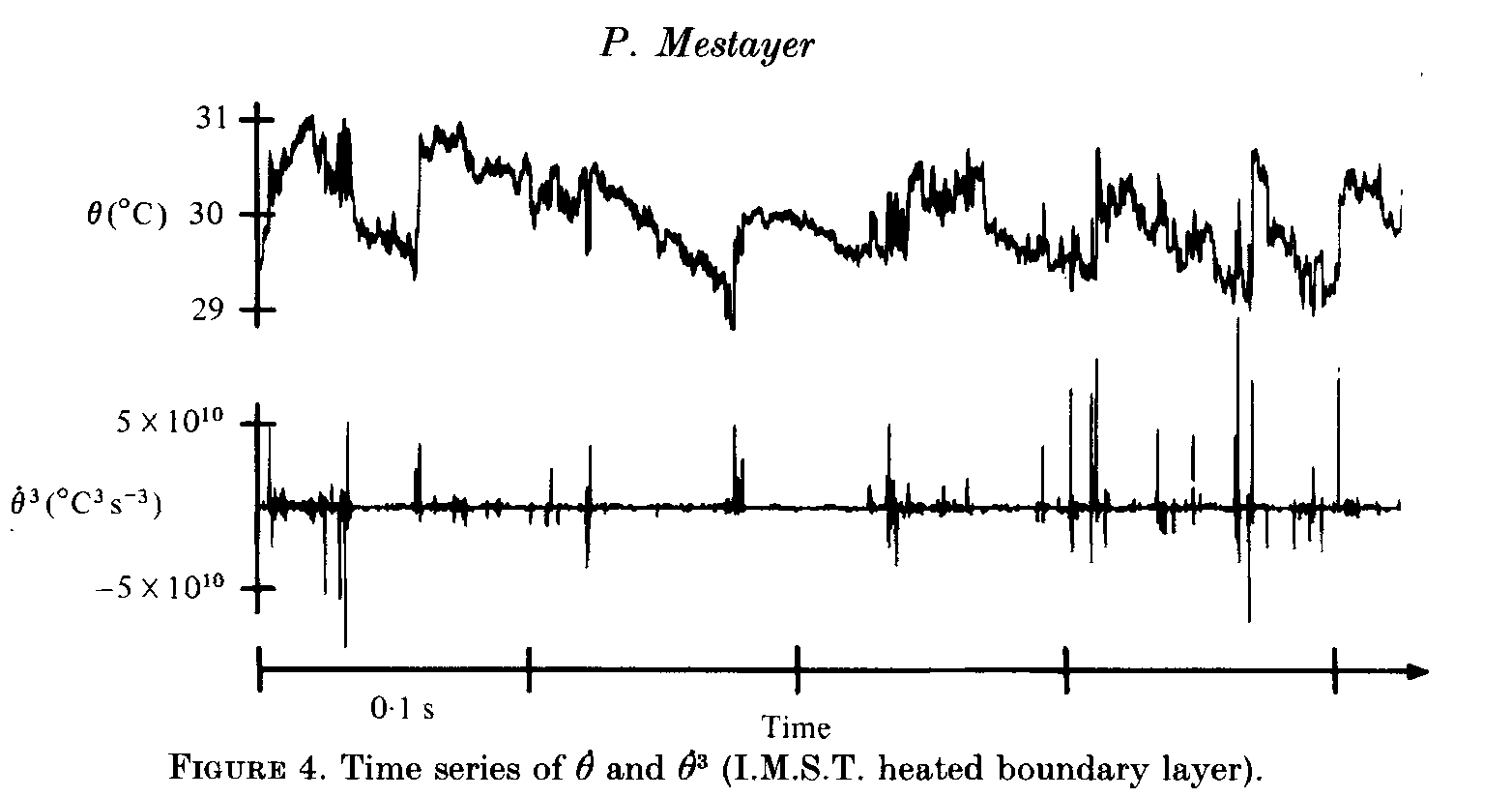}
\caption[Temperature Fluctuation]{
A plot from \cite{Mestayer1982} showing intermittent variations in temperature in a boundary layer.
}
 \label{fig:b_layer}
\end{figure}

Intermittency literally refers to an `on / off' nature - for example a usually quiet signal with occasional short bouts of noise. Sometimes this fleeting activity leaves the base signal unchanged, but it may be integral to the nature of a function --- an \corr{abstract} example of this is the `Cantor's function' or `Devil's Staircase' (Figure~\ref{fig:staircase}) which increases from 0 to 1 over the unit interval, but is constant almost everywhere \cite[\corr{except on the Cantor set, see e.g.}][]{Thomson2008}. In physical terms, an observable that when measured produces an intermittent signal usually has some kind of structure, \corr{as illustrated in figure}~\ref{fig:b_layer}. In turbulence this is somewhat contrary to the homogeneous, isotropic background to Kolmogorov's energy cascade hypothesis. Nonetheless, it is observed that at the outer scale the energetic motion injected into a turbulent system is often intermittent.

The classic example of this is a boundary layer - particularly relevant since the turbulent laminae where seeing arises can be thought of as a boundary layer between background fluids of different ambient temperatures \citep{Coulman1995}. Patches of concentrated vorticity can be seen to release from the boundary resulting in a highly non-uniform velocity field. There is also some experimental evidence to suggest that this intermittency persists into the inertial range, which is to say it is not dispersed by the turbulent mixing. However there have been no conclusive results on accurately characterising this intermittency in the inertial range of the velocity field, and there seems to be uncertainty over the significance.

\begin{figure}[htp]
 \includegraphics[width=\textwidth]{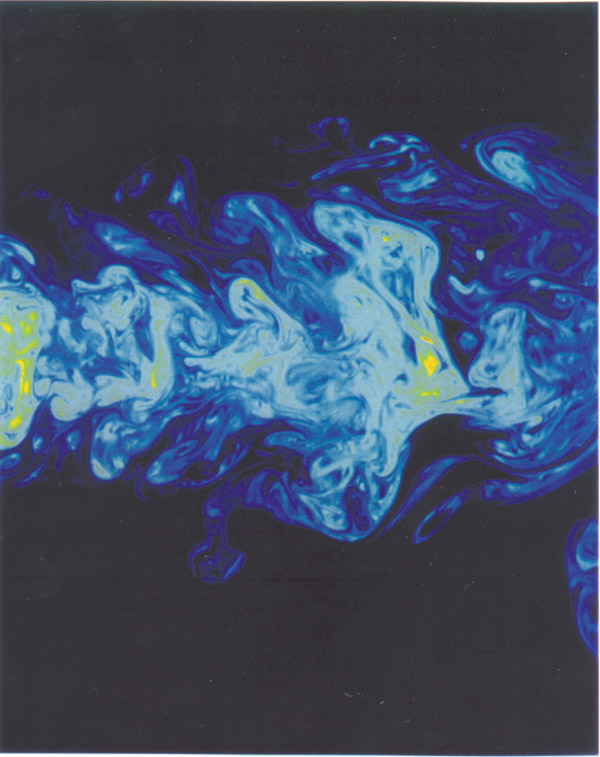}
\caption[Dye in a Jet]{Fluorescent dye in a turbulent jet. The dye distribution is significantly non-homogeneous and anisotropic. Reproduced from \cite{Sreenivasan1991}.}
 \label{fig:jet}
\end{figure}

However, passive scalars are now generally recognised to be less isotropic and homogeneous than the velocity field \citep{Sreenivasan1991}, particularly in shear layers. Intermittency produces significant deviations from the standard Gaussian Kolmogorov distribution in these observables, for example temperature fluctuations (Figure~\ref{fig:b_layer}). There are a few models for this fine structure of passive scalar fields, but none are singled out as the preferred approach. \cite{Shraiman2000} give a review of current work on scalar turbulence.

The extent to which intermittency effects can be observed at scales of interest to astronomy has been shown in simulations to have direct implications for both lucky imaging and adaptive optics, by \cite{Tubbs2006}, who proposes an adaptable generalised model of intermittency. In short, for a given mean level of seeing lucky imaging benefits from intermittency because it isolates the moments of good seeing and discards the poor ones. Similarly if intermittent effects are observed frequently then adaptive optics systems could benefit from continuous monitoring and  re-optimization of the system operation rates to achieve best correction at all times. If intermittency effects are extreme then it poses real problems for standard adaptive optics systems because the wavefront correction is performed in a feedback loop, with the potential for sudden large changes to upset the tracking and require repetition of the (often lengthy) target acquisition process known as ``closing the loop.''

Observational evidence for or against intermittency effects is sparse. It is well known that seeing can vary greatly on timescales of hours and minutes \citep[for example, see][]{Tokovinin2003} but without high temporal resolution measurements it is hard to determine whether these variations are due to large scale intermittency or simply a rapid but progressive change in the mean value of $r_0$. \cite{Baldwin2008} investigates this distinction, with the tentative conclusion that intermittency may vary with atmospheric conditions --- rapid step changes are observed in some data sets, while others are better characterised as Gaussian fluctuations about an evolving mean. In the absence of hard evidence I utilised the conventional model without intermittency effects for the simulations described in this thesis, but the subject merits further investigation since many expensive adaptive optics projects depend upon simulated data for design and optimization.

\subsubsection{Rolling phase screen generation}
Using the discrete Fourier transform phase-screen generation methods described above, long period simulations require large amounts of memory for allocating the phase screen arrays. This places a practical limit on the maximum simulation time, and also models a fixed mean intensity of turbulence for the duration of the simulation (i.e. fixed $r_0$). However, algorithms and code for producing dynamically updated phase screens have recently been developed, as reported in \cite{Assemat2006}. In short, the method uses an analysis of the covariance functions corresponding to the prescribed power spectrum model, in order to derive special matrix operators. These matrices may then be applied to a phase screen array and a vector of newly generated Gaussian pseudo-random values, to extend the phase screen along one edge. Considering a phase screen being blown past an aperture, one may then continually delete values from one side of the array while appending new values to the other side, producing a simulation of any 
desired timespan without large memory requirements. There is an additional advantage since the atmospheric parameters may be varied as the simulation progresses. A high degree of numerical precision is required to prevent divergent results due to round-off errors \citep{Basden2011}, but the technique has been successfully employed for some simulations, e.g. \cite{Basden2007}.

\subsubsection{Wavefront propagation methods}
If multiple atmospheric phase screens at various heights are modelled, the wavefront must be propagated between them.
The most commonly employed model for wavefront propagation is the so called ``geometric propagation'' method, where a uniform phase offset is added to the wavefront phase according to the distance travelled. While this is appropriate over relatively short distances or for wavefronts with only small phase perturbations (relative to a plane wave), if the phase perturbations are a significant proportion of a wavelength then diffraction effects may become significant, and a more sophisticated method may be required for accurate results. Examples include detailed simulations of defocused wavefronts, e.g. \cite{Guyon2010a}, or wavefronts focused through small lenslets, e.g. \cite{Goodwin2007}. For a full discussion of propagation algorithms the reader is referred to \cite{Papalexandris2000}.

\subsection{End-to-end Monte Carlo simulation packages}
\label{sec:MC_packages}
Of the Monte Carlo simulation packages, `CAOS' \cite[][IDL based ]{Carbillet2005}, and `YAO' \cite[][based on a free IDL-like language called `Yorick']{Dam2010},  are notable examples which are freely available. Other expansive end-to-end simulation packages exist, such as the ``Durham extremely large telescope adaptive optics simulation platform'' \citep{Basden2007} and the ``Octopus'' simulation code developed within the European Southern Observatories \citep{Montilla2010}, but these are generally utilised in-house and are not aimed at `casual' users.  Lucky imaging places some fairly strenuous requirements on the simulations package. It must be capable of a full time-evolving Monte-Carlo simulation, as opposed to an analytic model, in order to investigate the effects of atmospheric coherence times, etc. It must be able to provide a long period atmospheric phase screen, in order to prevent unrealistic repetition of point spread function behaviour. Finally, there is ongoing work to design a curvature sensor 
for use with a lucky imaging adaptive optics system. While this work is not discussed further here, the model relies upon Fresnel diffractive propagation effects, which are not modelled in many simulations, despite the fact that including these propagation effects produces significantly different results when compared to simple geometric propagation in some contexts \citep[see][]{Goodwin2007}.

For these reasons, and due to my familiarity with the C++ programming language, I chose to develop simulations using an open source C++ package called `Arroyo' \citep{Britton2004}. The package is well featured, but somewhat inaccessible to new users due to paucity of documentation and use of some rather esoteric programming idioms. During my use of the package I have made a number of alterations that I believe go some way to addressing these practical issues. I also modified sections of the code to allow for use of very large phase screens with pixel indices that may extend beyond the address capacity of a standard 32 bit integer variable, which is potentially of use on machines with RAM capacity of over 4 gigabytes.

\begin{figure}[htp]
\begin{center}

\subfigure[Original Arroyo code]{	
	\includegraphics[scale=0.5]{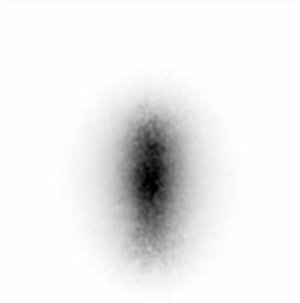}
	\label{subfig:elongated_arroyo_seeing}
}
\subfigure[Modified Arroyo code]{
	\includegraphics[scale=0.5]{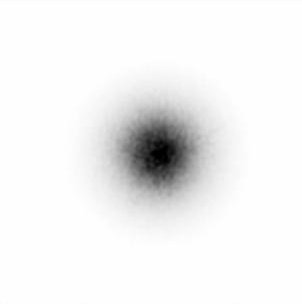}
	\label{subfig:fixed_arroyo_seeing}
}
\caption[Inadequate low spatial frequencies]{ \subref{subfig:elongated_arroyo_seeing}: Under certain conditions the original version of Arroyo produces phase screens which inadequately model low spatial frequencies in one direction, resulting in an unrealistically elongated long exposure seeing PSF. \subref{subfig:fixed_arroyo_seeing}: I modified the code to correct for this  --- details in text.}
\label{fig:bad_subharmonics}
\end{center}
\end{figure}

I also checked various aspects of the Arroyo simulated data to ensure that the simulated results were valid. While undertaking these tests I discovered that under certain conditions the low spatial frequencies are insufficiently reproduced in one direction even when subharmonic techniques are employed, due to use of long, thin `ribbon' phase screens. This was rectified by modifying the code to enforce a minimum phase screen pixel width in any direction equivalent to 50m, at the cost of a slight increase in computational memory requirements (Fig.~\ref{fig:bad_subharmonics}). 

I note that in theory, it should be relatively simple to integrate rolling phase screen functionality into the Arroyo package, which would allow for simulations of very long timespan and dynamically evolving atmospheric conditions. This potentially very useful enhancement is deferred to future work.

\section{Simulating photon shot noise and the EMCCD response}
\label{sec:EMCCD_sim}
For lucky imaging, the detector is crucial to the behaviour of the entire system. The detector characteristics not only determine the signal to noise of faint objects, but also influences the guide star registration process, and hence the resolution obtained. This, and the fact that I needed to develop and verify a range of detector analysis routines, meant that careful modelling of the detector was a necessity. Creating tools that would allow Monte-Carlo simulations required some work with pseudo-random number generation.

\begin{figure}[htp]
\begin{center}

\includegraphics[width=0.5\textwidth]{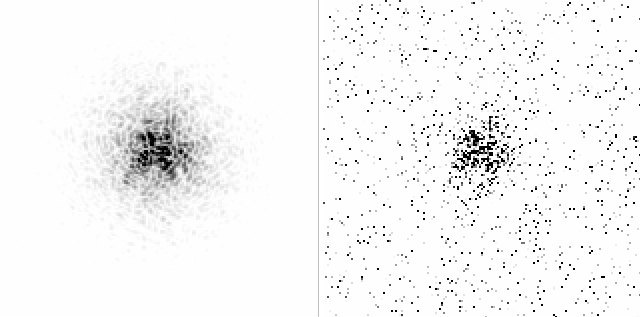}

\caption[Simulation of photon shot noise and detector response]{Simulation of photon shot noise and detector response. 
Left image depicts a typical intensity image output from Arroyo simulations, after binning to the desired pixel scale. 
Right image depicts the resulting EMCCD output in ADUs. Each pixel is sampled via Monte-Carlo methods from the detector output probability distribution corresponding to the number of photons captured by the pixel, which are previously generated via Monte-Carlo simulation of Poisson statistics.
}
\label{fig:simulation_detector}
\end{center}
\end{figure}

Photon shot noise and detector readout noise are easily modelled with widely available pseudo-random number generators conforming to Poissonian and Gaussian distributions. For conventional CCDs these often suffice. However, the electron multiplying CCD has an unusual ADU output probability distribution as detailed in Section~\ref{sec:pixel_PDFs}. To generate random numbers accordingly, the distribution first needs to be calculated, and then random numbers drawn from it. An early version of my code used a crude implementation of a ``discrete sequential search'' algorithm - essentially a random number is drawn from a uniform random number distribution, and the EMCCD distribution values are iterated through until the matching cumulative probability is reached. This has painfully slow performance for large simulations, in practice. Further investigation suggested the UNURAN package \citep{Hormann2000}, which provides ``black box'' tools for sampling pseudo-random numbers from any user generated distribution, and 
is also employed in the ROOT \citep{Brun1997} modelling package at CERN. Utilising the UNURAN package provides excellent performance, and the current revision of the code typically simulates $\sim$7 million pixel events per second on a single CPU core. Figure~\ref{fig:simulation_detector} illustrates the stages of the detector simulation process.

\section{Modelling lucky imaging at the Nordic Optical Telescope}
\label{sec:NOT_model}

\begin{figure}[htp]
\begin{center}
 \includegraphics[width=0.6\textwidth]{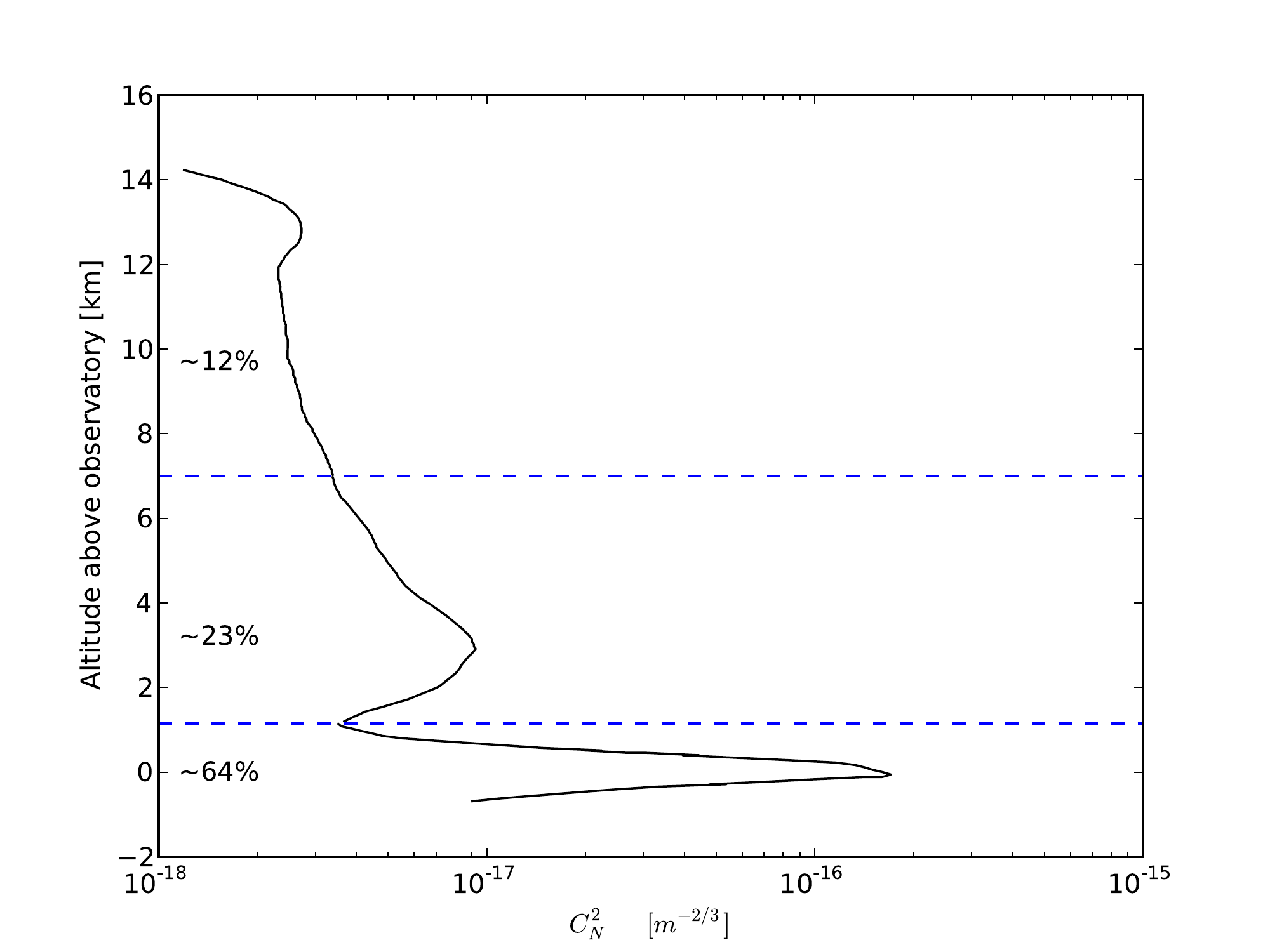}
\caption[Turbulence profile at the ORM]{A profile of the vertical structure of atmospheric turbulence, measured using generalised SCIDAR techniques at the Observatorio del Roque de los Muchachos. Specifically, this is a median profile obtained from 33854 individual profiles obtained in ``high resolution'' mode, i.e. with a vertical resolution $\Delta H(0) < 500$m --- see, e.g. \cite{Avila1997} for details of the technique, and \cite{Garcia-Lorenzo2011} for details of the observations. Note that the extension below the observatory level is an artefact of the measurement technique, and these values should be considered as contributing to the boundary layer at ground level. Horizontal lines represent partitioning into sections to be represented by individual layers in the simplified model of table~\ref{tab:atmos_model_layers}. Overlaid text denotes approximate percentage of total $C_N^2$ integral contained in each partition. For comparison with figures \ref{fig:Teide_profile} and \ref{fig:other_CN2_profiles} 
the observatory altitude of 2.4km should be added to the y-axis values.
}
\label{fig:ORM_profile}
\end{center}
\end{figure}

\begin{figure}[htp]
\begin{center}
 \includegraphics[width=0.8\textwidth]{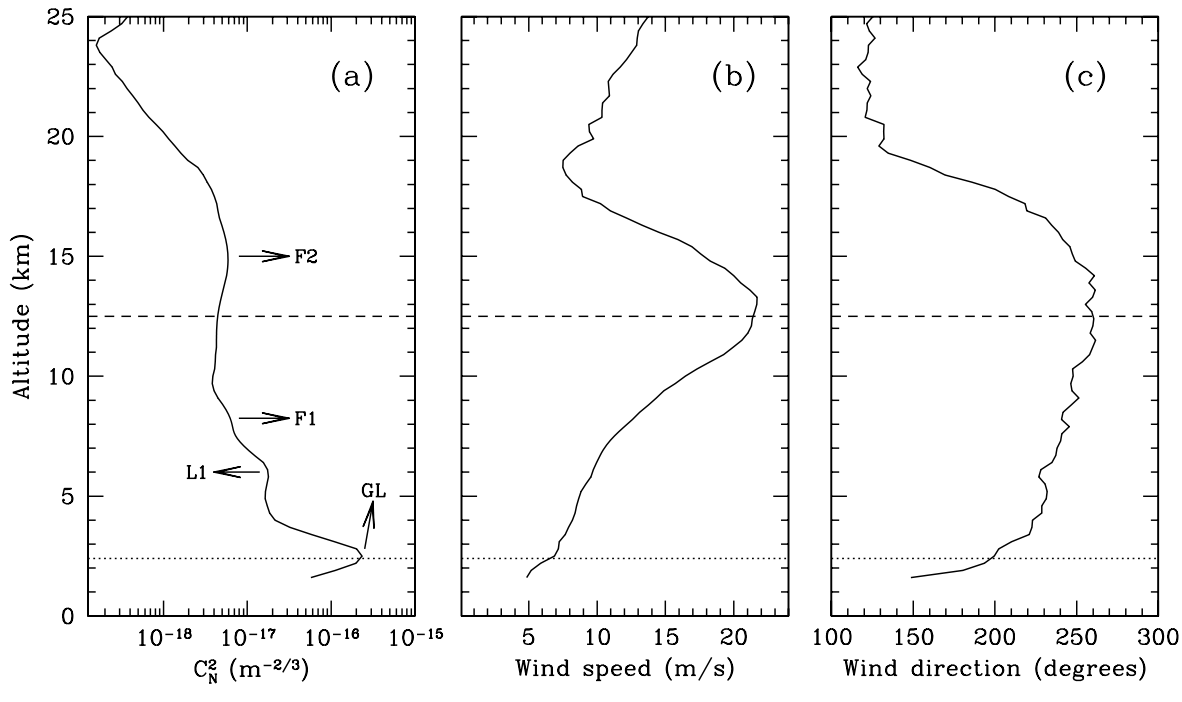}
\caption[Atmospheric profiles at the El Teide Observatory]{Vertical profiles of the turbulence (a), wind speed (b), and wind direction (c) observed at the El Teide observatory on Tenerife, approx. 100km from the Observatorio del Roque de los Muchachos. Vertical scale is altitude in km above sea level, horizontal dotted line denotes observatory altitude. Reproduced from \cite{Garcia-Lorenzo2009}.
}
\label{fig:Teide_profile}
\end{center}
\end{figure}

As astronomical observing techniques and equipment have progressed, methods of characterising observing sites and observing conditions have developed alongside. Since the late 1980's, differential image motion monitors \citep[DIMM's,][]{Pedersen1988} have enabled automatic monitoring of the seeing width, allowing observers to test prospective observatory sites, gain a uniform record of seeing conditions over long periods and optimise observation scheduling. 

\begin{figure}[htp]
\begin{center}

\subfigure[Mount Graham (summer)]{	
	\includegraphics[width=0.8\textwidth]{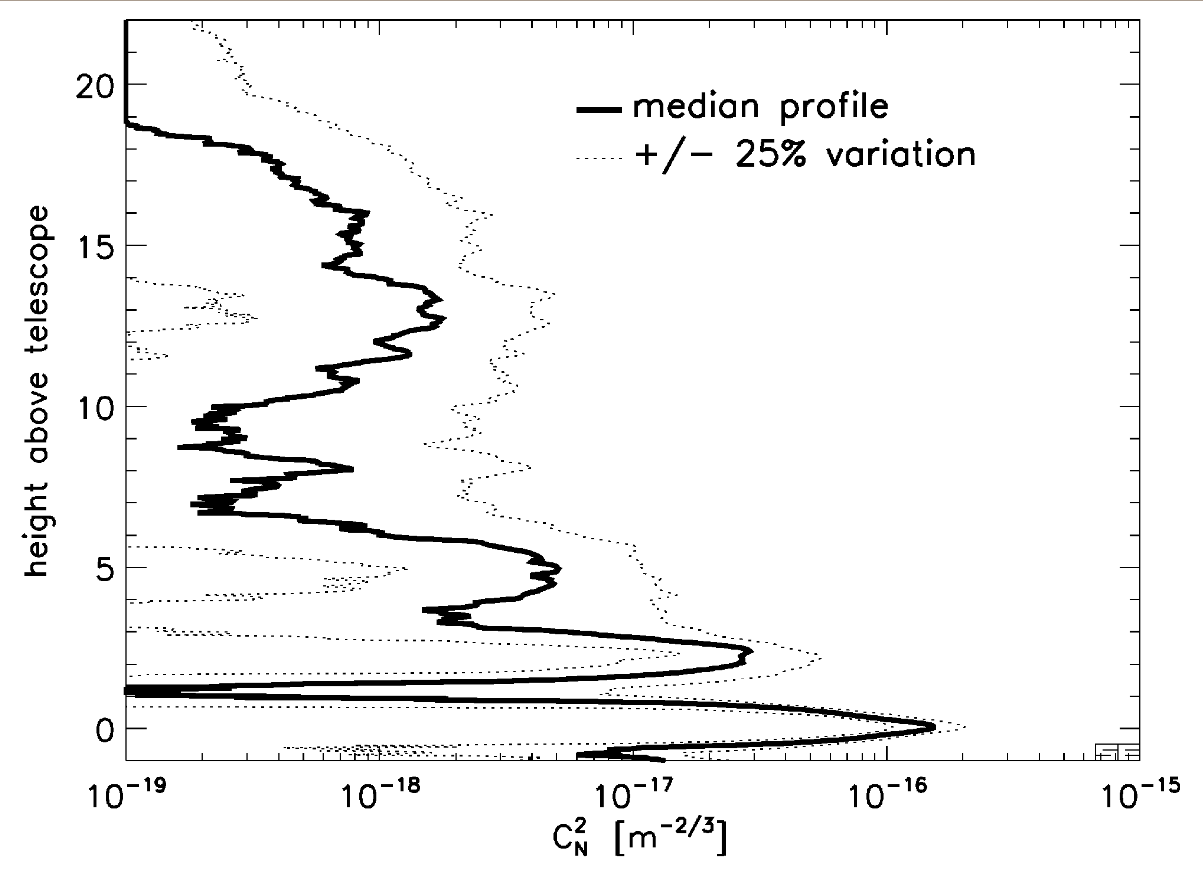}
	\label{subfig:MtGraham}
}
\subfigure[San Pedro Martir]{
	\includegraphics[width=0.8\textwidth]{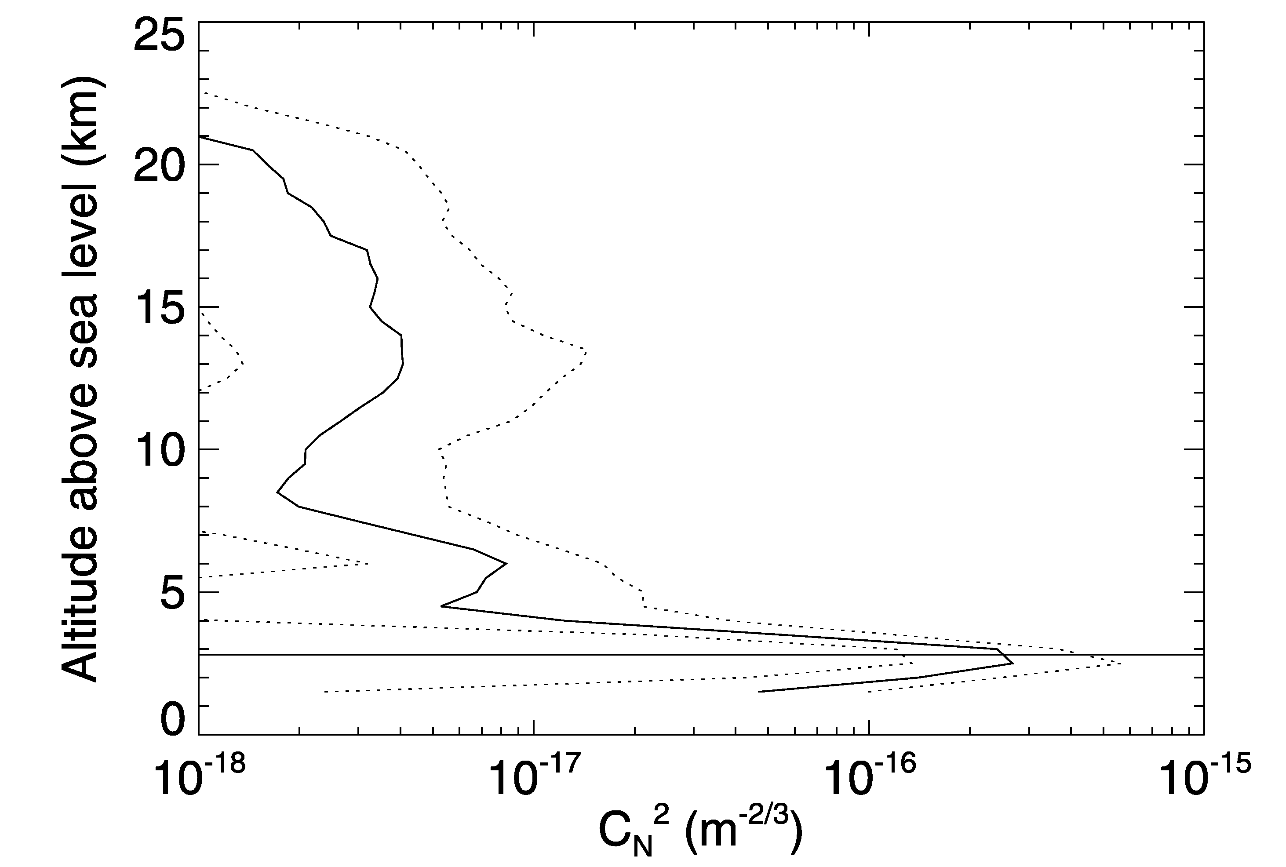}
	\label{subfig:SanPedro}
}

\caption[CN2 profiles at other locations]{
 Atmospheric turbulence profiles measured with generalised SCIDAR at other locations, both displaying the strong ground layer and tropospheric layer as described in the text.
 Figure~\subref{subfig:MtGraham} is the summer profile at Mount Graham, Arizona, reproduced from \cite{Egner2007}. 
 Figure~\subref{subfig:SanPedro} is a profile from San Pedro Martir, Mexico, observed over 15 nights in May 2000. Reproduced from \cite{Avila2006}.
 In both subfigures, the solid line represents the median profile while dashed lines represent the first and third quartiles.
}
\label{fig:other_CN2_profiles}
\end{center}
\end{figure}

As adaptive optics systems have become more widespread, more in depth knowledge of atmospheric conditions has become necessary in order to predict and optimise system behaviour. Real time correction techniques must be able to keep up with the changing atmospheric conditions, and so atmospheric coherence times have become important. More recently the vertical structure of atmospheric turbulence has come to the fore, since it is key to the performance of layer orientated adaptive optics techniques such as multi-conjugate AO \citep[][and references therein]{Campbell2010} and ground layer AO \citep{Tokovinin2004, Andersen2006}. 
Over the past decade a number of techniques have emerged to provide full vertical profiling, including `multi-aperture scintillation sensors'  \citep[MASS,][]{Tokovinin2007a}, `slope detection and ranging' \citep[SLODAR,][]{Wilson2002}, and generalized scintillation detection and ranging \citep[G-SCIDAR,][]{Fuchs1998}, but the data reduction methodologies are only now maturing, and only a handful of sites have been characterised using these new techniques over a significant length of time \cite[e.g.][]{Avila2006, Egner2007, Garcia-Lorenzo2011}. As one would expect, the long term (over many nights) average turbulence profiles observed vary from site to site and between seasons, although commonly detected features at sites of good seeing include a strong ground layer component, one or two peaks in the 3 to 5 km altitude region, and then another peak in the troposphere region of 8-15 km altitude --- see figures~\ref{fig:ORM_profile}, \ref{fig:Teide_profile} and \ref{fig:other_CN2_profiles}.

Fortunately, \cite{Garcia-Lorenzo2011} provide profiles for the Observatorio del Roque de los Muchachos. Their median high resolution turbulence profile is based largely on data taken between the months of June and August between 2004 and 2009 (i.e. under summer conditions). In theory, this provides an excellent match to the conditions under which the lucky imaging campaign of summer 2009 was undertaken, although the reality will of course depend upon the particular turbulence profiles of those nights --- indeed the observed seeing widths of the lucky imaging campaign range significantly about the seeing width of the median profile model. With the aim of producing simulated data broadly comparable to that of the observing campaign, I chose to use this data to inform my choice of atmospheric model.

When experimental data on the atmospheric turbulence profile \emph{is} available, one must carefully interpret the data, and decide how to proceed in modelling the turbulence profile computationally. There is experimental evidence from balloon based observations \citep{Vernin1994,Azouit2005} to support the use of thin layer approximations for regions of turbulence. The balloon data suggest there are often a few dominant layers, with many more adding smaller contributions to the total turbulence integral. When choosing how many layers to include in the simulation, a trade-off must be made between accurately modelling a detailed profile with many contributing layers, and the corresponding increase in computational requirements. For example, simulations of multi-conjugate adaptive optics require a large number of layers to be simulated, since too few would produce artificially high estimates of the image quality attainable, with phase perturbations from the few simulated turbulence layers all being perfectly 
corrected. Ground layer adaptive optics, which only aims to correct at one conjugation altitude, is often simulated with fewer layers \citep{Andersen2006}, since the most important property is the allocation of turbulent strength between low-altitude and high-altitude layers. Crucially, for short timescale techniques including lucky imaging, at least two layers should be simulated if there is any evidence for significant variance in the wind velocities, since the relative velocity of the wind layers \corr{along with the Fried coherence length determine the} atmospheric coherence time.

For my simulations, computational requirements were a limiting factor, since long timespan simulations were desired. As a result I chose a 3 layer model providing a crude approximation to the median high-resolution atmospheric profile measured at the Observatorio del Roque de los Muchachos, as described in \cite[][Figure~\ref{fig:ORM_profile}]{Garcia-Lorenzo2011}.  The layer properties are listed in table~\ref{tab:atmos_model_layers}.

Once the turbulent layers to model have been chosen, a realistic wind velocity profile is required. The combination of layer velocities determines the rate at which the instantaneous PSF evolves. This is typically measured by monitoring the autocorrelation properties of light intensity variation in the focal plane, as described in \cite{Scaddan1978,Fitzgerald2006}. To the best of my knowledge a full vertical profile of wind velocities is not available for the Observatorio del Roque de los Muchachos, however \cite{Garcia-Lorenzo2009} provide a profile for the nearby El Teide observatory (fig.~\ref{fig:Teide_profile}). The wind velocities chosen for my simulated model are based upon these data.

 \begin{table}
\caption{Layer properties for simple model of atmospheric conditions at Observatorio del Roque de los Muchachos.}
\label{tab:atmos_model_layers}
\begin{tabular}{lcccc}
\hline \hline 
Layer		& 		Height 				& $C_N^2$ portion  & Windspeed  & Wind direction    \\		
			&	[km above observatory]	&					&[m/s]		& [$^{\circ}$]  \\
\hline
Boundary layer	& 0							& 65\%			& 7			& 200\\
Lower free atmosphere	& 3					& 23\%			& 9			& 230\\
Upper free atmosphere	& 13				& 12\%			& 17			& 245\\

\hline

\end{tabular}
\end{table}

\section{Modelling hybrid adaptive optics systems}
One target of investigation with the simulations I developed is hybrid adaptive optics systems. While there has already been a proof of concept demonstration \citep{Law2009}, there are plenty of questions worth exploring, relating to expected performance over wider fields of view, use of different adaptive optics systems, and application to telescopes of different sizes. In this section I give a brief summary of the motivations for development of future hybrid systems, in terms of system performance --- a science case is discussed in Section~\ref{sec:lucky_AO_science}. I then give preliminary results that illustrate some of the performance gains which might be expected.

\subsection{Strehl ratio statistics behind adaptive optics}
As discussed in chapter~\ref{chap:frame_registration}, the instantaneous Strehl ratio as recorded in a very short exposure varies along with fluctuations in the seeing strength and (where relevant) errors in the adaptive optics corrections. For systems with no correction or partial AO, the histogram of these instantaneous Strehl ratios has a positive skew, i.e. a long tail of samples when the Strehl ratio is significantly better than average. Clearly by performing frame selection upon such a sample the mean Strehl ratio of the selected subsample may be much improved compared to the mean Strehl ratio of the full histogram.

\begin{figure}[htp]
\begin{center}
 \includegraphics[width=0.8\textwidth]{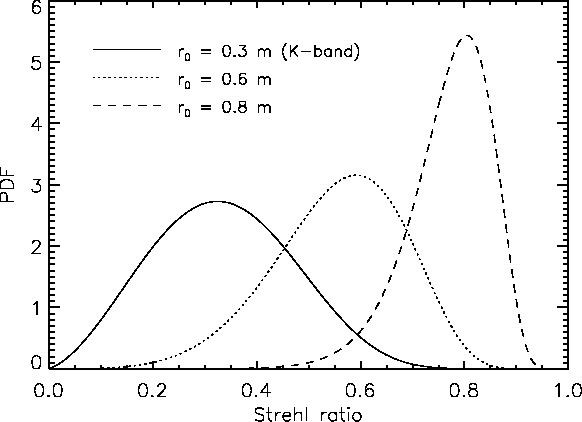}
\caption[Instantaneous Strehl probability distributions behind AO systems]{Instantaneous Strehl probability distributions behind AO systems, for low, intermediate and high Strehl regimes (solid, dotted and dashed lines respectively). Reproduced from~\cite{Gladysz2008}.
}
\label{fig:inst_Strehl_PDFs_gladysz}
\end{center}
\end{figure}

For high levels of adaptive optics correction, on-axis (i.e. near to the guide star), if a long exposure Strehl ratio of greater than say, 0.5 is achieved, this situation begins to change. The standard deviation of the instantaneous Strehl ratios is smaller, and the distribution becomes negatively skewed, as depicted in Figure~\ref{fig:inst_Strehl_PDFs_gladysz}. As a result, any benefits to be  gained from frame selection become diminished, and may be outweighed by the signal to noise ratio considerations of rejecting frames. For this reason, use of high speed imaging behind high Strehl systems for on-axis targets is only of use for specialized applications such as distinguishing faint companions from halo noise, e.g. \cite{Gladysz2008}.

However, current adaptive optics systems still struggle to produce long exposures of moderate Strehl ratio at visible wavelengths. Since this is the wavelength region where EMCCDs are most sensitive, it is an obvious target for application of lucky imaging techniques. Furthermore, for conventional single conjugate AO systems (as opposed to multi-conjugate AO, see Section~\ref{sec:AO_fov_intro}), the Strehl ratio always declines as the target separation from the guide star increases. Application of lucky imaging techniques to off-axis targets may significantly expand the effective isoplanatic patch size. Finally, for laser guide star systems, lucky imaging may offer greatly improved sky coverage. Laser guide stars do not provide tip-tilt information, and so a natural guide star, typically of magnitude $<14$ in V band, is still required to provide this information (further details in Section~\ref{sec:ao_sky_coverage}). The results of lucky imaging with guide stars of magnitude $18$ in \textit{i'} band on the 2.
5m Nordic Optical Telescope suggest that, with large aperture telescopes and partial correction from the AO system, guiding with much fainter natural guide stars may be possible.

\subsection{Preliminary investigations through simulation}
\label{sec:prelim_AO_sim}
An instrument concept that has recently been proposed \corr{by the Cambridge lucky-imaging group} is that of a hybrid lucky-imaging enhanced adaptive optics system for use on the 4.2 metre William Herschel Telescope, at the Observatorio del Roque de los Muchachos. I used the simulation tools to perform a preliminary analysis of what sort of performance might be expected from such an instrument, given different levels of adaptive optics correction.

For the analysis, I generated ten independent simulations of 1000 frames each, simulated at 20 frames per second. The wind model of table~\ref{tab:atmos_model_layers} was used, with a seeing width of 0.5 arcseconds at 500nm. An observation wavelength of 770nm was simulated, corresponding to the SDSS \textit{i'} band filter. This gives a $D/r_0$ ratio of around 12, which is nearly double the optimum for conventional lucky imaging \corr{--- hence conventional lucky imaging would perform very poorly}. The adaptive optics correction was simulated in an idealised fashion, implemented as a simple calculation of Zernike polynomials from the guide-star phase screen, up to a user specified number of Zernike modes. The combined Zernike polynomials were then subtracted from the phase screens for simulated targets across a field of view of 60 arcseconds. Three levels of correction were applied, 0 Zernike modes (regular seeing), 10 Zernike modes (all modes up to order 3), and 28 Zernike modes (all modes up to order 6). 
Detector and photon noise were not simulated, corresponding to performing the lucky guiding process upon bright stars of good signal to noise 
ratio.

\begin{figure}[htp]
\begin{center}
\subfigure[On axis]{	
	\includegraphics[width=0.7\textwidth]{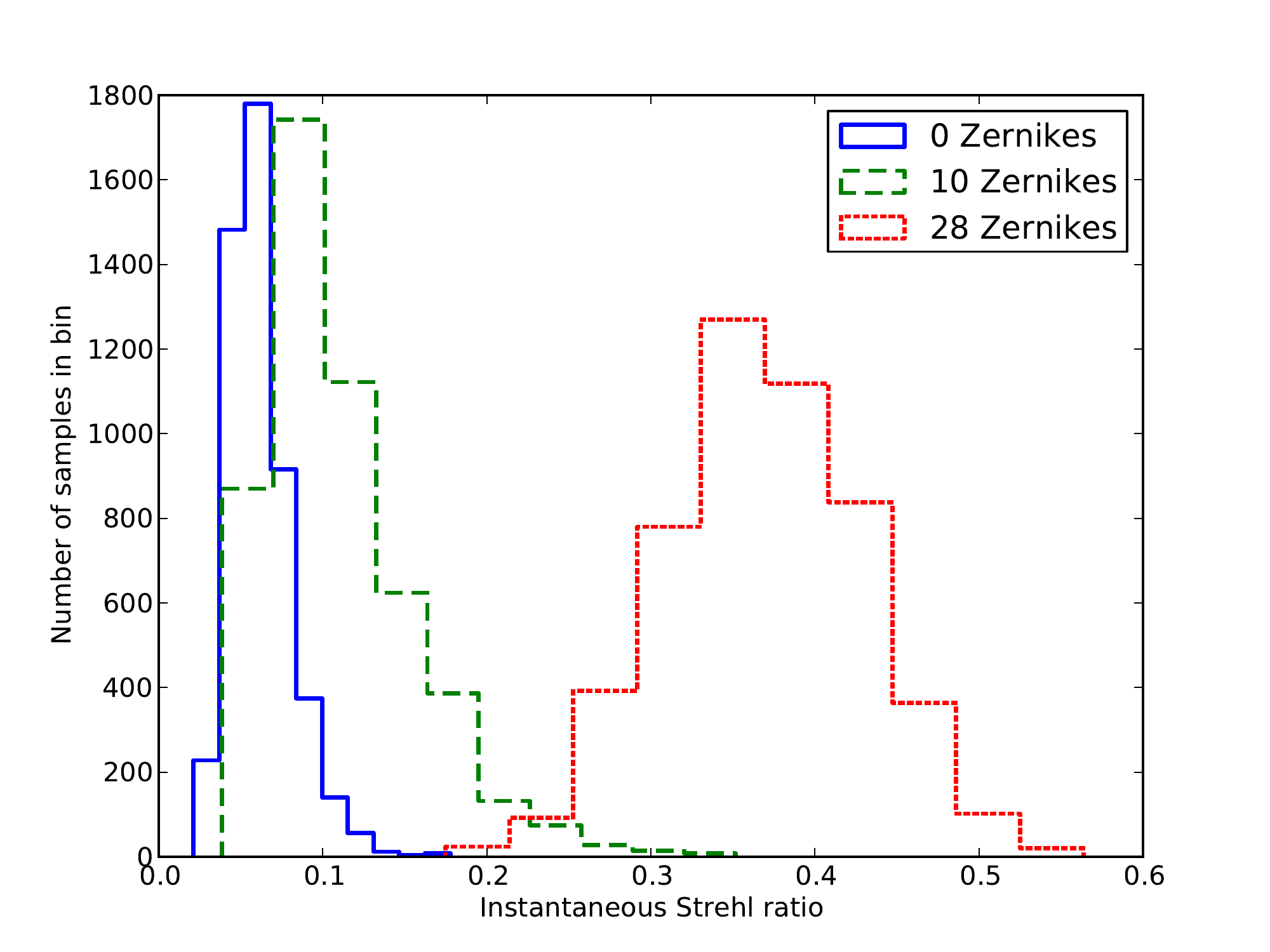}
	\label{subfig:on_axis}
}
\subfigure[Off axis, 30 arcseconds from guide star]{	
	\includegraphics[width=0.7\textwidth]{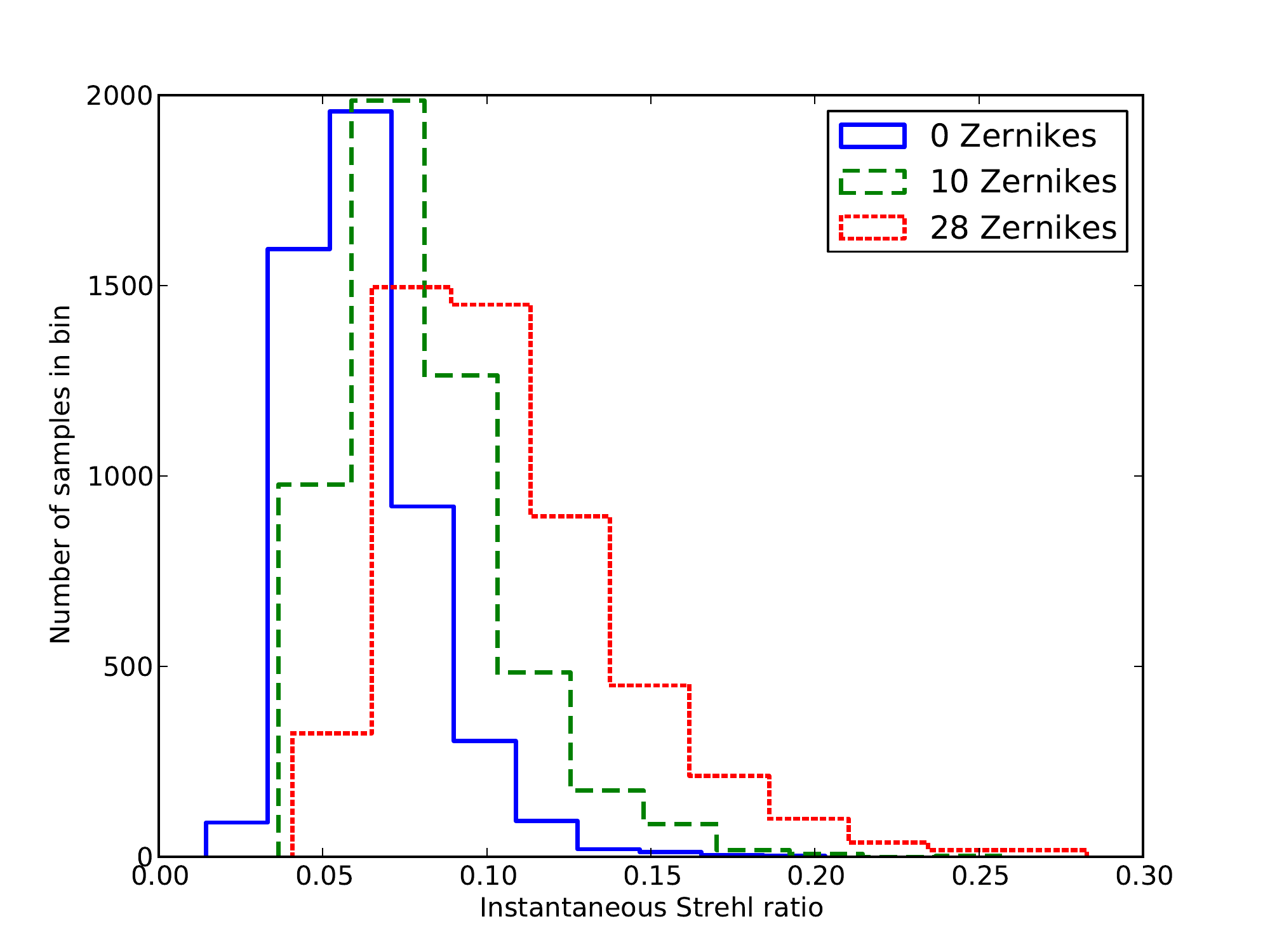}
	\label{subfig:off_axis}
}
 
\caption[Instantaneous Strehl probability distributions after removal of Zernike modes]{
Instantaneous Strehl probability distributions after simulation of adaptive optics by removal of Zernike modes from the phase screen. Clearly for the system parameters simulated, with $D/r_0 \approx 12$ , a fairly high number of Zernike modes must be corrected in order to obtain intermediate Strehl ratio. Even when this is the case, the off-axis Strehl distribution remains positively skewed, as seen in \subref{subfig:off_axis}. 
}
\label{fig:sim_inst_Strehl_PDFs}
\end{center}
\end{figure}

\begin{figure}[htp]
\begin{center}
\subfigure[10 Zernike modes corrected]{	
	\includegraphics[width=0.7\textwidth]{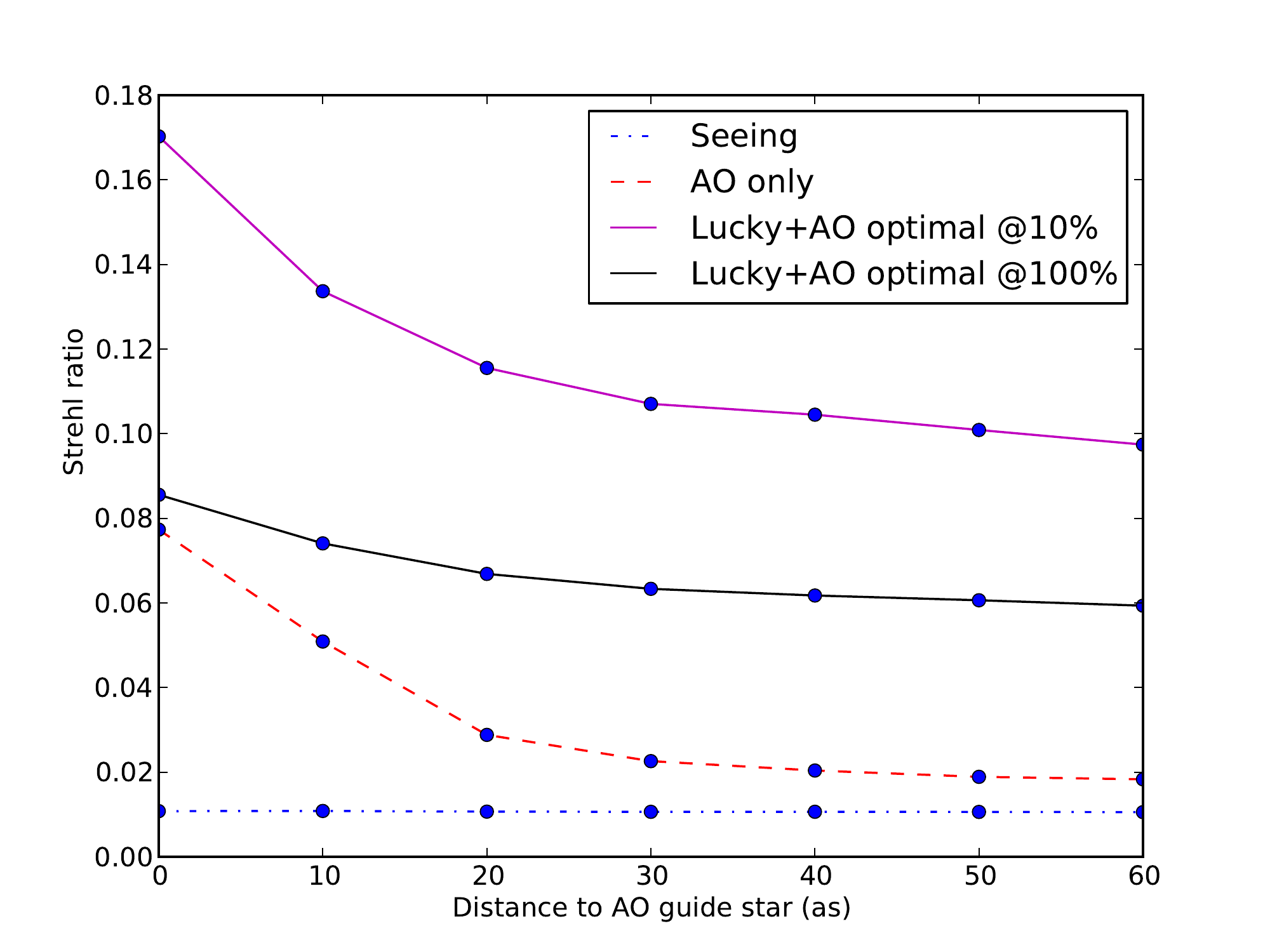}
	\label{subfig:10_zernikes_strehl}
}
\subfigure[28 Zernike modes corrected]{	
	\includegraphics[width=0.7\textwidth]{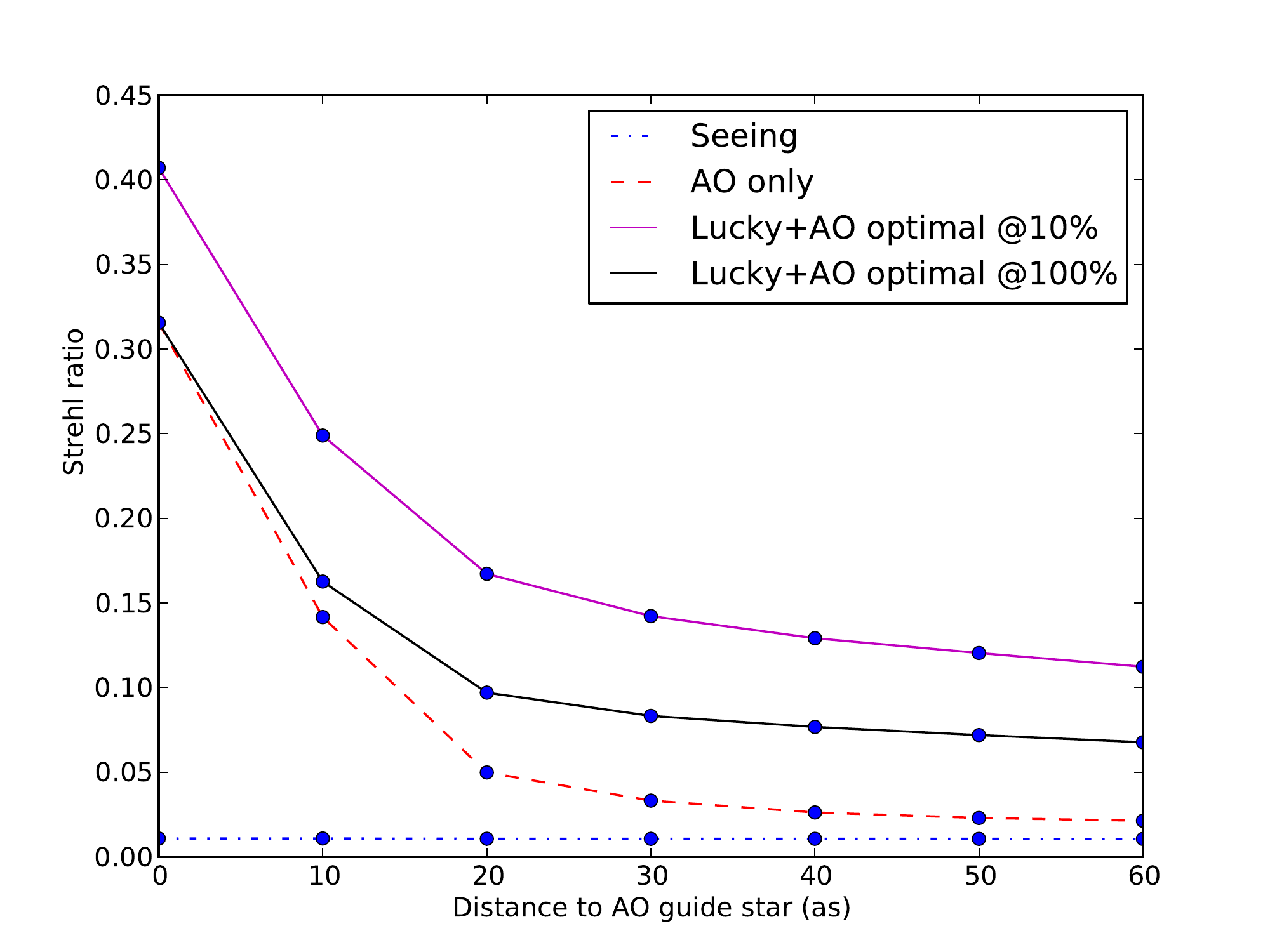}
	\label{subfig:28_zernikes_strehl}
}
\caption[Strehl ratios obtained with lucky imaging adaptive optics]{
Strehl ratios obtained across the field of view. Lines are plotted for no AO (conventional seeing exposure), AO only, and lucky imaging plus AO at 100\% and 10\% selection levels (lower and upper solid lines respectively). The lucky imaging values represent the ideal case of guiding using a separate bright guide star at each radius, rather than a single guide star across the entire field of view. Different levels of AO correction are shown in each subplot, as denoted by the sub-captions. On axis, frame selection significantly improves Strehl ratio in both cases. Off axis, the correction of local tip-tilt, or ``jitter'' as it is sometimes referred to in AO literature, causes a significant improvement even when 100\% of frames are selected.
}
\label{fig:lucky_AO_strehls}
\end{center}
\end{figure}

\begin{figure}[htp]
\begin{center}
\subfigure[10 Zernike modes corrected]{	
	\includegraphics[width=0.7\textwidth]{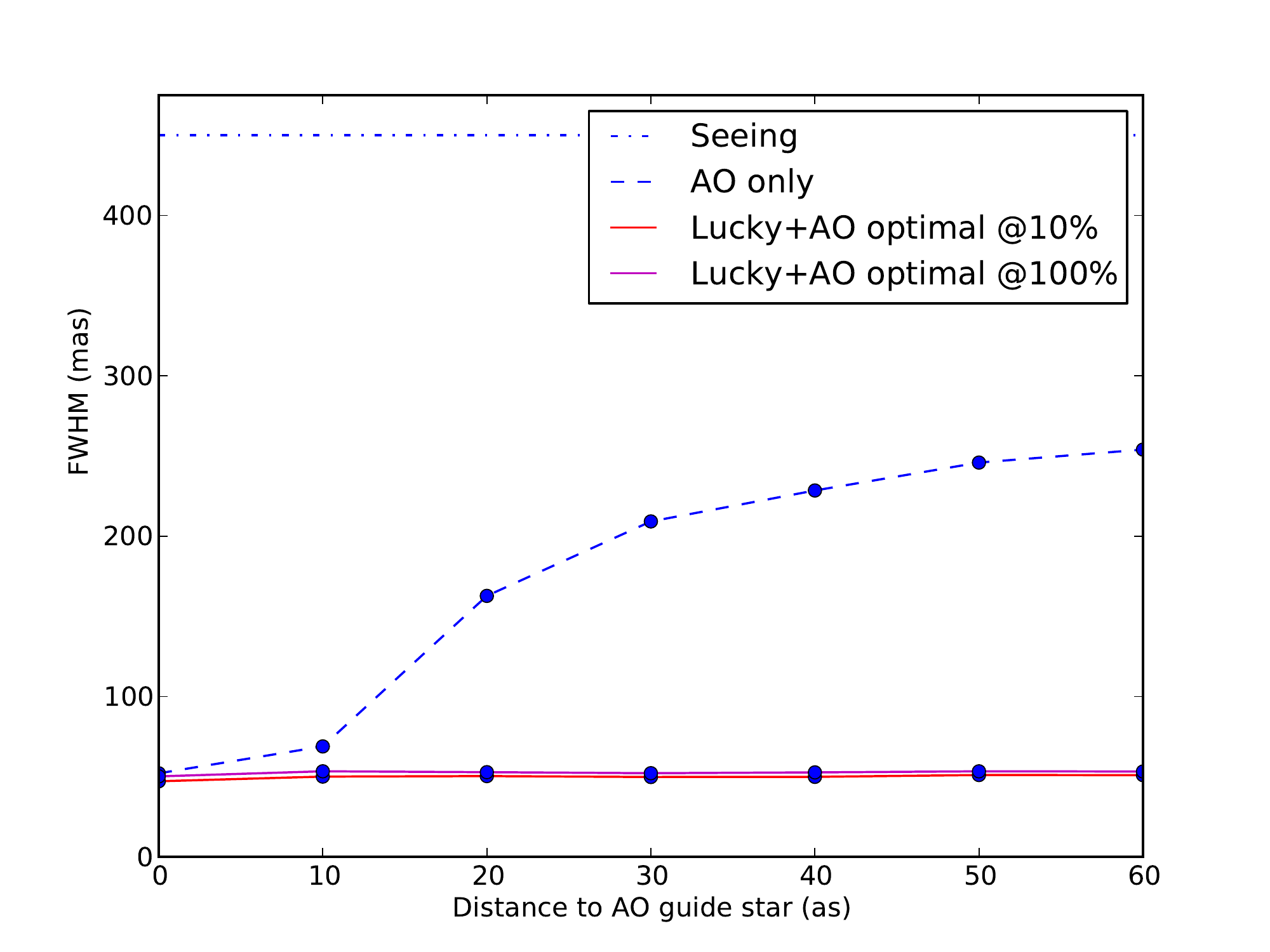}
	\label{subfig:10_zernikes_fwhm}
}
\subfigure[28 Zernike modes corrected]{	
	\includegraphics[width=0.7\textwidth]{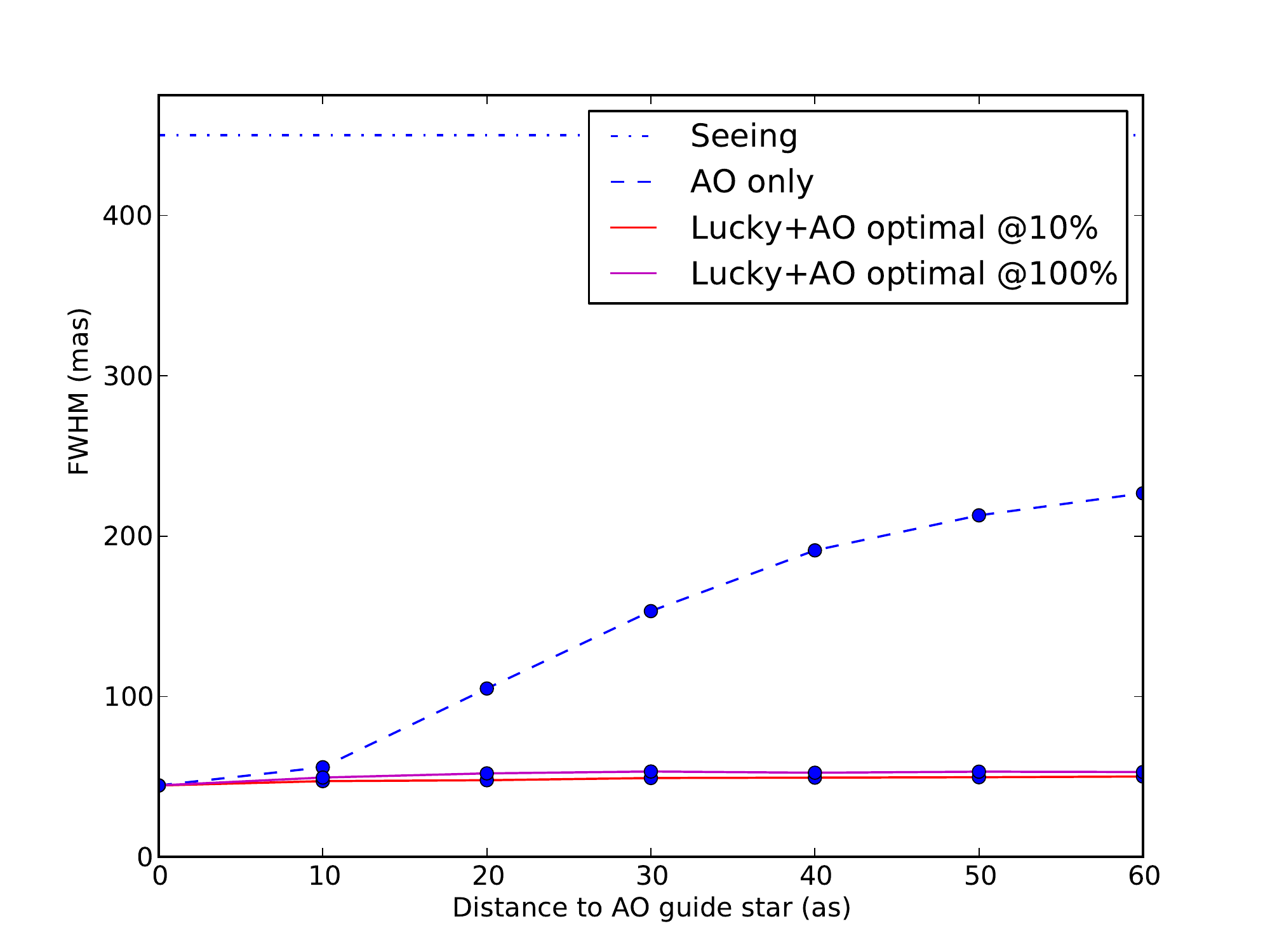}
	\label{subfig:28_zernikes_fwhm}
}
\caption[FWHMs obtained with lucky imaging adaptive optics]{
Full widths at half maximum obtained across the field of view. Plots depict FWHM for the same datasets as Figure~\ref{fig:lucky_AO_strehls}. Perhaps the most striking aspect here is that a diffraction limited FWHM may be obtained on a 4 metre class telescope across a wide field of view, even at 100\% frame selection.
}
\label{fig:lucky_AO_fwhms}
\end{center}
\end{figure}

Histograms of the instantaneous Strehl ratios resulting from the simulations are displayed in Figure~\ref{fig:sim_inst_Strehl_PDFs}, and clearly show how the off-axis Strehl distribution remains negatively skewed even for when moderate Strehl ratios are achieved on-axis. 
Figures~\ref{fig:lucky_AO_strehls} and \ref{fig:lucky_AO_fwhms} plot the image quality metrics obtained across the field of view, for AO and lucky imaging enhanced AO. The simulations clearly suggest that, at least under favourable conditions, lucky imaging techniques considerably enhance AO performance. The most intriguing plot is that depicting FWHM across the field. Using a lucky imaging stabilization of the adaptive optics frames, this is very much enhanced across the field of view. It should be pointed out that this result comes from simulating a local lucky imaging correction at each radius from the AO guide star --- that is, assuming a bright guide star could be utilised and the data reduced accordingly, for each point plotted. Crucially though, the AO corrected speckle patterns produce a diffraction limited PSF when re-aligned, even for a 4m class telescope.

\subsection{Future work}
A full investigation of lucky-imaging enhanced adaptive optics requires a full simulation of adaptive optics system components such as wavefront sensing and deformable mirrors, with detector and photon noise. Models for some of these components are implemented in Arroyo, but others require work. The simulations are also somewhat time consuming to compute. I have been in touch with Alistair Basden of the Durham AO simulation team, and procured a sample dataset of simulated post-AO short exposures, which may then be processed with the detector simulation package described above, and the standard lucky imaging pipeline. I hope to develop a full collaborative investigation in future, although I note that continued development of Arroyo remains valuable for small studies and development of novel instrument models and data reduction algorithms.
%
%

\chapter{Data reduction}
\label{chap:data_reduction}
This chapter describes the software I developed to reduce lucky imaging data. A general overview of the data reduction process is given, and some technical aspects of the software are highlighted. Finally, I describe the combinations of pre-existing astronomy software I have utilised to accomplish further tasks such as astrometric calibration.

\section{The lucky imaging pipeline}
\label{sec:pipeline}
The reduction of any particular dataset is currently a 2 stage process. In a first pass, calibration data is taken, the frames are cleaned of artefacts, cosmic rays are identified and guide star frames are registered. In the second pass this information is used to create drizzled (\cite{Fruchter2002}, also see Section~\ref{sec:image_formation_background}) science images at different selection levels. The drizzle implementation maintains two versions of the final image in parallel --- one thresholded image, one treated in the normal linear manner --- so both of these reduced outputs are produced in a single run. One C++ program deals with each of these stages, with frame registration data output into a formatted text file by the first. 

To preserve disk space and prevent data loss due to human error, the data is treated in an entirely read-only manner, to the extent that a data drive may be mounted in read-only mode. Input and output directories are designated in configuration files, allowing for easy re-reduction if the user wishes to try a reduction process with different parameters.
 
\subsection{Observation reduction preparation tool}
\begin{figure}[htp]
\begin{center}
\includegraphics[width=1.0\textwidth]{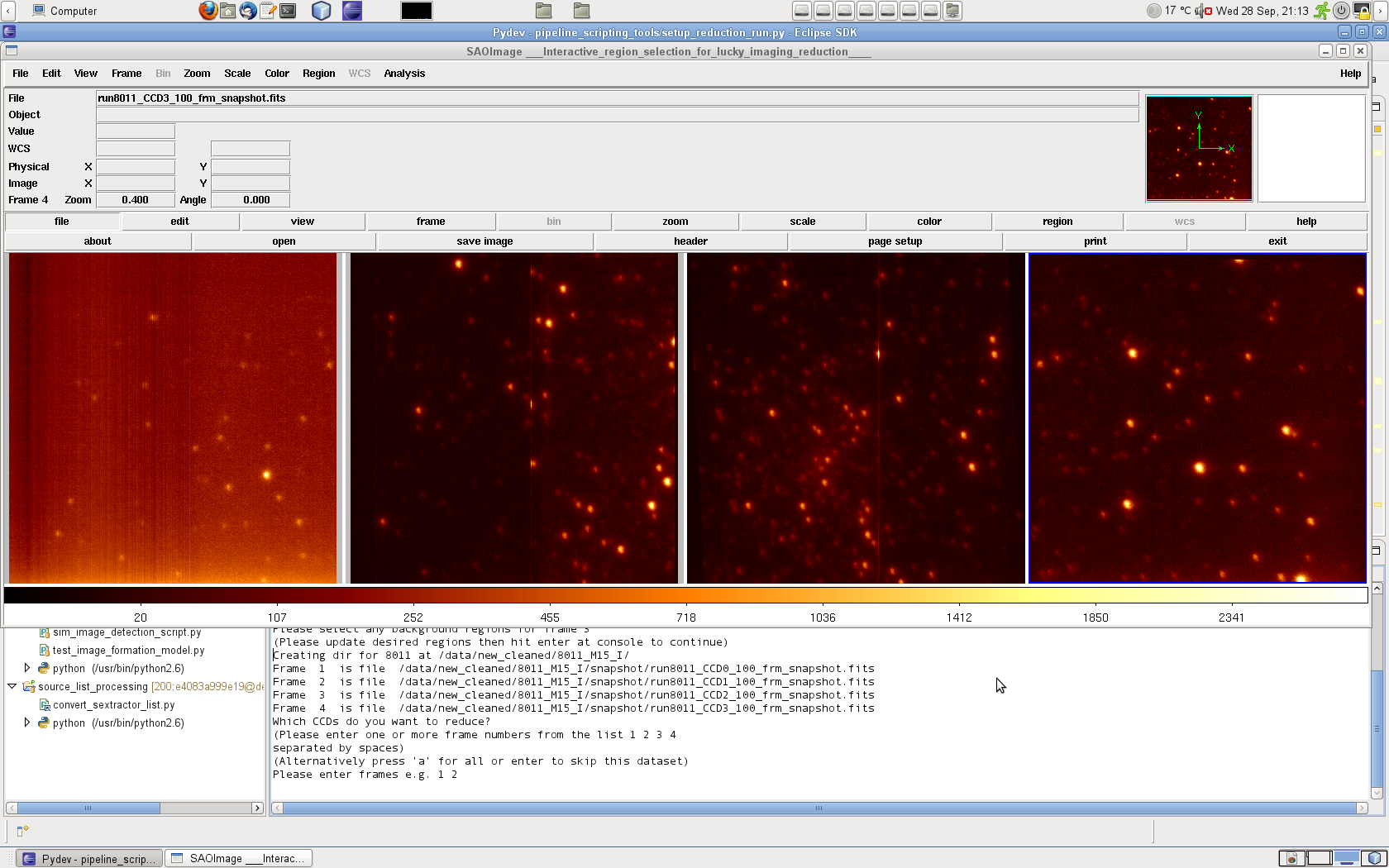}
\caption[Data reduction preparation.]{
Data reduction preparation. The process of preparing a data reduction run is largely automated, since many configuration files need creating and parameters selecting. This is a screenshot from one of the preparation scripts, which loads ``snapshots'' (averages created from the first hundred frames of an observation) and displays a tiled image of the multiple CCDs so that the user may choose which CCDs to reduce. Next, each selected CCD is displayed in turn and the user is prompted to select any guide stars and sky background regions.
}
\label{fig:reduction_prep}
\end{center}
\end{figure}

Before either of the main data reduction programs may be utilised, various configuration files must be written, and target regions selected. 
I created a pair of data-reduction preparation tools in order to ease the creation of these files, create output directory structures mirroring the input directories, and produce scripts which will automate reduction of many targets consecutively.

The run preparation tools are written in the Python programming language. Python allows easy alterations to the source code without recompilation, and enables use of many supporting packages, for example giving easy access to spreadsheet data, and interactivity with the DS9 image display tool for use with FITS format images \citep{Joye2011}.

The first stage of preparing for data reduction is to build a simple database containing information about where each dataset is located on the input drive, which pixel scale and wavelength was used for that observation, where the CCD calibration files are stored, and so on. I refer to this as ``localizing the observations spreadsheet.'' This process relies upon a careful naming convention, where each dataset folder is prefixed with an observation number, e.g. 7010 referring to the 10th observation on night 7 of an observing run (although more complex naming conventions could easily be encoded). The localization script takes as an input a spreadsheet with rows containing the id number and relevant information on filters, etc, for every observation. This may easily be recorded at the time of observation using any spreadsheet package, or created afterwards from observing logs.  The script scans a specified top-level directory for subdirectories prefixed with observation ID numbers, which are cross-referenced 
with the observing log spreadsheet. An extended spreadsheet is then output which has columns encoding the dataset locations, and specifies a ``camera configuration file'' for every observation, which is determined using a simple decision tree based on the filter, lens and so on. 

By automating the tedious process of dataset file location and parameter entry, the rest of the data reduction preparation proceeds much more easily. The user may specify observations for reduction simply by listing their observation ID numbers in a text file. Once a top level output directory has been specified, a second preparation script is employed for the purposes of snapshot creation and region selection. 

The first run of this second preparation script simply engages a cut-down version of the main pipeline program to produce `quick look snapshots,' average frames created by cleaning and averaging the first one hundred frames of each observation, which takes a few seconds per observation. 

The second run of the region selection script is more complex. For each observation in turn, an instance of the image display tool DS9 is initiated with the snapshot from each CCD of the camera mosaic loaded into a tiled image, with a preselected colour scheme and scaling (Figure~\ref{fig:reduction_prep}). The user is prompted to enter which CCDs of the mosaic require data reduction (possibly all of them). Each selected CCD snapshot is then redisplayed so that the user may select a guide star region and background region, if there are suitable regions on the CCD. This automation of file location, display and user prompting greatly speeds the reduction preparation process. Finally, a script is output which will consecutively call the main pipeline programs to reduce all the chosen datasets.

\subsection{First pass --- frame calibration and registration}
The first of the two main pipeline programs performs clean-up and registration of the short exposure images, and is run once for each of the datasets recorded from the multiple CCDs of the camera mosaic. 

The first action taken by the program is to attempt to load  bias calibration frames. Separate frames are stored to record the vertical and horizontal gradients in the bias, for the reasons covered in Section~\ref{sec:bias_pedestal}. If a horizontal gradient frame does not exist, it is generated by taking a column-by-column histogram of the first few thousand frames. An ``internally generated signal'' (see Section~\ref{sec:internal_signal}) frame is also loaded if one exists. Next, if a guide star is present on the CCD to be reduced, a cross correlation reference image is generated. The main reduction procedure then commences.

The various steps of the reduction process are implemented as independent ``filters'' which may be switched on or off depending upon reduction parameters. The filters interface with a multi-threading library (see Section~\ref{sec:pipeline_implementation_notes}) and may be parallel (in which case multiple instances proceed concurrently) or serial (in which case only one instance of the filter is ever in use).  The significant filters are ordered as follows:
\begin{itemize}
 \item Frame buffering: To ensure that data input-output speeds are maximised, I implemented a buffer filter. This does nothing except load the compressed data from disk into memory, leaving no ``dead time'' before the next data load request. This filter is serial, since the disk input speed is fastest when the data is accessed in a serial fashion.
 \item Decompression: This filter extracts the compressed data to an array of floating point numbers.
 \item Frame crop: Optionally, the raw data may be cropped to remove bad rows and columns at the edges.
 \item Bias drift tracking / uniform de-bias: If a sky background region has been defined, a histogram of every frame is taken from this region and used to estimate the bias pedestal of that frame. This value is stored (so it may be used again in the drizzling process), and then subtracted across the frame by the ``uniform debits filter.
 \item Bias gradient correction: Uses the pre-calibrated bias gradient frame to flatten the bias pedestal variations across the image.
 \item Frame summation. Records a sum of the debiased frames.
 \item Histogram recording: If a background region is designated, this builds a debiased frame from that region over the entire dataset. This ``cleaned'' histogram may then be used to estimate the bias pedestal with increased precision, estimate the electron multiplication gain level, and calibrate other detector characteristics.
 \item Cosmic ray detection and analysis filters: These are a series of filters which provide consecutively more stringent, and more time consuming, tests to determine whether a bright pixel is simply due to a bright stellar source, or a cosmic ray. The first filter simply uses a user-defined threshold to determine cosmic ray candidates. The second filter examines the candidate pixel to see if it is a sharp, local maxima or part of a bright region. The third detection filter compares this region of the frame to the same region in the previous frame, to see if the signal has increased dramatically or is roughly constant. Optionally, all candidate and confirmed cosmic ray pixel events may be output to file as 20x20 pixel images about the candidate pixels, which the user may visually inspect.

\item Cross correlation filter: Performs the frame registration procedures described in chapter~\ref{chap:frame_registration}. This is one of the most computationally intensive filters, but since it is implemented in a parallel fashion plenty of processing power may be applied (on a multi-core PC).
\end{itemize}

Finally, the list of recorded frame information is output as a formatted text file, encoding guide star locations and cross-correlation maxima, bias drift levels, any confirmed cosmic ray events, and the file location of each frame. A debiased and internally generated signal-subtracted average frame is also output to file.

\subsection{Second pass --- frame thresholding and recombination}
The second main pipeline program performs frame recombination, through use of the drizzle algorithm. This program needs to be able to reduce synchronised frames from the multiple CCDs, in order to produce a mosaic output from a single bright guide star located on any of the CCDs, which complicates matters somewhat.

The initialization procedures are similar to the registration program, with the addition of a frame list collation stage. A minor bug in the data acquisition program (since rectified) occasionally caused the loss of individual frames of data from the observation, and cosmic rays in the vicinity of the guide star may cause rejection of frames, so the program must be able to handle missing data. To solve this problem an intersection set of the frames which are present across all CCDs is determined. This section of the program also performs a cross-correlation analysis of the file timestamps, to ensure that the frame timestamps are in fact synchronised (again, minor bugs affected a few datasets from the 2009 observing run). 

Once the frame lists have been collated and validated, the frames are ranked by their cross-correlation maxima, which is used as an estimator of frame quality. If a background region was designated for creation of a dataset histogram, fitting procedures covered in Section~\ref{sec:gain_cal} are applied to estimate the electron multiplication gain setting.
The main procedure then begins. The initial steps are much the same as for the frame registration program, except that instead of loading all frames sequentially, only frames meeting the user defined selection requirements are loaded. Once the frames have been debiased, the following filters are applied:
\begin{itemize}
 \item Normalisation: If an estimate of the electron multiplication gain is available, the frames are normalised so that the pixel values are in units of photo-electrons per frame.
\item Thresholding: Optionally, if the frames are normalised, a thresholded copy of the data is created by applying a photon counting threshold across the image.
\item Dark signal subtraction: The pre-calibrated, normalised internal signal frame is used to subtract internally generated signal from the frames. 
\item Drizzle: Now fully calibrated, the frames can at last be realigned and combined according to the drizzle algorithm. Two output images are maintained, one containing weighted sums, while the other is a weight map. Pixels in the region of cosmic ray events or bad columns have zero weights assigned. This is the most computationally intensive filter in this stage of the pipeline. To improve distribution of the processing load over multiple CPU cores, if data from multiple CCDs are being reduced, then a pair of drizzle output images is maintained for each CCD. The drizzle output frames are carefully pre-aligned to the output mosaic pixels, so that creation of the output mosaic from the final images is a simple process of concatenating the images.
\end{itemize}

Finally, the drizzled image pairs of weighted sums and weight maps are used to output reduced images representing the weighted average of the input pixel values. If a thresholded image was produced, an image with good signal to noise across a wide range of light levels may be created by combining it with the linear image as described in Section~\ref{sec:combining_thresholded_data}.

If multiple frame selection cut-off levels are specified, the main procedure runs repeatedly until all the desired reduced images have been generated.

\section{Technical aspects}
\subsection{Data storage considerations}

EMCCDs by design are well suited to imaging at high frame rates and low light levels, where a conventional CCD would have a readout noise level far above the signal. The increasing pixel array sizes and readout rates are enabling application to an ever greater range of problems. As a result it is feasible and sometimes desirable to build an EMCCD imaging system with extremely high data acquisition requirements. As an example, the 4 CCD camera used by the Cambridge Lucky imaging group in summer 2009 produced $\sim4.5$ megapixels of data per exposure at a frame rate of 21 Hz, or 188 MB/sec if the data is stored naively at 16 bits per pixel. As a result, the step up in computational resources required for both storage and processing from conventional imaging is a large one. Sustained data write rates of 100MB/sec are about the limit of what can be achieved with current off-the-shelf computing hardware (circa 2009-2011), which leaves the user with 3 choices: high-end data storage solutions, real-time processing, 
or compression. 

Fortunately, in the case of lucky imaging the data generally lends itself easily to compression, enough that data storage and offline data processing was still possible for the 2009 mosaic camera. The pixel data output from our camera is represented by 14 bits per pixel, representing integer values from 0 to 16384. For a typical lucky imaging astronomical exposure the majority of these pixels do not represent photo-electron events, i.e. they have a value drawn from a Gaussian distribution as described in Section~\ref{sec:pixel_PDFs}. With typical readout noise, all but a negligible proportion of these pixels are within say, 255 data numbers of the mean and as such only require 8 bits to encode.

\cite{Law2007} introduced a simple compression system whereby these ``dark'' pixels are represented by one byte, and pixels outside the offset range -128 to 127 were represented by 3 bytes --- 1 key byte followed by 2 data bytes. The current version of the Cambridge lucky group's data acquisition software takes this concept one stage further; varying storage sizes at the bit, rather than byte, level. For each exposure a histogram is generated and analysed. Depending upon the value range of the dark pixels, the majority of the pixels in an image will then be represented by a smaller number of bits, typically 5 or 6 bits, while pixels with high values (usually due to a photo-electron event) are represented by a key number of the same bit length as a ``dark'' pixel, followed by the full 14 data bits. 

The pixel representations are then combined into a conventional stream of 8-bit bytes using bitwise operations. Higher compression rates could be achieved by using the Rice algorithm as implemented in the CFITSIO library \citep{Pence1999}, but any algorithm must be low enough in complexity to keep up with the data rate while allowing sufficient remaining CPU time to deal with all other aspects of the data acquisition process. This simple algorithm is relatively easy to implement and optimise, and serves well in practice, though future systems may require further work on this aspect.

\subsection{Notes on pipeline implementation}
\label{sec:pipeline_implementation_notes}
The particular needs of a lucky imaging data reduction pipeline --- high performance, custom decompression, large but fairly homogeneous data sets --- resulted in the decision to write a pipeline from scratch, rather than attempt scripting via IRAF, IDL or some other high level language. The code base has been written from the ground up in C++ with the goals of being both object orientated and easily separable into coherent modules, which allows for rapid development, testing and debugging (indeed each section of the library on which the pipeline is built includes a full suite of `unit tests' \citep{Hamill2004}, which may be invoked after compilation of the code to ensure that updates do not break any functionality. The code base is now quite well featured, including reusable modules for drizzling, interpolation, convolution, data calibration, PSF characterisation, etc, and will hopefully be useful for further astronomical image data reduction. I have summarised a few key features of the implementation below,
 but the interested reader is referred to further notes and the source code, available at the author's homepage\footnote{www.ast.cam.ac.uk/$\sim$ts337}.

It became apparent early in the development that a lot of code complexity arose due to the complicated interplay of different co-ordinate systems and indexes used for the different purposes of pixel indexing, describing sub-pixel locations, describing mosaic locations, and for dealing with pixels at different scales (e.g. when creating an interpolated or drizzled image). To alleviate confusion, different C++ classes
\footnote{For an introduction to C++, especially classes and templates, I recommend \cite{Koenig2000}.}
 are used to represent each type of co-ordinate system --- thereby causing code to throw errors at compile time if the wrong co-ordinate type is used for a particular purpose. However, since the co-ordinate systems are all \emph{functionally} the same, they are actually implemented using a single, template class. 

The class used to represent an image carries information on its outline position and pixel scale relative to some global co-ordinate system, thereby allowing easy conversion between different frames of reference. This works rather nicely in practice, creating code that is shorter, easier to read and maintain, and no less efficient if coded carefully. In fact, the addition of a pixel iteration class ensures pixel values are always accessed in the same order they are stored in memory, to improve caching and hence increase performance when writing code which loops over pixel arrays. 

These features, together with template classes representing image arrays and many subroutines go some way towards creating a high performance, compact, shorthand language that may be used to solve data reduction problems quickly and neatly, in pure C++ with few dependencies. While C++ certainly has drawbacks, it does match this sort of challenge well. The interested reader is referred to \cite{Heinzl2007} for further discussion on this topic.

In terms of overall efficiency, the code has been both optimised and multi-threaded. Optimisation was undertaken using the C++ profiling tool set Valgrind \citep{Nethercote2007} to identify bottlenecks and memory cache misses (and ensure no memory leaks). 

The different stages of the data reduction process have been multi-threaded using the open source Intel ``Thread Building Blocks'' package \citep{Reinders2007}. This includes a set of features for pipeline functionality  which map exceedingly well to the demands of processing lucky imaging data, allowing for simultaneous data access and processing across multiple threads with a relatively simple interface. As a result, on our `off-the-shelf' reduction workstation (running with 2 quad core Intel Xeon 5530 processors) the pipeline will process data as fast as it can be read from the disk, typically $\sim$80MB/sec of compressed data - it is therefore capable of reducing the data nearly as fast as the current camera can produce it. 

Since the data reduction process is limited by data access rates, and the datasets fill a considerable amount of disk space, it makes sense to keep the data in compressed form. The data is most compressible in the original raw integer format, i.e. before subtraction of floating point bias frame pixel values. Therefore, keeping it compressed means reapplying the calibration process whenever the frame is loaded, for example in the second pass of the reduction process when the reduced image is being created. While this requires slightly more processor time, the routines are well optimized and data throughput rates are still the limiting factor.

\section{Astrometric Calibration}
\label{sec:astrometric_cals}
With any astronomical instrument, astrometric calibration is essential to determine the pixel scale and sky orientation of the images recorded. One simple, often used method of astrometric calibration is simply to observe a ``calibrator binary,'' a pair of well separated bright stars with accurately known parameters. Calibrating a mosaic camera is more complex. If any astrometric measurements are to be made across different sub-fields of the mosaic, then the offset between the CCD sub-fields must also be carefully measured.

In order to cross-calibrate the 4 CCDs of the Cambridge lucky imaging camera, we observed several crowded fields of globular cluster regions. These regions give many stars across all 4 CCDs which may be used as calibration points.

In theory it is possible to calibrate the relative pixel sizes and pixel offsets between camera sub-fields simply by observing the same field of stars at different offsets --- certain pairs of stars are picked out, their parameters are measured across one sub-field, then the mosaic field of view is shifted and the pair parameters are measured again. One example of this is the Hubble Space Telescope calibration program described in \cite{Casertano2001}. However such a task is challenging, requiring many observations and  careful analysis.

A much simpler alternative is to use a pre-existing catalogue of star positions, if one of sufficient accuracy exists, that provides reference points for stars across the entire field of view. Then the subfields may be calibrated with respect to their on-sky properties, and the inter-CCD offsets can be inferred.

When looking for a way to calibrate our globular cluster fields it became apparent that no high-resolution catalogues existed. While the fields are full of stars that are above, for example, the USNO-B catalogue faint limit, the fields were too crowded and saturated to provide source resolution in the data used to create such catalogues. The 2MASS catalogue \citep{Skrutskie2006} was the only source of reference points available. While I was able to use 2MASS data to produce approximate calibrations the density of catalogue entries is relatively low compared to the number of stars visible in a lucky imaging observation of a globular cluster field --- typically ten or so catalogue entries on a field of a hundred visible stars. To make matters worse, it was often clear that a single entry in the 2MASS catalogue was really a mid-point between 2 stars unresolved in the 2MASS data. Initial calibrations suggested discrepancies between calibrations as large as half an arcsecond, which is significant when each sub 
field is only 30 arcseconds across.

\subsection{Creating an astrometric calibration catalogue from HST archive data}
In order to obtain a high resolution astrometric catalogue, I turned to archival data from the Hubble Space Telescope Wide Field / Planetary Camera 2. (HST WFPC2). There is an ongoing effort to catalogue this archival data \citep{Casertano2010}, but the default source extraction parameters are not optimized for crowded fields, and it was necessary to create the source catalogues myself.

\subsubsection{HSTphot}
The specialized tool of choice for performing source extraction on HST WFPC2 data is HSTphot \citep{Dolphin2000}. HSTphot uses a semi-analytical pre-determined PSF model and detailed information on the WFPC2 chipset to analyse raw archival data. As a result HSTphot can perform source extraction with excellent faint limits, good source de-blending, and astrometric accuracies down to around 1/100th of a pixel. After extensive experimentation I was able to produce source catalogues with excellent depth from the WFPC2 data. However, the program is not without its drawbacks, requiring a rather laborious series of data reduction steps to run, and lacking routines for converting pixel locations into sky co-ordinates. I was able to overcome these issues somewhat by writing auxiliary tools in python, but the final obstacle was that of saturated sources. By default HSTphot entirely rejects sources which have saturated pixels at the core. This often means that the stars which may be located with best accuracy in the 
lucky imaging data are not represented in the HSTphot catalogue, which makes it unsuitable for observations taken in poor seeing or over a short time.

\subsubsection{S-Extractor}
\begin{figure}[htp]
\begin{center}
\includegraphics[width=1.0\textwidth]{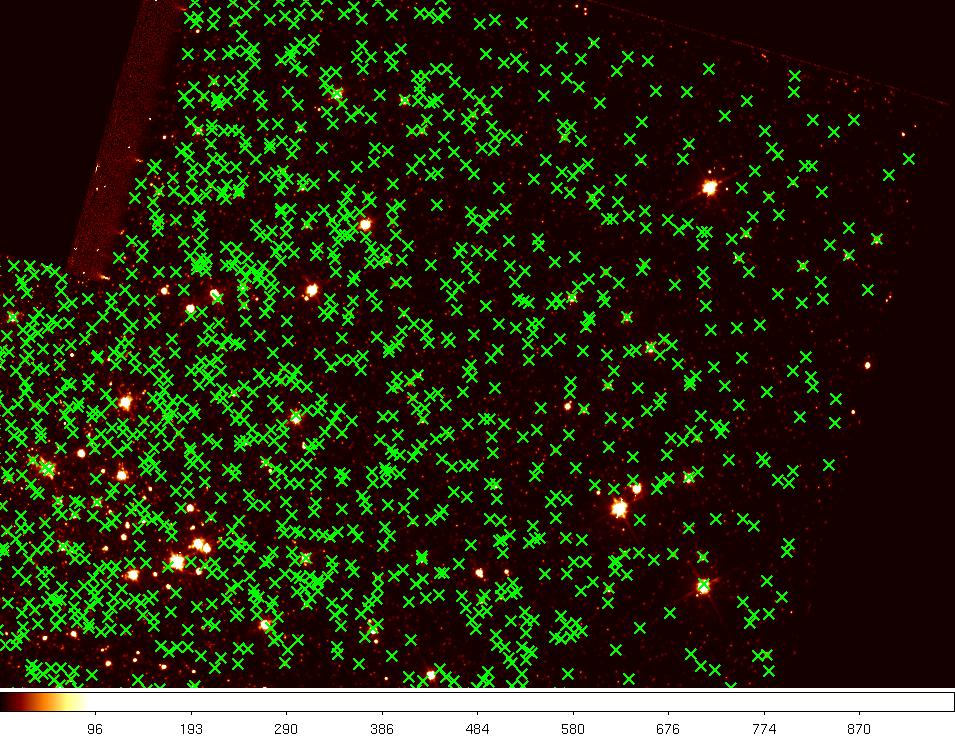}
\caption[Astrometric catalogue created using S-Extractor and archival HST data]{
An astrometric catalogue for the globular cluster M13 created using S-Extractor and archival HST data, after removal of bad sources (details in text). Green crosses depict identified source locations. A high density of catalogue sources is achieved, which is ideal for astrometric calibration of the lucky imaging mosaic through observations of globular clusters. A representative sample of the source extractions has been visually verified.
}
\label{fig:HST_catalog}
\end{center}
\end{figure}

I eventually settled upon the well known S-Extractor \citep{Bertin1996} as my tool of choice for creating WFPC2 catalogue data. Initial trials suggested S-Extractor would not perform well with crowded WFPC2 data, but further investigation of the parameters resulted in decent results. Since the algorithm used by S-Extractor does not model the PSF, the catalogues produced are not as photometrically deep or astrometrically precise as HSTphot, but it can be used to give locations of sufficient bright and intermediate sources with sub-pixel accuracy that the astrometric calibrations should be reasonably accurate. It is also simpler to run, using pre-reduced `science' images rather than the archival raw-data. One problem was that the `science image' mosaics are embedded in a larger blank frame with low signal regions around the edges due to the drizzling process. As a result, S-Extractor often produces false sources in these edge regions. I wrote a Python script to examine the source catalogues produced and 
compare them with the `weights' image which accompanies the science image, rejecting any sources that lie on a region of low weight and therefore low signal to noise. Sources at low signal level in the main image region were also rejected. Figure~\ref{fig:HST_catalog} depicts catalogue positions after rejection of bad sources.

\subsection{Calibration procedure}
Having created a reference catalogue, I required a method of matching the images to known positions. 

The first step was to extract source locations from the images. To do this I wrote a short program to look for local maxima which are above a user specified threshold in order to identify bright star locations, then fit a Gaussian profile to 9 pixels about each bright pixel to determine a sub-pixel position and interpolated peak pixel value.

\begin{figure}[htp]
\begin{center}
\includegraphics[width=1.0\textwidth]{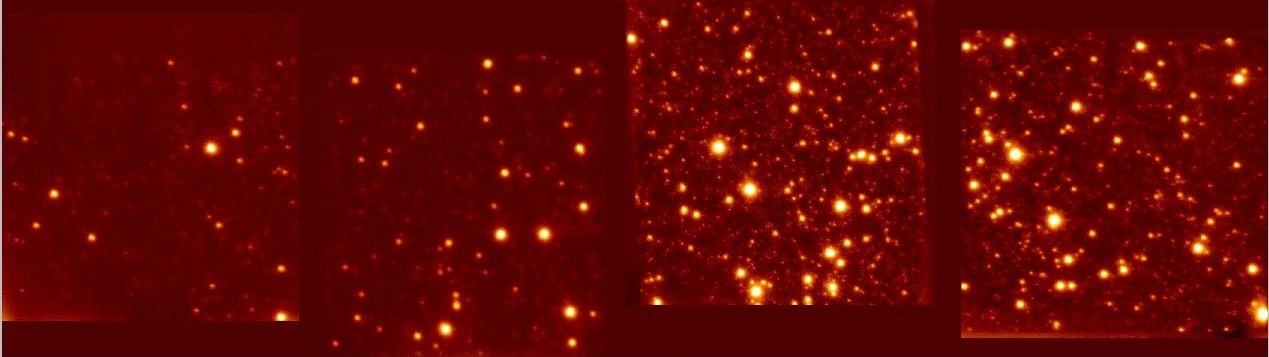}
\caption[A drizzled mosaic of M13, created using the astrometrically calibrated inter-CCD spacings]{
A drizzled mosaic, created using the astrometrically calibrated inter-CCD spacings. Field of view size is approximately 30 by 120 arcseconds. 
}
\label{fig:drizzled_mosaic}
\end{center}
\end{figure}

Second, I required a fitting routine to match the image catalogues to the reference catalogue. There is at least one publicly available tool for this task, ``Astromatic'' \citep{Bertin2010}, but it is focused on large surveys and only provides calibration against pre-existing standard catalogues. I resorted to implementing my own fitting routine, which varied pixel scale, central pixel sky location, and orientation. Each field is first calibrated approximately by eye, then the image catalogue positions are repeatedly converted to sky positions and compared with the reference catalogue while the parameters are varied by the fitting algorithm. While undoubtedly a crude solution compared to specialized tools such as Astromatic, this served well enough and seems to produce good fits. Figure~\ref{fig:drizzled_mosaic} displays the final drizzled mosaic produced using the astrometric calibrations.

Multiple astrometric calibrations using observations of different targets, and from different observing nights, revealed discrepancies in the calibrated inter-CCD spacings of up to a quarter of an arcsecond. However, since the sample size is small and the calibration procedure is not yet extensively tested, it is impossible to be sure whether the discrepancies are due to real movement in the instrument, or simply calibration errors.

%

\chapter{Conclusion}
\label{chap:conclusion}

In conclusion, the specialised hardware and software required for lucky imaging techniques are now sufficiently mature that it should start becoming less of a specialist technique, and more of a mainstream observing tool, either where `better than seeing' observations are required, or under favourable conditions, for diffraction-limited imaging at medium class telescopes. 

In this dissertation I have demonstrated that, through careful calibration and use of data thresholding techniques, the faint limit achieved \corr{in lucky imaging} is comparable to conventional imaging if a moderate pixel angular width is employed (Section~\ref{sec:general_science}). This situation will improve as development of electron multiplying CCDs continues \corr{and levels of noise due to clock-induced charge get lower --- since 2009 the Cambridge group have achieved just this using updated clock-signal-generation electronics}. The concentration of stellar light compared to conventional imaging should result in much better faint limits under observing conditions \corr{with high sky background levels, such as during a full moon}. Previous estimations of high image quality over a large field of view have been reconfirmed using the mosaic CCD camera, with a factor of 2 improvement in full width at half maximum demonstrated even at a radius of 30 arcseconds from the guide star. More sophisticated data 
reduction techniques may widen this patch size even further.

I believe we are already seeing a wider interest in lucky imaging. It is telling that while I was only aware of the Cambridge lucky imaging group when I began my PhD, there are now at least 2 other active instrumentation groups \corr{performing binarity surveys} with lucky imaging techniques \citep{Hormuth2008b,Oscoz2008}. No doubt this is in part due to a growing recognition of the potential of electron multiplying CCDs \citep{Daigle2010,Tulloch2011}, but also there is a growing awareness and understanding of how ground based astronomical imaging systems behave on short timescales. The astronomical community is now beginning to consider how we might exploit that knowledge \corr{to design adaptive optics systems employing exposure selection shutters controlled by wavefront sensors,  or fast tip-tilt-mirror image stabilisation} \cite[see, e.g.][among many others]{Keremedjiev2011,Gladysz2010,Males2010}.

Definitively staking out the observational programs that will benefit from lucky imaging seems to me to be a pressing project. I hope that the analytical and computational models I have developed will be of some use in this. With mosaic EMCCD cameras providing wide fields of view and excellent faint limits, at cheap cost and without moving parts, automated high resolution survey programs seem like a natural target. Detecting and measuring gravitational microlensing events in crowded fields is one example of a science driver to which the technique may be well suited.

While I have developed what are hopefully robust and reusable software components, this is likely to be an ever evolving aspect of the lucky imaging technique, as it is for most of astronomy. Larger mosaic cameras, and more complex image reduction algorithms will likely require judicious use of the latest computational tools such as general purpose graphical processing units, to ensure that data may be processed in a timely fashion.

Finally, probably the most exciting aspect of lucky imaging is the potential of hybrid adaptive optics systems. The experimental set-up described in \cite{Law2009} achieved the highest resolution astronomical images ever obtained with \emph{direct} imaging,%
\footnote{\corr{As opposed to aperture synthesis methods.}} 
with full width at half maximum measurements as small as 42 milliarcseconds. Application of lucky imaging techniques behind the latest generation of adaptive optics systems will push the resolution limit even further, and should enable a much improved effective isoplanatic patch size (cf. Section~\ref{sec:prelim_AO_sim}). Combined with laser guide stars or novel curvature wavefront sensors \corr{which may utilise fainter natural guide stars} \citep{Guyon2010a}, the sky coverage of adaptive optics systems may be greatly improved.

The Hubble space telescope revealed a wealth of new science by improving angular resolution in astronomy. Who knows what we will find lurking at the next resolution limit.


\appendix

%


\bibliographystyle{astronmb}
\setlength{\bibsep}{0pt}

\bibliography{lucky_refs}

\begin{thebibliography}{}

\bibitem[\protect\astroncite{Aime et~al.}{1986}]{Aime1986}
Aime, C., Borgnino, J., Martin, F., Petrov, R., and Ricort, G.: 1986,
\newblock Contribution to the space-time study of stellar speckle patterns,
\newblock {\em Journal of the Optical Society of America A} {\bf 3}, 1001

\bibitem[\protect\astroncite{Andersen et~al.}{2006}]{Andersen2006}
Andersen, D.~R., Stoesz, J., Morris, S., Lloyd-Hart, M., Crampton, D.,
  Butterley, T., Ellerbroek, B., Jolissaint, L., Milton, N.~M., Myers, R.,
  Szeto, K., Tokovinin, A., V{\'{e}}ran, J.-P., and Wilson, R.: 2006,
\newblock Performance modeling of a wide-field ground-layer adaptive optics
  system,
\newblock {\em PASP} {\bf 118}, 1574

\bibitem[\protect\astroncite{Anderson and King}{2000}]{Anderson2000}
Anderson, J. and King, I.~R.: 2000,
\newblock Toward high-precision astrometry with {WFPC2. I.} deriving an
  accurate point-spread function,
\newblock {\em PASP} {\bf 112}, 1360

\bibitem[\protect\astroncite{{Ass{\'e}mat} et~al.}{2007}]{Ass'emat2007}
{Ass{\'e}mat}, F., {Gendron}, E., and {Hammer}, F.: 2007,
\newblock {The FALCON concept: multi-object adaptive optics and atmospheric
  tomography for integral field spectroscopy - principles and performance on an
  8-m telescope},
\newblock {\em MNRAS} {\bf 376}, 287

\bibitem[\protect\astroncite{Assemat et~al.}{2006}]{Assemat2006}
Assemat, F., Wilson, R., and Gendron, E.: 2006,
\newblock Method for simulating infinitely long and non stationary phase
  screens with optimized memory storage,
\newblock {\em Optics Express} {\bf 14}, 988

\bibitem[\protect\astroncite{Avila et~al.}{2006}]{Avila2006}
Avila, R., Carrasco, E., Ibañez, F., Vernin, J., Prieur, J.-L., and Cruz,
  D.~X.: 2006,
\newblock Generalized scidar measurements at san pedro mártir. ii. wind
  profile statistics,
\newblock {\em PASP} {\bf 118}, 503

\bibitem[\protect\astroncite{Avila et~al.}{1997}]{Avila1997}
Avila, R., Vernin, J., and Masciadri, E.: 1997,
\newblock Whole atmospheric-turbulence profiling with generalized scidar,
\newblock {\em Applied Optics} {\bf 36}, 7898

\bibitem[\protect\astroncite{{Azouit} and {Vernin}}{2005}]{Azouit2005}
{Azouit}, M. and {Vernin}, J.: 2005,
\newblock {Optical Turbulence Profiling with Balloons Relevant to Astronomy and
  Atmospheric Physics},
\newblock {\em PASP} {\bf 117}, 536

\bibitem[\protect\astroncite{Bailyn}{1995}]{Bailyn1995}
Bailyn, C.~D.: 1995,
\newblock Blue stragglers and other stellar anomalies:implications for the
  dynamics of globular clusters,
\newblock {\em Annual Review of Astronomy and Astrophysics} {\bf 33}, 133

\bibitem[\protect\astroncite{Bakos et~al.}{2004}]{Bakos2004}
Bakos, G., Noyes, R.~W., Kovács, G., Stanek, K.~Z., Sasselov, D.~D., and
  Domsa, I.: 2004,
\newblock Wide-field millimagnitude photometry with the hat: A tool for
  extrasolar planet detection,
\newblock {\em Publications of the Astronomical Society of the Pacific} {\bf
  116}, 266

\bibitem[\protect\astroncite{Baldwin et~al.}{2001}]{Baldwin2001}
Baldwin, J.~E., Tubbs, R.~N., Cox, G.~C., Mackay, C.~D., Wilson, R.~W., and
  Andersen, M.~I.: 2001,
\newblock Diffraction-limited 800 nm imaging with the 2.56 m nordic optical
  telescope,
\newblock {\em AAP} {\bf 368}, L1

\bibitem[\protect\astroncite{{Baldwin} et~al.}{2008}]{Baldwin2008}
{Baldwin}, J.~E., {Warner}, P.~J., and {Mackay}, C.~D.: 2008,
\newblock {The point spread function in Lucky Imaging and variations in seeing
  on short timescales},
\newblock {\em AAP} {\bf 480}, 589

\bibitem[\protect\astroncite{Basden}{2011}]{Basden2011}
Basden, A.: 2011,
\newblock {\em Private Communication},
\newblock High numerical precision required for rolling phase screen generation

\bibitem[\protect\astroncite{Basden et~al.}{2007}]{Basden2007}
Basden, A., Butterley, T., Myers, R., and Wilson, R.: 2007,
\newblock Durham extremely large telescope adaptive optics simulation platform,
\newblock {\em Applied Optics} {\bf 46}, 1089

\bibitem[\protect\astroncite{Basden et~al.}{2003}]{Basden2003}
Basden, A.~G., Haniff, C.~A., and Mackay, C.~D.: 2003,
\newblock Photon counting strategies with low-light-level {CCDs},
\newblock {\em MNRAS} {\bf 345}, 985

\bibitem[\protect\astroncite{Batalha et~al.}{2012}]{Batalha2012}
Batalha, N.~M., Rowe, J.~F., Bryson, S.~T., Barclay, T., Burke, C.~J.,
  Caldwell, D.~A., Christiansen, J.~L., Mullally, F., Thompson, S.~E., Brown,
  T.~M., Dupree, A.~K., Fabrycky, D.~C., Ford, E.~B., Fortney, J.~J.,
  Gilliland, R.~L., Isaacson, H., Latham, D.~W., Marcy, G.~W., Quinn, S.,
  Ragozzine, D., Shporer, A., Borucki, W.~J., Ciardi, D.~R., Gautier,
  Thomas~N., I., Haas, M.~R., Jenkins, J.~M., Koch, D.~G., Lissauer, J.~J.,
  Rapin, W., Basri, G.~S., Boss, A.~P., Buchhave, L.~A., Charbonneau, D.,
  Christensen-Dalsgaard, J., Clarke, B.~D., Cochran, W.~D., Demory, B.-O.,
  Devore, E., Esquerdo, G.~A., Everett, M., Fressin, F., Geary, J.~C.,
  Girouard, F.~R., Gould, A., Hall, J.~R., Holman, M.~J., Howard, A.~W.,
  Howell, S.~B., Ibrahim, K.~A., Kinemuchi, K., Kjeldsen, H., Klaus, T.~C., Li,
  J., Lucas, P.~W., Morris, R.~L., Prsa, A., Quintana, E., Sanderfer, D.~T.,
  Sasselov, D., Seader, S.~E., Smith, J.~C., Steffen, J.~H., Still, M., Stumpe,
  M.~C., Tarter, J.~C., Tenenbaum, P., Torres, G., Twicken, J.~D., Uddin, K.,
  Van~Cleve, J., Walkowicz, L., and Welsh, W.~F.: 2012,
\newblock {\em Planetary Candidates Observed by Kepler, III: Analysis of the
  First 16 Months of Data}

\bibitem[\protect\astroncite{Bergfors et~al.}{2010}]{Bergfors2010}
Bergfors, C., Brandner, W., Janson, M., Daemgen, S., Geissler, K., Henning, T.,
  Hippler, S., Hormuth, F., Joergens, V., and Köhler, R.: 2010,
\newblock Lucky imaging survey for southern m dwarf binaries,
\newblock {\em Astronomy and Astrophysics} {\bf 520}, 54

\bibitem[\protect\astroncite{Bertin}{2010}]{Bertin2010}
Bertin, E.: 2010,
\newblock in {\em JENAM 2010, Joint European and National Astronomy Meeting},
  p. 227

\bibitem[\protect\astroncite{Bertin and Arnouts}{1996}]{Bertin1996}
Bertin, E. and Arnouts, S.: 1996,
\newblock Sextractor: Software for source extraction.,
\newblock {\em Astronomy and Astrophysics Supplement Series} {\bf 117}, 393

\bibitem[\protect\astroncite{Borucki et~al.}{2010}]{Borucki2010}
Borucki, W.~J., Koch, D., Basri, G., Batalha, N., Brown, T., Caldwell, D.,
  Caldwell, J., Christensen-Dalsgaard, J., Cochran, W.~D., DeVore, E., Dunham,
  E.~W., Dupree, A.~K., Gautier, T.~N., Geary, J.~C., Gilliland, R., Gould, A.,
  Howell, S.~B., Jenkins, J.~M., Kondo, Y., Latham, D.~W., Marcy, G.~W.,
  Meibom, S., Kjeldsen, H., Lissauer, J.~J., Monet, D.~G., Morrison, D.,
  Sasselov, D., Tarter, J., Boss, A., Brownlee, D., Owen, T., Buzasi, D.,
  Charbonneau, D., Doyle, L., Fortney, J., Ford, E.~B., Holman, M.~J., Seager,
  S., Steffen, J.~H., Welsh, W.~F., Rowe, J., Anderson, H., Buchhave, L.,
  Ciardi, D., Walkowicz, L., Sherry, W., Horch, E., Isaacson, H., Everett,
  M.~E., Fischer, D., Torres, G., Johnson, J.~A., Endl, M., MacQueen, P.,
  Bryson, S.~T., Dotson, J., Haas, M., Kolodziejczak, J., Van~Cleve, J.,
  Chandrasekaran, H., Twicken, J.~D., Quintana, E.~V., Clarke, B.~D., Allen,
  C., Li, J., Wu, H., Tenenbaum, P., Verner, E., Bruhweiler, F., Barnes, J.,
  and Prsa, A.: 2010,
\newblock Kepler planet-detection mission: Introduction and first results,
\newblock {\em Science} {\bf 327}, 977

\bibitem[\protect\astroncite{Britton}{2004}]{Britton2004}
Britton, M.~C.: 2004,
\newblock in {\em Modeling and Systems Engineering for Astronomy}, Vol. 5497 of
  {\em Proc. SPIE}, pp 290--300

\bibitem[\protect\astroncite{Brun and Rademakers}{1997}]{Brun1997}
Brun, R. and Rademakers, F.: 1997,
\newblock Root -- an object oriented data analysis framework,
\newblock {\em Nuclear Instruments and Methods in Physics Research Section A:
  Accelerators, Spectrometers, Detectors and Associated Equipment} {\bf
  389(1-2)}, 81

\bibitem[\protect\astroncite{Caccia et~al.}{1987}]{Caccia1987}
Caccia, J.~L., Azouit, M., and Vernin, J.: 1987,
\newblock Wind and c(n)-squared profiling by single-star scintillation
  analysis,
\newblock {\em Applied Optics} {\bf 26}, 1288

\bibitem[\protect\astroncite{Campbell et~al.}{2010}]{Campbell2010}
Campbell, M.~A., Evans, C.~J., Mackey, A.~D., Gieles, M., Alves, J., Ascenso,
  J., Bastian, N., and Longmore, A.~J.: 2010,
\newblock {VLT-MAD} observations of the core of 30 {D}oradus,
\newblock {\em MNRAS} {\bf 405}, 421

\bibitem[\protect\astroncite{Canales and Cagigal}{1999}]{Canales1999a}
Canales, V.~F. and Cagigal, M.~P.: 1999,
\newblock Rician distribution to describe speckle statistics in adaptive
  optics,
\newblock {\em Applied Optics} {\bf 38}, 766

\bibitem[\protect\astroncite{Carbillet et~al.}{2005}]{Carbillet2005}
Carbillet, M., Vérinaud, C., Femenía, B., Riccardi, A., and Fini, L.:
  2005,
\newblock Modelling astronomical adaptive optics - i. the software package
  caos,
\newblock {\em Monthly Notices of the Royal Astronomical Society} {\bf 356},
  1263

\bibitem[\protect\astroncite{Casertano et~al.}{2010}]{Casertano2010}
Casertano, S., Lindsay, K., and Stankiewicz, M.: 2010,
\newblock in {\em Bulletin of the American Astronomical Society}, Vol. 215, p.
  512

\bibitem[\protect\astroncite{Casertano and Wiggs}{2001}]{Casertano2001}
Casertano, S. and Wiggs, M.~S.: 2001,
\newblock {\em An Improved Geometric Solution for WFPC2},
\newblock Technical report, Space Telescope Science Institute

\bibitem[\protect\astroncite{Charbonneau et~al.}{2004}]{Charbonneau2004}
Charbonneau, D., Brown, T.~M., Dunham, E.~W., Latham, D.~W., Looper, D.~L.,
  Mandushev, G., and Deming, D.: 2004,
\newblock in S.~S. Holt (ed.), {\em The Search for Other Worlds}, Vol. 713, pp
  151--160

\bibitem[\protect\astroncite{{Christou}}{1991}]{Christou1991}
{Christou}, J.~C.: 1991,
\newblock {Image quality, tip-tilt correction, and shift-and-add infrared
  imaging},
\newblock {\em PASP} {\bf 103}, 1040

\bibitem[\protect\astroncite{Committee et~al.}{2010}]{Decadal2010}
Committee, N.~R., Astrophysics, and for a Decadal Survey~of Astronomy, C.:
  2010,
\newblock {\em New Worlds, New Horizons in Astronomy and Astrophysics},
\newblock The National Academies Press

\bibitem[\protect\astroncite{Corrsin}{1951}]{Corrsin1951}
Corrsin, S.: 1951,
\newblock On the spectrum of isotropic temperature fluctuations in an isotropic
  turbulence,
\newblock {\em J. Appl. Phys.} {\bf 22}, 469

\bibitem[\protect\astroncite{Coulman et~al.}{1995}]{Coulman1995}
Coulman, C.~E., Vernin, J., and Fuchs, A.: 1995,
\newblock Optical seeing-mechanism of formation of thin turbulent laminae in
  the atmosphere,
\newblock {\em Applied Optics} {\bf 34}, 5461

\bibitem[\protect\astroncite{Daemgen et~al.}{2009}]{Daemgen2009}
Daemgen, S., Hormuth, F., Brandner, W., Bergfors, C., Janson, M., Hippler, S.,
  and Henning, T.: 2009,
\newblock Binarity of transit host stars. implications for planetary
  parameters,
\newblock {\em AAP} {\bf 498}, 567

\bibitem[\protect\astroncite{Daigle et~al.}{2010}]{Daigle2010}
Daigle, O., Quirion, P.-O., and Lessard, S.: 2010,
\newblock in {\em Society of Photo-Optical Instrumentation Engineers (SPIE)
  Conference Series}, Vol. 7742, p.~2

\bibitem[\protect\astroncite{{Dantowitz} et~al.}{2000}]{Dantowitz2000}
{Dantowitz}, R.~F., {Teare}, S.~W., and {Kozubal}, M.~J.: 2000,
\newblock {Ground-based High-Resolution Imaging of Mercury},
\newblock {\em AJ} {\bf 119}, 2455

\bibitem[\protect\astroncite{Davies and Hansen}{1998}]{Davies1998}
Davies, M.~B. and Hansen, B. M.~S.: 1998,
\newblock Neutron star retention and millisecond pulsar production in globular
  clusters,
\newblock {\em Monthly Notices of the Royal Astronomical Society} {\bf 301}, 15

\bibitem[\protect\astroncite{Davies et~al.}{2008}]{Davies2008}
Davies, R., Rabien, S., Lidman, C., Le~Louarn, M., Kasper, M.,
  Förster~Schreiber, N.~M., Roccatagliata, V., Ageorges, N., Amico, P., Dumas,
  C., and Mannucci, F.: 2008,
\newblock Laser guide star adaptive optics without tip-tilt,
\newblock {\em The Messenger} {\bf 131}, 7

\bibitem[\protect\astroncite{{Devaney}}{2007}]{Devaney2007}
{Devaney}, N.: 2007,
\newblock in G. Cheriaux, C.~J. Hooker, and M. Stupka (eds.), {\em Adaptive
  Optics for Laser Systems and Other Applications.}, Vol. 6584 of {\em
  Proceedings of the SPIE}

\bibitem[\protect\astroncite{Dhillon et~al.}{2007}]{Dhillon2007a}
Dhillon, V.~S., Marsh, T.~R., Stevenson, M.~J., Atkinson, D.~C., Kerry, P.,
  Peacocke, P.~T., Vick, A. J.~A., Beard, S.~M., Ives, D.~J., Lunney, D.~W.,
  McLay, S.~A., Tierney, C.~J., Kelly, J., Littlefair, S.~P., Nicholson, R.,
  Pashley, R., Harlaftis, E.~T., and O'Brien, K.: 2007,
\newblock Ultracam: an ultrafast, triple-beam ccd camera for high-speed
  astrophysics,
\newblock {\em MNRAS} {\bf 378}, 825

\bibitem[\protect\astroncite{Diolaiti et~al.}{2000}]{Diolaiti2000}
Diolaiti, E., Bendinelli, O., Bonaccini, D., Close, L., Currie, D., and
  Parmeggiani, G.: 2000,
\newblock Analysis of isoplanatic high resolution stellar fields by the
  {StarFinder} code,
\newblock {\em AAPS} {\bf 147}, 335

\bibitem[\protect\astroncite{Dolphin}{2000}]{Dolphin2000}
Dolphin, A.~E.: 2000,
\newblock {WFPC2} stellar photometry with {HSTPHOT},
\newblock {\em PASP} {\bf 112}, 1383

\bibitem[\protect\astroncite{Durant et~al.}{2011}]{Durant2011}
Durant, M., Shahbaz, T., Gandhi, P., Cornelisse, R., Muñoz-Darias, T.,
  Casares, J., Dhillon, V., Marsh, T., Spruit, H., O'Brien, K., Steeghs, D.,
  and Hynes, R.: 2011,
\newblock High time resolution optical/x-ray cross-correlations for x-ray
  binaries: anticorrelations and rapid variability,
\newblock {\em Monthly Notices of the Royal Astronomical Society} {\bf 410},
  2329

\bibitem[\protect\astroncite{{E2V Technologies}}{2005}]{e2v2005}
{E2V Technologies}: 2005,
\newblock {\em CCD201-20 Back Illuminated 2-Phase IMO Series Electron
  Multiplying CCD Sensor},
\newblock Technical report, E2V

\bibitem[\protect\astroncite{Egner et~al.}{2007}]{Egner2007}
Egner, S.~E., Masciadri, E., and McKenna, D.: 2007,
\newblock Generalized scidar measurements at mount graham,
\newblock {\em PASP} {\bf 119}, 669

\bibitem[\protect\astroncite{Esposito et~al.}{2010}]{Esposito2010}
Esposito, S., Riccardi, A., Fini, L., Puglisi, A.~T., Pinna, E., Xompero, M.,
  Briguglio, R., Quirós-Pacheco, F., Stefanini, P., Guerra, J.~C., Busoni,
  L., Tozzi, A., Pieralli, F., Agapito, G., Brusa-Zappellini, G., Demers, R.,
  Brynnel, J., Arcidiacono, C., and Salinari, P.: 2010,
\newblock in {\em Society of Photo-Optical Instrumentation Engineers (SPIE)
  Conference Series}, Vol. 7736, p.~7

\bibitem[\protect\astroncite{Fitzgerald and Graham}{2006}]{Fitzgerald2006}
Fitzgerald, M.~P. and Graham, J.~R.: 2006,
\newblock Speckle statistics in adaptively corrected images,
\newblock {\em ApJ} {\bf 637}, 541

\bibitem[\protect\astroncite{{Fried}}{1965}]{Fried1965}
{Fried}, D.~L.: 1965,
\newblock Statistics of a geometric representation of wavefront distortion,
\newblock {\em Journal of the Optical Society of America (1917-1983)} {\bf 55},
  1427

\bibitem[\protect\astroncite{Fried}{1978}]{Fried1978}
Fried, D.~L.: 1978,
\newblock Probability of getting a lucky short-exposure image through
  turbulence,
\newblock {\em J. Opt. Soc. Am.} {\bf 68(12)}, 1651

\bibitem[\protect\astroncite{Fruchter and Hook}{2002}]{Fruchter2002}
Fruchter, A.~S. and Hook, R.~N.: 2002,
\newblock Drizzle: A method for the linear reconstruction of undersampled
  images,
\newblock {\em PASP} {\bf 114}, 144

\bibitem[\protect\astroncite{Fuchs et~al.}{1998}]{Fuchs1998}
Fuchs, A., Tallon, M., and Vernin, J.: 1998,
\newblock Focusing on a turbulent layer: Principle of the ``generalized
  scidar'',
\newblock {\em Publications of the Astronomical Society of the Pacific} {\bf
  110}, 86

\bibitem[\protect\astroncite{García-Lorenzo et~al.}{2009}]{Garcia-Lorenzo2009}
García-Lorenzo, B., Eff-Darwich, A., Fuensalida, J.~J., and Castro-Almazán,
  J.: 2009,
\newblock Adaptive optics parameters connection to wind speed at the teide
  observatory,
\newblock {\em Monthly Notices of the Royal Astronomical Society} {\bf 397},
  1633

\bibitem[\protect\astroncite{García-Lorenzo and
  Fuensalida}{2011}]{Garcia-Lorenzo2011}
García-Lorenzo, B. and Fuensalida, J.~J.: 2011,
\newblock Atmospheric optical turbulence at the roque de los muchachos
  observatory: data base and recalibration of the generalized scidar data,
\newblock {\em Monthly Notices of the Royal Astronomical Society} {\bf 1}, 1124

\bibitem[\protect\astroncite{Giersz and Heggie}{2009}]{Giersz2009}
Giersz, M. and Heggie, D.~C.: 2009,
\newblock Monte carlo simulations of star clusters - vi. the globular cluster
  ngc 6397,
\newblock {\em Monthly Notices of the Royal Astronomical Society} {\bf 395},
  1173

\bibitem[\protect\astroncite{Gladysz et~al.}{2008}]{Gladysz2008}
Gladysz, S., Christou, J., Law, N., Dekany, R., Redfern, M., and Mackay, C.:
  2008,
\newblock in {\em Society of Photo-Optical Instrumentation Engineers (SPIE)
  Conference Series}, Vol. 7015, p.~66

\bibitem[\protect\astroncite{Gladysz et~al.}{2010a}]{Gladysz2010}
Gladysz, S., Yaitskova, N., and Christou, J.~C.: 2010a,
\newblock in {\em Society of Photo-Optical Instrumentation Engineers (SPIE)
  Conference Series}, Vol. 7736, p.~50

\bibitem[\protect\astroncite{Gladysz et~al.}{2010b}]{Gladysz2010b}
Gladysz, S., Yaitskova, N., and Christou, J.~C.: 2010b,
\newblock Statistics of intensity in adaptive-optics images and their
  usefulness for detection and photometry of exoplanets,
\newblock {\em Journal of the Optical Society of America A} {\bf 27}, 64

\bibitem[\protect\astroncite{Goodman}{1975}]{Goodman1975}
Goodman, J.: 1975,
\newblock in {\em Laser Speckle and Related Phenomena}, Vol.~9 of {\em Topics
  in Applied Physics}, pp 9--75, Springer Berlin / Heidelberg,
\newblock 10.1007/BFb0111436

\bibitem[\protect\astroncite{Goodwin et~al.}{2007}]{Goodwin2007}
Goodwin, M., Jenkins, C., and Lambert, A.: 2007,
\newblock Improved detection of atmospheric turbulence with slodar,
\newblock {\em Optics Express} {\bf 15}, 14844

\bibitem[\protect\astroncite{Gratadour et~al.}{2005}]{Gratadour2005}
Gratadour, D., Mugnier, L.~M., and Rouan, D.: 2005,
\newblock Sub-pixel image registration with a maximum likelihood estimator.
  application to the first adaptive optics observations of arp 220 in the l'
  band,
\newblock {\em AAP} {\bf 443}, 357

\bibitem[\protect\astroncite{{Gullieuszik} et~al.}{2008}]{Gullieuszik2008}
{Gullieuszik}, M., {Greggio}, L., {Held}, E.~V., {Moretti}, A., {Arcidiacono},
  C., {Bagnara}, P., {Baruffolo}, A., {Diolaiti}, E., {Falomo}, R., {Farinato},
  J., {Lombini}, M., {Ragazzoni}, R., {Brast}, R., {Donaldson}, R., {Kolb}, J.,
  {Marchetti}, E., and {Tordo}, S.: 2008,
\newblock {Resolving stellar populations outside the Local Group: MAD
  observations of UKS 2323-326},
\newblock {\em AAP} {\bf 483}, L5

\bibitem[\protect\astroncite{Guyon}{2010}]{Guyon2010a}
Guyon, O.: 2010,
\newblock High sensitivity wavefront sensing with a nonlinear curvature
  wavefront sensor,
\newblock {\em Publications of the Astronomical Society of the Pacific} {\bf
  122}, 49

\bibitem[\protect\astroncite{Hamill}{2004}]{Hamill2004}
Hamill, P.: 2004,
\newblock {\em Unit test frameworks},
\newblock O'Reilly

\bibitem[\protect\astroncite{Hardy}{1998}]{Hardy1998}
Hardy, J.~W.: 1998,
\newblock {\em Adaptive Optics for Astronomical Telescopes},
\newblock Oxford University Press

\bibitem[\protect\astroncite{Harris}{1996}]{Harris1996}
Harris, W.~E.: 1996,
\newblock A catalog of parameters for globular clusters in the milky way,
\newblock {\em The Astronomical Journal} {\bf 112}, 1487

\bibitem[\protect\astroncite{Harris}{2010}]{Harris2010}
Harris, W.~E.: 2010,
\newblock {\em A New Catalog of Globular Clusters in the Milky Way}

\bibitem[\protect\astroncite{Heggie and Giersz}{2008}]{Heggie2008}
Heggie, D.~C. and Giersz, M.: 2008,
\newblock Monte carlo simulations of star clusters - v. the globular cluster
  m4,
\newblock {\em Monthly Notices of the Royal Astronomical Society} {\bf 389},
  1858

\bibitem[\protect\astroncite{Heinzl}{2007}]{Heinzl2007}
Heinzl, R.: 2007,
\newblock {\em Ph.D. thesis}, Vienna University of Technology Institute for
  Microelectronics

\bibitem[\protect\astroncite{Hormann and Leydold}{2000}]{Hormann2000}
Hormann, W. and Leydold, J.: 2000,
\newblock in {\em Proceedings of the 32nd conference on Winter simulation}, pp
  675--682, Society for Computer Simulation International, Orlando, Florida

\bibitem[\protect\astroncite{Hormuth}{2008}]{Hormuth2008}
Hormuth, F.: 2008,
\newblock in {\em Society of Photo-Optical Instrumentation Engineers (SPIE)
  Conference Series}, Vol. 7014, p. 136

\bibitem[\protect\astroncite{Hormuth et~al.}{2008a}]{Hormuth2008a}
Hormuth, F., Brandner, W., Hippler, S., and Henning, T.: 2008a,
\newblock Astralux - the calar alto 2.2-m telescope lucky imaging camera,
\newblock {\em ArXiv e-prints} {\bf 0807}, 504

\bibitem[\protect\astroncite{Hormuth et~al.}{2008b}]{Hormuth2008b}
Hormuth, F., Hippler, S., Brandner, W., Wagner, K., and Henning, T.: 2008b,
\newblock in {\em Ground-based and Airborne Instrumentation for Astronomy II.},
  Vol. 7014 of {\em Proc. SPIE}, p. 138

\bibitem[\protect\astroncite{Howell}{2000}]{Howell2000}
Howell, S.~B.: 2000,
\newblock {\em Handbook of CCD Astronomy}, Chapt.~4, p.~60,
\newblock Cambridge University Press

\bibitem[\protect\astroncite{Hufnagel}{1966}]{Hufnagel1966}
Hufnagel, R.: 1966,
\newblock in {\em Restoration of Atmospherically Degraded Images}, Vol.~3,
  p.~11

\bibitem[\protect\astroncite{Hurley et~al.}{2007}]{Hurley2007}
Hurley, J.~R., Aarseth, S.~J., and Shara, M.~M.: 2007,
\newblock The core binary fractions of star clusters from realistic
  simulations,
\newblock {\em The Astrophysical Journal} {\bf 665}, 707

\bibitem[\protect\astroncite{Hut et~al.}{1992}]{Hut1992}
Hut, P., McMillan, S., Goodman, J., Mateo, M., Phinney, E.~S., Pryor, C.,
  Richer, H.~B., Verbunt, F., and Weinberg, M.: 1992,
\newblock Binaries in globular clusters,
\newblock {\em Publications of the Astronomical Society of the Pacific} {\bf
  104}, 981

\bibitem[\protect\astroncite{Hut et~al.}{1991}]{Hut1991}
Hut, P., Murphy, B.~W., and Verbunt, F.: 1991,
\newblock The formation rate of low-mass x-ray binaries in globular clusters,
\newblock {\em Astronomy and Astrophysics} {\bf 241}, 137

\bibitem[\protect\astroncite{Ivanova et~al.}{2005}]{Ivanova2005}
Ivanova, N., Fregeau, J.~M., Rasio, F.~A., and Stairs, I.~H.: 2005,
\newblock in F.~A. Rasio (ed.), {\em Binary Radio Pulsars}, Vol. 328, p. 231

\bibitem[\protect\astroncite{Jerram et~al.}{2001}]{Jerram2001}
Jerram, P., Pool, P.~J., Bell, R., Burt, D.~J., Bowring, S., Spencer, S.,
  Hazelwood, M., Moody, I., Catlett, N., Heyes, P.~S., Canosa, J., and Sampat,
  N.: 2001,
\newblock in M.~M. Blouke (ed.), {\em Society of Photo-Optical Instrumentation
  Engineers (SPIE) Conference Series}, Vol. 4306, pp 178--186

\bibitem[\protect\astroncite{Johansson and Gavel}{1994}]{Johansson1994}
Johansson, E.~M. and Gavel, D.~T.: 1994,
\newblock in J.~B. Breckinridge (ed.), {\em Society of Photo-Optical
  Instrumentation Engineers (SPIE) Conference Series}, Vol. 2200, pp 372--383

\bibitem[\protect\astroncite{Jolissaint et~al.}{2006}]{Jolissaint2006}
Jolissaint, L., Véran, J.-P., and Conan, R.: 2006,
\newblock Analytical modeling of adaptive optics: foundations of the phase
  spatial power spectrum approach,
\newblock {\em Journal of the Optical Society of America A} {\bf 23}, 382

\bibitem[\protect\astroncite{Joye et~al.}{2011}]{Joye2011}
Joye, W., Accomazzi, A., Mink, D.~J., and Rots, A.~H.: 2011,
\newblock in I.~N. Evans (ed.), {\em Astronomical Data Analysis Software and
  Systems XX}, Vol. 442, p. 633

\bibitem[\protect\astroncite{Keremedjiev and
  Eikenberry}{2011}]{Keremedjiev2011}
Keremedjiev, M. and Eikenberry, S.~S.: 2011,
\newblock A comparison between lucky imaging and speckle stabilization for
  astronomical imaging,
\newblock {\em Publications of the Astronomical Society of the Pacific} {\bf
  123}, 213

\bibitem[\protect\astroncite{Knigge et~al.}{2009}]{Knigge2009}
Knigge, C., Leigh, N., and Sills, A.: 2009,
\newblock A binary origin for `blue stragglers' in globular clusters,
\newblock {\em Nature} {\bf 457}, 288

\bibitem[\protect\astroncite{Koenig and Moo}{2000}]{Koenig2000}
Koenig, A. and Moo, B.~E.: 2000,
\newblock {\em Accelerated C++ - Practical Programming by Example},
\newblock Addison-Wesley

\bibitem[\protect\astroncite{Krauss and Chaboyer}{2003}]{Krauss2003}
Krauss, L.~M. and Chaboyer, B.: 2003,
\newblock Age estimates of globular clusters in the milky way: Constraints on
  cosmology,
\newblock {\em Science} {\bf 299}, 65

\bibitem[\protect\astroncite{Labeyrie}{1970}]{Labeyrie1970}
Labeyrie, A.: 1970,
\newblock Attainment of diffraction limited resolution in large telescopes by
  fourier analysing speckle patterns in star images,
\newblock {\em Astronomy and Astrophysics} {\bf 6}, 85

\bibitem[\protect\astroncite{Lane et~al.}{1992}]{Lane1992}
Lane, R.~G., Glindemann, A., and Dainty, J.~C.: 1992,
\newblock Simulation of a kolmogorov phase screen,
\newblock {\em Waves in Random Media} {\bf 2}, 209

\bibitem[\protect\astroncite{Lantz et~al.}{2008}]{Lantz2008}
Lantz, E., Blanchet, J.-L., Furfaro, L., and Devaux, F.: 2008,
\newblock Multi-imaging and {B}ayesian estimation for photon counting with
  {EMCCDs},
\newblock {\em MNRAS} {\bf 386}, 2262

\bibitem[\protect\astroncite{Law}{2008}]{Law2008a}
Law, N.: 2008,
\newblock in {\em Adaptive Optics Systems}, Vol. 7015 of {\em Proc. SPIE},
  p.~67

\bibitem[\protect\astroncite{Law}{2007}]{Law2007}
Law, N.~M.: 2007,
\newblock {\em Ph.D. thesis}, University of Cambridge

\bibitem[\protect\astroncite{Law et~al.}{2008}]{Law2008}
Law, N.~M., Hodgkin, S.~T., and Mackay, C.~D.: 2008,
\newblock The luckycam survey for very low mass binaries - ii. 13 new m4.5-m6.0
  binaries,
\newblock {\em Monthly Notices of the Royal Astronomical Society} {\bf 384},
  150

\bibitem[\protect\astroncite{Law et~al.}{2005}]{Law2005}
Law, N.~M., Hodgkin, S.~T., Mackay, C.~D., and Baldwin, J.~E.: 2005,
\newblock Ten new very low-mass close binaries resolved in the visible,
\newblock {\em Astronomische Nachrichten} {\bf 326}, 1024

\bibitem[\protect\astroncite{Law et~al.}{2006}]{Law2006}
Law, N.~M., Mackay, C.~D., and Baldwin, J.~E.: 2006,
\newblock Lucky imaging: high angular resolution imaging in the visible from
  the ground,
\newblock {\em AAP} {\bf 446}, 739

\bibitem[\protect\astroncite{Law et~al.}{2009}]{Law2009}
Law, N.~M., Mackay, C.~D., Dekany, R.~G., Ireland, M., Lloyd, J.~P., Moore,
  A.~M., Robertson, J.~G., Tuthill, P., and Woodruff, H.~C.: 2009,
\newblock Getting lucky with adaptive optics: Fast adaptive optics image
  selection in the visible with a large telescope,
\newblock {\em The Astrophysical Journal} {\bf 692}, 924

\bibitem[\protect\astroncite{Le~Louarn}{2010}]{LeLouarn2010}
Le~Louarn, M.: 2010,
\newblock in {\em Society of Photo-Optical Instrumentation Engineers (SPIE)
  Conference Series}, Vol. 7736, p.~41

\bibitem[\protect\astroncite{Lodieu et~al.}{2009}]{Lodieu2009}
Lodieu, N., Zapatero~Osorio, M.~R., and Martín, E.~L.: 2009,
\newblock Lucky imaging of {M} subdwarfs,
\newblock {\em Astronomy and Astrophysics} {\bf 499}, 729

\bibitem[\protect\astroncite{Lopez}{1992}]{Lopez1992}
Lopez, B.: 1992,
\newblock How to monitor optimum exposure times for high resolution imaging
  modes?,
\newblock {\em AAP} {\bf 253}, 635

\bibitem[\protect\astroncite{Lopez and Sarazin}{1993}]{Lopez1993}
Lopez, B. and Sarazin, M.: 1993,
\newblock The {ESO} atmospheric temporal coherence monitor dedicated to high
  angular resolution imaging,
\newblock {\em AAP} {\bf 276}, 320

\bibitem[\protect\astroncite{Mackay et~al.}{2010}]{Mackay2010}
Mackay, C., Staley, T.~D., King, D., Suess, F., and Weller, K.: 2010,
\newblock in {\em Society of Photo-Optical Instrumentation Engineers (SPIE)
  Conference Series}, Vol. 7742, p.~1

\bibitem[\protect\astroncite{Mackay}{1986}]{Mackay1986}
Mackay, C.~D.: 1986,
\newblock Charge-coupled devices in astronomy,
\newblock {\em Annual Review of Astronomy and Astrophysics} {\bf 24}, 255

\bibitem[\protect\astroncite{Mackay et~al.}{2001}]{Mackay2001}
Mackay, C.~D., Tubbs, R.~N., Bell, R., Burt, D.~J., Jerram, P., Moody, I.,
  Canosa, J., and Sampat, N.: 2001,
\newblock in M.~M. Blouke (ed.), {\em Society of Photo-Optical Instrumentation
  Engineers (SPIE) Conference Series}, Vol. 4306, pp 289--298

\bibitem[\protect\astroncite{{Maire} et~al.}{2008}]{Maire2008}
{Maire}, J., {Ziad}, A., {Borgnino}, J., and {Martin}, F.: 2008,
\newblock {Comparison between atmospheric turbulence models by angle-of-arrival
  covariance measurements},
\newblock {\em MNRAS} {\bf 386}, 1064

\bibitem[\protect\astroncite{Males et~al.}{2010}]{Males2010}
Males, J.~R., Close, L.~M., Kopon, D., Gasho, V., and Follette, K.: 2010,
\newblock in {\em Society of Photo-Optical Instrumentation Engineers (SPIE)
  Conference Series}, Vol. 7736, p. 198

\bibitem[\protect\astroncite{Marcy and Butler}{1995}]{Marcy1995}
Marcy, G.~W. and Butler, R.~P.: 1995,
\newblock in {\em Bulletin of the American Astronomical Society}, Vol. 187, p.
  1379

\bibitem[\protect\astroncite{Matsuo et~al.}{1985}]{Matsuo1985}
Matsuo, K., Teich, M., and Saleh, B.: 1985,
\newblock Noise properties and time response of the staircase avalanche
  photodiode,
\newblock {\em Electron Devices, IEEE Transactions on} {\bf 32(12)}, 2615

\bibitem[\protect\astroncite{Meekins et~al.}{1984}]{Meekins1984}
Meekins, J.~F., Wood, K.~S., Hedler, R.~L., Byram, E.~T., Yentis, D.~J., Chubb,
  T.~A., and Friedman, H.: 1984,
\newblock Millisecond variability of {Cygnus X-1},
\newblock {\em The Astrophysical Journal} {\bf 278}, 288

\bibitem[\protect\astroncite{{Mestayer}}{1982}]{Mestayer1982}
{Mestayer}, P.: 1982,
\newblock {Local isotropy and anisotropy in a high-Reynolds-number turbulent
  boundary layer},
\newblock {\em Journal of Fluid Mechanics} {\bf 125}, 475

\bibitem[\protect\astroncite{Montilla et~al.}{2010}]{Montilla2010}
Montilla, I., Béchet, C., Lelouarn, M., Correia, C., Tallon, M., Reyes, M.,
  and Thiébaut, Ã.: 2010,
\newblock in {\em Adaptative Optics for Extremely Large Telescopes}, p. 3002

\bibitem[\protect\astroncite{Motch et~al.}{1982}]{Motch1982}
Motch, C., Ilovaisky, S.~A., and Chevalier, C.: 1982,
\newblock Discovery of fast optical activity in the {X}-ray source {GX 339-4},
\newblock {\em Astronomy and Astrophysics} {\bf 109}, L1

\bibitem[\protect\astroncite{Naylor}{1998}]{Naylor1998}
Naylor, T.: 1998,
\newblock An optimal extraction algorithm for imaging photometry,
\newblock {\em MNRAS} {\bf 296}, 339

\bibitem[\protect\astroncite{Nethercote and Seward}{2007}]{Nethercote2007}
Nethercote, N. and Seward, J.: 2007,
\newblock Valgrind: a framework for heavyweight dynamic binary instrumentation,
\newblock {\em SIGPLAN Not.} {\bf 42(6)}, 89

\bibitem[\protect\astroncite{{Nieto} et~al.}{1987}]{Nieto1987}
{Nieto}, J.-L., {Llebaria}, A., and {di Serego Alighieri}, S.: 1987,
\newblock {Photon-counting detectors in time-resolved imaging mode - Image
  recentring and selection algorithms},
\newblock {\em AAP} {\bf 178}, 301

\bibitem[\protect\astroncite{{Noll}}{1976}]{Noll1976}
{Noll}, R.~J.: 1976,
\newblock {Zernike polynomials and atmospheric turbulence.},
\newblock {\em Journal of the Optical Society of America (1917-1983)} {\bf 66},
  207

\bibitem[\protect\astroncite{Obukhov}{1949}]{Obukhov1949}
Obukhov, A.~M.: 1949,
\newblock Structure of the temperature field in a turbulent flow,
\newblock {\em Izv. Akad. Nauk. S.S.S.R, Ser. Geograf. Geofiz.} {\bf 13}, 58

\bibitem[\protect\astroncite{Orosz et~al.}{2011}]{Orosz2011}
Orosz, J.~A., McClintock, J.~E., Aufdenberg, J.~P., Remillard, R.~A., Reid,
  M.~J., Narayan, R., and Gou, L.: 2011,
\newblock The mass of the black hole in {Cygnus X-1},
\newblock {\em The Astrophysical Journal} {\bf 742}, 84

\bibitem[\protect\astroncite{Oscoz et~al.}{2008}]{Oscoz2008}
Oscoz, A., Rebolo, R., López, R., Pérez-Garrido, A., Pérez, J.~A.,
  Hildebrandt, S., Rodríguez, L.~F., Piqueras, J.~J., Villó, I., González,
  J.~M., Barrena, R., Gómez, G., García, A., Montañés, P., Rosenberg, A.,
  Cadavid, E., Calcines, A., Díaz-Sánchez, A., Kohley, R., Martín, Y.,
  Peñate, J., and Sánchez, V.: 2008,
\newblock in {\em Ground-based and Airborne Instrumentation for Astronomy II.},
  Vol. 7014 of {\em Proc. SPIE}, p. 137

\bibitem[\protect\astroncite{Oskinova et~al.}{2012}]{Oskinova2012}
Oskinova, L.~M., Feldmeier, A., and Kretschmar, P.: 2012,
\newblock Clumped stellar winds in supergiant high-mass x-ray binaries: X-ray
  variability and photoionization,
\newblock {\em Monthly Notices of the Royal Astronomical Society} {\bf 421},
  2820

\bibitem[\protect\astroncite{Papalexandris and
  Redding}{2000}]{Papalexandris2000}
Papalexandris, M.~V. and Redding, D.~C.: 2000,
\newblock Calculation of diffraction effects on the average phase of an optical
  field,
\newblock {\em J. Opt. Soc. Am. A} {\bf 17(10)}, 1763

\bibitem[\protect\astroncite{Park and Schowengerdt}{1983}]{Park1983}
Park, S.~K. and Schowengerdt, R.~A.: 1983,
\newblock Image reconstruction by parametric cubic convolution,
\newblock {\em Computer Vision, Graphics, and Image Processing} {\bf 23(3)},
  258

\bibitem[\protect\astroncite{Pedersen et~al.}{1988}]{Pedersen1988}
Pedersen, H., Rigaut, F., and Sarazin, M.: 1988,
\newblock Seeing measurements with a differential image motion monitor.,
\newblock {\em The Messenger} {\bf 53}, 8

\bibitem[\protect\astroncite{Pence et~al.}{1999}]{Pence1999}
Pence, W., Plante, R.~L., and Roberts, D.~A.: 1999,
\newblock in D.~M. Mehringer (ed.), {\em Astronomical Data Analysis Software
  and Systems VIII}, Vol. 172, p. 487

\bibitem[\protect\astroncite{Pollacco et~al.}{2006}]{Pollacco2006}
Pollacco, D.~L., Skillen, I., Collier~Cameron, A., Christian, D.~J., Hellier,
  C., Irwin, J., Lister, T.~A., Street, R.~A., West, R.~G., Anderson, D.,
  Clarkson, W.~I., Deeg, H., Enoch, B., Evans, A., Fitzsimmons, A., Haswell,
  C.~A., Hodgkin, S., Horne, K., Kane, S.~R., Keenan, F.~P., Maxted, P. F.~L.,
  Norton, A.~J., Osborne, J., Parley, N.~R., Ryans, R. S.~I., Smalley, B.,
  Wheatley, P.~J., and Wilson, D.~M.: 2006,
\newblock The wasp project and the superwasp cameras,
\newblock {\em Publications of the Astronomical Society of the Pacific} {\bf
  118}, 1407

\bibitem[\protect\astroncite{Poyneer et~al.}{2009}]{Poyneer2009}
Poyneer, L., van Dam, M., and Véran, J.-P.: 2009,
\newblock Experimental verification of the frozen flow atmospheric turbulence
  assumption with use of astronomical adaptive optics telemetry,
\newblock {\em Journal of the Optical Society of America A} {\bf 26}, 833

\bibitem[\protect\astroncite{Poyneer}{2003}]{Poyneer2003}
Poyneer, L.~A.: 2003,
\newblock Scene-based {S}hack-{H}artmann wave-front sensing: analysis and
  simulation,
\newblock {\em Applied Optics} {\bf 42}, 5807

\bibitem[\protect\astroncite{{Racine}}{2006}]{Racine2006}
{Racine}, R.: 2006,
\newblock {The Strehl Efficiency of Adaptive Optics Systems},
\newblock {\em PASP} {\bf 118}, 1066

\bibitem[\protect\astroncite{Reinders}{2007}]{Reinders2007}
Reinders, J.: 2007,
\newblock {\em Intel threading building blocks},
\newblock O'Reilly \& Associates, Inc., Sebastopol, CA, USA, first edition

\bibitem[\protect\astroncite{Remillard and McClintock}{2006}]{Remillard2006}
Remillard, R.~A. and McClintock, J.~E.: 2006,
\newblock X-ray properties of black-hole binaries,
\newblock {\em Annual Review of Astronomy and Astrophysics} {\bf 44}, 49

\bibitem[\protect\astroncite{Richer et~al.}{2004}]{Richer2004}
Richer, H.~B., Fahlman, G.~G., Brewer, J., Davis, S., Kalirai, J., Stetson,
  P.~B., Hansen, B. M.~S., Rich, R.~M., Ibata, R.~A., Gibson, B.~K., and Shara,
  M.: 2004,
\newblock Hubble space telescope observations of the main sequence of m4,
\newblock {\em The Astronomical Journal} {\bf 127}, 2771

\bibitem[\protect\astroncite{{Roddier}}{1981}]{Roddier1981}
{Roddier}, F.: 1981,
\newblock The effects of atmospheric turbulence in optical astronomy,
\newblock {\em Progress in Optics} {\bf 19}, 281

\bibitem[\protect\astroncite{{Roddier} et~al.}{1982}]{Roddier1982a}
{Roddier}, F., {Gilli}, J.~M., and {Lund}, G.: 1982,
\newblock On the origin of speckle boiling and its effects in stellar speckle
  interferometry,
\newblock {\em Journal of Optics} {\bf 13}, 263

\bibitem[\protect\astroncite{Rogalski}{2005}]{Rogalski2005}
Rogalski, A.: 2005,
\newblock {HgCdTe} infrared detector material: history, status and outlook,
\newblock {\em Reports on Progress in Physics} {\bf 68(10)}, 2267

\bibitem[\protect\astroncite{Ross}{2009}]{Ross2009}
Ross, T.~S.: 2009,
\newblock Limitations and applicability of the maréchal approximation,
\newblock {\em Applied Optics} {\bf 48}, 1812

\bibitem[\protect\astroncite{Rothman et~al.}{2009}]{Rothman2009}
Rothman, J., de~Borniol, E., Bisotto, S., Mollard, L., and Guellec, F.: 2009,
\newblock in {\em Proceedings of the Quantum of Quasars workshop.}, p.~9

\bibitem[\protect\astroncite{Sarazin et~al.}{2002}]{Sarazin2002}
Sarazin, M., Tokovinin, A., Ragazzoni, R., Esposito, S., and Hubin, N.: 2002,
\newblock in E. Vernet (ed.), {\em European Southern Observatory Conference and
  Workshop Proceedings}, Vol.~58, p. 321

\bibitem[\protect\astroncite{Scaddan and Walker}{1978}]{Scaddan1978}
Scaddan, R.~J. and Walker, J.~G.: 1978,
\newblock Statistics of stellar speckle patterns,
\newblock {\em Applied Optics} {\bf 17}, 3779

\bibitem[\protect\astroncite{Shannon}{1998}]{Shannon1998}
Shannon, C.: 1998,
\newblock Communication in the presence of noise,
\newblock {\em Proceedings of the IEEE DOI - 10.1109/JPROC.1998.659497} {\bf
  86(2)}, 447

\bibitem[\protect\astroncite{Shraiman and Siggia}{2000}]{Shraiman2000}
Shraiman, B.~I. and Siggia, E.~D.: 2000,
\newblock Scalar turbulence,
\newblock {\em Nature} {\bf 405(6787)}, 639

\bibitem[\protect\astroncite{Skrutskie et~al.}{2006}]{Skrutskie2006}
Skrutskie, M.~F., Cutri, R.~M., Stiening, R., Weinberg, M.~D., Schneider, S.,
  Carpenter, J.~M., Beichman, C., Capps, R., Chester, T., Elias, J., Huchra,
  J., Liebert, J., Lonsdale, C., Monet, D.~G., Price, S., Seitzer, P., Jarrett,
  T., Kirkpatrick, J.~D., Gizis, J.~E., Howard, E., Evans, T., Fowler, J.,
  Fullmer, L., Hurt, R., Light, R., Kopan, E.~L., Marsh, K.~A., McCallon,
  H.~L., Tam, R., Van~Dyk, S., and Wheelock, S.: 2006,
\newblock The two micron all sky survey (2mass),
\newblock {\em The Astronomical Journal} {\bf 131}, 1163

\bibitem[\protect\astroncite{Sreenivasan}{1991}]{Sreenivasan1991}
Sreenivasan, K.~R.: 1991,
\newblock {On local isotropy of passive scalars in turbulent shear flows},
\newblock {\em Royal Society of London Proceedings Series A} {\bf 434}, 165

\bibitem[\protect\astroncite{Staley and Mackay}{2010}]{Staley2010a}
Staley, T.~D. and Mackay, C.~D.: 2010,
\newblock in {\em Ground-based and Airborne Instrumentation for Astronomy III},
  No. 7735 in Proc. SPIE

\bibitem[\protect\astroncite{Stetson}{1987}]{Stetson1987}
Stetson, P.~B.: 1987,
\newblock {DAOPHOT} - a computer program for crowded-field stellar photometry,
\newblock {\em PASP} {\bf 99}, 191

\bibitem[\protect\astroncite{Takeda et~al.}{2009}]{Takeda2009}
Takeda, G., Kita, R., and Rasio, F.~A.: 2009,
\newblock in {\em IAU Symposium}, Vol. 253, pp 181--187

\bibitem[\protect\astroncite{Tatarski}{1961}]{Tatarski1961}
Tatarski, V.~I.: 1961,
\newblock {\em Wave Propagation in a turbulent medium},
\newblock McGraw-Hill

\bibitem[\protect\astroncite{Thomas et~al.}{2006}]{Thomas2006}
Thomas, S., Fusco, T., Tokovinin, A., Nicolle, M., Michau, V., and Rousset, G.:
  2006,
\newblock Comparison of centroid computation algorithms in a {S}hack-{H}artmann
  sensor,
\newblock {\em MNRAS} {\bf 371}, 323

\bibitem[\protect\astroncite{Thomson}{2008}]{Thomson2008}
Thomson, B. S. . J. B. B. A. M.~B.: 2008,
\newblock {\em Elementary Real Analysis},
\newblock CreateSpace

\bibitem[\protect\astroncite{Tokovinin}{2002}]{Tokovinin2002}
Tokovinin, A.: 2002,
\newblock From differential image motion to seeing,
\newblock {\em Publications of the Astronomical Society of the Pacific} {\bf
  114}, 1156

\bibitem[\protect\astroncite{Tokovinin}{2004}]{Tokovinin2004}
Tokovinin, A.: 2004,
\newblock Seeing improvement with ground-layer adaptive optics,
\newblock {\em Publications of the Astronomical Society of the Pacific} {\bf
  116}, 941

\bibitem[\protect\astroncite{{Tokovinin} et~al.}{2003}]{Tokovinin2003}
{Tokovinin}, A., {Baumont}, S., and {Vasquez}, J.: 2003,
\newblock {Statistics of turbulence profile at Cerro Tololo},
\newblock {\em MNRAS} {\bf 340}, 52

\bibitem[\protect\astroncite{Tokovinin and Kornilov}{2007}]{Tokovinin2007a}
Tokovinin, A. and Kornilov, V.: 2007,
\newblock Accurate seeing measurements with mass and dimm,
\newblock {\em Monthly Notices of the Royal Astronomical Society} {\bf 381},
  1179

\bibitem[\protect\astroncite{{Troy} et~al.}{2000}]{Troy2000}
{Troy}, M., {Dekany}, R.~G., {Brack}, G., {Oppenheimer}, B.~R., {Bloemhof},
  E.~E., {Trinh}, T., {Dekens}, F.~G., {Shi}, F., {Hayward}, T.~L., and
  {Brandl}, B.: 2000,
\newblock in P.~L. Wizinowich; (ed.), {\em Adaptive Optical Systems
  Technology}, Vol. 4007 of {\em Presented at the Society of Photo-Optical
  Instrumentation Engineers (SPIE) Conference}, pp 31--40

\bibitem[\protect\astroncite{Trujillo et~al.}{2001}]{Trujillo2001}
Trujillo, I., Aguerri, J. A.~L., Cepa, J., and Gutiérrez, C.~M.: 2001,
\newblock The effects of seeing on {S\'ersic profiles} - {II.} the {Moffat}
  {PSF},
\newblock {\em MNRAS} {\bf 328}, 977

\bibitem[\protect\astroncite{Tubbs}{2003}]{Tubbs2003}
Tubbs, R.~N.: 2003,
\newblock {\em Ph.D. thesis}, University of Cambridge

\bibitem[\protect\astroncite{Tubbs}{2006}]{Tubbs2006}
Tubbs, R.~N.: 2006,
\newblock in B.~L. Ellerbroek and D.~B. Calia (eds.), {\em Advances in Adaptive
  Optics II}, Vol. 6272 of {\em Proc. SPIE}

\bibitem[\protect\astroncite{Tubbs et~al.}{2002}]{Tubbs2002}
Tubbs, R.~N., Baldwin, J.~E., Mackay, C.~D., and Cox, G.~C.: 2002,
\newblock Diffraction-limited {CCD} imaging with faint reference stars,
\newblock {\em AAP} {\bf 387}, L21

\bibitem[\protect\astroncite{Tulloch}{2004}]{Tulloch2004}
Tulloch, S.~M.: 2004,
\newblock in {\em Society of Photo-Optical Instrumentation Engineers (SPIE)
  Conference Series}, Vol. 5492, pp 604--614

\bibitem[\protect\astroncite{Tulloch and Dhillon}{2011}]{Tulloch2011}
Tulloch, S.~M. and Dhillon, V.~S.: 2011,
\newblock On the use of electron-multiplying {CCDs} for astronomical
  spectroscopy,
\newblock {\em Monthly Notices of the Royal Astronomical Society} {\bf 411},
  211

\bibitem[\protect\astroncite{Tyson}{2000}]{Tyson2000}
Tyson, R.~K.: 2000,
\newblock {\em Introduction to Adaptive Optics},
\newblock SPIE Press

\bibitem[\protect\astroncite{van Dam et~al.}{2010}]{Dam2010}
van Dam, M.~A., Hinz, P.~M., Codona, J.~L., Hart, M., Garcia-Rissmann, A.,
  Johns, M.~W., Shectman, S.~A., Bouchez, A.~H., McLeod, B.~A., and Rigaut, F.:
  2010,
\newblock in {\em Society of Photo-Optical Instrumentation Engineers (SPIE)
  Conference Series}, Vol. 7736, p. 136

\bibitem[\protect\astroncite{van~der Klis}{2000}]{Klis2000}
van~der Klis, M.: 2000,
\newblock Millisecond oscillations in x-ray binaries,
\newblock {\em Annual Review of Astronomy and Astrophysics} {\bf 38}, 717

\bibitem[\protect\astroncite{Vernin and Munoz-Tunon}{1994}]{Vernin1994}
Vernin, J. and Munoz-Tunon, C.: 1994,
\newblock Optical seeing at {La Palma Observatory}. 2: Intensive site testing
  campaign at the {Nordic Optical Telescope},
\newblock {\em AAP} {\bf 284}, 311

\bibitem[\protect\astroncite{Westphal et~al.}{1968}]{Westphal1968}
Westphal, J.~A., Sandage, A., and Kristian, J.: 1968,
\newblock Rapid changes in the optical intensity and radial velocities of the
  {X}-ray source {SCO X-1},
\newblock {\em The Astrophysical Journal} {\bf 154}, 139

\bibitem[\protect\astroncite{Wilson}{2002}]{Wilson2002}
Wilson, R.~W.: 2002,
\newblock {SLODAR}: measuring optical turbulence altitude with a
  {Shack-Hartmann} wavefront sensor,
\newblock {\em Monthly Notices of the Royal Astronomical Society} {\bf 337},
  103

\bibitem[\protect\astroncite{Winn}{2010}]{Winn2010}
Winn, J.~N.: 2010,
\newblock {\em Transits and Occultations}

\end{thebibliography}

\end{document}